\pdfoutput=1			
\documentclass[showpacs,showkeys]{revtex4}
\usepackage{graphicx}
\def\lsim{\mathrel{\rlap{\lower4pt\hbox{\hskip1pt$\sim$}}
    \raise1pt\hbox{$<$}}}         
\def\gsim{\mathrel{\rlap{\lower4pt\hbox{\hskip1pt$\sim$}}
    \raise1pt\hbox{$>$}}}         
\newcommand\T{\rule{0mm} {3.0ex}}
\newcommand\B{\rule[-1.6ex]{0mm}{0.0ex}}

\begin{document}

\title{The Form of the Effective Interaction in Harmonic-Oscillator-Based Effective Theory}

\author{W. C. Haxton}

\affiliation{Institute for Nuclear Theory and Department of Physics, \\
University of Washington, Seattle, WA 98195, USA}

\begin{abstract}
I explore the form of the effective interaction in harmonic-oscillator-based effective
theory (HOBET) in leading-order (LO) through next-to-next-to-next-to-leading order (N$^3$LO).
As the included space in a HOBET (as in the shell model) is defined by the oscillator energy,
both long-distance (low-momentum) and short-distance (high-momentum) degrees of
freedom reside in the high-energy excluded space.  A HOBET effective interaction is
developed in which a short-range contact-gradient expansion, free of operator mixing and
corresponding to a systematic expansion in nodal quantum numbers, is combined with
an exact summation of the relative kinetic energy.   By this means the very strong
coupling of the included ($P$) and excluded ($Q$) spaces by the kinetic energy is
removed.  One finds a simple and rather surprising
result, that the interplay of $QT$ and $QV$ is governed by a single parameter $\kappa$,
the ratio of an observable, the binding energy $|E|$, to a parameter in the effective
theory, the oscillator energy $\hbar \omega$.  Once the functional dependence on $\kappa$
is identified, the remaining order-by-order subtraction 
of the short-range physics residing in $Q$ becomes systematic and rapidly converging.
Numerical calculations are used to demonstrate how well the resulting expansion
reproduces the running of $H^{eff}$ from high
scales to a typical shell-model scale of 8$\hbar \omega$.  At N$^3$LO various global 
properties of $H^{eff}$ are reproduced to a typical accuracy of 0.01\%, or about 1 keV,
at 8$\hbar \omega$.
Channel-by-channel variations in convergence rates are
similar to those found in effective field theory approaches.

The state-dependence of the effective interaction has been a troubling problem in nuclear
physics, and is embodied in the energy dependence of $H^{eff}(|E|)$ in the Bloch-Horowitz
formalism.  It is shown that almost all of this state dependence is also
extracted in the procedures followed here, isolated in the analytic dependence 
of $H^{eff}$ on $\kappa$.  Thus there exists a simple, Hermitian $H^{eff}$ that can be
use in spectral calculations.

The existence of a systematic operator expansion for $H^{eff}$, depending on a series of short-range
constants augmented by $\kappa$, will be important to future efforts to determine the HOBET interaction directly from experiment, rather than from an underlying NN potential.   
\end{abstract}

\email{haxton@phys.washington.edu}

\maketitle

\section{Introduction}
In nuclear physics one often faces the problem of determining long-wavelength
properties of nuclei, such as binding energies, radii, or responses to low-momentum
probes.  One approach would be to evaluate the relevant operators between exact nuclear
wave functions obtained from solutions of the many-body Schroedinger equation.  
Because the NN potential is strong, characterized by anomalously large NN
scattering lengths, and highly repulsive at very short distances,  this task becomes
exponentially more difficult as the nucleon number increases.  Among available quasi-exact
methods, the variational and Green's function Monte Carlo work of the Argonne group has perhaps set
the standard \cite{argonne}, yielding accurate results throughout most of the $1p$ shell.  

Effective theory (ET) potentially offers an alternative, a method that limits the numerical difficulty
of a calculation by restricting it to a finite Hilbert space (the $P$- or
``included"-space), while correcting the bare Hamiltonian $H$ (and other operators) for the
effects of the $Q$- or ``excluded"-space. 
Calculations using the effective Hamiltonian $H^{eff}$ within $P$ reproduce
the results using $H$ within $P+Q$, over the domain of overlap. 
That is, the effects of $Q$ on $P$-space 
calculations are absorbed into $P(H^{eff}-H)P$.

One interesting challenge for ET is the case of a $P$-space basis of harmonic
oscillator (HO) Slater determinants.  This is a special basis for nuclear physics
because of center-of-mass separability:  if all Slater determinants containing
up to $N$ oscillator quanta are retained, $H^{eff}$ will be translationally invariant (assuming
$H$ is).  Such bases are also important because of powerful shell-model (SM) techniques that 
have been developed for iterative diagonalization and for evaluating inclusive responses.  The larger
$P$ can be made, the smaller the effects of $H^{eff}-H$.  If one could fully develop
harmonic-oscillator based effective theory (HOBET), it would provide a prescription for
eliminating the SM's many uncontrolled approximations, while retaining the 
model's formidable numerical apparatus.

The long-term goal is a HOBET resembling standard effective field theories (EFTs)  \cite{weinberg,savage}.  That is, for a given choice of $P$, the effective interaction would be a sum of a long-distance
``bare" interaction whose form would be determined by chiral symmetry, augmented by 
some general effective interaction that accounts for the excluded Q space.  That effective
interaction would be expanded systematically and in some natural way, with the parameters
governing the strength of successive terms
determined by directly fitting to experiment.  There would be no need to introduce or integrate
out any high-momentum NN potential, an unnecessary intermediate
effective theory between QCD and the SM scale.

One prerequisite for such an approach is the demonstration that a systematic expansion for
the HOBET effective interaction exists.  This paper explores this issue, making use of numerically
generated effective interaction matrix elements for the deuteron, obtained by solving
the Bloch-Horowitz (BH) equation for the Argonne $v_{18}$ potential, an example of a
potential with a relatively
hard core ($\lsim$ 2 GeV) \cite{av18}.
The BH $H^{eff}$ is a Hermitian but energy-dependent
Hamiltonian satisfying
\begin{eqnarray}
H^{eff} = H &+& H {1 \over E - Q H} Q H \nonumber \\
H^{eff} |\Psi_P \rangle = E |\Psi_P \rangle ~~~&&~~~ |\Psi_P \rangle
 = (1-Q) |\Psi \rangle.
 \label{BH}
 \end{eqnarray}
Here $H$ is the bare Hamiltonian and $E$ and $\Psi$ are the exact eigenvalue and wave function
(that is, the solution of the Schroedinger equation in the full $P+Q$ space).   
$E$ is negative for a bound state.  Because $H^{eff}$
depends on the unknown exact eigenvalue $E$, Eqs. (\ref{BH}) must be solved self-consistently,
state by state,
a task that in practice proves to be relatively straightforward.   If this is done, the $P$-space
eigenvalue will be the exact energy $E$ and the $P$-space wave function $\Psi_P$ will be the
restriction of the exact wave function $\Psi$ to $P$.  This implies a nontrivial normalization
and nonorthogonality of the restricted ($P$-space) wave functions.  
If $P$ is enlarged, new components are added to the existing ones,
and for a sufficiently large $P$ space, the norm approaches ones.  This convergence is slow
for potentials like $v_{18}$, with many shells being required before norms near one are
achieved \cite{song,luu}.  Observables calculated with the restricted wave
functions and the appropriate effective operators are independent of the choice of
$P$, of course.  All of these properties follow from physics encoded in $H^{eff}$.
 
In HOBET $P$ and thus $H^{eff}$ are functions of the oscillator parameter $b$ and the
number of included HO quanta $\Lambda_P$.  In this paper I study the behavior 
of matrix elements 
$\langle \alpha | H^{eff} | \beta \rangle$ generated for the Argonne $v_{18}$ potential, as both $b$
and  $\Lambda_P$ are varied.  In particular, $\Lambda_P$ is allowed to run from
very high values to the ``shell-model" scale of 8 $\hbar \omega$, in order
to test whether the physics above a specified scale can be efficiently absorbed
into the coefficients of some systematic expansion, e.g., one
analogous to the contact-gradient expansions employed in EFTs
(which are generally formulated in plane wave bases).  
There are reasons the HOBET effective interaction could prove more
complicated:
\begin{itemize}   
\item An effective theory defined by a subset of HO Slater determinants
is effectively an expansion around a typical momentum scale $q \sim 1/b$.  That is, the $P$-space
omits both long-wavelength and short-wavelength degrees of freedom. The former are
connected with the overbinding of the HO, while the latter are due to absence
in $P$ of the strong, short-range NN interaction.  As
any systematic expansion of the effective interaction must simultaneously address
both problems, the form of the effective interaction cannot be as simple as a
contact-gradient expansion (which would be appropriate if the missing physics were only
short-ranged).
\item The relative importance of the missing long-wavelength and short-wavelength excitations
is governed by the binding energy, $|E|$, with the former increasing as $|E| \rightarrow 0$.
These long-range interactions allow nuclear states to de-localize, minimizing the kinetic
energy.  But nuclei are weakly bound -- binding energies are very small compared
to the natural scales set by the scalar and vector potentials in nuclei.  One concludes that
the effective interaction must depend delicately on $|E|$.  
\item  An effective theory is generally considered successful if it can reproduce the lowest energy
excitations in $P$.   But one asks for much more when one seeks to accurately represent
the effective interaction, which governs all of the spectral properties within $P$.  The HO appears
to be an especially difficult case in which to attempt such a representation.
The kinetic energy operator  in the HO has strong off-diagonal components which
raise or lower the nodal quantum number, and thus connect
Slater determinants containing $\Lambda_P$ quanta with those containing $\Lambda_P \pm 2$.
This means that $P$ and $Q$ are strongly coupled through low-energy excitations, a 
situation that is usually problematic for an effective theory.
\end{itemize}

All of these problems involve the interplay, governed by $|E|$, of $QT$ (delocalization)
and $QV$ (corrections for short-range repulsion).
The explicit energy dependence
of the BH equation proves to be
a great advantage in resolving the problems induced by this interplay,
leading to a natural factorization of the long- and short-range contributions to the
effective interaction, and thereby to a successful systematic representation of the effective
interaction.  (Conversely, techniques such as Lee-Suzuki \cite{suzuki} will intermingle these effects
in a complex way and obscure the underlying simplicity of the effective interaction.)
The result is an  energy-dependent
contact-gradient expansion at N$^3$LO that
reproduces the entire effective interaction to an accuracy of about a few keV.   The contact-gradient expansion is defined in a way that is appropriate to the HO, eliminating operator mixing and
producing a simple dependence on nodal quantum numbers.  The coefficients
in the expansion play the role of generalized Talmi integrals.

The long-range physics residing in $Q$ can be isolated analytically and
expressed in terms of a single parameter, $\kappa = \sqrt{2 |E|/\hbar \omega}$,  remarkably
the ratio of an observable ($|E|$) to a parameter one chooses in defining the ET.  The 
dependence of $H^{eff}$ on $\kappa$ is determined by summing $QT$ to
all orders.  The resulting
$H^{eff}$ is defined by $\kappa$ and by the coefficients of the short-ranged expansion.

This
same parameter governs almost all of the state dependence that enters when
one seeks to describe multiple states.   Thus it appears that there is a systematic,
rapidly converging representation for $H^{eff}$ in HOBET that could be used to
describe a set of nuclear states.  The short-range parameters in that representation
are effectively state-independent, as the state-dependence usually attacked with
techniques like Lee-Suzuki is isolated in $\kappa$.

\section{Long- and short-wavelength separations in $H^{eff}$}
In Refs. \cite{song,luu} a study was done of the evolution of matrix elements
$\langle \alpha | H^{eff} | \beta \rangle$, for the deuteron and for $^3$He/$^3$H, 
from the $\Lambda_P \rightarrow \infty$ limit, where $H^{eff} \rightarrow H$, down to
$\Lambda_P$ characteristic of the shell model (SM), e.g., small $P$ spaces with 4, 6, or 8
$\hbar \omega$ excitations, relative to the naive $1s$-shell ground state.
As noted above, this definition of $P$ in terms of the total quanta in  HO Slater
determinants maintains center-of-mass separability and thus 
leads to an $H^{eff}$ that is translationally invariant, just like $H$.  Indeed, the HO
basis is the only set of compact wave functions with this attractive
property.

\begin{figure}
\begin{center}
\includegraphics[width=10cm]{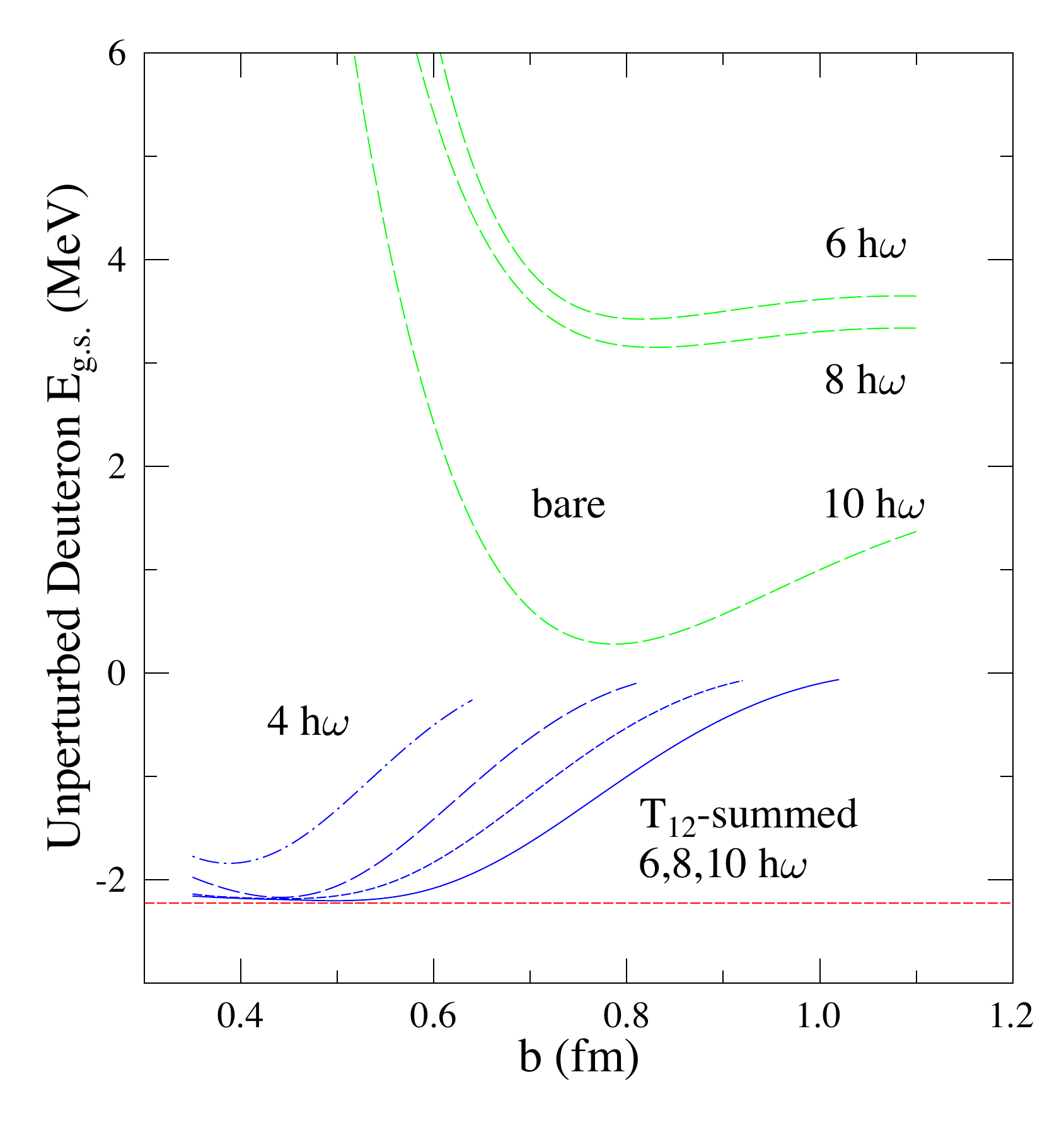}
\end{center}
\caption{(Color online) Deuteron ground-state convergence for ``bare" calculations
in small $P$-spaces, which omit all effects
due to the multiple scattering of $V$ in $Q$.  The three curves on the upper right
were calculated from the standard BH equation,  which identifies the bare interaction
as $P(T+V)P$.  These calculations fail
to bind the deuteron, even with $\Lambda_P=10$, for all values of the HO size
parameter $b$: the $P$-space estimate for $V$ is poor if $b$ is much above 1fm,
while the estimate for $T$ is poor if $b$ is below that value.    The lower four curves were evaluated
for the bare interaction of the reordered BH given by Eq. (\ref{BHnew}), which incorporates
the long-range effects of $QT$ to all orders, building in the correct asymptotic form of
the wave function.   This allows one to reduce $b$ to small values,
pulling most of the effects of $V$ into $P$, without distorting the long-distance behavior
of the wave function or, therefore, the estimate for $T$.  Rather remarkably, this bare calculation reproduces the correct
binding energy for $P$ spaces as small as $\Lambda_P$=6.  That is, by the combination of
the summation of $QT$ to all orders and the adjustment of $b$ to an optimal value 
characteristic of the hard core radius of $v_{18}$, the effective interaction contribution
can be driven to such small values that it can be ignored.}
\label{fig_1}
\end{figure}

But this choice leads to a more complicated ET, as $P$ excludes both 
short-distance and long-distance  components of wave functions.  This problem was first
explored in connection with the nonperturbative behavior of $H^{eff}$: the need 
to simultaneously correct for the missing long- and short-distance behavior of 
$\Psi_P$ is the reason one cannot tune $P$ to make
$H^{eff}$ converge rapidly.   For example, while it is possible to ``pull" more of the
missing short-range physics into $P$ by choosing a small $b$, this adjustment 
produces a more compact state with very large $Q$-space corrections to the kinetic 
energy.  Conversely, one can tune $b$ to large values
to improve the description of the nuclear tail, but at the cost of missing even more of
the short-range physics.  At no value of $b$ are both problems handled well: Fig. \ref{fig_1}
shows that a poor minimum is reached at some intermediate $b$, with a $10 \hbar \omega$
``bare" calculation failing to bind the deuteron.

The solution found to this hard-core correlation/extended state quandary is an
a priori treatment of the overbinding of the harmonic oscillator.  The BH equation is
rewritten in a form that allows the relative kinetic energy operator to be
summed to all orders.  (This form was introduced in the first of Refs. \cite{luu}; a
detailed derivation can be found in the Appendix of the third of these references.  The 
kinetic energy sum can be done analytically for calculations performed in a
Jacobi basis.)  This reordered BH equation has the form
\begin{equation}
H^{eff} = H + HQ {1 \over E-QH} QH =
{E \over E -TQ} \left[ T -T {Q \over E}T + V + V {1 \over E-QH} QV \right] {E \over E-QT}
\label{BHnew}
\end{equation}  
where the bare $H$ is the sum of the relative kinetic energy and a two-body interaction
\begin{equation}
H = {1 \over 2} \sum_{i,j=1}^A \left(T_{ij} + V_{ij} \right), \mathrm{~~~with~~} T_{ij} = {({\bf p}_i-{\bf p}_j)^2 \over 2 A M}.
\end{equation}
This effective interaction is to be evaluated between a finite basis of Slater determinants 
$| \alpha \rangle \in P$, which is equivalent to evaluating the Hamiltonian
\begin{equation}
\widetilde{H}^{eff} \equiv T -T {Q \over E}T + V + V {1 \over E-QH} QV
\end{equation}
between the states
\begin{equation}
|\widetilde{\alpha} \rangle \equiv {E \over E-QT} |\alpha \rangle
\end{equation}
By summing $QT$ to all orders, the proper behavior at large $r$ can be built in, which
then allows $b$ to be adjusted, without affecting
the long-wavelength properties of the wave function.  Fig. \ref{fig_1}, from Ref. \cite{luu}, shows that the
resulting decoupling of the long- and short-wavelength physics can greatly improve
convergence: a ``bare" 6 $\hbar \omega$ calculation that neglects all contributions of $QV$
gives an excellent binding energy.  This decoupling
of $QV$ and $QT$ is also important
in finding a systematic expansion for $H^{eff}$.

This reorganization produces an $H^{eff}$ with
three terms operating between HO Slater determinants,
\begin{eqnarray}
\langle \alpha | T {E \over E-QT} | \beta \rangle=\langle \alpha | {E \over E-TQ} T | \beta \rangle~&\stackrel{\mathrm{nonedge}}{\longrightarrow}&~\langle \alpha | T | \beta \rangle \nonumber \\
\langle \alpha | {E \over E-TQ} V {E \over E-QT} | \beta \rangle~&\stackrel{\mathrm{nonedge}}{\longrightarrow}&~\langle \alpha | V | \beta \rangle \nonumber \\
\langle \alpha | {E \over E-TQ} V {1 \over E-QH} QV {E \over E-QT} | \beta \rangle~&\stackrel{\mathrm{nonedge}}{\longrightarrow}&~\langle \alpha | V {1 \over E-QH} QV | \beta \rangle.
\label{wh:eq4}
\end{eqnarray}
The ladder properties of $QT$ make
$E/(E-QT)$ the identity operator except when it acts on an $|\alpha \rangle$
with energy $\Lambda_P\hbar \omega$  or $(\Lambda_P-1) \hbar \omega$.  These are
called the edge states.   For nonedge states, the new grouping
of terms in $H^{eff}$ 
reduces to the expressions on the right-hand side of Eq. \ref{wh:eq4}, the
conventional components of
$H^{eff}$.  Thus the summation over $QT$ alters only a subset of the matrix elements
of $H^{eff}$, while leaving other states unaffected.  

\begin{figure}
\begin{center}
\includegraphics[width=10cm]{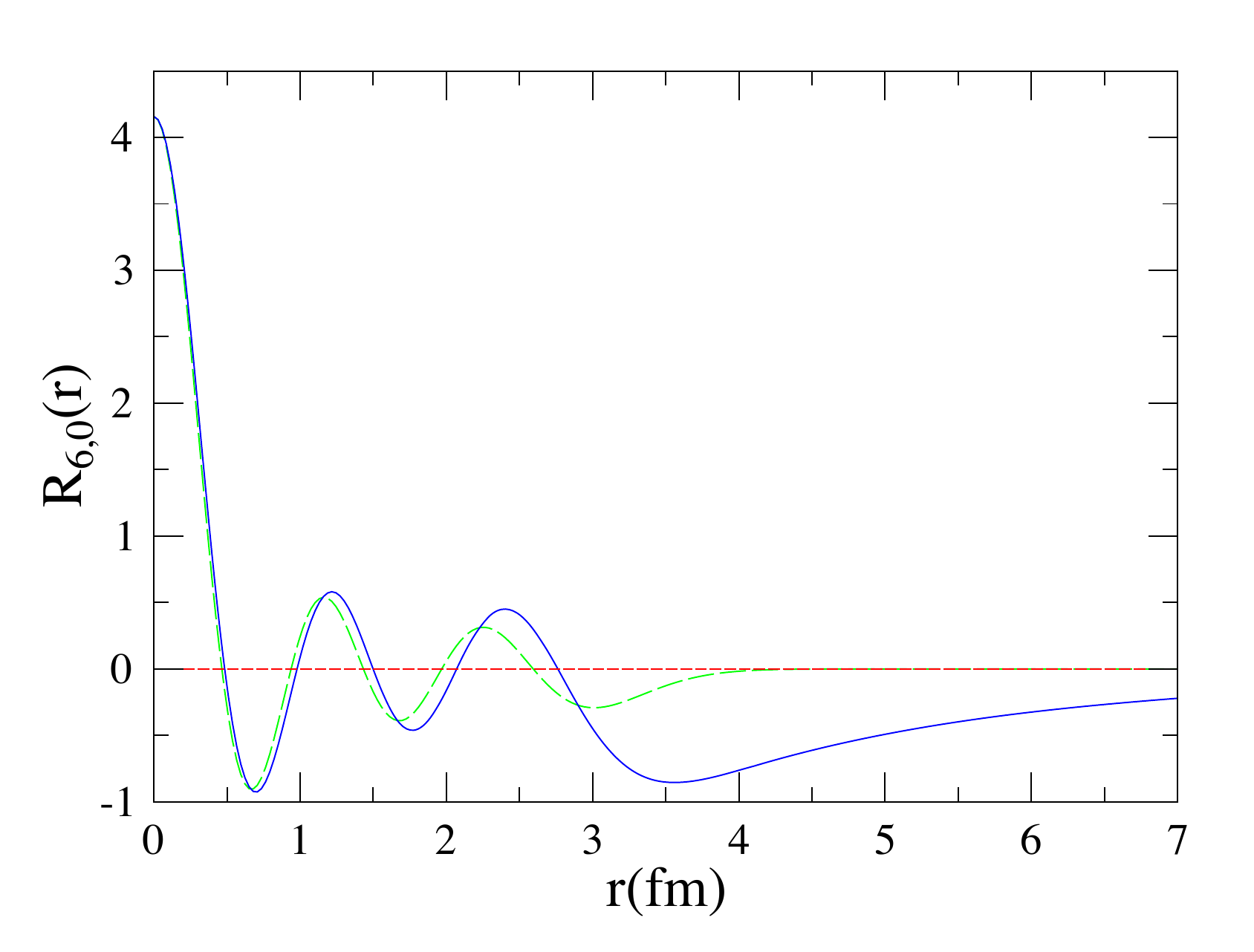}
\end{center}
\caption{(Color online) A comparison of the radial wave functions for the HO state $|n l\rangle$ (dashed) 
and for the extended
state $(E/E-QT) |n l\rangle$ (solid), for $(n,l)=(6,0)$
in a $\Lambda_P=10$ deuteron calculation.  The extended tail of the
latter is apparent.  Note that the normalization of the extended
state has been adjusted to match that of $|nl\rangle$ at $r$=0, in order to show that the
shapes differ only at large $r$.  Thus a depletion of the extended state at small $r$
is not apparent in this figure.}
\label{fig_2}
\end{figure}

Figure \ref{fig_2} shows the extended tail of the relative two-particle wave function that is induced by $E/(E-QT)$ acting on an edge HO state \cite{luu}.  As 
will become apparent from later expressions, this tail has the proper exponential fall-off,
\begin{equation}
\sim {e^{-\kappa r} \over \kappa r}
\end{equation}
where $\kappa= \sqrt{2|E|/\hbar \omega}$ and $r=|\vec{r}_1-\vec{r}_2|/\sqrt{2}b$ is the
dimensionless Jacobi coordinate, not the Gaussian tail of the HO.   At small $r$ the wave function
is basically unchanged (apart from normalization).

\section{The HOBET Effective Interaction}
Contact-gradient expansions 
are used in approaches like EFT to correct for the exclusion of short-range (high-momentum)
interactions.  The most general scalar  interaction is constructed, consistent with Hermiticity, 
parity conservation, and time-reversal invariance, as an expansion in the momentum.  Such
an interaction for the two-nucleon system, expanded to order N$^3$LO (or up to six gradients),
is shown in Table \ref{table:1}.  (Later these operators will be slightly modified for HOBET.)

\begin{table}
\centering
\caption{Contact-gradient expansion for relative-coordinate two-particle matrix elements. Here
$\stackrel{\rightarrow}{D^2_M} =  (\stackrel{\rightarrow}{\nabla} \otimes \stackrel{\rightarrow}{\nabla})_{2M}$,
$\stackrel{\rightarrow}{D^0_0} =  [ (\sigma(1) \otimes \sigma(2))_2 \otimes  D^2]_{00}$, 
 $\stackrel{\rightarrow}{F^3_M} =  (\stackrel{\rightarrow}{\nabla} \otimes \stackrel{\rightarrow}{D^2})_{3M}$, 
$\stackrel{\rightarrow}{F^1_M}= [ (\sigma(1) \otimes \sigma(2))_2 \otimes  F^3]_{1M}$,
$\stackrel{\rightarrow}{G^4_M} =  (\stackrel{\rightarrow}{D^2} \otimes \stackrel{\rightarrow}{D^2})_{4M},$
$\stackrel{\rightarrow}{G^2_M} =  [(\sigma(1) \otimes \sigma(2))_2 \otimes G^4]_{2M}$,
and the scalar product of tensor operators is defined as $A^J \cdot B^J = \sum_{M=-J}^{M=J} (-1)^M A^J_M B^J_{-M}$.}
\label{table:1}
{\footnotesize
\begin{tabular}{|c||c|c|c|c|}
\hline
Transitions & LO & NLO & NNLO & N$^3$LO \\ \hline
${}^3S_1 \leftrightarrow {}^3S_1$ \T & $a_{LO}^{3S1} \delta({\bf r})$ & $a_{NLO}^{3S1} (\stackrel{\leftarrow}{\nabla^2} \delta({\bf r}) + \delta({\bf r}) \stackrel{\rightarrow}{\nabla^2})$ & $a_{NNLO}^{3S1, 22} \stackrel{\leftarrow}{\nabla^2} \delta({\bf r}) \stackrel{\rightarrow}{\nabla^2}$ & $a_{N^3LO}^{3S1, 42} (\stackrel{\leftarrow}{\nabla^4} \delta({\bf r}) \stackrel{\rightarrow}{\nabla^2}+ \stackrel{\leftarrow}{\nabla^2} \delta({\bf r}) \stackrel{\rightarrow}{\nabla^4})$ \\
or ${}^1S_0 \leftrightarrow {}^1S_0$ \B & & & $a_{NNLO}^{3S1, 40} (\stackrel{\leftarrow}{\nabla^4} \delta({\bf r}) + \delta({\bf r}) \stackrel{\rightarrow}{\nabla^4})$ &
$a_{N^3LO}^{3S1, 60} (\stackrel{\leftarrow}{\nabla^6} \delta({\bf r}) + \delta({\bf r}) \stackrel{\rightarrow}{\nabla^6}$) \\ \hline
${}^3S_1 \leftrightarrow {}^3D_1$  \T & & $a_{NLO}^{SD} (\delta({\bf r}) \stackrel{\rightarrow}{D^0}+  \stackrel{\leftarrow}{D^0} \delta({\bf r}))$ & $a_{NNLO}^{SD, 22} (\stackrel{\leftarrow}{\nabla^2} \delta({\bf r}) \stackrel{\rightarrow}{D^0} + \stackrel{\leftarrow}{D^0} \delta({\bf r})\stackrel{\rightarrow}{\nabla^2} )$ &  $a_{N^3LO}^{SD, 42} ( \stackrel{\leftarrow}{\nabla^4} \delta({\bf r}) \stackrel{\rightarrow}{D^0} +\stackrel{\leftarrow}{D^0} \delta({\bf r})\stackrel{\rightarrow}{\nabla^4})$\\
 & & & $a_{NNLO}^{SD, 04} (\delta({\bf r}) \stackrel{\rightarrow}{\nabla^2} \stackrel{\rightarrow}{D^0} + \stackrel{\leftarrow}{D^0}\stackrel{\leftarrow}{\nabla^2} \delta({\bf r}))$ & $a_{N^3LO}^{SD, 24} ( \stackrel{\leftarrow}{\nabla^2}\delta({\bf r}) \stackrel{\rightarrow}{\nabla^2} \stackrel{\rightarrow}{D^0} +\stackrel{\leftarrow}{D^0} \stackrel{\leftarrow}{\nabla^2} \delta({\bf r})\stackrel{\rightarrow}{\nabla^2} )$\\
 \B & & & &  $a_{N^3LO}^{SD, 06} ( \delta({\bf r}) \stackrel{\rightarrow}{\nabla^4} \stackrel{\rightarrow}{D^0} +\stackrel{\leftarrow}{D^0} \stackrel{\leftarrow}{\nabla^4} \delta({\bf r}))$\\ \hline
${}^1D_2 \leftrightarrow {}^1D_2$ \T & & & $a_{NNLO}^{1D2}  \stackrel{\leftarrow}{D^2} \cdot \delta({\bf r})\stackrel{\rightarrow}{D^2}$ &  $a_{N^3LO}^{1D2}  (\stackrel{\leftarrow}{D^2} \stackrel{\leftarrow}{\nabla^2} \cdot \delta({\bf r})\stackrel{\rightarrow}{D^2} + \stackrel{\leftarrow}{D^2} \cdot \delta({\bf r}) \stackrel{\rightarrow}{\nabla^2} \stackrel{\rightarrow}{D^2})$ \\
or ${}^3 D_J \leftrightarrow {}^3 D_J$\B & & & & \\ \hline
${}^3D_3 \leftrightarrow {}^3G_3$ \T & & & &  $a_{N^3LO}^{DG} ( \stackrel{\leftarrow}{D^2} \cdot \delta({\bf r})\stackrel{\rightarrow}{G^2} + \stackrel{\leftarrow}{G^2} \cdot \delta({\bf r}) \stackrel{\rightarrow}{D^2})$  \B \\ \hline
${}^1P_1 \leftrightarrow {}^1P_1$ \T& & $a_{NLO}^{1P1} \stackrel{\leftarrow}{\nabla^{}} \cdot \delta({\bf r}) \stackrel{\rightarrow}{\nabla^{}}$ & $a_{NNLO}^{1P1} (\stackrel{\leftarrow}{\nabla^{}} \stackrel{\leftarrow}{\nabla^2} \cdot \delta({\bf r}) \stackrel{\rightarrow}{\nabla^{}} + \stackrel{\leftarrow}{\nabla^{}} \cdot \delta({\bf r}) \stackrel{\rightarrow}{\nabla^2} \stackrel{\rightarrow}{\nabla^{}})$ & $a_{N^3LO}^{1P1,33} \stackrel{\leftarrow}{\nabla^{}} \stackrel{\leftarrow}{\nabla^2} \cdot \delta({\bf r}) \stackrel{\rightarrow}{\nabla^2} \stackrel{\rightarrow}{\nabla^{}} $ \\
or ${}^3P_J \leftrightarrow {}^3P_J \B$ & & & & $a_{N^3LO}^{1P1,51} (\stackrel{\leftarrow}{\nabla^{}} \stackrel{\leftarrow}{\nabla^4} \cdot \delta({\bf r}) \stackrel{\rightarrow}{\nabla^{}} + \stackrel{\leftarrow}{\nabla^{}} \cdot \delta({\bf r}) \stackrel{\rightarrow}{\nabla^4} \stackrel{\rightarrow}{\nabla^{}})$ \\ \hline
${}^3P_2 \leftrightarrow {}^3F_2$ \T & & & $a_{NNLO}^{PF} (\stackrel{\leftarrow}{\nabla^{}} \cdot \delta({\bf r}) \stackrel{\rightarrow}{F^1} + \stackrel{\leftarrow}{F^1} \cdot \delta({\bf r}) \stackrel{\rightarrow}{\nabla^{}})$
& $a_{N^3LO}^{PF, 33} (\stackrel{\leftarrow}{\nabla^{}} \stackrel{\leftarrow}{\nabla^2} \cdot \delta({\bf r}) \stackrel{\rightarrow}{F^1} + \stackrel{\leftarrow}{F^1} \cdot \delta({\bf r}) \stackrel{\rightarrow}{\nabla^2} \stackrel{\rightarrow}{\nabla^{}}) $ \\
\B & & & & $a_{N^3LO}^{PF, 1 5} (\stackrel{\leftarrow}{\nabla^{}}  \cdot \delta({\bf r}) \stackrel{\rightarrow}{\nabla^2}\stackrel{\rightarrow}{F^1} + \stackrel{\leftarrow}{F^1} \stackrel{\leftarrow}{\nabla^2} \cdot \delta({\bf r})  \stackrel{\rightarrow}{\nabla^{}})$ \\ \hline
${}^1F_3 \leftrightarrow {}^1F_3$ \T & & & & $a_{N^3LO}^{1F3} \stackrel{\leftarrow}{F^3}  \cdot \delta({\bf r}) \stackrel{\rightarrow}{F^3} $ \\
or  ${}^3F_J \leftrightarrow {}^3F_J$\B & & & & \\
\hline \hline
\end{tabular}}
\end{table}

The ``data" for testing such an expansion for HOBET are
deuteron matrix elements $\langle \alpha | P(H^{eff}-H)P | \beta \rangle$ evaluated
as in Refs. \cite{song,luu} for $v_{18}$.  I take an  8$\hbar \omega$ $P$-space ($\Lambda_{P}$ = 8).
The evolution of the matrix elements will be followed as
contributions from scattering in $Q$ are integrated out progressively, starting with the highest
energy contributions.  To accomplish this, the
contribution to  $H^{eff}$
coming from excitations in $Q$ up to a scale $\Lambda > \Lambda_{P}$ is defined as
$H^{eff}(\Lambda)$, obtained by explicitly summing over all states in $Q$ 
up to that scale:
\begin{equation}
H^{eff}(\Lambda) \equiv H +  H {1 \over E-Q_\Lambda H} Q_\Lambda H~~~~Q_\Lambda \equiv
\sum_{\alpha=\Lambda_P+1}^\Lambda |\alpha \rangle \langle \alpha|~~~~Q_{\Lambda_P} \equiv 0.
\end{equation}
Thus $H^{eff} = H^{eff} (\Lambda \rightarrow \infty)$ and $H^{eff}(\Lambda_P)=H$.
The quantity 
\begin{equation}
\Delta(\Lambda) \equiv H^{eff}-H^{eff}(\Lambda) =
H {1 \over E-Q H} QH - H {1 \over E-Q_\Lambda H} Q_\Lambda H
\label{delta}
\end{equation}
represents the contributions to
$H^{eff}$ involving excitations in $Q$ above the scale $\Lambda$.  For $\Lambda >> \Lambda_{P}$, one expects $\Delta(\Lambda)$ 
to be small and well represented by a LO
interaction.  As $\Lambda$ runs to values closer to $\Lambda_{P}$, one would
expect to find that NLO, NNLO, N$^3$LO, .... contributions become successively more
important.   If one could formulate
some expansion that continues to accurately reproduce the various matrix
elements of $\Delta(\Lambda)$ as
$\Lambda \rightarrow \Lambda_P$, then a successful expansion for the
HOBET effective interaction $\Delta(\Lambda_P) = H^{eff}-H$ would be in hand.  

\begin{figure}
\begin{minipage}{0.5\linewidth}
\begin{center}
\includegraphics[width=7.5cm]{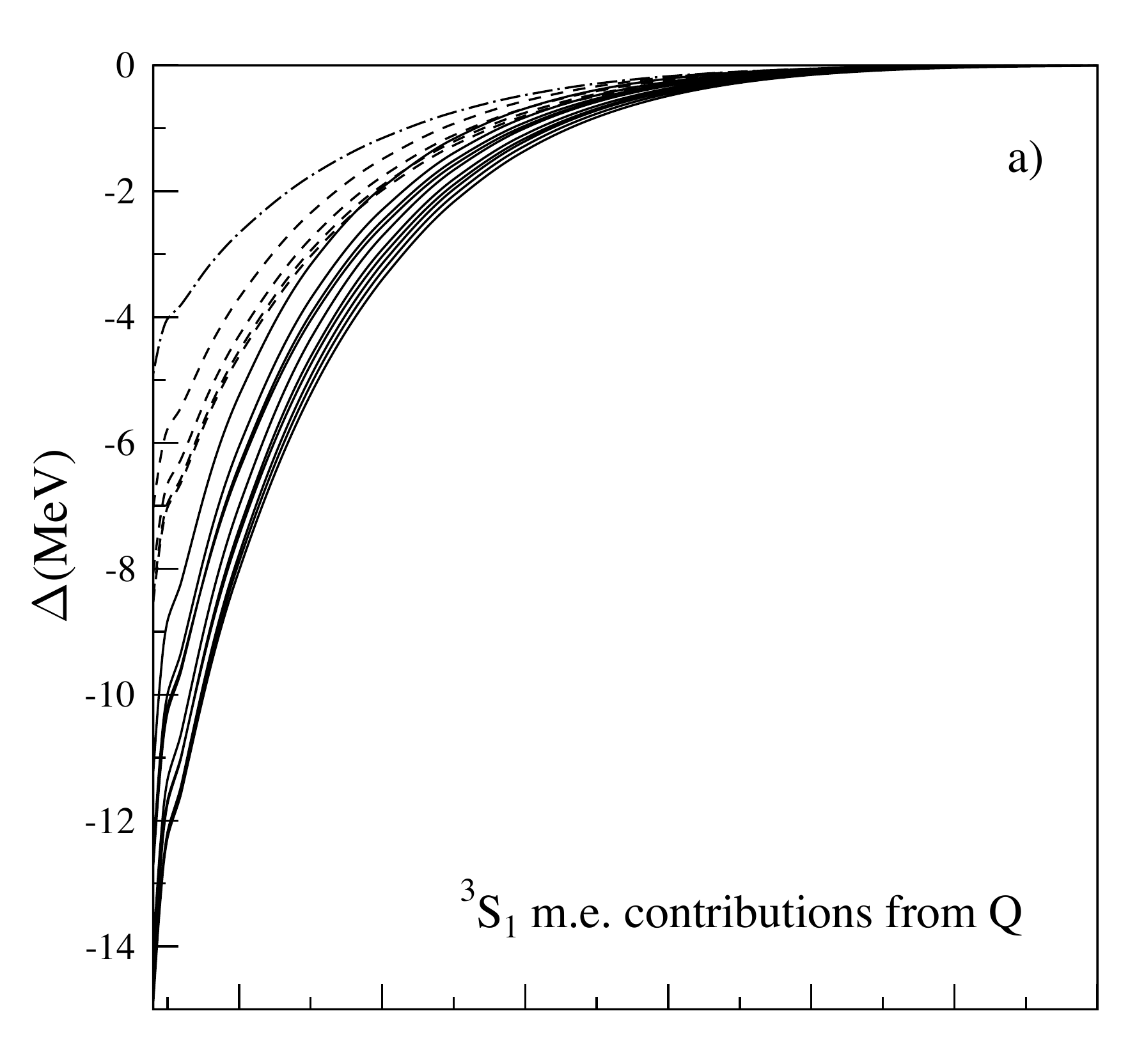}
\includegraphics[width=7.5cm]{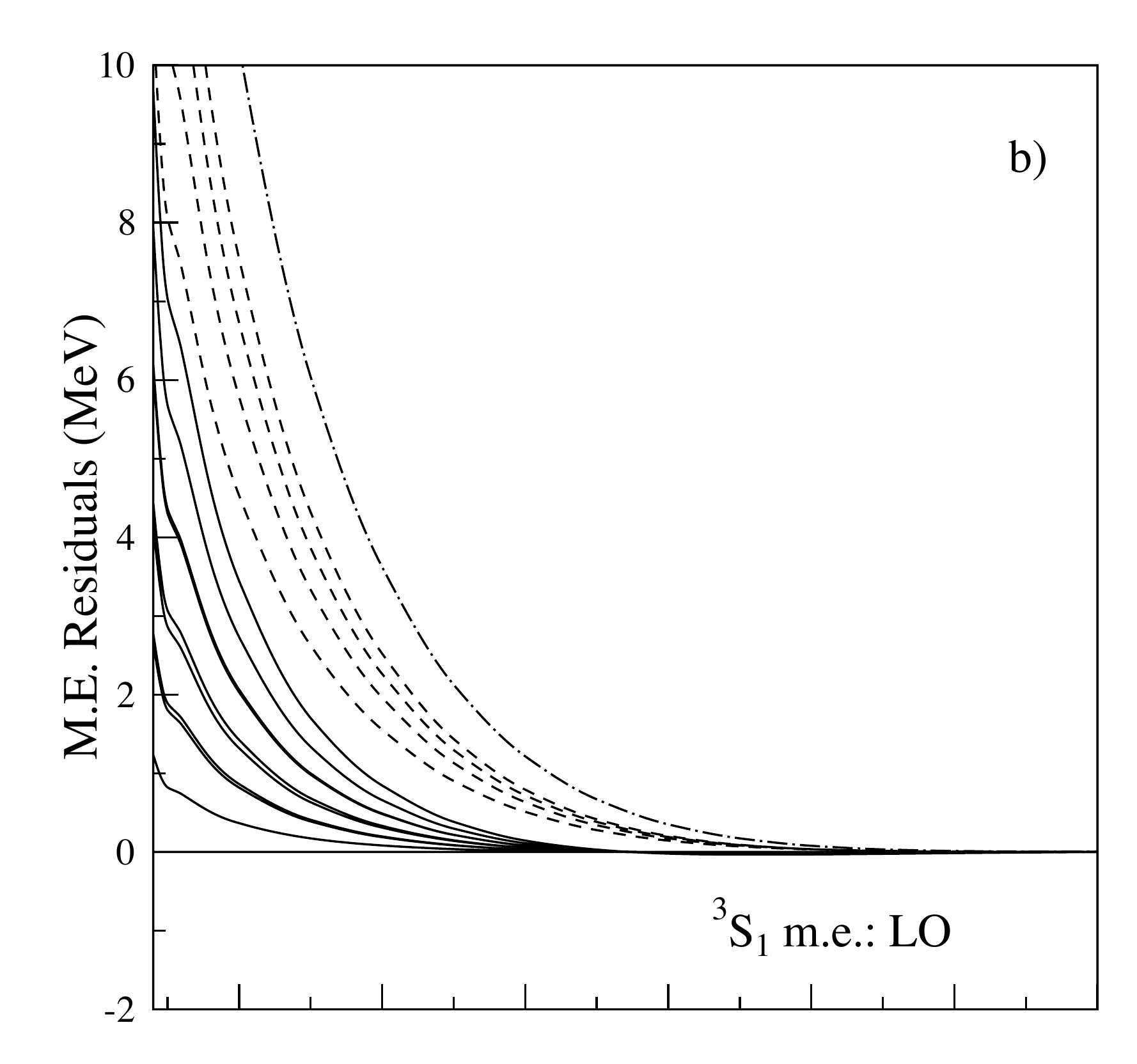}
\includegraphics[width=7.5cm]{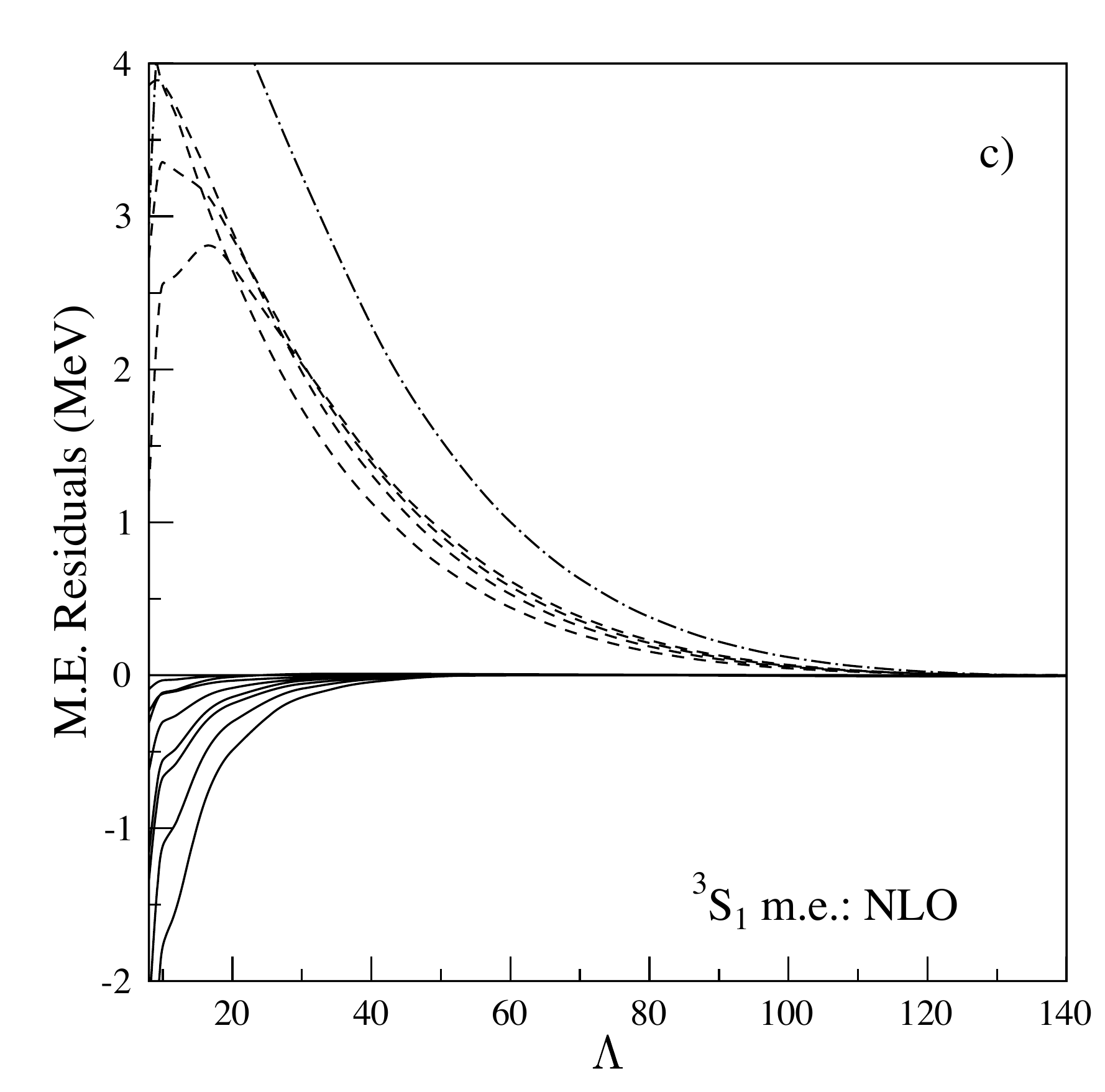}
\end{center}
\end{minipage}%
\begin{minipage}{0.5\linewidth}
\begin{center}
\includegraphics[width=7.5cm]{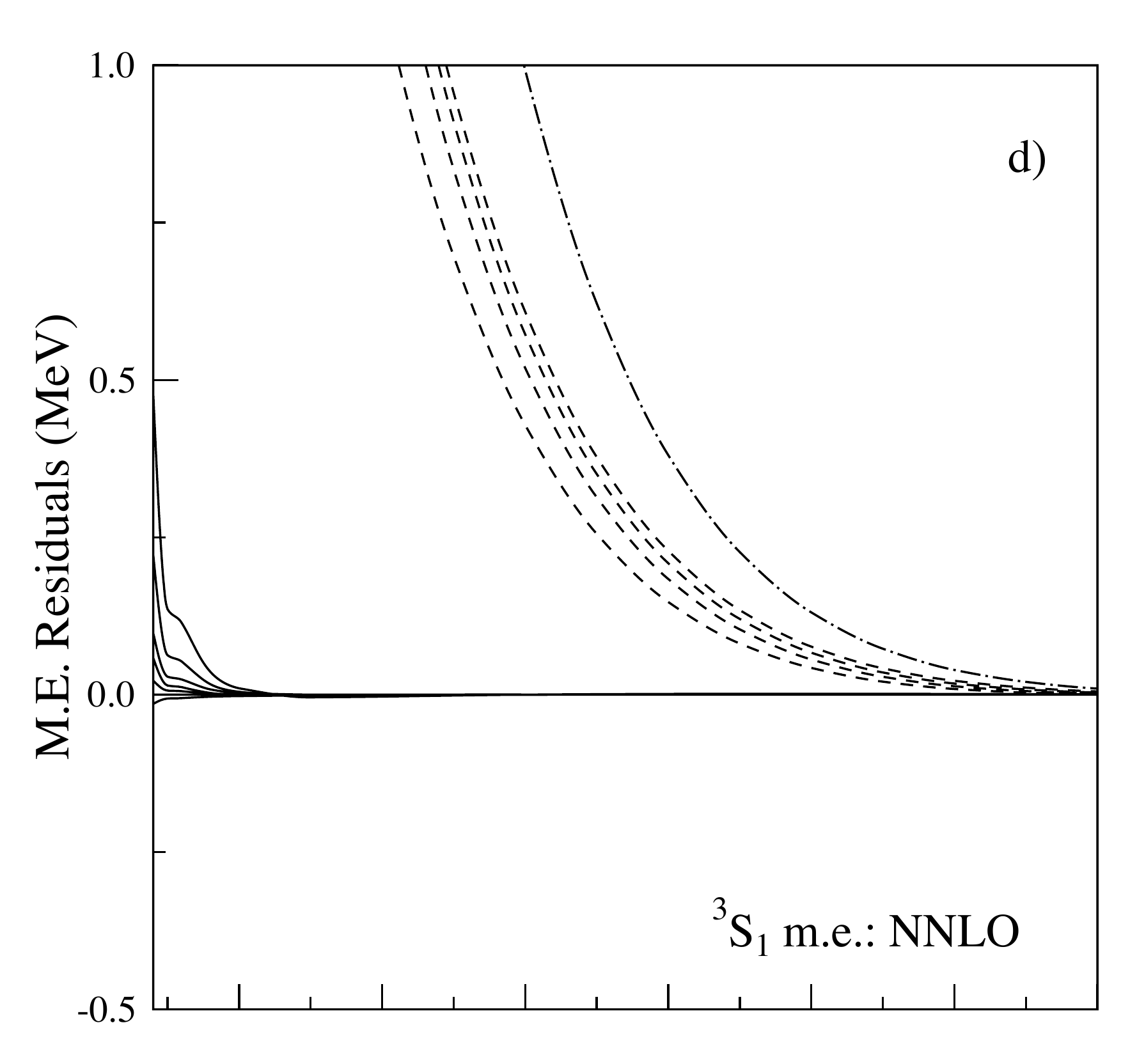}
\includegraphics[width=7.5cm]{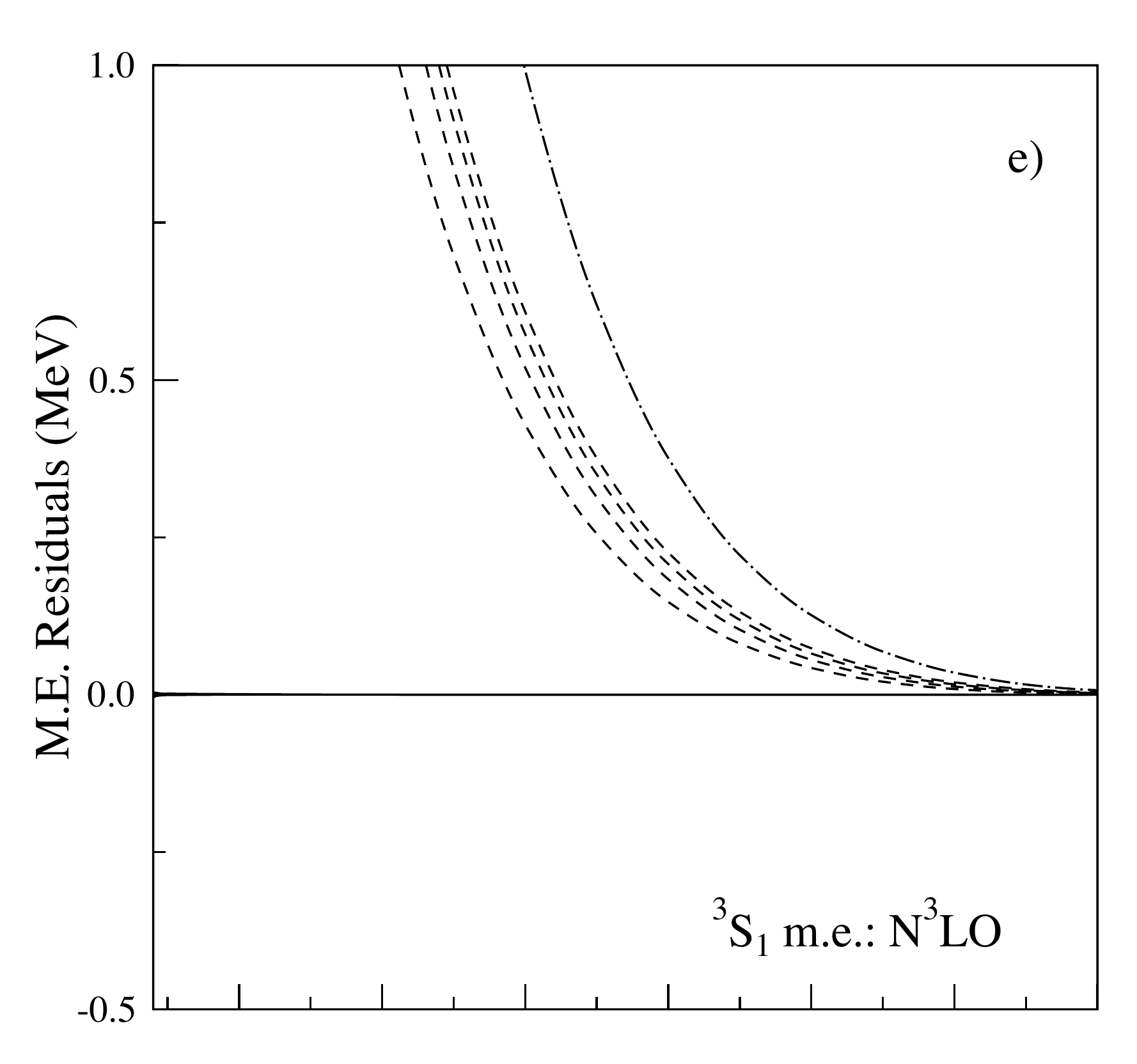}
\includegraphics[width=7.5cm]{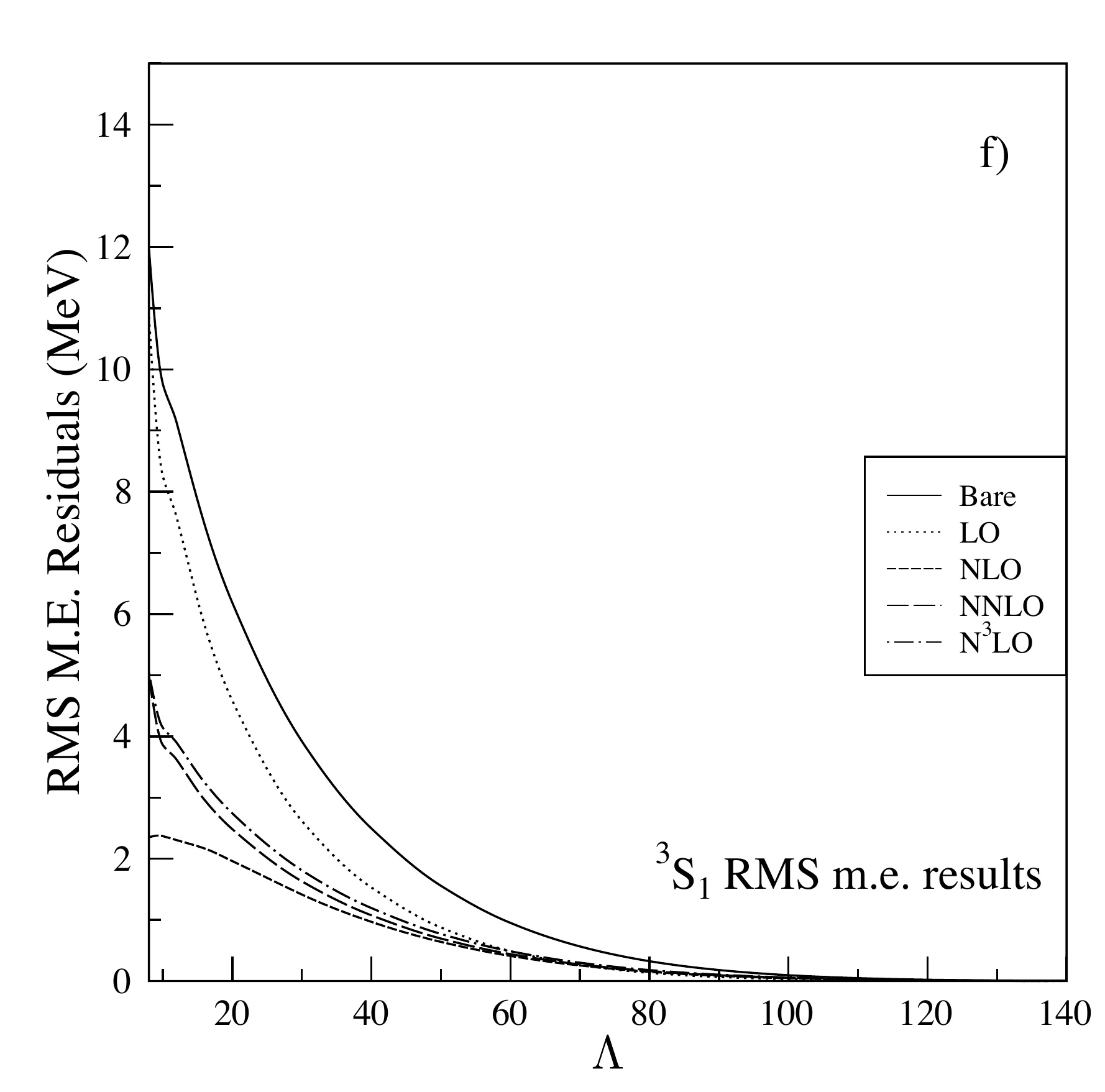}
\end{center}
\end{minipage}
\caption{In a) the contributions to $H^{eff}-H$ from excitations in $Q$ above $\Lambda$ are plotted
for a calculation with $\Lambda_P=8$ and $b$=1.7 fm.  Each line describes the running
of one of the 15 independent $P$-space
matrix elements $\langle n^\prime l^\prime=0 | H^{eff}-H | n l=0 \rangle$, $1 \leq n \leq n^\prime \leq 5$.
Ten of the matrix elements are between nonedge states (solid), four connect the $n^\prime=5$
edge state to the $n$=1,2,3,4
nonedge states (dashed), and one is the diagonal $n^\prime=n=5$ edge-edge case (dot dashed).
b)-e) show the residuals for naive LO, NLO, NNLO, and N$^3$LO fits (see text).  f) shows the RMS
deviation for the set of $P$-space matrix elements.
The expected systematic improvement with increasing  order is apparent only for matrix elements between nonedge states.}
\label{fig_simple}
\end{figure}

Figure \ref{fig_simple}a is a plot of $\Delta(\Lambda)$ for the 15 $^3S_1$ matrix elements in the chosen P-space.  
For typical matrix elements $\Delta(\Lambda_P)=H^{eff}-H
\sim$ -12 MeV -- a great deal of the deuteron binding comes from the Q-space. 
Five of the matrix elements involve bra or ket edge states.
The evolution of these contributions with $\Lambda$ appears to be less regular
than is observed for nonedge-state matrix elements.

One can test whether the results shown in Fig. \ref{fig_simple}a can be reproduced in a contact-gradient
expansion.  At each $\Lambda$ the coefficients
$a_{LO}^{3S1}(\Lambda)$, $a_{NLO}^{3S1}(\Lambda)$, etc., would be determined
from the lowest-energy ``data,"
those matrix elements
$\langle \alpha | \Delta(\Lambda) | \beta \rangle$ carrying the fewest HO quanta.
Thus, in LO,  $a_{LO}^{3S1}(\Lambda)$
would be determined from the $(n^\prime,n)=(1,1)$ matrix element.  The remaining 
14 $P$-space matrix elements are then predicted, not fit;  in NNLO four coefficients would
be determined from the (1,1), (1,2), (1,3), and (2,2) matrix elements, and
eleven predicted.
Figures \ref{fig_simple}b-d show the residuals -- the differences between the predicted and calculated matrix elements. 
For successive LO, NLO, and NNLO calculations, the scale at which
residuals in $\Delta$ are significant, say greater than 10 keV, is brought down successively,
e.g., from an initial $\sim 100 \hbar \omega$, to $\sim 60 \hbar \omega$ (LO), to $\sim 30 \hbar \omega$
(NLO), and finally to $\sim 20 \hbar \omega$ (NNLO), except for matrix elements involving edge
states.  There the improvement is not significant, with noticeable deviations remaining at 
$\sim 100 \hbar \omega$ even at NNLO.  This irregularity indicates a flaw in the underlying
physics of this approach -- specifically the use of a short-range expansion for $H^{eff}$
when important contributions to $H^{eff}$ are coming from long-range interactions in $Q$. 
So this must be fixed.

\subsection{The contact-gradient expansion for HOBET}
The gradient with respect to the dimensionless coordinate $\vec{r} \equiv (\vec{r}_1-\vec{r}_2) / b\sqrt{2}$ is denoted by $\overrightarrow{\nabla}$. 
The coefficients $a_{LO}, a_{NLO}, ...$  in Table \ref{table:1} then carry the dimensions of MeV.

The contact-gradient expansion defined in Table \ref{table:1} is that commonly used in plane-wave bases,
where one expands around $\vec{k}=0$ with
\begin{equation}
\left. \overrightarrow{\nabla}^2 \exp{i \vec{k} \cdot \vec{r}} ~\right|_{\vec{k}=0} = 0. 
\end{equation}
HOBET begins with a lowest-energy $1s$ Gaussian wave packet with a characteristic
momentum $\sim 1/b$.  An analogous definition of gradients
such that
\begin{equation}
\overrightarrow{\nabla}^2 \psi_{1s}(b) = 0 
\end{equation}
is obtained by redefining each operator appearing in Table \ref{table:1} by  
\begin{equation}
O \rightarrow \bar{O} \equiv e^{r^2/2} O e^{r^2/2}.
\label{bar}
\end{equation}
The gradients appearing in the operators of Table \ref{table:1}
then act on polynomials in $r$.  This leads to two
attractive properties.  First is the removal of operator mixing.  Once 
$a_{LO}^{3S1}$ is fixed in LO to the $(n^\prime,n)=(1,1)$ matrix element, this quantity remains fixed in NLO, NNLO, etc.  Higher-order terms make no contributions to this matrix element.
Similarly, $a_{NLO}$, once fixed to the $(1,2)$ matrix element, is unchanged
in NNLO.  That is, the NLO results contain the LO results, and so on.
Second, this definition gives the HOBET effective interaction a simple
dependence on nodal quantum numbers,
\begin{equation}
\overrightarrow{\nabla}^2 \sim -4 (n-1)~~~~~~~~\overrightarrow{\nabla}^4 \sim16 (n-1)(n-2).
\end{equation}
(The Appendix describes this expansion in some detail.)  In each channel, this dependence 
agrees with the plane-wave result in lowest contributing order, but otherwise differs in
terms of relative order $1/n$.  This HO form of the contact-gradient expansion is connected
with standard Talmi integrals \cite{talmi}, generalized for nonlocal potentials, e.g.,
\begin{eqnarray}
a_{LO} &\sim&  \int^\infty_0 \int^\infty_0 e^{-r_1^2}  \left[V(r_1,r_2)\right]  e^{-r_2^2} r_1^2 r_2^2 dr_1 dr_2 \nonumber \\
a_{NLO} &\sim& \int^\infty_0 \int^\infty_0 e^{-r_1^2}  \left[r_1^2 V(r_1,r_2) \right] e^{-r_2^2} r_1^2 r_2^2 dr_1 dr_2 = \int^\infty_0 \int^\infty_0 e^{-r_1^2} \left[ V(r_1,r_2)  r_2^2 \right] e^{-r_2^2} r_1^2 r_2^2 dr_1 dr_2 
\end{eqnarray}
and so on.
 
\subsection{Identifying terms with the contact-gradient expansion}
The next question is the association of the operators in Table \ref{table:1} with an appropriate
set of terms in $H^{eff}-H$, so that the difficulties apparent in Fig. \ref{fig_simple} are avoided.
The reorganized BH equation of Eq. (\ref{BHnew})
\begin{eqnarray}
H^{eff} & = &
{E \over E -TQ} \left[ T -T {Q \over E}T + V + V {1 \over E-QH} QV \right] {E \over E-QT}  \nonumber \\
& \rightarrow &{E \over E -TQ} \biggl[ T -T {Q \over E}T + V + \sum_{ i=LO, NLO, ...} \bar{O}^i \biggr] {E \over E-QT}
\label{final}
\end{eqnarray}
isolates
$V (E-QH)^{-1} QV$, a term that is sandwiched between short-range operators that scatter
to high-energy states: one anticipates this term can be successfully represented by a short-range expansion
like the contact-gradient expansion.  This identification is made here and tested later in this
paper.

This reorganization
only affects the edge-state matrix elements, clearly.   As the process of fitting coefficients
uses matrix elements of low $(n^\prime,n)$, none of which involves edge states, the 
coefficients are unchanged.  But every matrix element involving edge states now includes
the effects of rescattering by $QT$ to all orders.  Thus a procedure for 
evaluating these matrix elements is needed.  

\subsection{Matrix element evaluation}
There are several alternatives for evaluating Eq. (\ref{final}) for edge states.  One of these 
exploits the tri-diagonal form of $QT$ for the deuteron.  If $|n l \rangle$ is an edge state in
$P$ then
\begin{equation}
{E \over E-QT} |n~ l \rangle = |n~ l \rangle + {1 \over E-QT} QT |n~ l \rangle = |n ~l \rangle 
+ \sqrt{n(n+l+1/2)} {1 \over -\kappa^2 - {2 Q T \over \hbar \omega}} |n+1 ~l \rangle
\label{st1}
\end{equation}
where $E<0$ for a bound state.  The dimensionless parameter $\kappa=\sqrt{2|E| \over \hbar \omega}$ depends on the ratio of the binding energy $|E|$
to the HO energy scale.  Note that the second vector on the right in Eq. (\ref{st1}) lies entirely in $Q$.  Now
\begin{eqnarray}
{2 \over \hbar \omega}QT |n+1~ l\rangle &=& (2n+l+3/2) |n+1~l\rangle + \sqrt{(n+1)(n+l+3/2)} |n+2~l\rangle \nonumber \\
{2 \over \hbar \omega}QT |n+2~l\rangle &=& \sqrt{(n+1)(n+l+3/2)} |n+1~l \rangle +
(2n+l+7/2) |n+2~l\rangle + \sqrt{(n+2)(n+l+5/2)} |n+3~l\rangle \nonumber \\
{2 \over \hbar \omega}QT |n+3~l\rangle &=& \sqrt{(n+2)(n+l+5/2)} |n+2~l\rangle+
(2n+l+11/2) |n+3~l\rangle+\sqrt{(n+3)(n+l+7/2)} |n+4~l \rangle \nonumber \\
{2 \over \hbar \omega}QT | n+4~l\rangle &=& ...
\end{eqnarray}
So the operator $2 QT/\hbar \omega$ in the basis $\{ |n+i~l\rangle, i=1, 2, ...$ \} has the form
\begin{equation}
{2 \over \hbar \omega} QT=\left( \begin{array}{ccccc} \alpha_1 & \beta_1 & 0 & 0 & \\ \beta_1 & \alpha_2 &
\beta_2 & 0 & \\ 0 & \beta_2 & \alpha_3 & \beta_3 & \cdots \\ 0 & 0 & \beta_3 & \alpha_4 & \\
 & & \vdots & & \end{array} \right)
 \end{equation}
 where
 \begin{equation}
 \alpha_i = \alpha_i(n,l) = 2n + 2i + l-1/2,~~~~\beta_i =\beta_i(n,l) = \sqrt{(n+i)((n+i+l+1/2)}~.
 \end{equation}
 
As is well known, if this representation of the operator $2 QT/\hbar \omega$ is truncated after
$k$ steps, the $2k-1$ nonzero coefficients \{$\alpha_i,\beta_i$\} determine the $2k-1$ operator moments
of the starting vector $|n+1~ l\rangle$, 
\begin{equation}
\langle n+1~l | \left( {2 QT \over \hbar \omega} \right)^i |n+1~l \rangle,~~~i=1,....,2k-1
\end{equation}
A standard formula exists \cite{haydock} for the moments expansion of the Green's function acting on the first vector $|n+1~l\rangle$ of such a
tri-diagonal matrix, allowing us to write
 \begin{equation}
\sqrt{n(n+l+1/2)} {1 \over -\kappa^2 - {2 QT \over \hbar \omega}} |n+1~ l\rangle= \widetilde{g}_1(-\kappa^2;n,l) |n+1~l\rangle + \widetilde{g}_2(-\kappa^2;n,l) |n+2~l\rangle + \widetilde{g}_3(-\kappa^2;n,l) |n+3~l \rangle + \cdots
 \end{equation}
 The coefficients \{$\widetilde{g}_i$\} can be obtained from an auxiliary set of continued fractions \{$g_i^\prime$\}
 that are determined by downward recursion
 \begin{eqnarray}
 g_k^\prime(-\kappa^2;n,l) &\equiv& {1 \over -\kappa^2 - \alpha_k(n,l)} \nonumber \\
 g_{i-1}^\prime(-\kappa^2;n,l) &=& {1 \over -\kappa^2 - \alpha_{i-1}(n,l) - \beta_{i-1}(n,l)^2 g_i^\prime(-\kappa^2;n,l)},~~i=k,....2
 \end{eqnarray}
 From these continued fractions the needed coefficients can be computed from the algebraic relations
 \begin{eqnarray}
 \widetilde{g}_1(-\kappa^2;n,l) &=& \sqrt{n(n+l+1/2)} g_1^\prime(-\kappa^2;n,l) \nonumber \\
 \widetilde{g}_i(-\kappa^2;n,l) &=& \widetilde{g}_{i-1}(-\kappa^2;n,l) \beta_{i-1} g_i^\prime(-\kappa^2;n,l),~~ i=2,...,k
 \end{eqnarray}
 Defining $\widetilde{g}_0(-\kappa^2;n,l) \equiv 1$ it follows
 \begin{eqnarray}
 {E \over E-QT} |n~l> &=& \sum_{i=0}^{k \rightarrow \infty} \widetilde{g}_i(-\kappa^2;n,l) |n+i~l\rangle, ~~\mathrm{edge~state} \nonumber \\
 &=& |n~l>,~~~~~~~~~~~~~~~~\mathrm{otherwise}
 \label{gf}
 \end{eqnarray}
 where it is understood that $k$ is made large enough so that the moments expansion for
 the Green's function is accurate throughout the region in coordinate space where
 $E/(E-QT) |n~l\rangle$ is needed.  Note that the first line of Eq. (\ref{gf}) can be viewed as the
 general result if one defines
 \begin{equation}
 g_i(-\kappa^2;n,l) \equiv 0,~~i=1,...k,~~\mathrm{if}~|n~l\rangle ~\mathrm{is~not~an~edge~state}.
 \end{equation}
 (For $A \ge$ 3 one would be treating the 3($A$-1)-dimensional HO, with the role of
 the spherical harmonics replaced by the corresponding hyperspherical harmonics.)
 Eq. (\ref{gf}) can now be used to evaluate the various terms in Eq. (\ref{final}). \\
 
 \noindent
 {\it Matrix elements for the contact-gradient operators:} The matrix elements have the general form
\begin{equation}
\langle n' l' | {E \over E-TQ} \bar{O} {E \over E-QT} | n l \rangle = \sum_{i,j=0} \widetilde{g}_j(-\kappa^2;n',l') \widetilde{g}_i(-\kappa^2;n,l) \langle n'+j~ l | \bar{O} |n+i~ l \rangle
\end{equation}
where $\bar{O}$ is formed from gradients acting on the bra and ket, evaluated at $\vec{r}$=0.
The general matrix element (any partial wave) is worked out in the Appendix.  For example,
one needs for $S$-wave channels the relation 
 \begin{equation}
 (\vec{\nabla}^2)^p e^{r^2/2} R_{nl=0}(r) Y_{00}(\Omega_r) \big|_{\vec{r} \rightarrow 0} = (-4)^p ~{(n-1)! \over (n-1-p)!} ~ {1 \over \pi} \left[ {\Gamma(n+1/2) \over
 (n-1)!}\right]^{1/2} 
  \end{equation}
from which it follows
 \begin{eqnarray}
&& \langle n^\prime (l^\prime=0~S=1) J=1 |{E \over E -TQ} \biggl[ \sum_{ i=LO, ...,N^3LO} \bar{O}^{3S1,i} \biggr] {E \over E-QT} | n (l=0~S=1) J=1 \rangle = \nonumber \\
 &&{2 \over \pi^2}  \sum_{i,j=0} \widetilde{g}_j(-\kappa^2;n^\prime,l^\prime=0) \widetilde{g}_i(-\kappa^2;n,l=0) \left[ {\Gamma{n^\prime+j +1/2} \Gamma{n+i+1/2} \over (n^\prime+j-1)! (n+i-1)!} \right] \biggl[ a_{LO}^{3S1}
 -4\left((n^\prime+j-1)+(n+i-1)\right) a_{NLO}^{3S1}   \nonumber \\
&& +16\left\{(n^\prime+j-1)(n+i-1)a_{NNLO}^{3S1,22} + \left((n^\prime+j-1)(n^\prime+j-2)+
 (n+i-1)(n+i-2)\right)a_{NNLO}^{3S1,40}\right\} \nonumber \\
&&   -64 \left\{ (n^\prime+j-1)(n+i-1) \bigl( (n^\prime+j-2)+(n+i-2) \bigr) a_{N^3LO}^{3S1,42}  \right. \nonumber \\
&& + \left.  \left((n^\prime+j-1)(n^\prime+j-2)(n^\prime+j-3)+(n+i-1)(n+i-2)(n+i-3)\right) a_{N^3LO}^{3S1,60} \right\} \biggr].
 \end{eqnarray}
 
In the case of nonedge states,  $\widetilde{g}_i\equiv 0$ except for the case of $\widetilde{g}_0 \equiv 1$.  Thus it is apparent
that the net consequence of the rearrangement of the BH equation and the identification of the
contact-gradient expansion with $V(E-QH)^{-1}QV$, is effectively a renormalization of the coefficients of that expansion
for the edge HO states.  That renormalization is governed by $\kappa^2 =2|E|/\hbar \omega$, e.g., 
\begin{eqnarray}
a_{LO}(n',l',n,l) \rightarrow a_{LO}^\prime(E;n',l',n,l) = a_{LO}(n',l',n,l)  \sum_{i,j=0} \widetilde{g}_j(-\kappa^2;n',l') \widetilde{g}_i(-\kappa^2;n,l) \nonumber \\
\times \left[ {\Gamma(n'+j+1/2)
\Gamma(n+i+1/2) \over \Gamma(n'+1/2) \Gamma(n+1/2)} \right]^{1/2}
\left[{(n'-1)! (n-1)! \over (n'+j-1)! (n+i-1)!} \right]^{1/2} .
\label{primes}
\end{eqnarray}
This renormalization is large, typically a reduction in strength by a factor of 2-4, 
for $|E|$=2.224 MeV,  and also remains substantial for more deeply bound systems, as 
will be illustrated later.  (The binding energy for this purpose is defined relative to the lowest particle
breakup channel, the first extended state.)
The effects encoded into $|\widetilde{\alpha} \rangle$  by summing $QT$ to all orders are 
nontrivial:  they depend on a nonperturbative strong interaction parameter $|E|$ as well
as $QT$, and they alter
effective matrix elements of the strong potential.    For a given choice of $\Lambda_P$,
the renormalization depends
on a single parameter, $2|E|/\hbar \omega$, not on $|E|$ or $b$ separately.
In the plane-wave limit $b \rightarrow \infty$, this parameter is driven to $\infty$,
so that  $a_{LO}^\prime \rightarrow a_{LO}$.  No renormalization is required in this limit.
The dependence on $|E|$ is discussed in more detail later, including its connection
to the state-dependence
inherent in effective theory. \\

\noindent
{\it Matrix elements of the relative kinetic energy:}  The relative kinetic energy operator couples
$P$ and $Q$ via strong matrix elements that grow as $n$.  As Ref. \cite{luu} discusses, this coupling
causes difficulties with perturbative expansions in $H$ even in the case of $P$ spaces that
contain almost all of the wave function (e.g., $\Lambda_P
\sim$ 70).  There is always a portion of the wave function 
tail at large $r$ that is nonperturbative, involving matrix elements of $T$ that
exceed $\Lambda_P \hbar \omega/2$.

The kinetic energy contribution is
\begin{equation}
\langle \alpha | T+T {1 \over E-QT} QT | \beta \rangle = \langle \widetilde{\alpha} | T-T{Q \over E} T | \widetilde{\beta} \rangle = \langle \alpha |T |\widetilde{\beta} \rangle = \langle \widetilde{\alpha} | T | \beta \rangle
\end{equation}
where the last two terms show that the transformation to states $|\widetilde{\alpha} \rangle =
E/(E-QT) |\alpha \rangle$ reduces the calculation of the rescattering to that of a bare matrix element.
It follows from this expression
\begin{equation}
\langle n^\prime~l | T + T {1 \over E-QT} QT |n~l \rangle = \langle n^\prime~l | T | n~l \rangle + { \hbar \omega \over 2} \delta_{n^\prime n} \sqrt{n(n+l+1/2)} \widetilde{g}_1(-\kappa^2;n,l).
\end{equation}
Thus, rescattering via $QT$ alters the diagonal matrix element of the effective interaction
for edge states, as determined by
$\widetilde{g}_1(-\kappa^2; n,l)$.\\

\noindent
{\it Matrix elements of the bare potential:}  The $P$-space matrix element of $V$ becomes
$\langle \widetilde{\alpha} | V | \widetilde{\beta} \rangle$ which, as is illustrated in Fig. \ref{fig_2}, involves
an integral over a wave function that, apart from normalization, differs from the 
HO only in the tail, where the potential is weak.  It can be evaluated by generating the 
wave functions $| \widetilde{\alpha} \rangle$ and $|\widetilde{\beta} \rangle$ as HO expansions,
\begin{equation}
\left[ \sum_{j=0} \widetilde{g}_j(-\kappa^2;n^\prime,l^\prime) \langle n^\prime+j~l^\prime | \right]
~V~\left[ \sum_{i=0} |n+i~l \rangle \widetilde{g}_i(-\kappa^2;n,l) \right] = \sum_{i,j=0}\widetilde{g}_j(-\kappa^2;n^\prime,l^\prime) \widetilde{g}_i(-\kappa^2;n,l) \langle n^\prime+j~l^\prime | 
~V~ |n+i~l \rangle .
\label{barev}
\end{equation}
though the alternative Green's function expression, discussed below, is simpler. \\

\noindent
{\it Use of the free Green's function:} 
An alternative to an expansion in an HO basis is generation of  $|\widetilde{\alpha}\rangle $ with
the free (modified Helmholtz) Green's function.  For any P-space state $|n~l\rangle$,
\begin{equation}
(E-QT) |\widetilde{\alpha}\rangle = E |\alpha \rangle~\Rightarrow~ (E-T) |\widetilde{\alpha}\rangle = E |\alpha \rangle - PT |\widetilde{\alpha}\rangle
\end{equation}
That is, both $E-QT$ and $E-T$ project $|\widetilde{\alpha}\rangle$ back into the $P$-space.  The
free Green's function equation can be written
\begin{eqnarray}
(E-T) |\widetilde{\alpha} \rangle &=& P \left[ E - T {E \over E-QT} \right] P | \alpha \rangle \nonumber \\
&=& \left[ P {1 \over E-T} P \right]^{-1} | \alpha \rangle.
\label{gf2} 
\end{eqnarray}
Either of the driving terms on the right-hand side is easy to manipulate.  The second expression 
requires inversion of a P-space matrix, one most easily calculated
in momentum space, as the HO is its own Fourier transform and as the resulting momentum-space integrals can be done in
closed form.  This form was used in 
the three-body calculations of Ref. \cite{tom}.

Here I will use the first expression above, rewriting  the right-hand-side driving term
in terms of $ |\alpha_{nlm_l} \rangle$,
\begin{eqnarray}
&&P \left[ E - T {E \over E-QT} \right] P |\alpha \rangle = {\hbar \omega \over 2} \biggl[ \biggl(-\kappa^2 - (2n+l-1/2) - \widetilde{g}_1(-\kappa^2;n,l) \sqrt{n(n+l+1/2)} \biggr) |n~l~m_l>  \nonumber \\
&& -  \sqrt{(n-1)(n+l-1/2)} |n-1~l~m_l\rangle - \sqrt{n(n+l+1/2)} P |n+1~l ~m_l\rangle \biggr] \equiv {\hbar \omega \over 2} |\alpha_{nlm_l} \rangle,
\label{greens}
\end{eqnarray}
where the driving term has been kept general, valid for either edge or nonedge states: the latter can be a
helpful numerical check, verifying that a HO wave function is obtained, for such cases, from
the expression below.  For an edge state, $\widetilde{g}_1$ is nonzero and $P|n+1~l\rangle \equiv 0$;
for a nonedge state, $\widetilde{g}_1=0$ and $P$=1.   Labeling the corresponding edge state
as  $|\widetilde{\alpha}_{nlm_l} \rangle$, 
\begin{eqnarray}
\langle \vec{r} |\widetilde{\alpha}_{nlm_l} \rangle &=& \int d^3 \vec{r}^{~\prime} {1 \over 4 \pi |\vec{r}-\vec{r}^{~\prime}|} e^{-\kappa |\vec{r}-\vec{r}^{~\prime}|} ~ \langle \vec{r}^{~\prime}|\alpha_{nlm_l}> 
\nonumber \\
&=& -Y_{lm}(\Omega_r) \biggl[ {1 \over \sqrt{r}}~ I_{l+1/2}(\kappa r) \int_r^\infty d^3\vec{r}^{~\prime} (r^\prime)^{3/2} K_{l+1/2}(\kappa r^\prime)~\langle \vec{r}^{~\prime}|\alpha_{nlm_l}> \nonumber \\
&&+ {1 \over \sqrt{r}} ~K_{l+1/2}(\kappa r) \int_0^r d^3\vec{r}^{~\prime} (r^\prime)^{3/2} I_{l+1/2}(\kappa r^\prime)~\langle \vec{r}^{~\prime}|\alpha_{nlm_l}>
\label{eq:partial-wave}
\end{eqnarray}
where $I$ and $K$ denote the standard modified Bessel functions.
By expressing the HO radial wave functions in terms of the underlying Laguerre polynomials and integrating the polynomials term by term, 
alternative expressions are obtained for the various quantities previously expressed as expansions
in the $\widetilde{g}_i$.  This is detailed in the Appendix. 
One finds, for example, 
\begin{eqnarray}
\langle \vec{r}=0|\widetilde{\alpha}_{nlm_l} \rangle &=& \delta_{l,0} ~\delta_{m_l,0} ~ \sqrt{{(n-1)! \Gamma(n+1/2) \over 2 \pi}} ~\sum_{k=0}^{n} {(-2)^k \over k! (n-k)! \Gamma(k+3/2)}~ \times \nonumber \\
&&\biggl[(n-k)(\kappa^2 + 3n-3/2-k 
+\widetilde{g}_1(-\kappa^2;n,0) \sqrt{n(n+1/2)}) 
+ P[n+1,l=0] n(n+1/2) \biggr] ~ \times \nonumber \\
&& \biggl[-\sqrt{2}~ \kappa ~\Gamma(k+3/2)~ {}_1F_1[k+3/2;3/2;\kappa^2/2] + k!~ {}_1F_1[k+1;1/2;\kappa^2/2] \biggr]
\end{eqnarray}
where ${}_1F_1$ is Kummer's confluent hypergeometric function, and where
$P[n+1,l=0] =1$ if $|n+1~l\rangle$ is in $P$, and 0 otherwise. Similar expressions can be
derived to handle all of the operators $\bar{O}$ appearing in the contact-gradient expansion
(see the Appendix).

\subsection{Numerical Tests}
In this subsection channel-by-channel N$^3$LO results are presented for $H^{eff}$ 
based on Eqs. (\ref{bar}) and (\ref{final}),
which isolate a short-range operator that plausibly can be
accurately and systematically expanded via contact-gradient operators.  

For $\Lambda_P=8$, the fitting procedure determines all N$^3$LO
coefficients from nonedge matrix elements, 
leaving all edge matrix elements and a substantial set of nonedge matrix
elements unconstrained.  Thus one can use these matrix elements to test whether the
expansion systematically accounts for the ``data," the set of numerically generated $v_{18}$ matrix elements of $H^{eff}$.  One test is the running of the results
as a function of $\Lambda$: a systematic progression through LO, NLO, etc., operators should
be observed as $\Lambda$ is lowered to the SM scale.  A second test is the ``Lepage plot" \cite{lepage},
which displays residual errors in matrix elements: if the improvement is systematic, these
residual errors should reflect the nodal-quantum-number dependence of the operators that would
correct these results, in next order.

Eq. (\ref{final}) includes ``bare" terms -- the matrix elements $\langle \alpha | T | \widetilde{\beta} \rangle$
and $\langle \widetilde{\alpha} | V | \widetilde{\beta} \rangle$ -- and a term involving repeated
scattering by $H$ in $Q$, but sandwiched between the short-range operator $QV$.
To test the dependence on $\Lambda$, the rescattering term is decomposed in the manner of Eq. (\ref{delta}),
\[
\Delta_{QT}(\Lambda) = {E \over E-TQ} \left[V {1 \over E-QH} QV - V{1 \over E-Q_\Lambda H}
Q_\Lambda V \right] {E \over E-QT}, \]
to isolate the contribution of scattering above the
scale $\Lambda.$  $\Delta_{QT}(\Lambda)$ is evaluated numerically for $v_{18}$ at each required $\Lambda.$
The long-wavelength summation is always done to
all orders -- the running with $\Lambda$ thus reflects the behavior
of the short-range piece, $V (E-QH)^{-1} QV$.
The full $P$-space effective interaction is obtained as $\Lambda \rightarrow \Lambda_P$.  

\begin{figure}
\begin{minipage}{0.5\linewidth}
\begin{center}
\includegraphics[width=8cm]{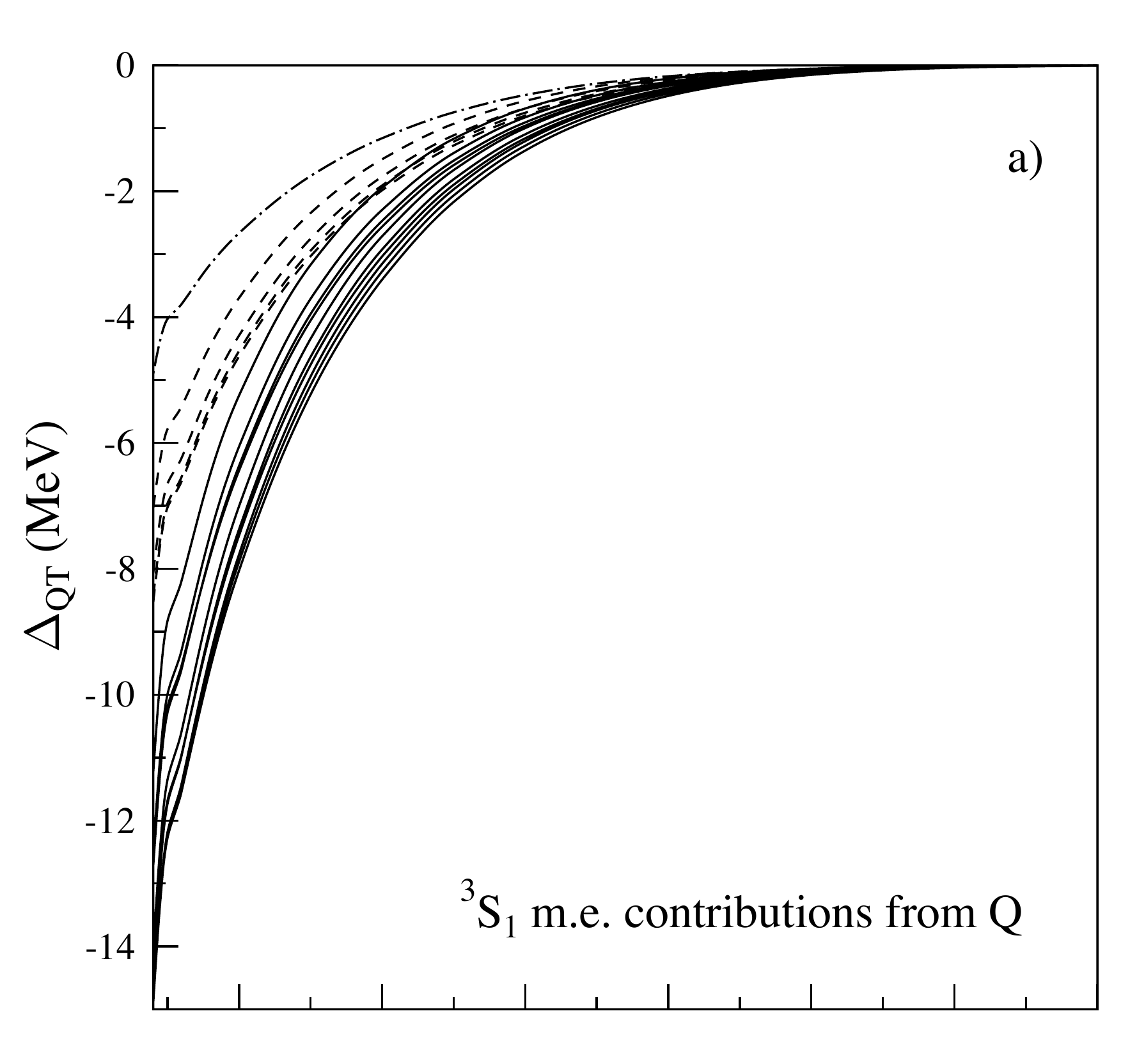}
\includegraphics[width=8cm]{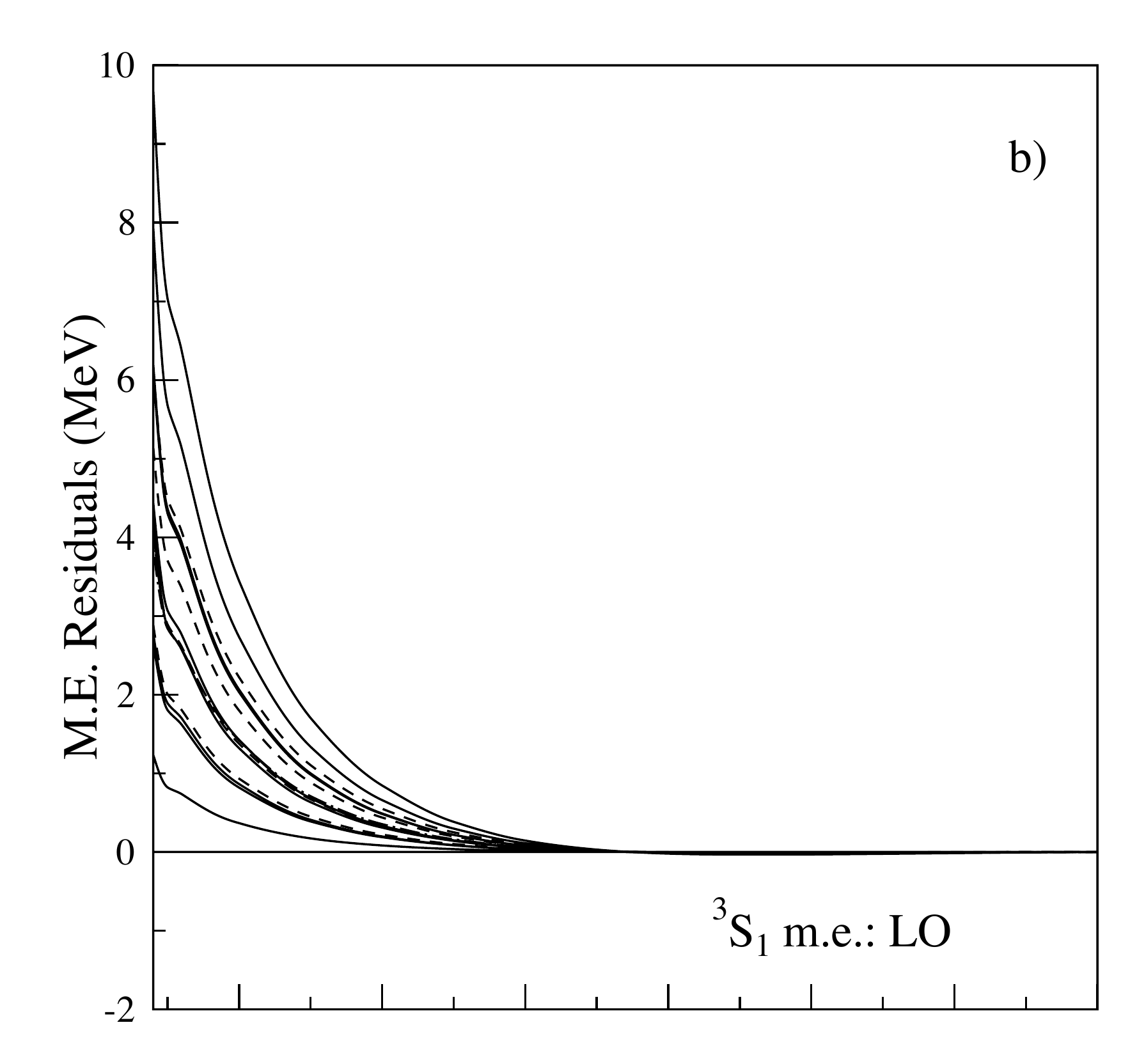}
\includegraphics[width=8cm]{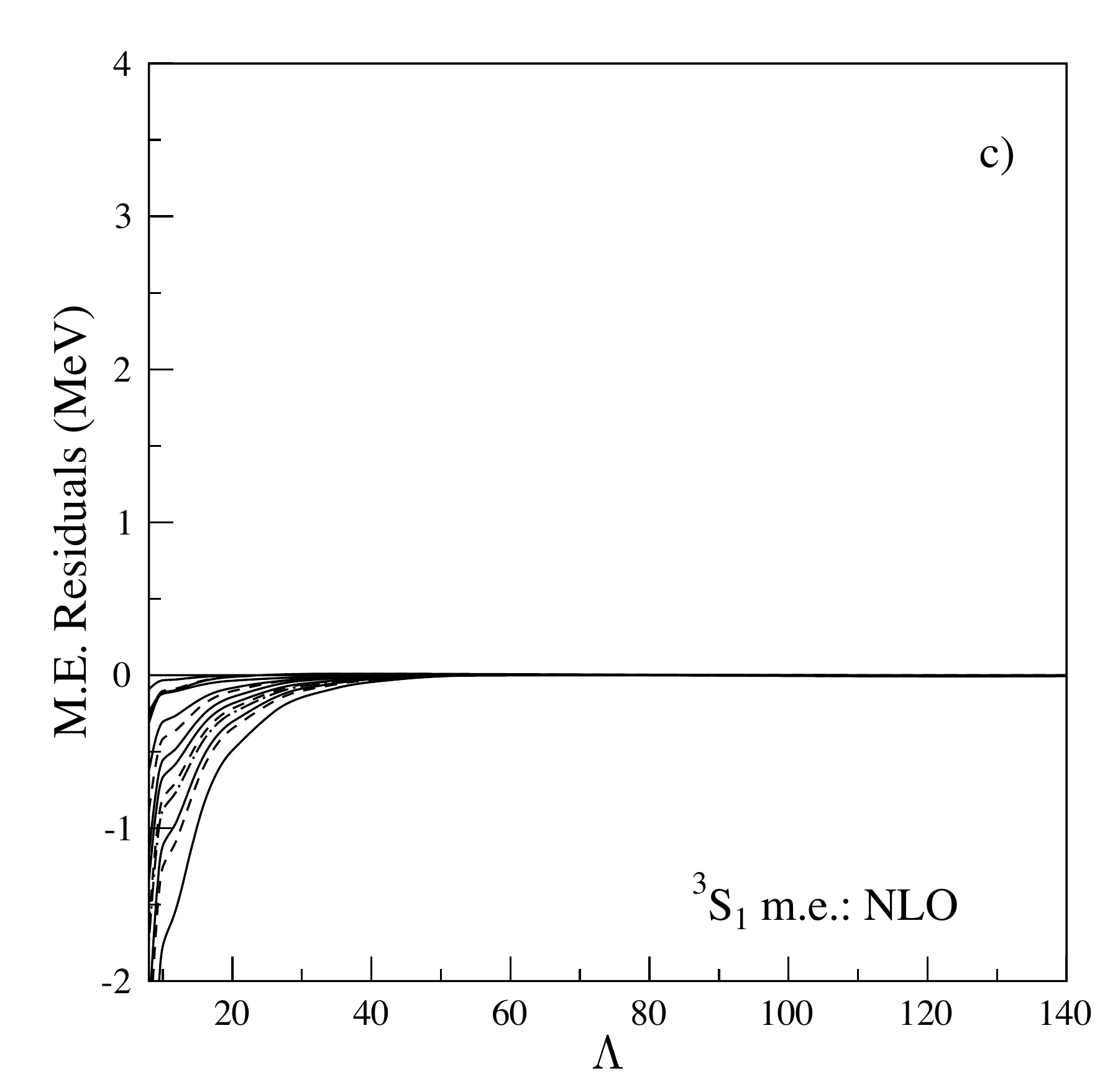}
\end{center}
\end{minipage}%
\begin{minipage}{0.5\linewidth}
\begin{center}
\includegraphics[width=8cm]{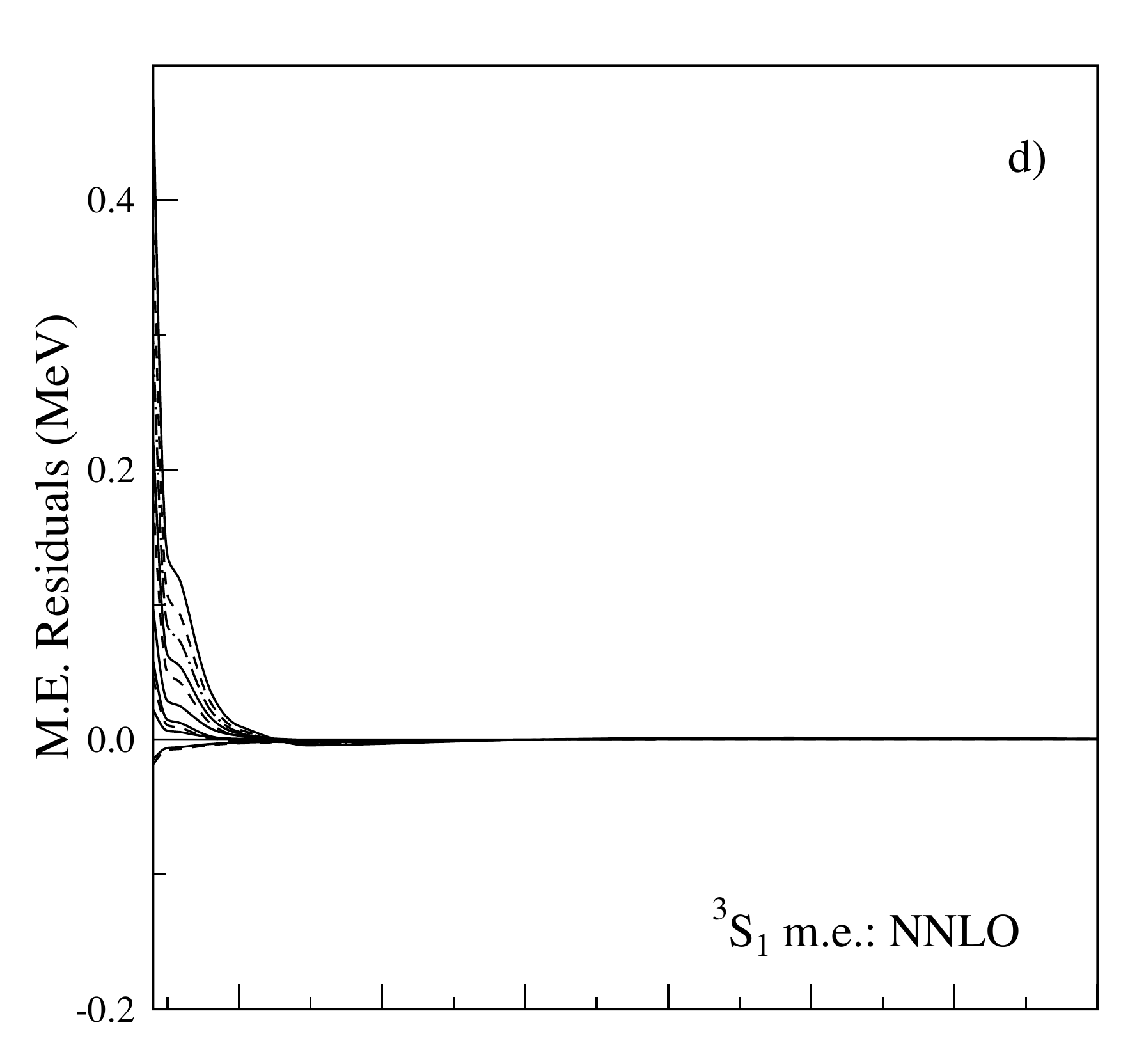}
\includegraphics[width=8cm]{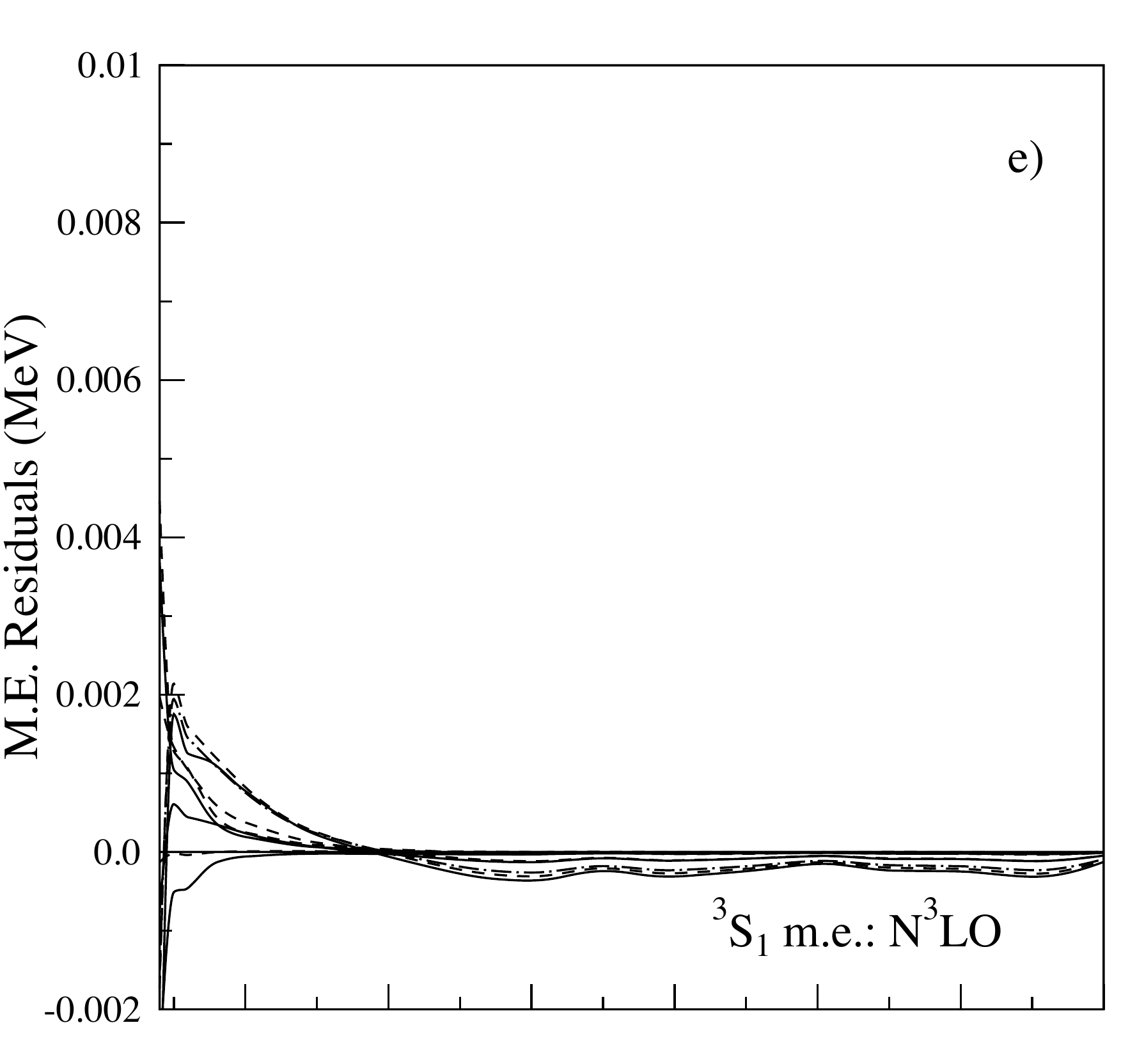}
\includegraphics[width=8cm]{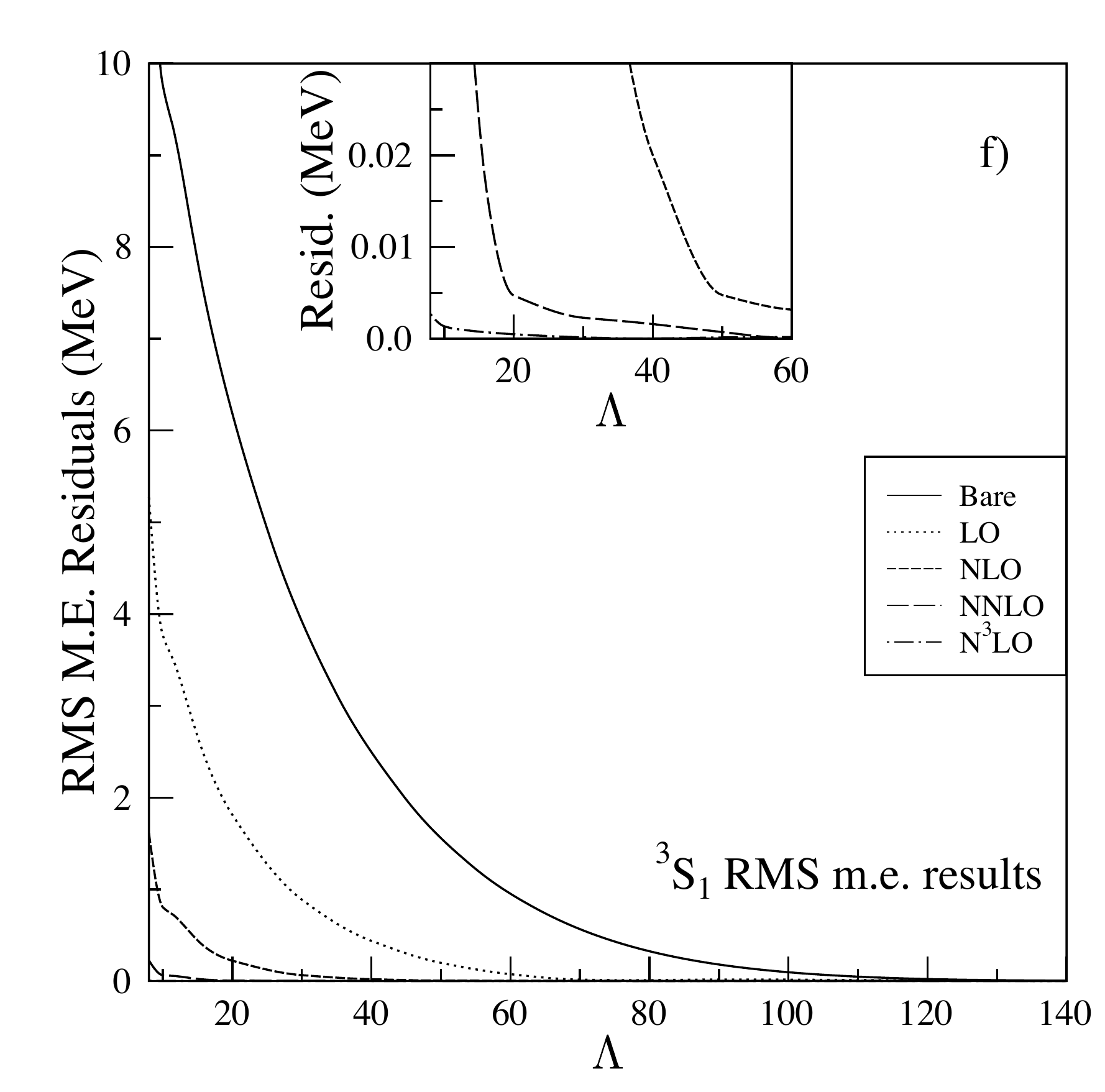}
\end{center}
\end{minipage}
\caption{As in Fig. \ref{fig_simple}, but for the $QT$-summed reordering of $H^{eff}$.  The contributions to the
effective interaction  from excitations in $Q$ above $\Lambda$, denoted $\Delta_{QT}(\Lambda)$ in
the text,  are plotted.  Each line gives the running of a $P$-space
matrix element.
b)-e) show the residuals for  LO, NLO, NNLO, and N$^3$LO fits (see text).  f) shows the RMS
deviation for the set of $P$-space matrix elements.  The improvement with increasing order is 
systematic and rapid: at N$^3$LO the RMS deviation for unconstrained matrix elements as
$\Lambda \rightarrow \Lambda_P$ is about 3 keV.  That is, the entire effective interaction is
reproduced to a few parts in 10$^4$. }
\label{fig_3s1}
\end{figure}

\begin{figure}
\begin{minipage}{0.5\linewidth}
\begin{center}
\includegraphics[width=8cm]{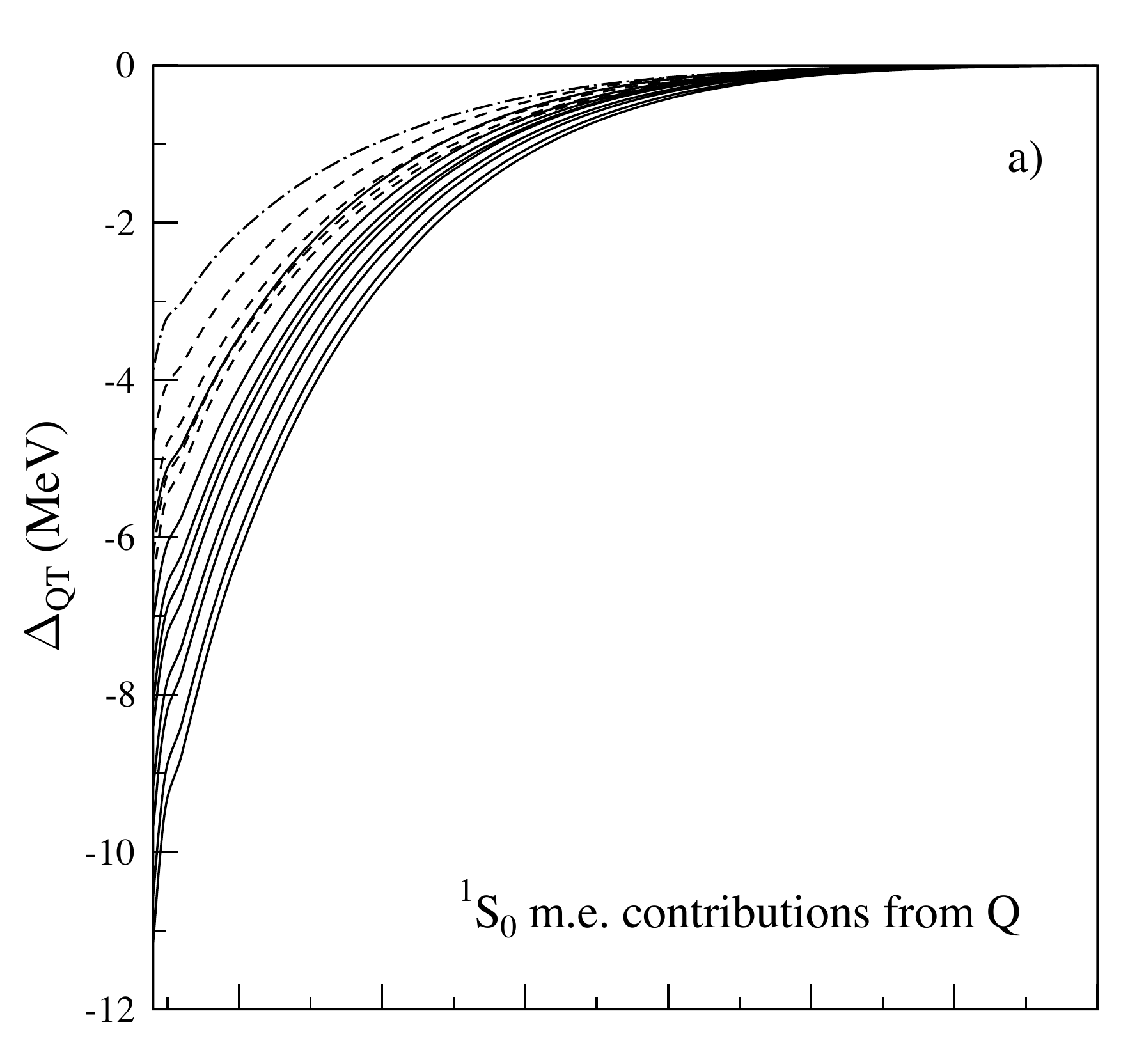}
\includegraphics[width=8cm]{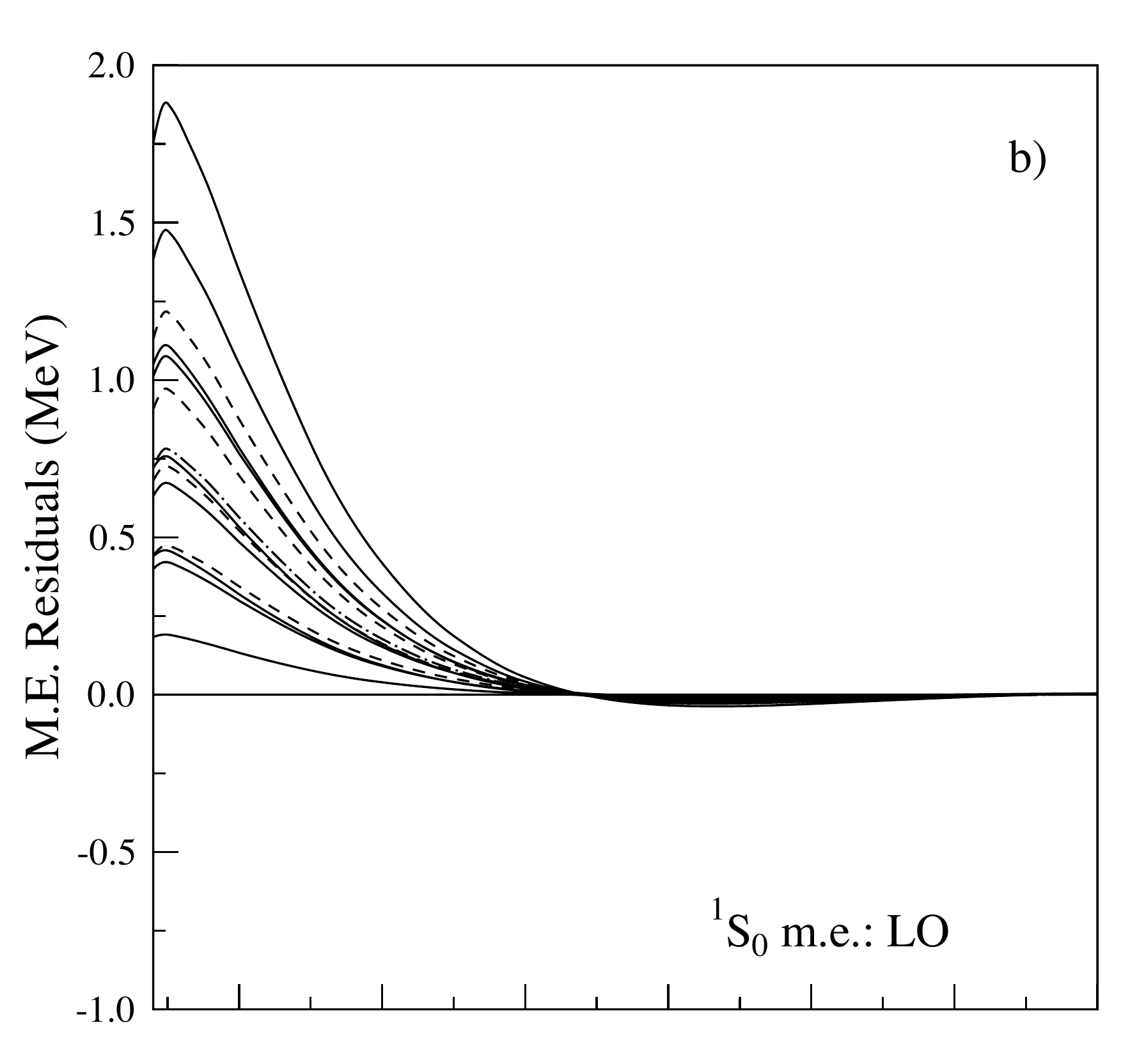}
\includegraphics[width=8cm]{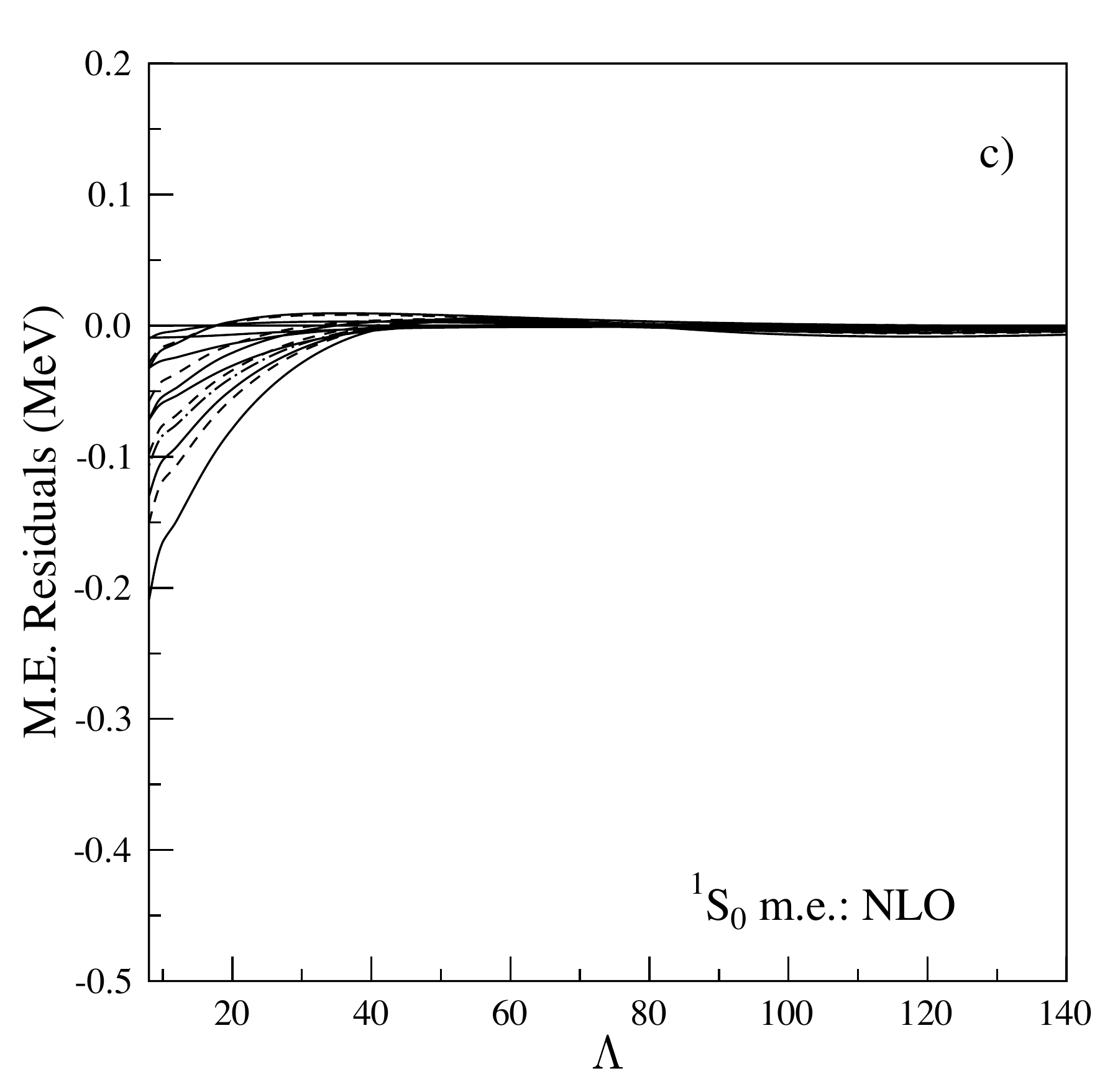}
\end{center}
\end{minipage}%
\begin{minipage}{0.5\linewidth}
\begin{center}
\includegraphics[width=8cm]{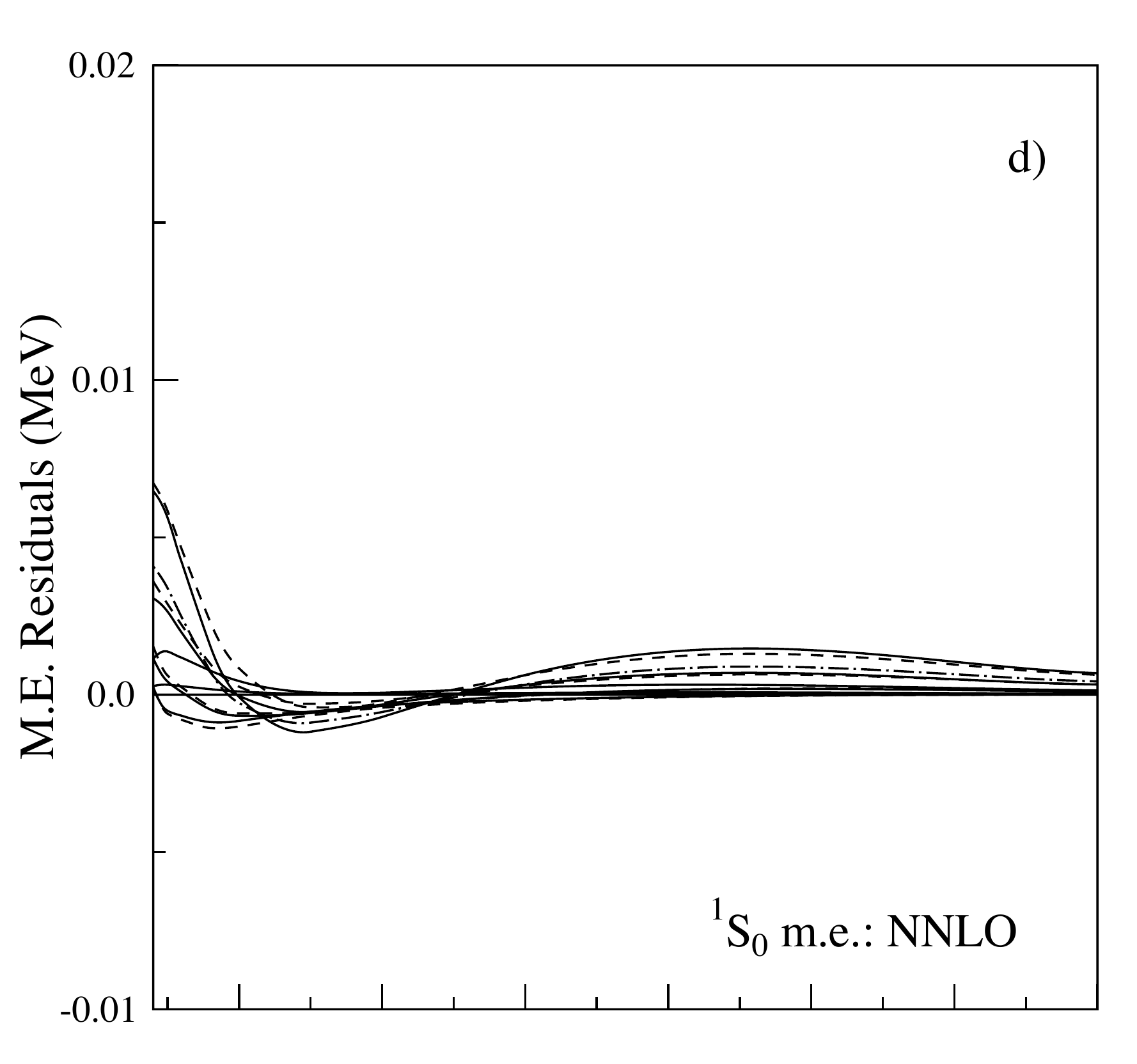}
\includegraphics[width=8cm]{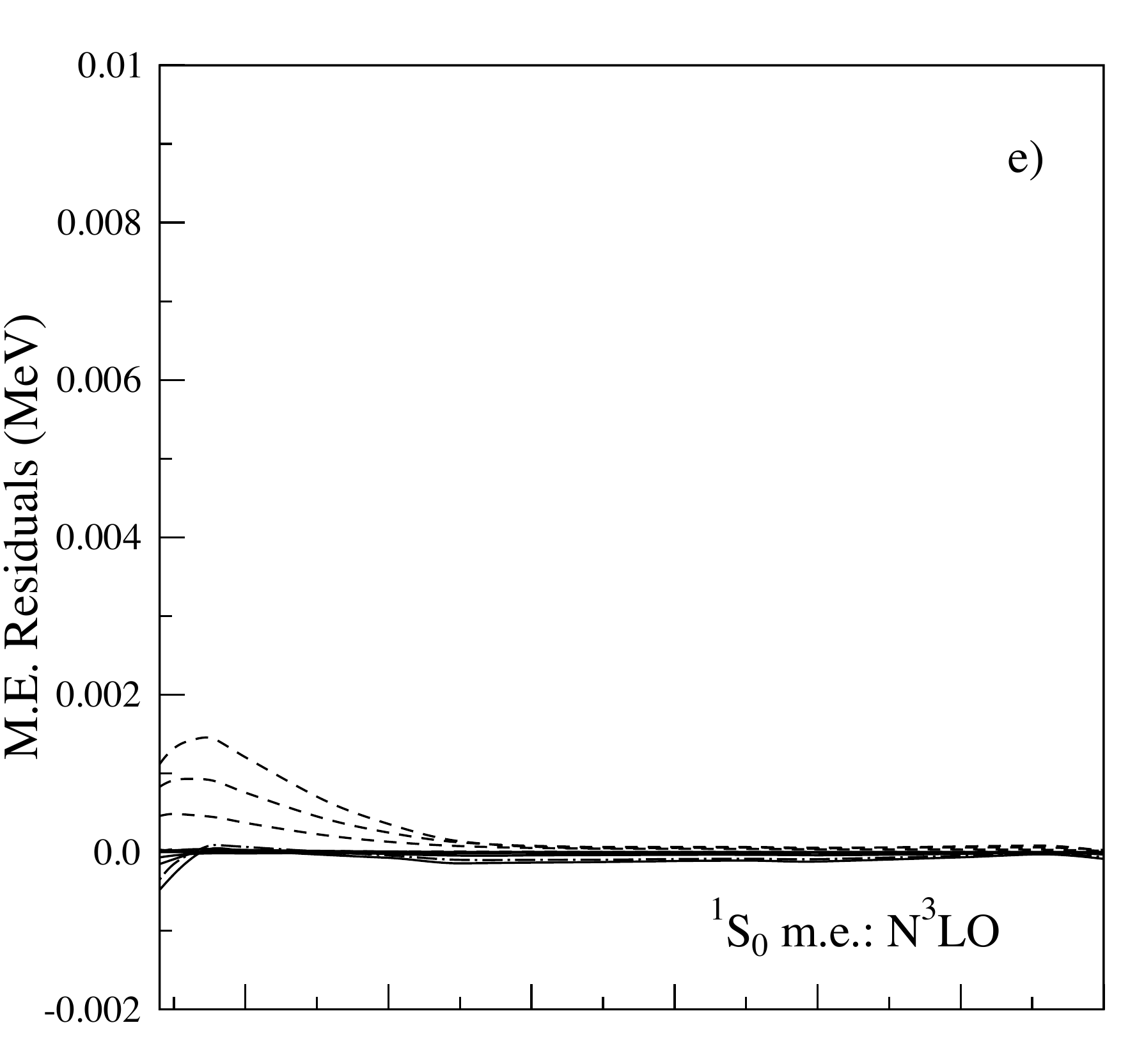}
\includegraphics[width=8cm]{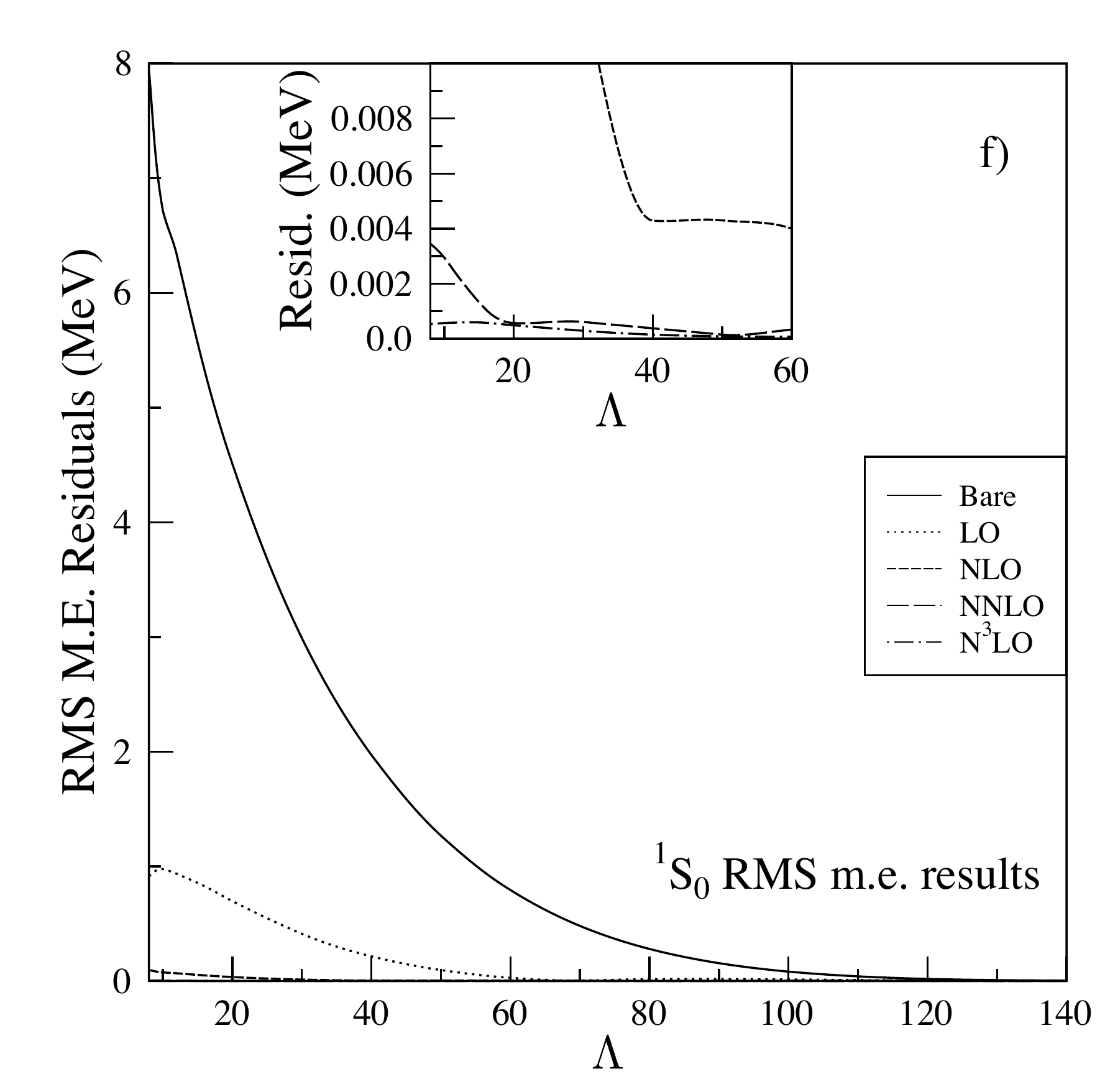}
\end{center}
\end{minipage}
\caption{As in Fig. \ref{fig_3s1}, but for the $^1S_0$ channel.  The N$^3$LO results are seen
to reproduce the entire effective interaction to the accuracy of about a keV, or one part in
$10^4$. }
\label{fig_1s0}
\end{figure}

As outlined before, coefficients are fitting to the longest wavelength information.
For example, in $S$ channels, $a_{LO}$
is fixed to the $(n^\prime,n)$ = (1,1)  matrix element; the absence of operator mixing then
guarantees this coefficient remains fixed, as higher order terms
are evaluated.  The single $a_{NLO}$ coefficient is fixed to (2,1) (or equivalently (1,2)) ; $a_{NNLO}^{22}$ and $a_{NNLO}^{40}$ are determined from (2,2) and (3,1); and finally
$a_{N^3LO}^{42}$ and $a_{N^3LO}^{60}$ are fixed to (3,2) and (4,1).    So at N$^3$LO there 
are a total of 6 parameters.  This procedure
is repeated for a series of $\Lambda$ ranging from 140 to $\Lambda_P$=8.  The results in each
order, and the improvement order by order, are thus obtained as a function of $\Lambda$.
$P$ contains 15 independent matrix elements in the ${}^3S_1-{}^3S_1$ channel,
nine of which play no role in the fitting: these test whether the improvement is
systematic.

Figures \ref{fig_3s1} and \ref{fig_1s0} show the results for ${}^3S_1-{}^3S_1$ and ${}^1S_0-{}^1S_0$.
Panel a) shows the evolution of the matrix elements $\langle \alpha | \Delta_{QT}(\Lambda)
| \beta \rangle$ for each of the 15 independent matrix elements.   Matrix
elements involving only nonedge states,  a single edge state, or two edge states
are denoted by solid, dashed, and dash-dotted lines, respectively.   Progressively more
binding is recovered as $\Lambda \rightarrow \Lambda_P$.   In
the $^3S_1-{}^3S_1$ case, the contribution at $\Lambda_P$ is
$\sim$ 12-14 MeV for nonedge matrix elements, $\sim$ 7-8 MeV for 
matrix elements with one edge state, and $\sim$ 5 MeV for the $\langle n=5 l=0 | \Delta_{QT}(\Lambda_P)
| n=5 l=0 \rangle$ double-edge matrix element.  

Panels b)-e) show the residuals -- the difference
between the matrix elements of $\Delta_{QT}(\Lambda_P)$ and those of the contact-gradient
potential of Eq. (\ref{final}) -- from LO through N$^3$LO.   The trajectories correspond to
the unconstrained matrix elements (14 in LO, 9 in N$^3$LO): the fitted matrix elements 
produce the horizontal line at 0.  Unlike the naive approach in Fig. \ref{fig_simple}, the improvement
is now systematic in all matrix elements.   In the $^3S_1-{}^3S_1$ channel, a LO treatment 
effectively removes all contributions in $Q$ above $\Lambda \sim$ 60; NLO lowers this
scale to $\sim$ 40, and NNLO is $\sim$ 20.  The magnitude of N$^3$LO
residuals at $\Lambda_P$  is typically
$\lsim$ 2 keV -- the entire effective interaction can be  represented by Eq. (\ref{final}) to
an accuracy of about 0.01\%.  Panel f) shows the root mean square (RMS) deviation among the
unconstrained matrix elements, and the rapid order-by-order improvement.

The pattern repeats in the $^1S_0-{}^1S_0$ channel, where the convergence (in terms
of the size of the residuals) is somewhat faster.  The
N$^3$LO RMS deviation among the unconstrained matrix elements at $\Lambda_P$ is
$\sim$ 0.5 keV.

\begin{figure}
\begin{minipage}{0.5\linewidth}
\begin{center}
\includegraphics[width=8cm]{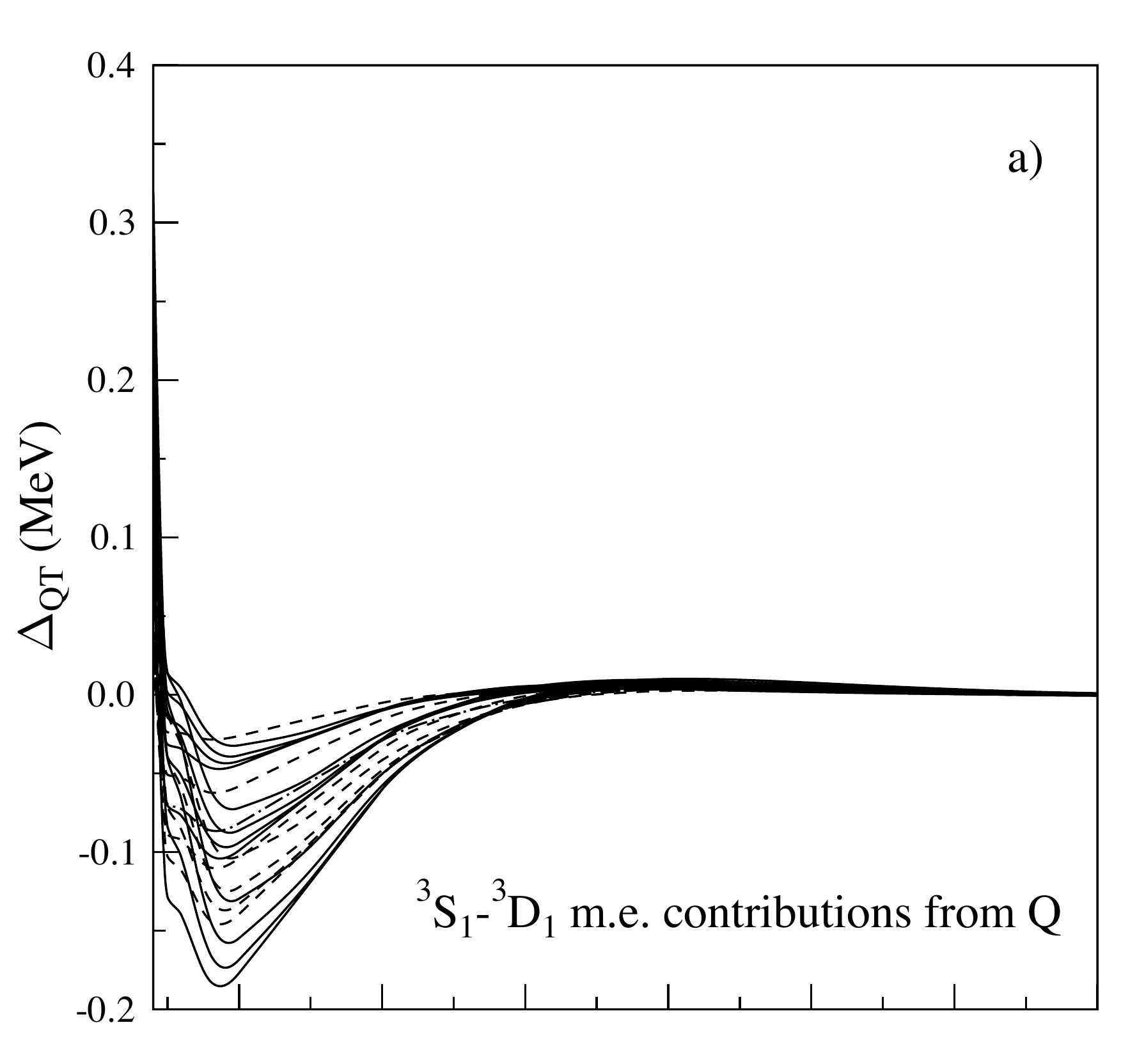}
\includegraphics[width=8cm]{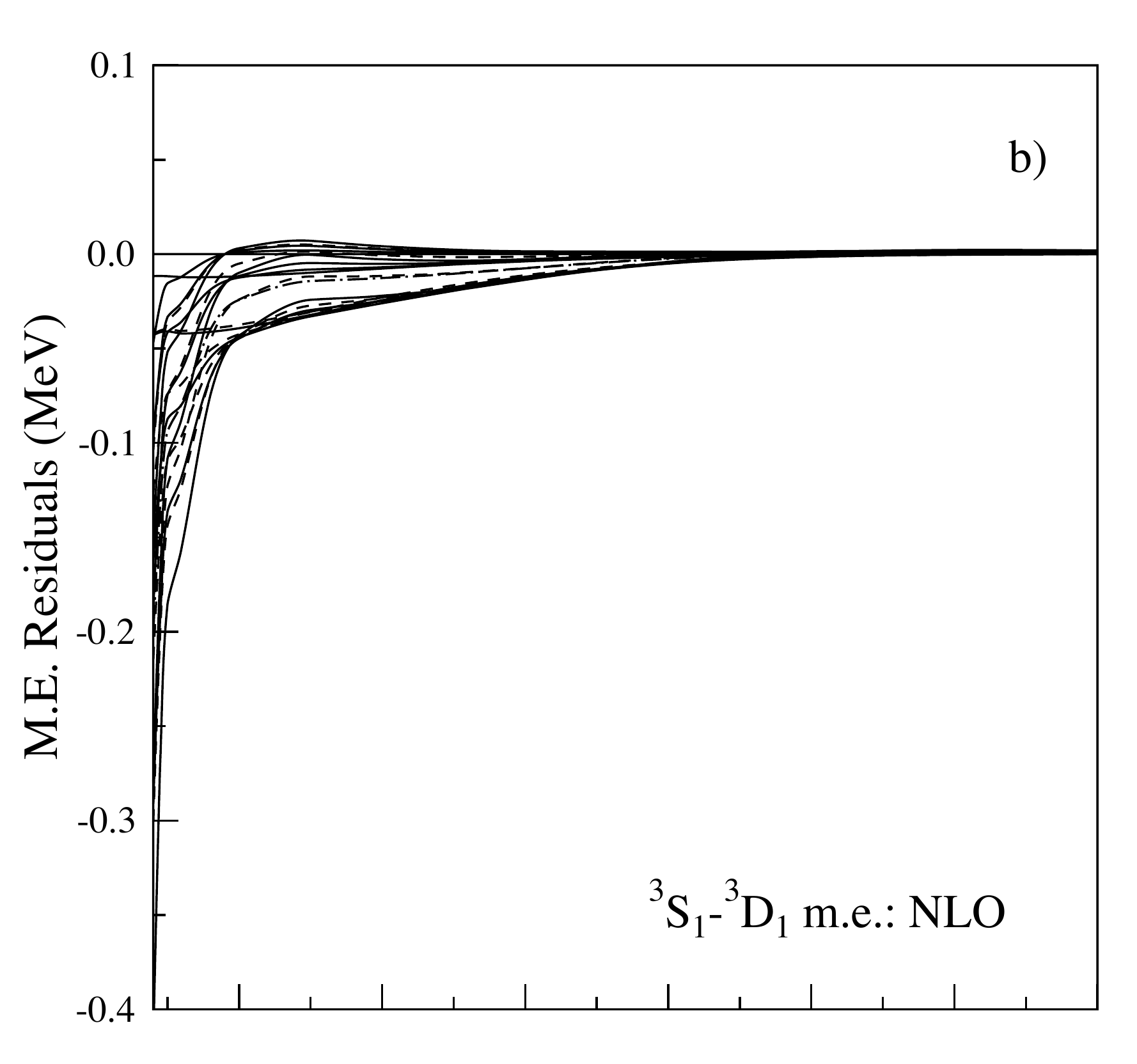}
\end{center}
\caption{As in Figs. \ref{fig_3s1} and \ref{fig_1s0}, but for the $^3S_1-{}^3D_1$ channel.  As with the cases
described before, the N$^3$LO results remain accurate at the few keV level, as the
integration is brought down to the shell-model scale, $\Lambda \rightarrow \Lambda_P$. }
\label{fig_sd}
\end{minipage}%
\begin{minipage}{0.5\linewidth}
\begin{center}
\includegraphics[width=8cm]{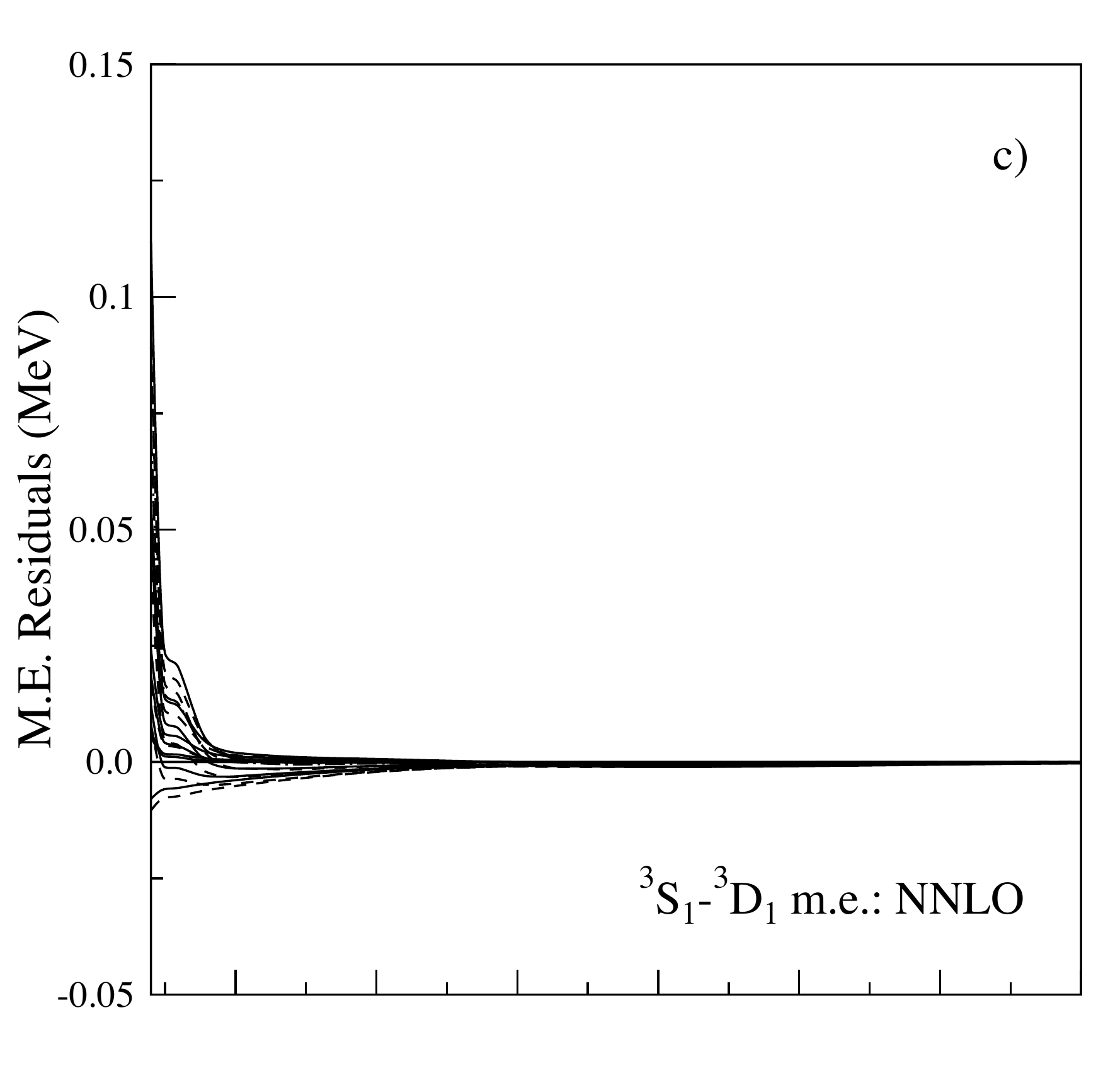}
\includegraphics[width=8cm]{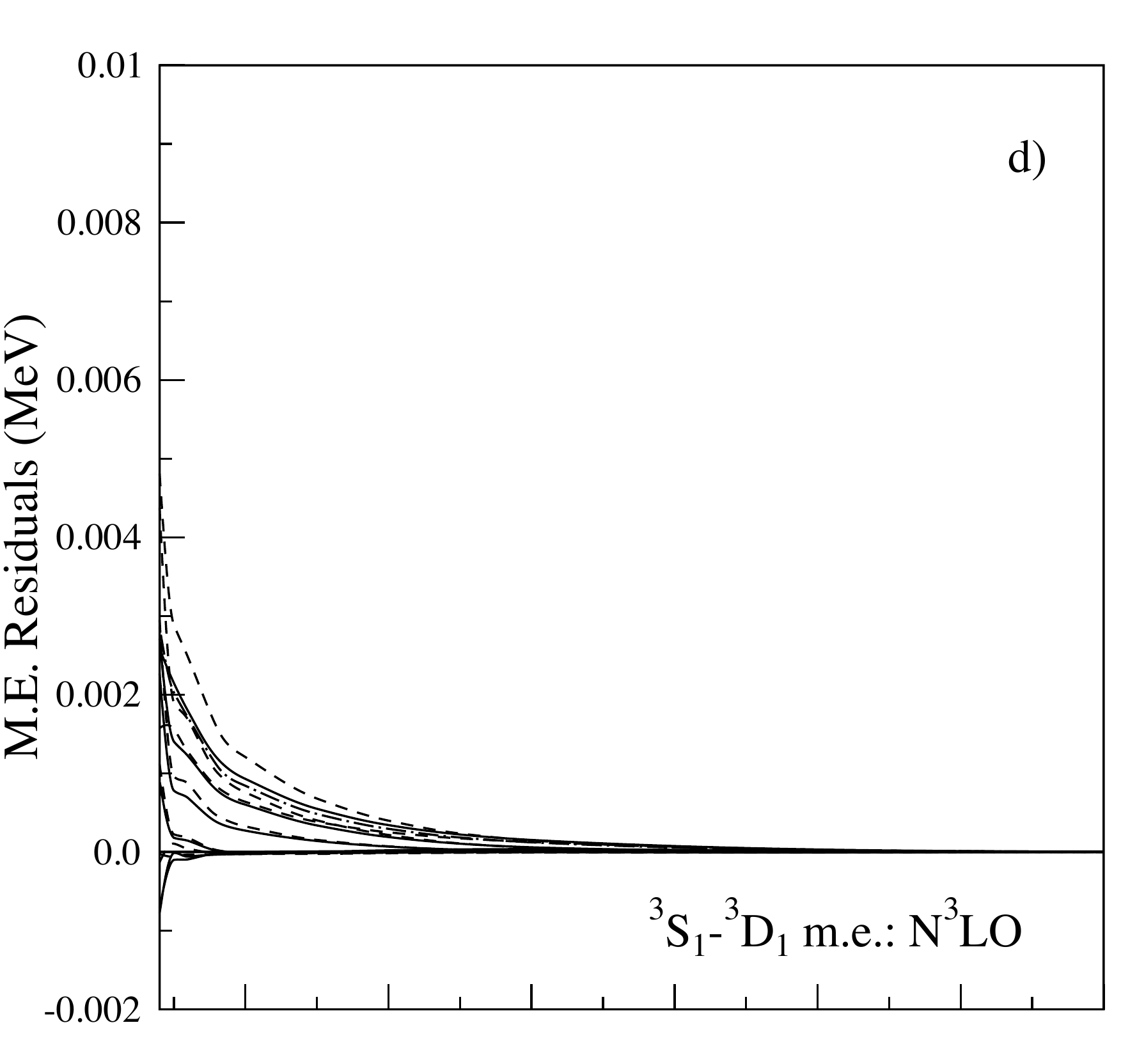}
\includegraphics[width=8cm]{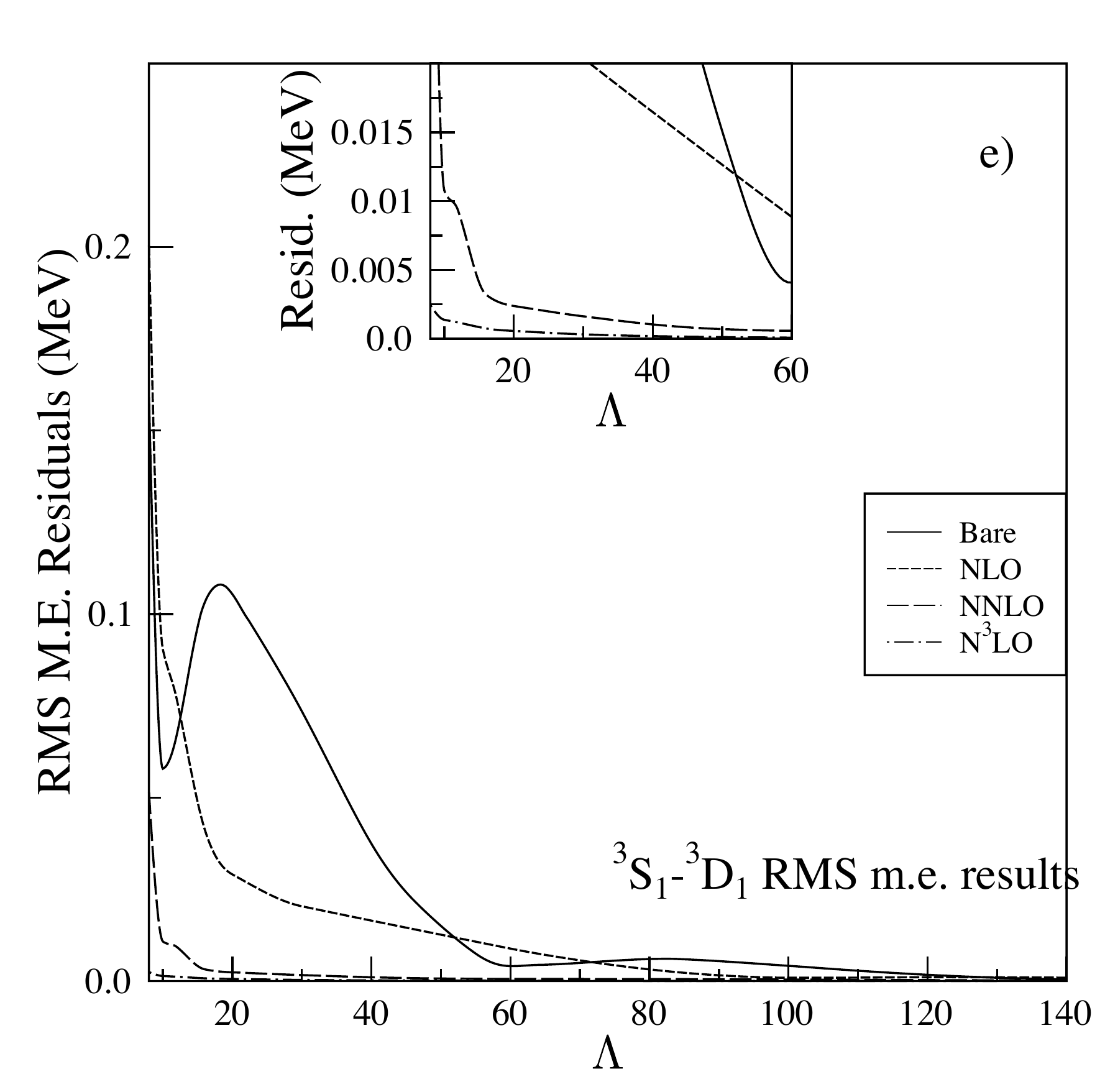}
\end{center}
\end{minipage}
\end{figure}

\begin{figure}
\begin{minipage}{0.5\linewidth}
\begin{center}
\includegraphics[width=8cm]{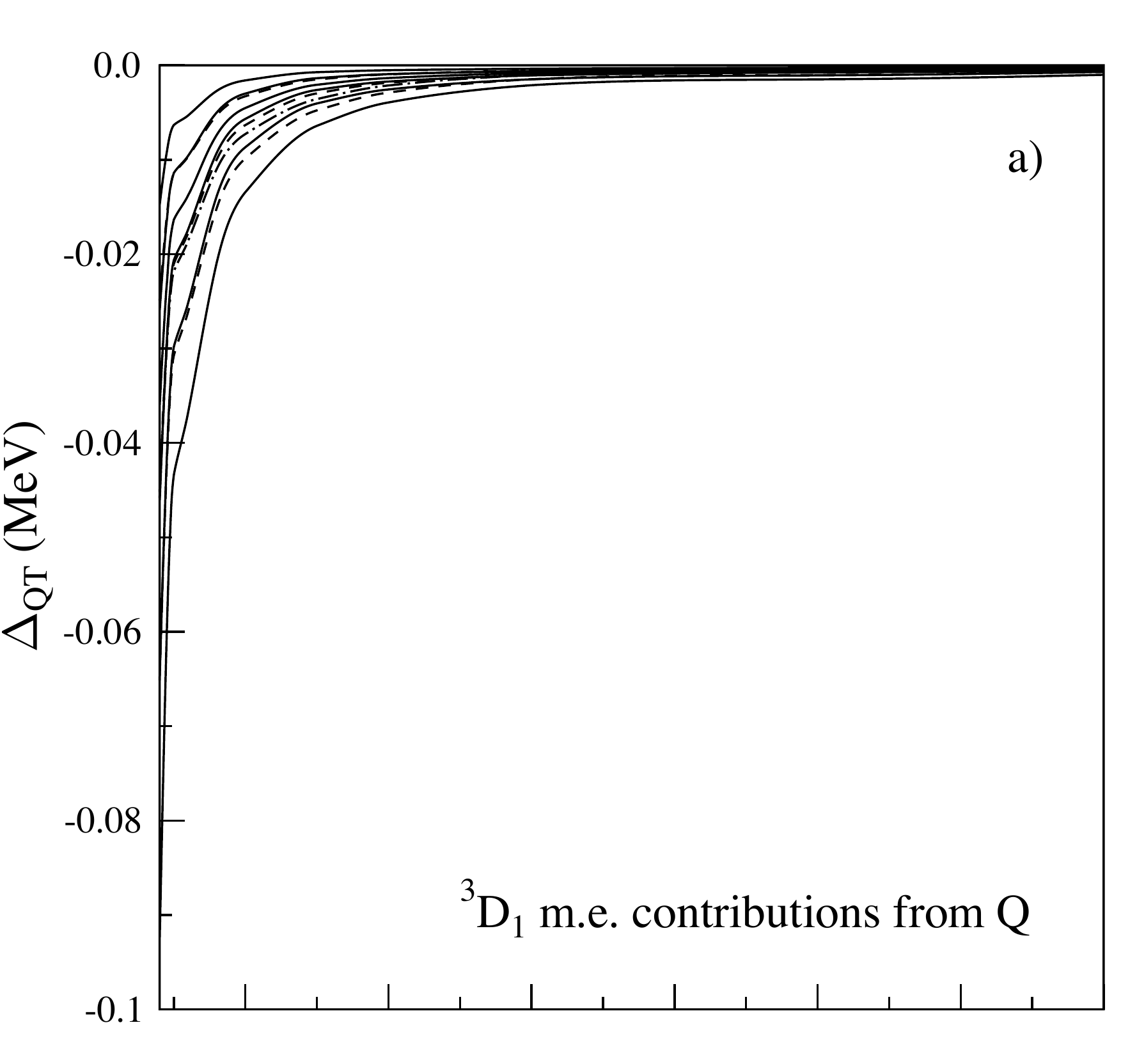}
\includegraphics[width=8cm]{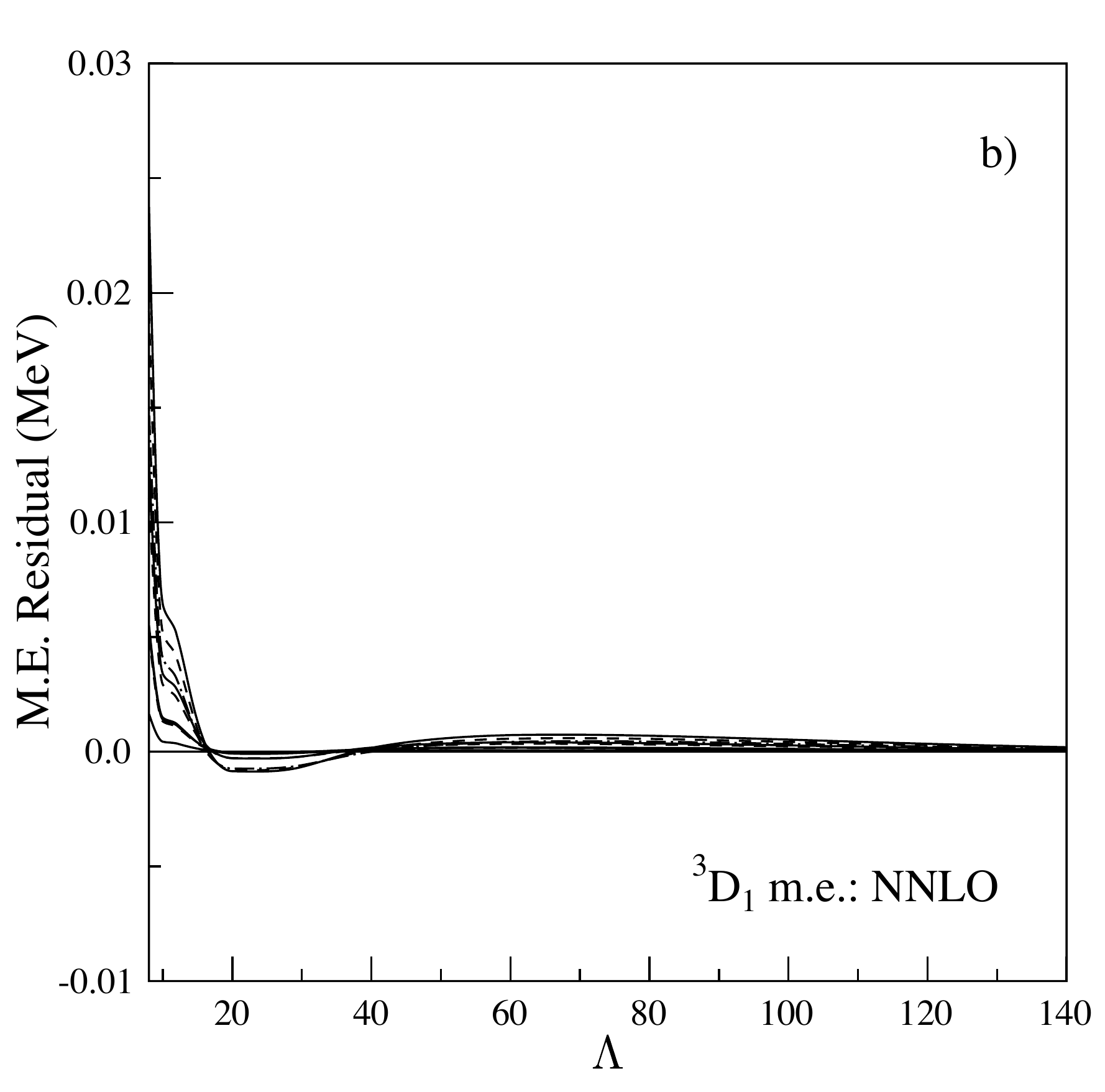}
\end{center}
\end{minipage}%
\begin{minipage}{0.5\linewidth}
\begin{center}
\includegraphics[width=8cm]{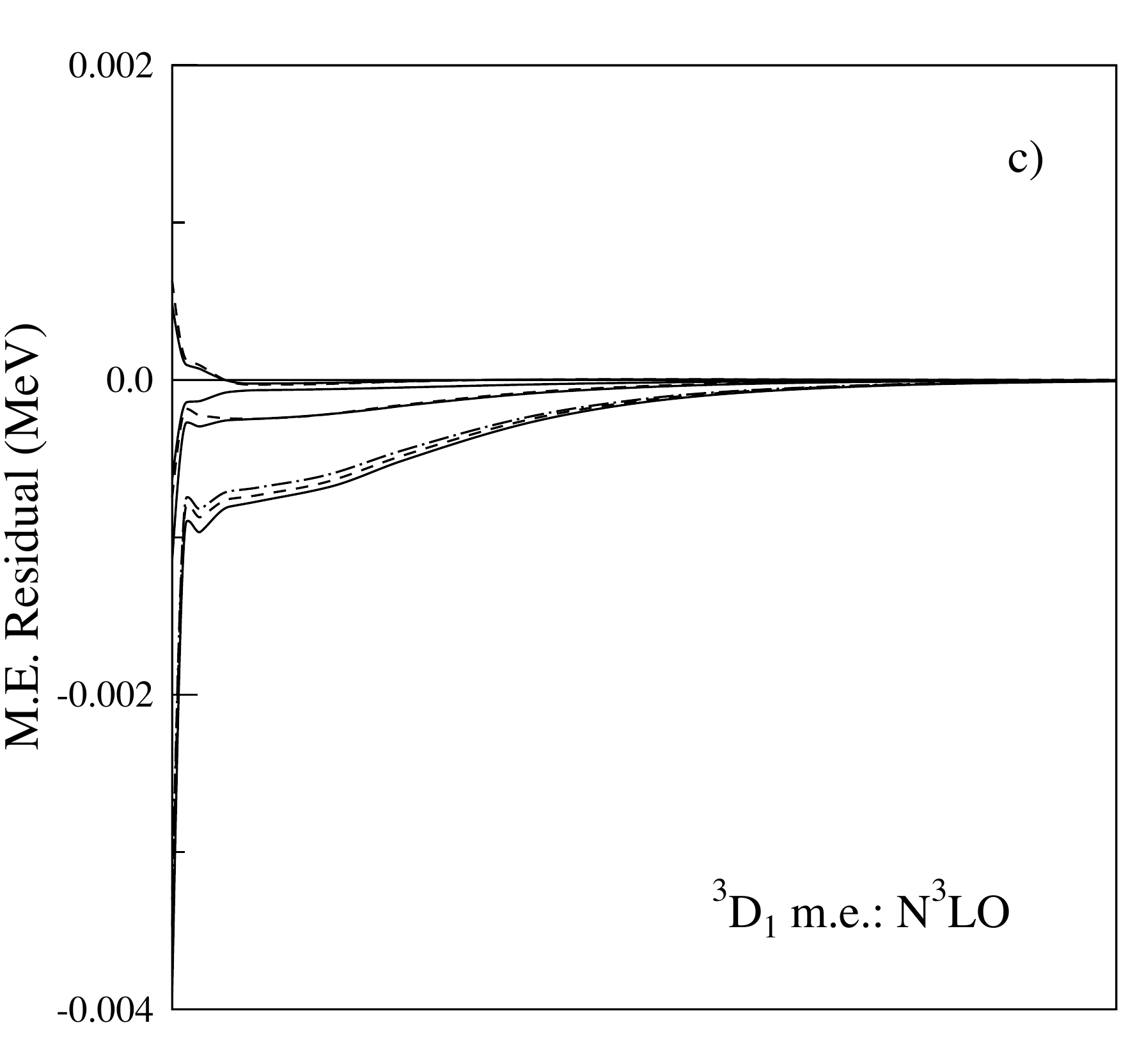}
\includegraphics[width=8cm]{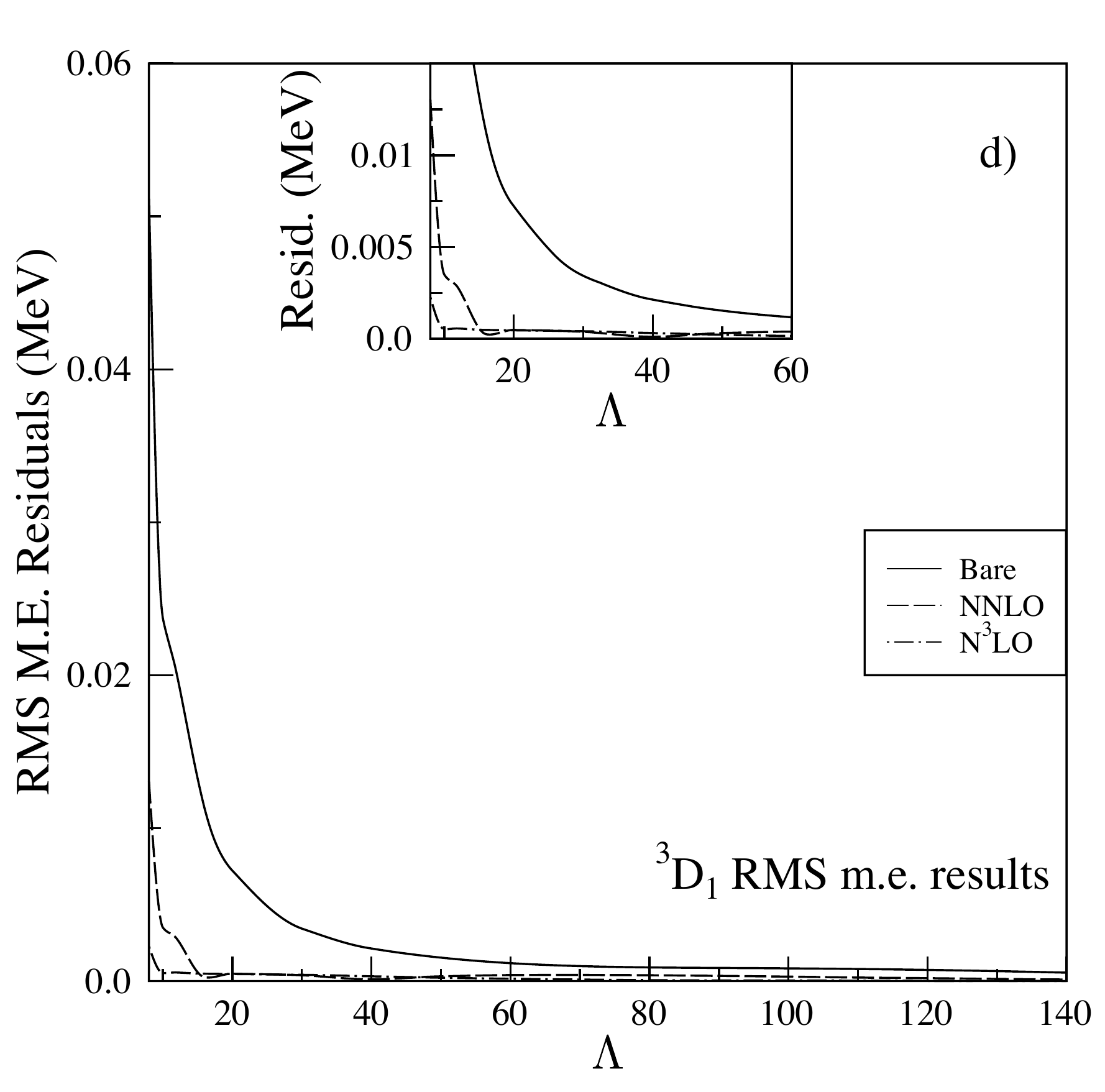}
\end{center}
\end{minipage}
\caption{As in Fig. \ref{fig_3s1}, but for the $^3D_1$ channel.   }
\label{fig_3d1}
\end{figure}

\begin{figure}
\begin{minipage}{0.5\linewidth}
\begin{center}
\includegraphics[width=8cm]{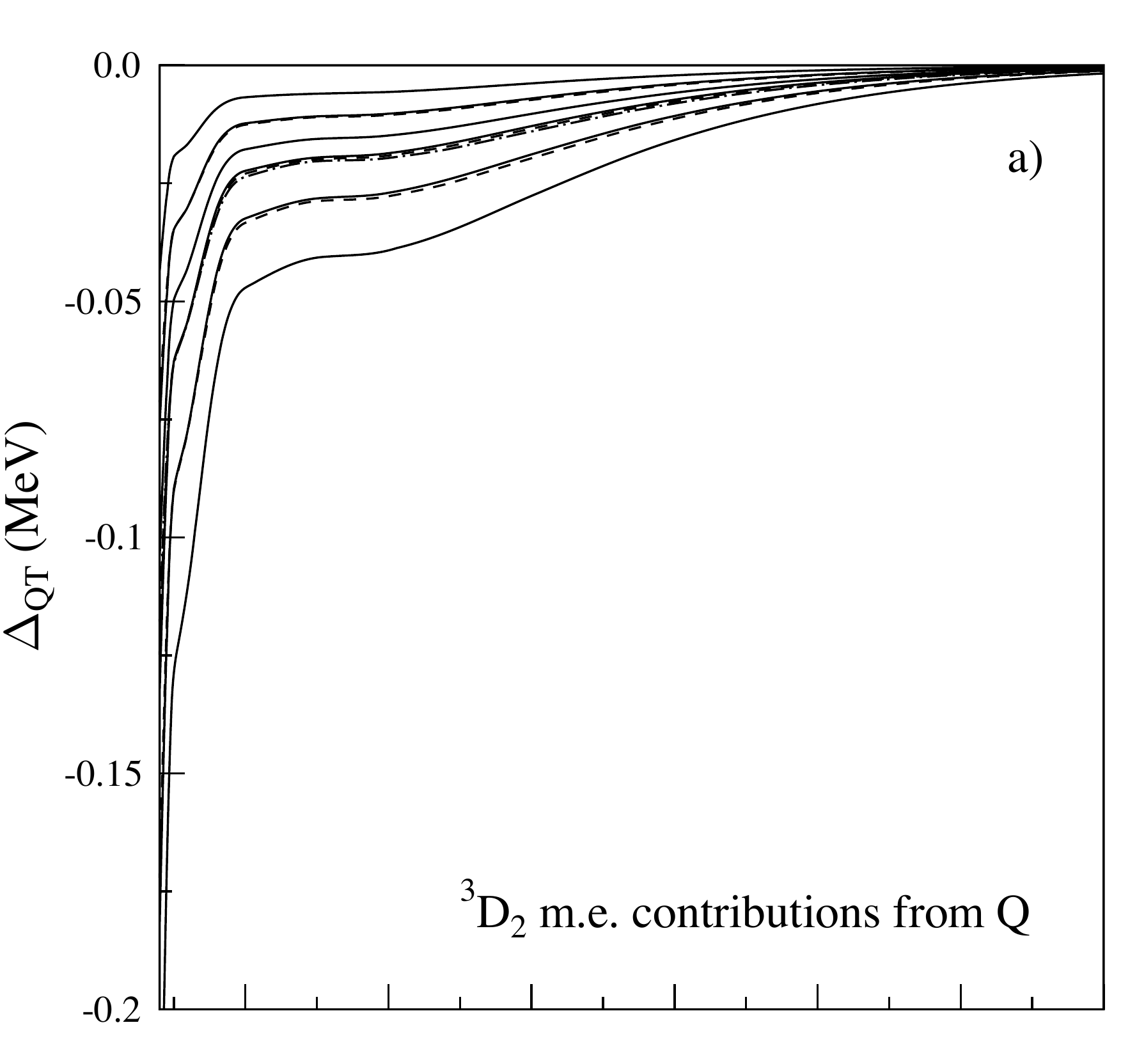}
\includegraphics[width=8cm]{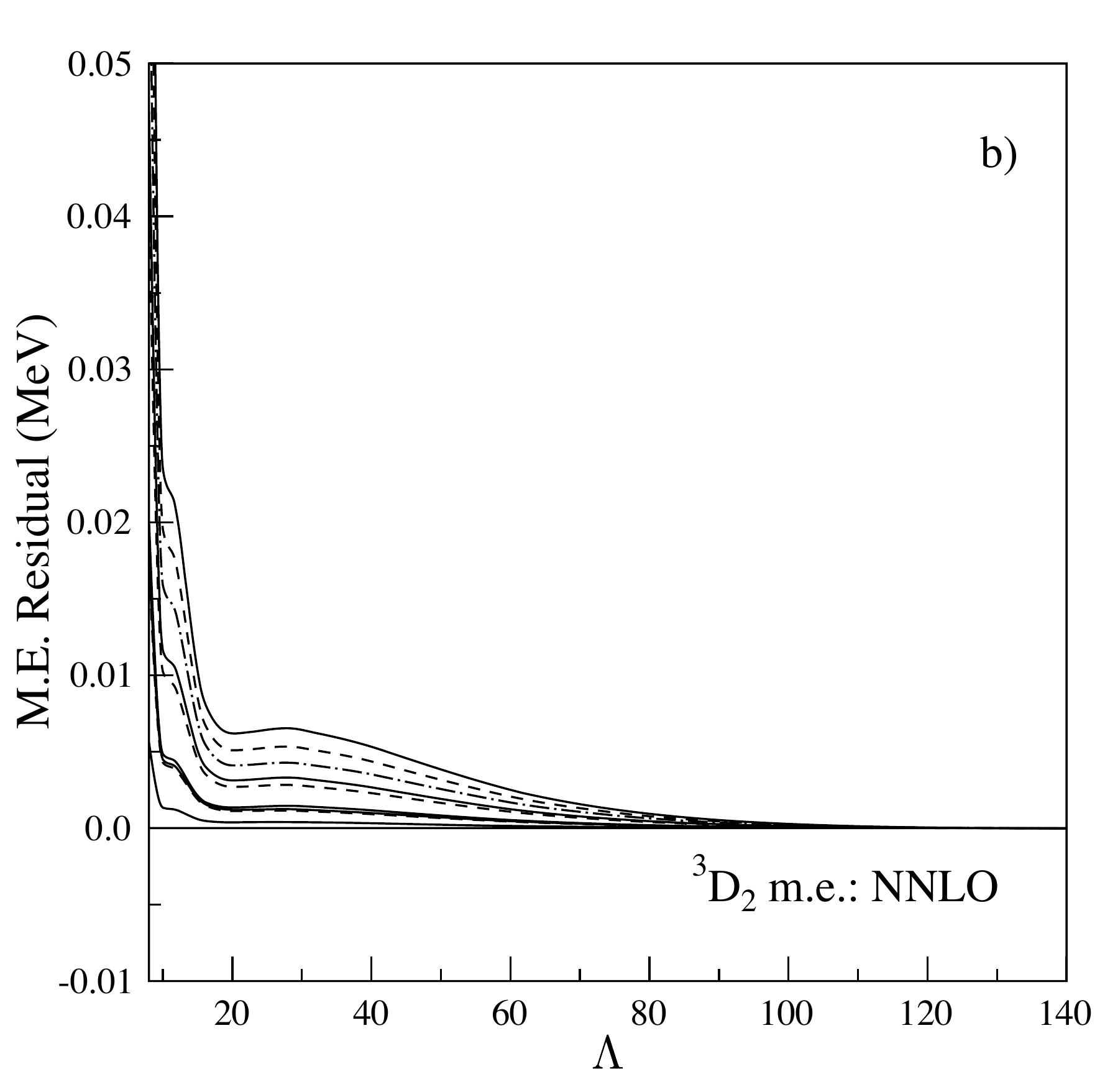}
\end{center}
\end{minipage}%
\begin{minipage}{0.5\linewidth}
\begin{center}
\includegraphics[width=8cm]{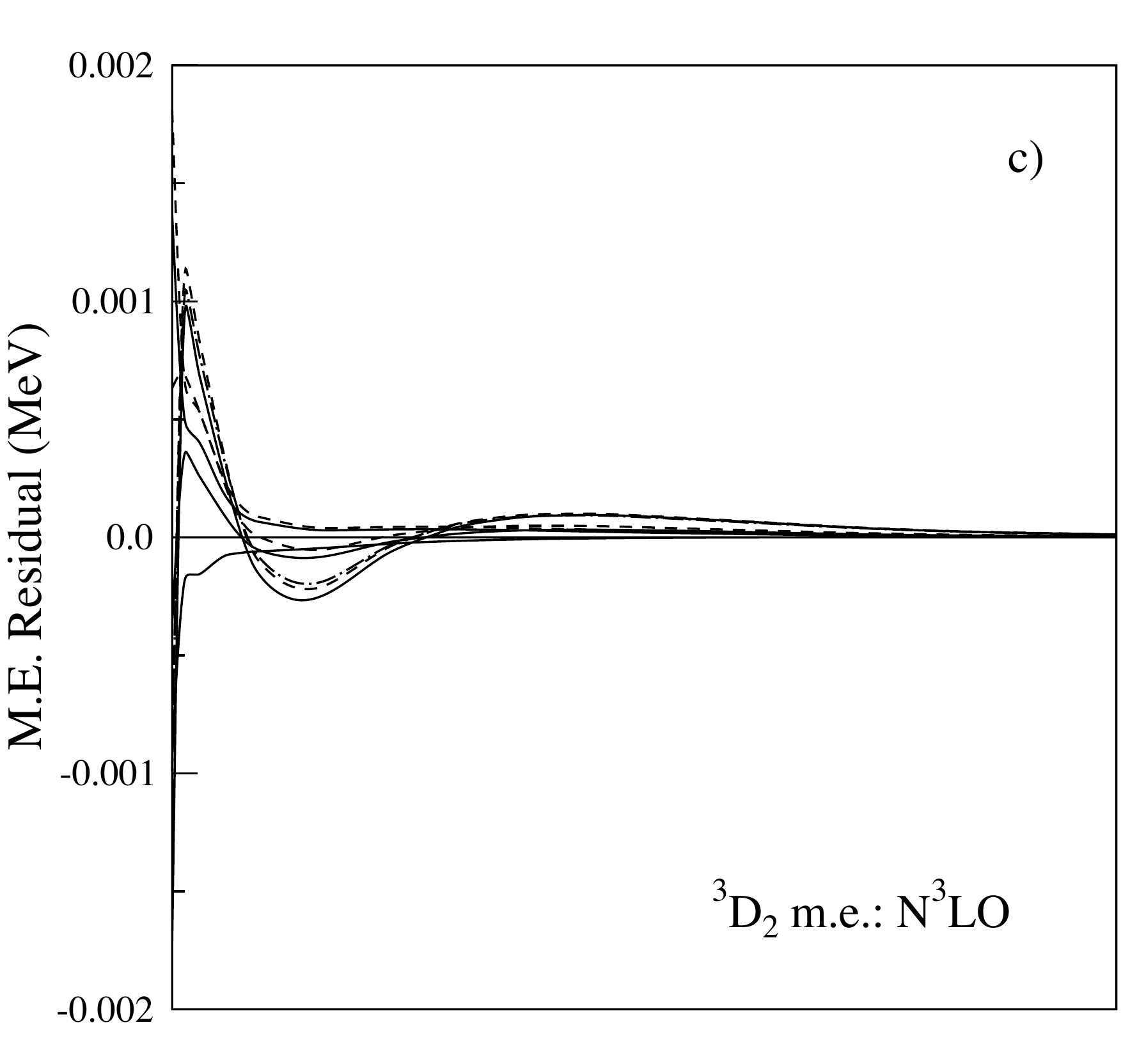}
\includegraphics[width=8cm]{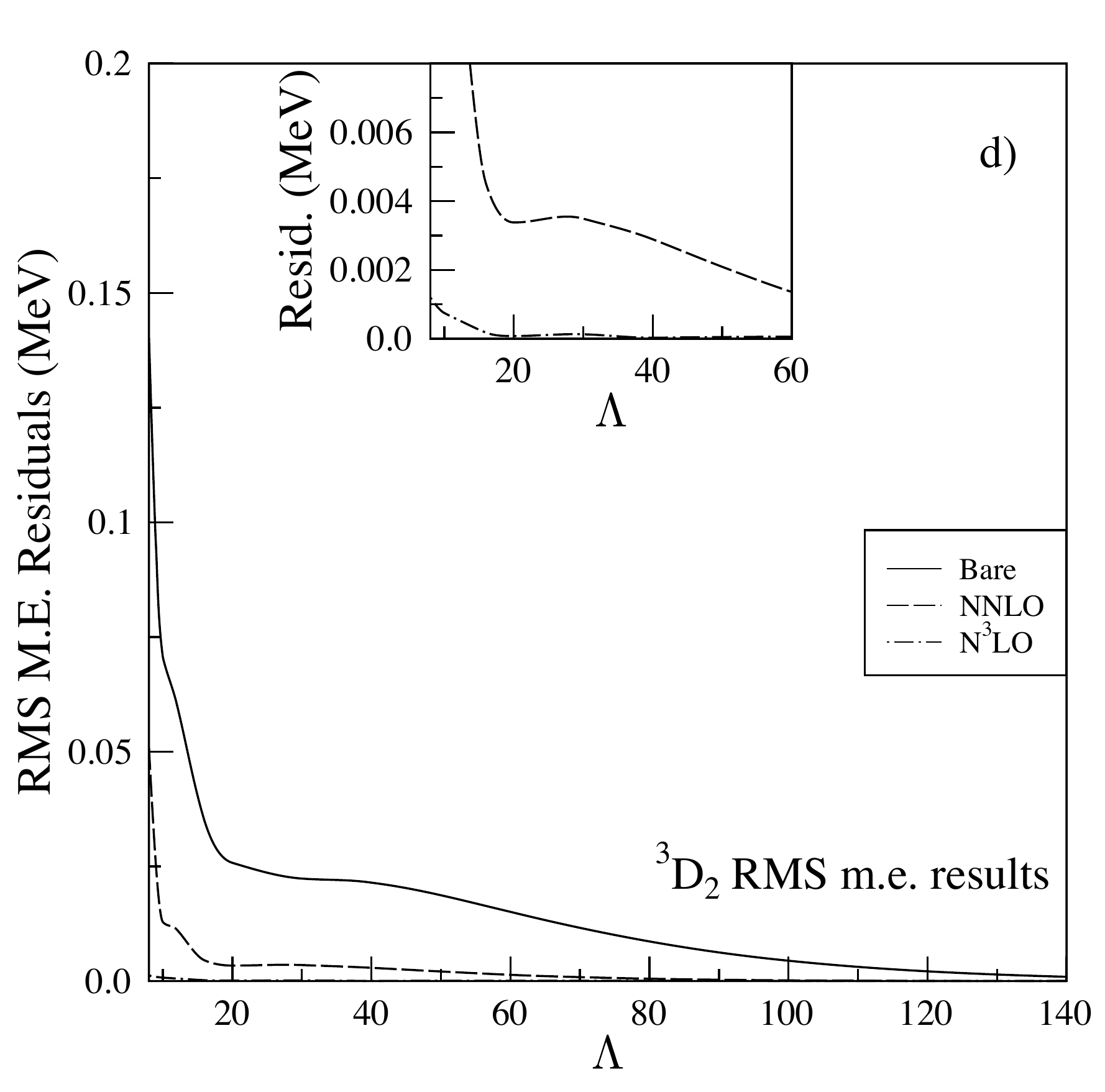}
\end{center}
\end{minipage}
\caption{As in Fig. \ref{fig_3s1}, but for the $^3D_2$ channel.   }
\label{fig_3d2}
\end{figure}

\begin{figure}
\begin{minipage}{0.5\linewidth}
\begin{center}
\includegraphics[width=8cm]{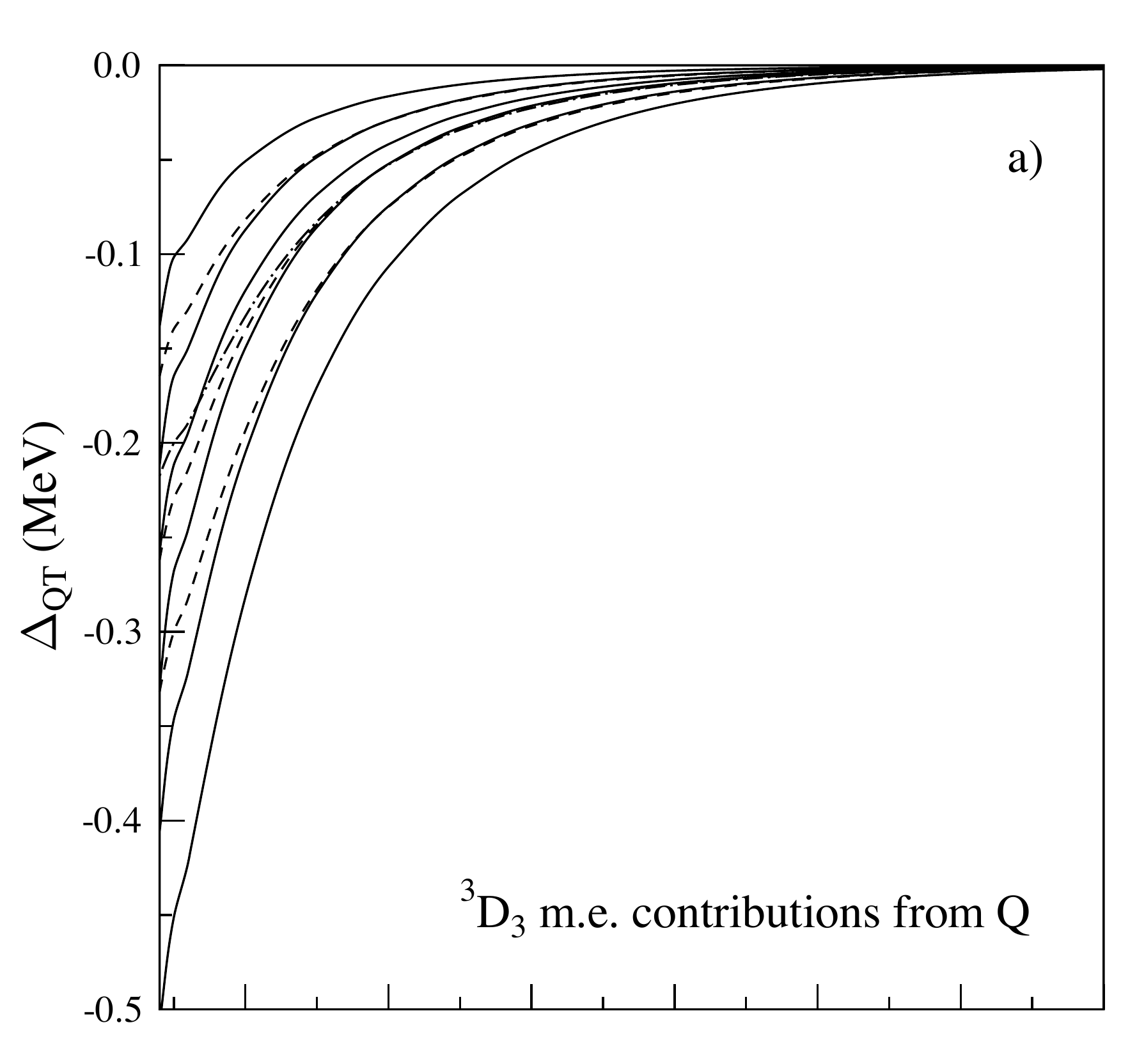}
\includegraphics[width=8cm]{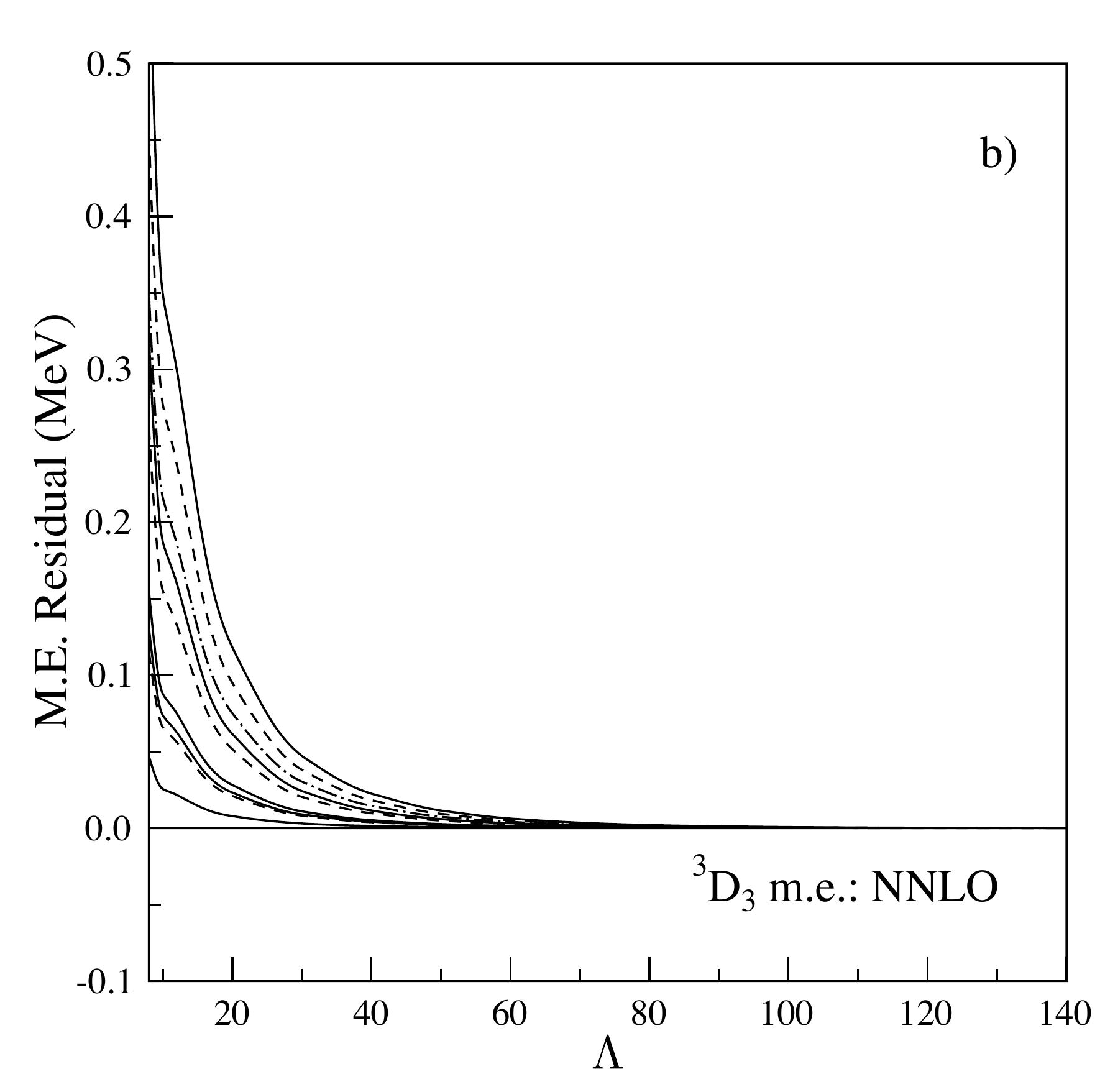}
\end{center}
\caption{As in Fig. \ref{fig_3d1} and \ref{fig_3d2},  but for the $^3D_3$ channel.   This ``stretched"
configuration generates much larger residuals than the other $l=2$ channels.  
Consequently a calculation to N$^4$LO would be needed to reduce typical matrix element
errors to $\sim$ 10 keV, in the limit $\Lambda \rightarrow \Lambda_P$.}
\label{fig_3d3}
\end{minipage}%
\begin{minipage}{0.5\linewidth}
\begin{center}
\includegraphics[width=8cm]{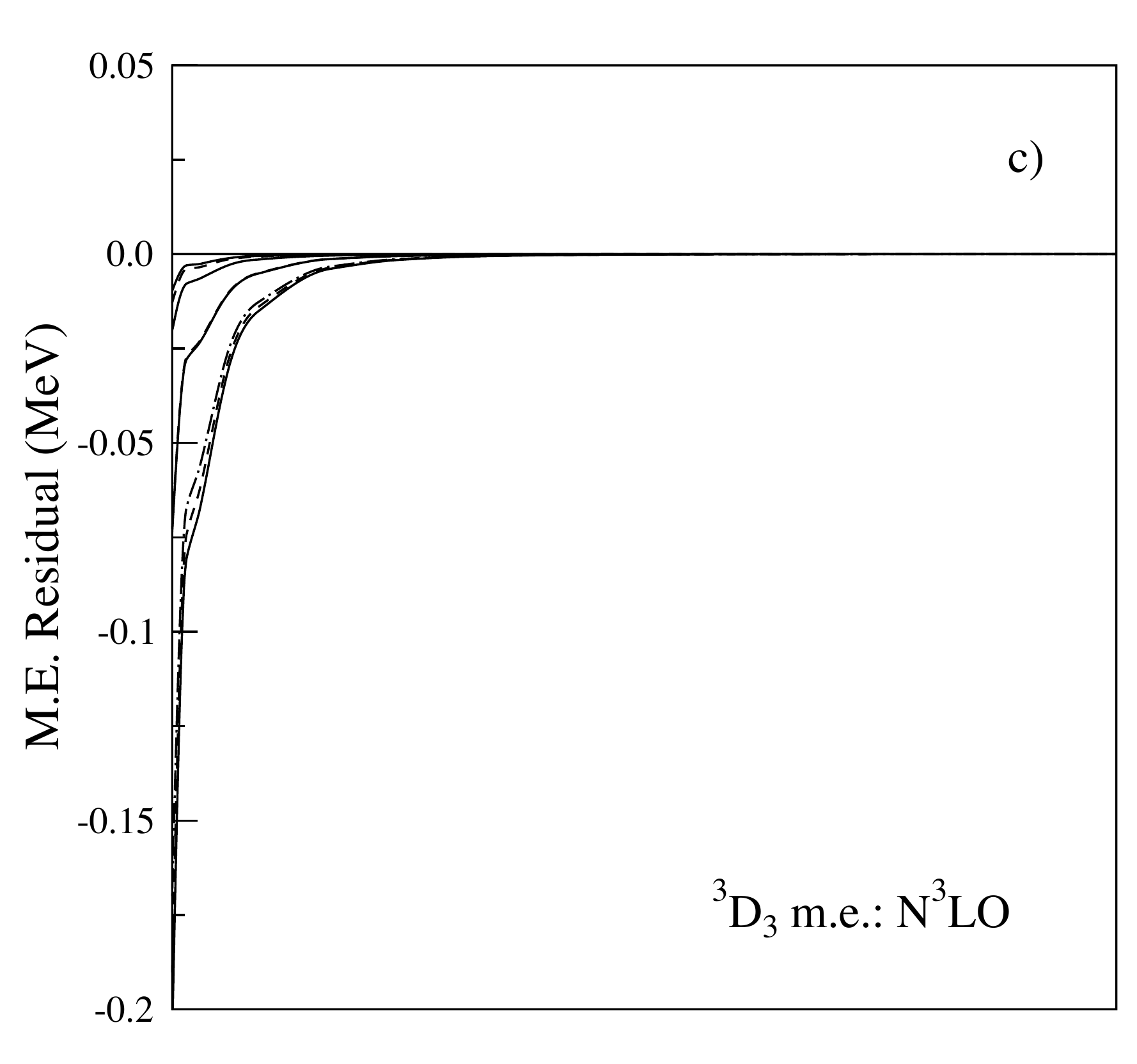}
\includegraphics[width=8cm]{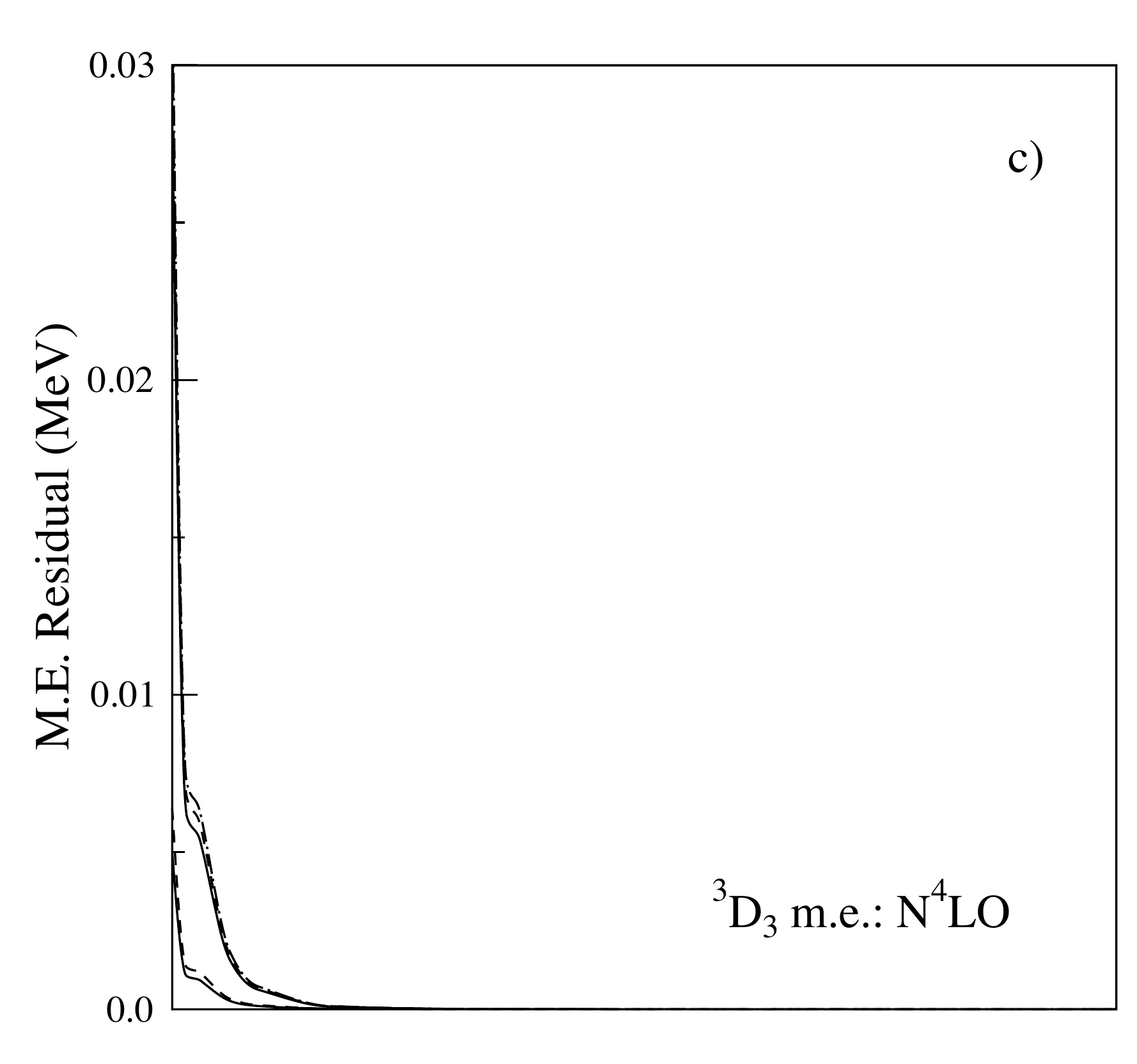}
\includegraphics[width=8cm]{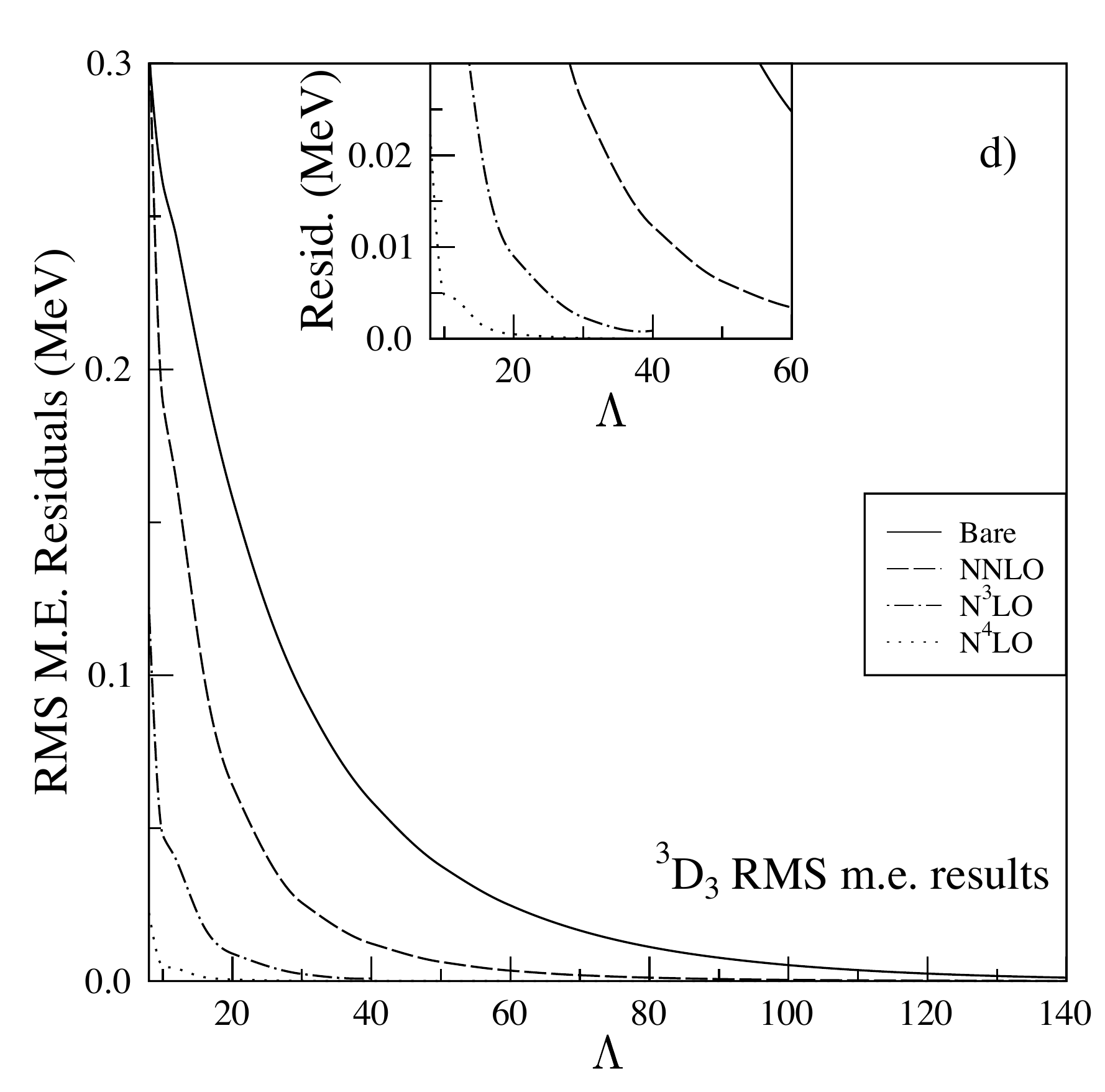}
\end{center}
\end{minipage}
\end{figure}

\begin{figure}
\begin{minipage}{0.5\linewidth}
\begin{center}
\includegraphics[width=8cm]{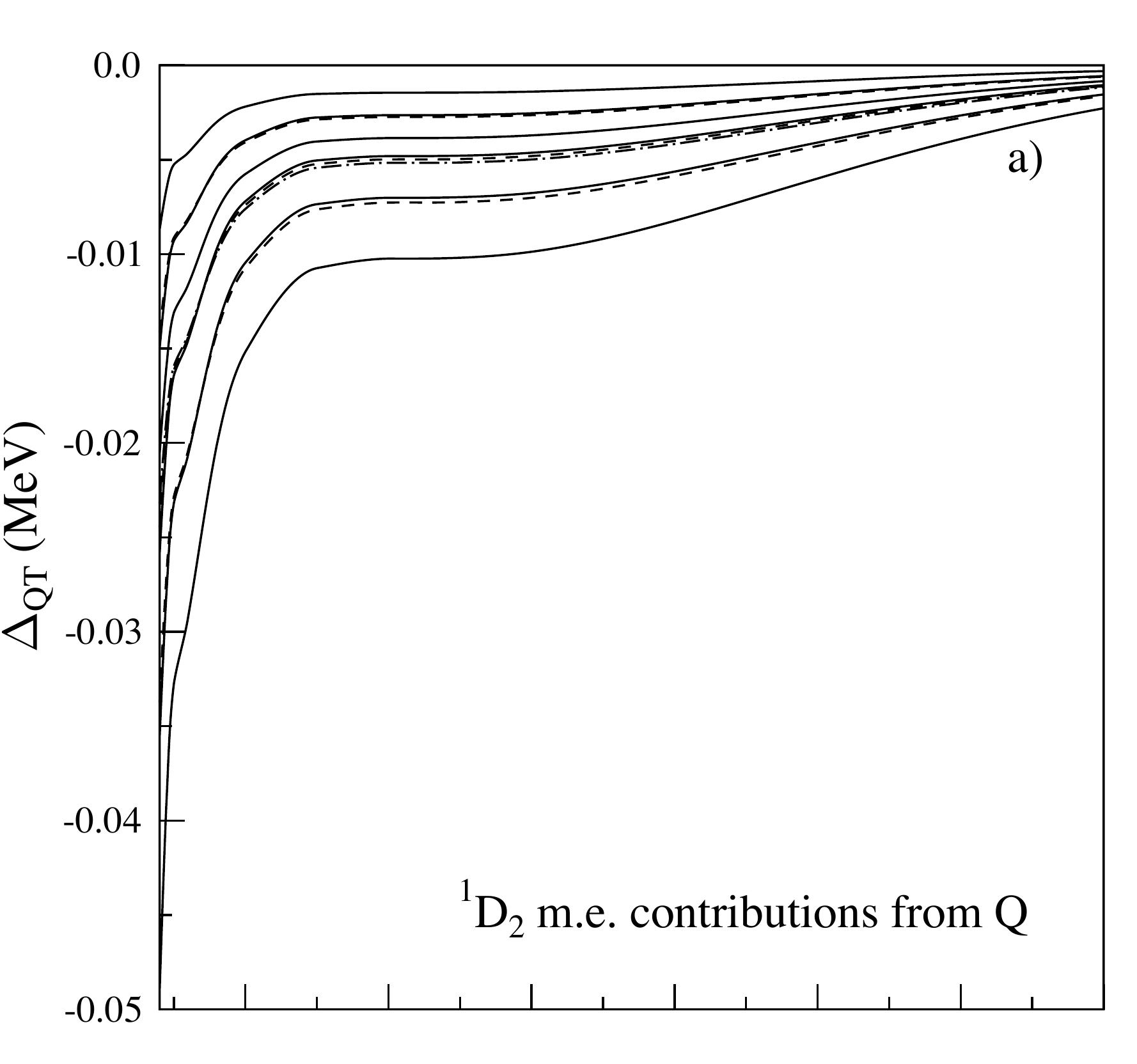}
\includegraphics[width=8cm]{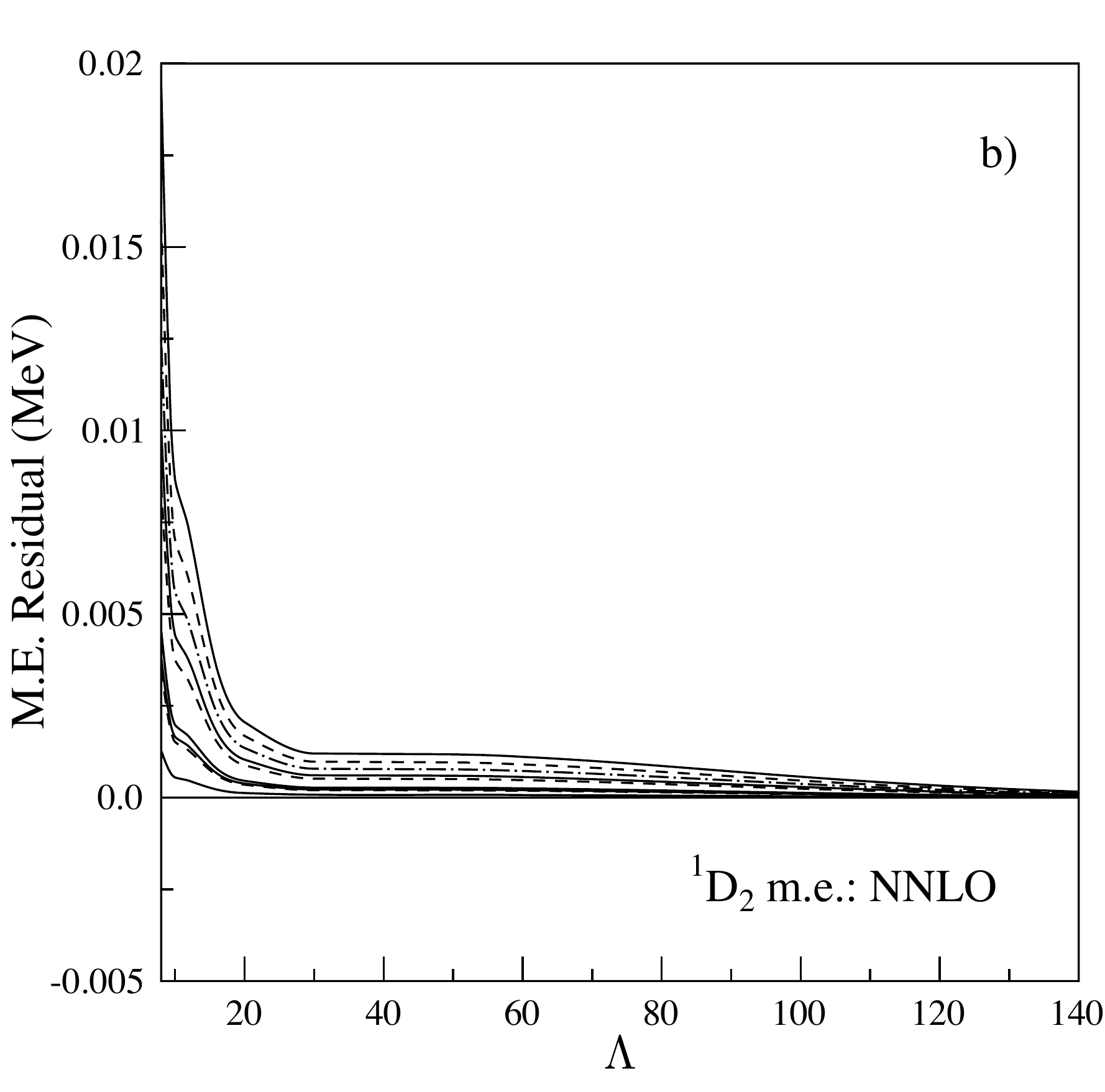}
\end{center}
\end{minipage}%
\begin{minipage}{0.5\linewidth}
\begin{center}
\includegraphics[width=8cm]{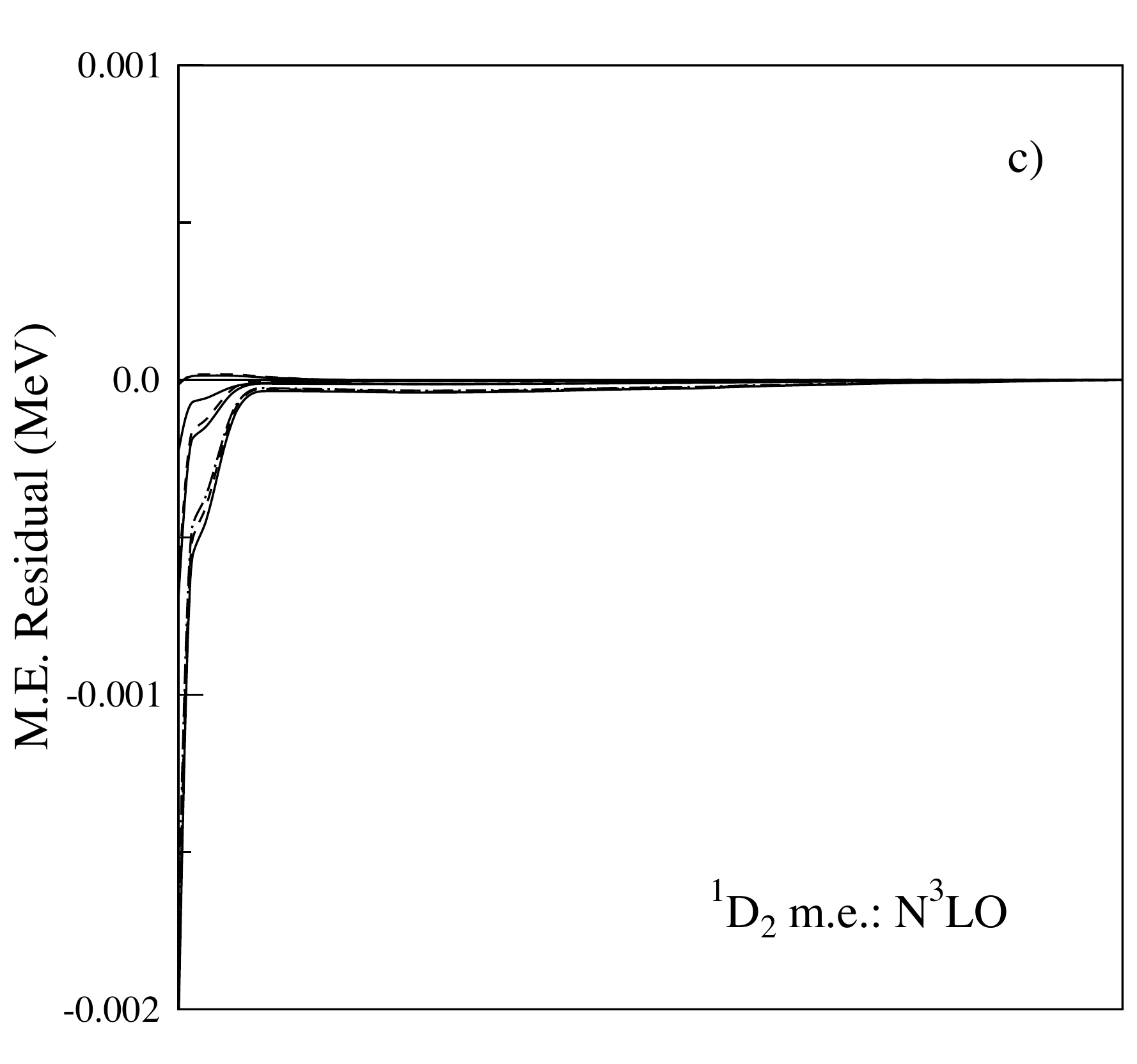}
\includegraphics[width=8cm]{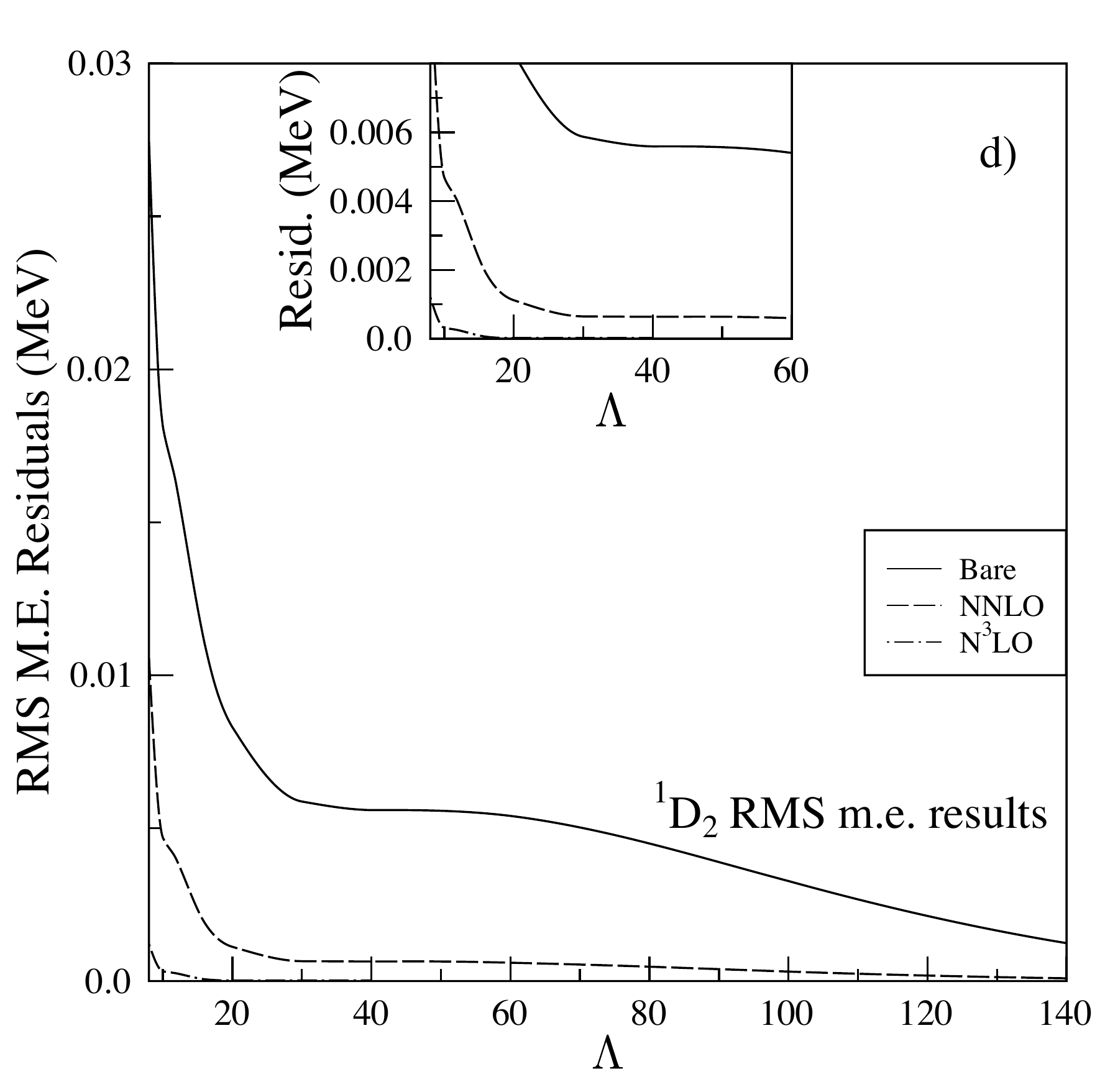}
\end{center}
\end{minipage}
\caption{As in Fig. \ref{fig_3s1}, but for the $^1D_2$ channel.   }
\label{fig_1d2}
\end{figure}

\begin{figure}
\begin{minipage}{0.5\linewidth}
\begin{center}
\includegraphics[width=8cm]{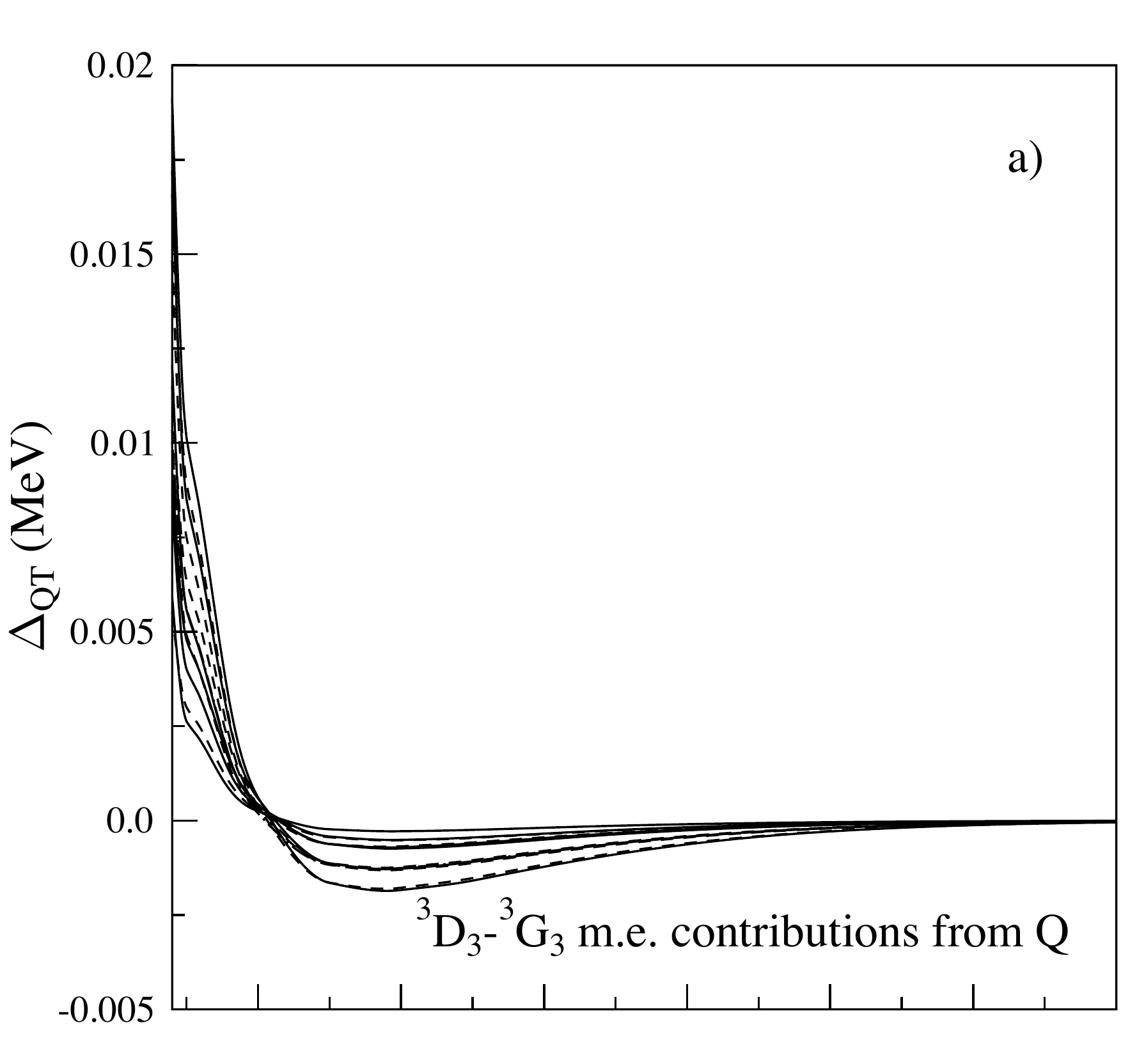}
\includegraphics[width=8cm]{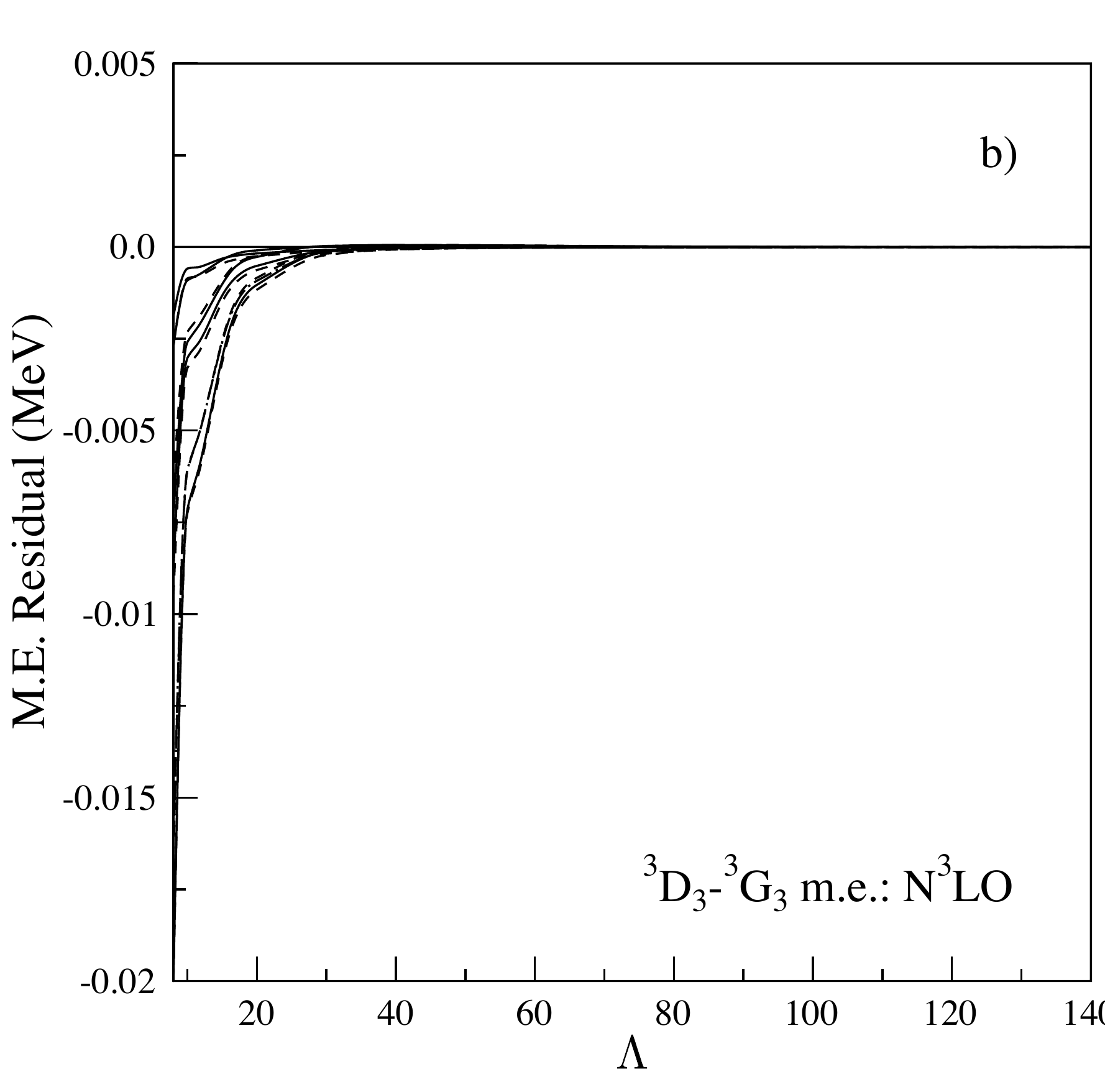}
\end{center}
\end{minipage}%
\begin{minipage}{0.5\linewidth}
\begin{center}
\includegraphics[width=8cm]{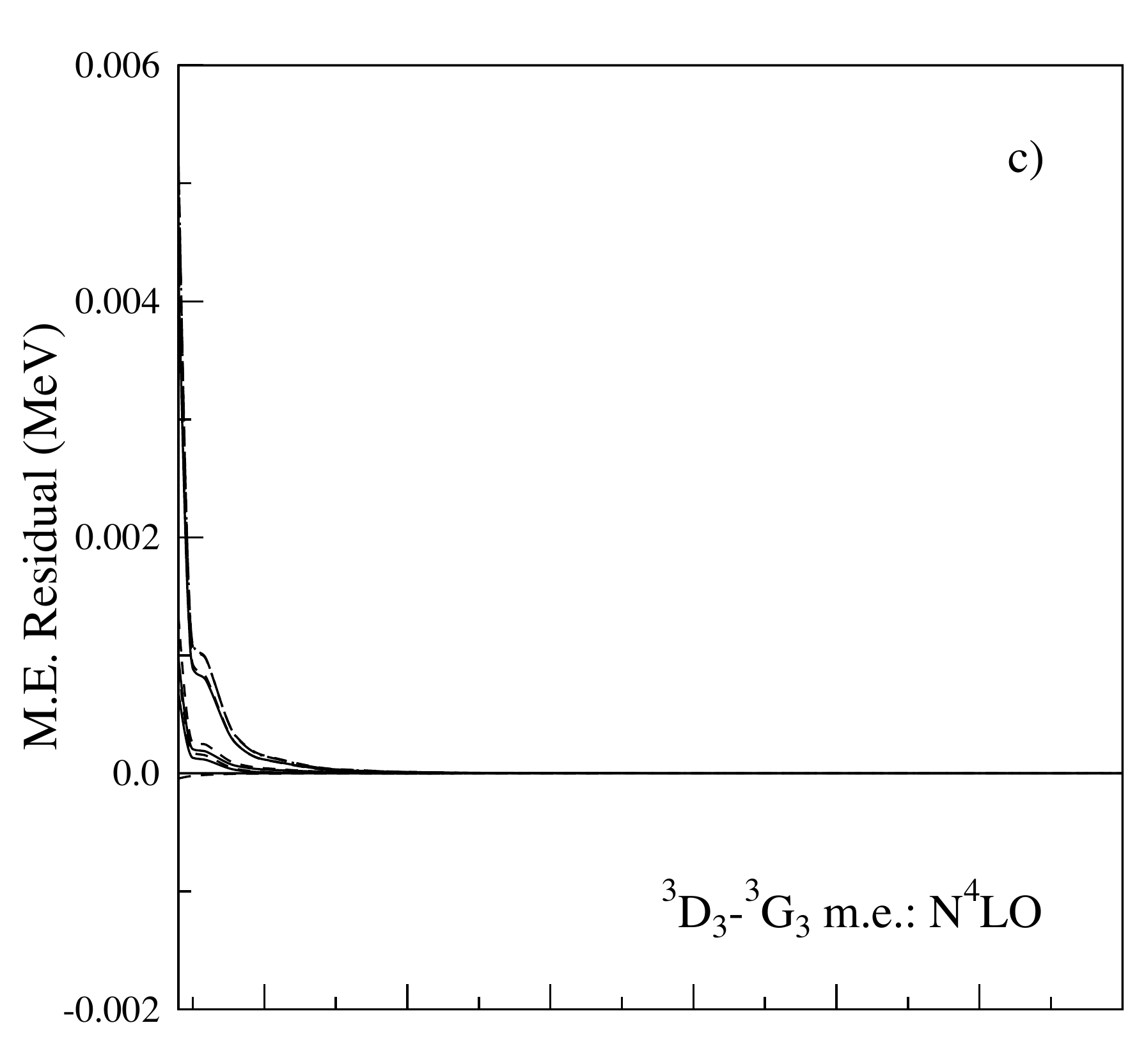}
\includegraphics[width=8cm]{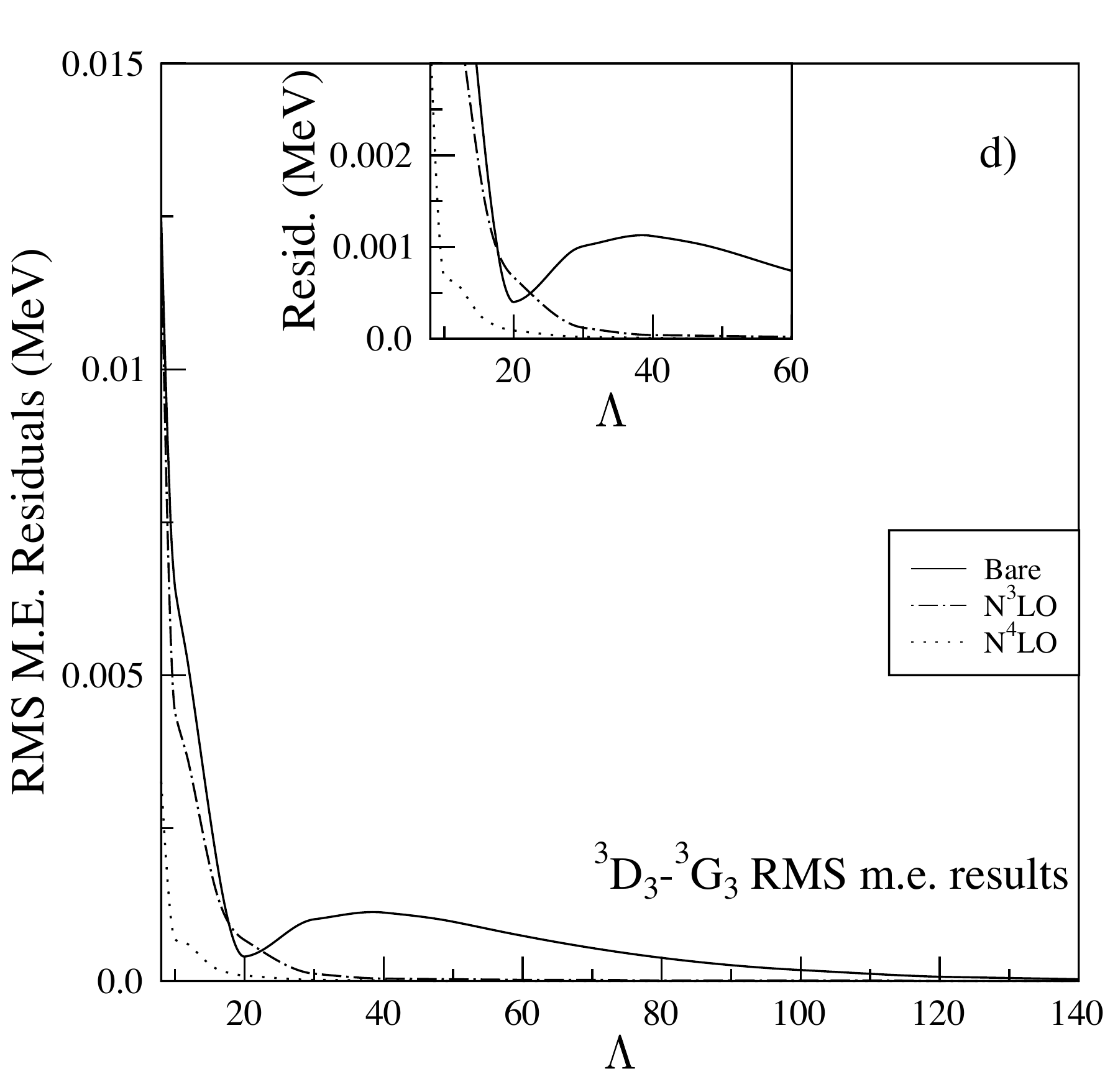}
\end{center}
\end{minipage}
\caption{As in Fig. \ref{fig_3s1}, but for the $^3D_3-{}^3G_3$ channel.  The N$^4$LO
contribution is also shown. }
\label{fig_dg}
\end{figure}

The remaining positive-parity channels that can be constrained at N$^3$LO are given
in Figs. \ref{fig_sd} through \ref{fig_dg}: ${}^3S_1-{}^3D_1$ (leading order contribution NLO);
${}^1D_2-{}^1D_2$, ${}^3D_1-{}^3D_1$, ${}^3D_2-{}^3D_2$, and ${}^3D_3-{}^3D_3$
(NNLO); and ${}^3D_3-{}^3G_3$ (N$^3$LO).   

Table \ref{table:2} gives the resulting fitted couplings at N$^3$LO for all contributing
channels, along with numerical results for the root-mean-square $Q$-space contributions
to $\Delta_{QT}(\Lambda_P)$ and the root-mean-square residuals (the deviation between
the contact-gradient prediction and $\Delta_{QT}(\Lambda_P)$ for the remaining
unconstrained effective interactions matrix elements).  The quality of the agreement found
in the $^1S_0$ and $^3S_1$channels is generally typical -- residuals at the few kilovolt
level -- though there are some exceptions and some general patterns that emerge.
One of these is the tendency of the triplet channels with spin and angular momentum aligned
($^3S_1-{}^3S_1$, $^3P_2-{}^3P_2$, $^3D_3-{}^3D_3$, and $^3F_4-{}^3F4$) to exhibit larger
residuals than the remaining $S, P, D$ and $F$ channels, respectively.  The
$^3D_3-{}^3D_3$, which has contributions at NNLO and N$^3$LO,
stands out as the most difficult channel, with a residual of 122 keV, one to two
orders of magnitude greater than the typical scale of N$^3$LO residuals.

\begin{table*}
\begin{center}
\caption{The effective interaction for LO through N$^3$LO, with $\Lambda_P=8$ and $b$=1.7 fm.$^\dagger$}
\label{table:2}
\begin{tabular}{||l||c||c||c|c||c|c|c||c|c||}
        \hline
Channel & \multicolumn{7}{c||}{Couplings (MeV)} & $\langle$M.E.$\rangle_{RMS}$ (MeV) & $\langle$Resid.$\rangle_{RMS}$ (keV) \\ 
\hline \hline
 & $a_{LO}^S$ & $a_{NLO}^S$ & $a_{NNLO}^{S,22}$ & $a_{NNLO}^{S,40}$ & $a_{N^3LO}^{S,42}$ &
 $a_{N^3LO}^{S,60}$ & & & \\
 ${}^1S_0-{}^1S_0$ & -32.851 & -2.081E-1 & -2.111E-3 & -1.276E-3 & -7.045E-6 & -1.8891E-6  & & 7.94 & 0.53 \\
 ${}^3S_1-{}^3S_1$ &  -62.517 & -1.399 & -5.509E-2 & -1.160E-2 & -5.789E-4 & -1.444E-4 & & 11.97 & 2.71\\  
 \hline
 & & $a_{NLO}^{SD}$ & $a_{NNLO}^{SD,22}$ & $a_{NNLO}^{SD,04}$ & $a_{N^3LO}^{SD,42}$ &
 $a_{N^3LO}^{SD,24}$ & $a_{N^3LO}^{SD,06}$ & & \\
 ${}^3S_1 - {}^3D_1$ & &2.200E-1 & 1.632E-2 & 2.656E-2 & 2.136E-4 & 3.041E-4 & -1.504E-4 & 0.160 & 2.45 \\ \hline
  & & & $a_{NNLO}^D$ & & $a_{N^3LO}^D$ & & & & \\
  ${}^1D_2 -{}^1D_2$ & & & -6.062E-3 & & -1.189E-4 & & & 0.027 & 1.21 \\
  ${}^3D_1-{}^3D_1$ & & & -1.034E-2 & & -1.532E-4 & & & 0.051 & 2.27 \\
  ${}^3D_2-{}^3D_2$ & & & -3.048E-2 & & -5.238E-4 & & & 0.141& 1.20 \\
  ${}^3D_3-{}^3D_3$ & & & -9.632E-2 & & -4.355E-3 & & & 0.303 & 122$^\ddagger$ \\ \hline
  & & & & & $a_{N^3LO}^{SD}$ & & &  &\\
  ${}^3D_3-{}^3G_3$ & & & & & 3.529E-4 & &  & 0.012 & 12.2$^\ddagger$ \\ \hline
   & & $a_{NLO}^P$ & $a_{NNLO}^P$ &  &$a_{N^3LO}^{P,33}$ & $a_{N^3LO}^{P,51}$ & & &  \\
   ${}^1P_1-{}^1P_1$ & & -8.594E-1 & -7.112E-3 &  & -6.822E-5 & 1.004E-5 & & 0.694 & 0.11 \\
   ${}^3P_0-{}^3P_0$ & & -1.641 & -1.833E-2 & & -2.920E-4 & -1.952E-4 & & 1.283 & 2.26 \\
   ${}^3P_1-{}^3P_1$ & & -1.892 & -1.588E-2 & & -1.561E-4 & -6.737E-6 & & 1.526 & 0.08 \\
   ${}^3P_2-{}^3P_2$ & & -4.513E-1 & -1.257E-2 & & -5.803E-4 & -1.421E-4 & & 0.285 & 5.61 \\ \hline
   & & & $a_{NNLO}^{PF}$ & & $a_{N^3LO}^{PF,33}$ & $a_{N^3LO}^{PF,15}$ & & & \\
   ${}^3P_2-{}^3F_2$ & & & -4.983E-3 & & 1.729E-5 & -5.166E-5 & & 0.034 & 1.43 \\ \hline
   & & & & & $a_{N^3LO}^F$ & & &  & \\
   ${}^1F_3-{}^1F_3$ & & & & & -3.135E-4 & & & 0.007 & 1.03 \\
   ${}^3F_2-{}^3F_2$ & & & & & -8.537E-4 & & & 0.020 & 2.34 \\
   ${}^3F_3-{}^3F_3$ & & & & & -2.647E-4 & & & 0.006 & 0.61 \\
   ${}^3F_4-{}^3F_4$ & & & & & -5.169E-4 & & & 0.008 & 6.23 \\
 \hline  \hline
\end{tabular}
\end{center}
\flushleft{$^\dagger$ The appropriate LO, NLO, and NNLO interactions are obtained by truncating the
table at the desired order.  \\
$^\ddagger$ An $N^4LO$ calculation in the ${}^3D_3-{}^3D_3$ channel
yields $a_{N^4LO}^{3D3,44}$=-2.510E-4 MeV
and $a_{N^4LO}^{3D3,62}$ = -7.550E-5 MeV, and reduces $\langle \mathrm{Resid.} \rangle_{RMS}$ to 22.3 keV; and in the $^3D_3-{}^3G_3$ channel yields
$a_{N^4LO}^{DG,44}$ = -2.141E-5 MeV and $a_{N^4LO}^{DG,26}$ = 1.180E-5 MeV and
reduces $\langle \mathrm{Resid.} \rangle_{RMS}$ to 3.26 keV.}
\end{table*}

\begin{figure}
\begin{minipage}{0.5\linewidth}
\begin{center}
\includegraphics[width=8cm]{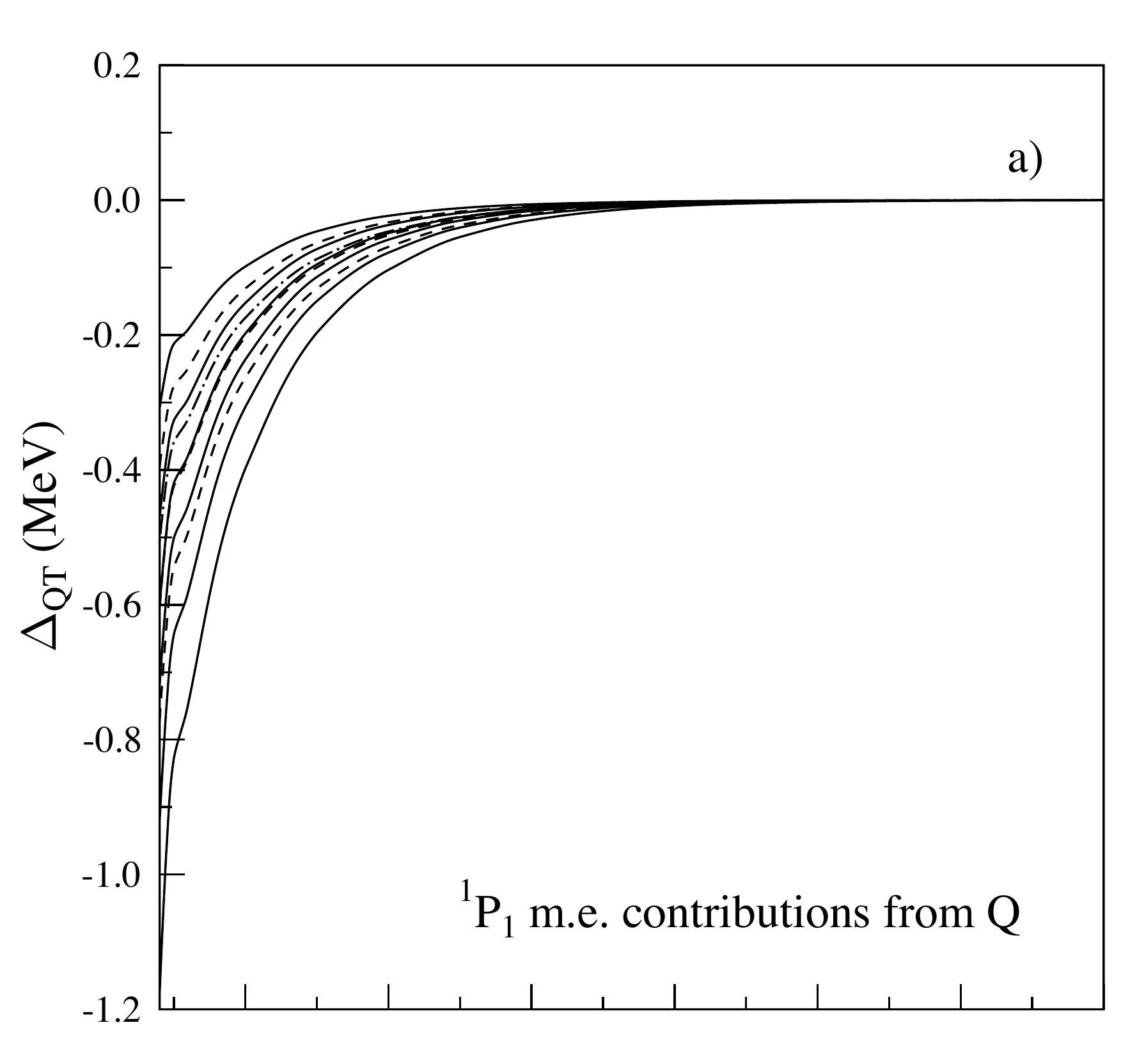}
\includegraphics[width=8cm]{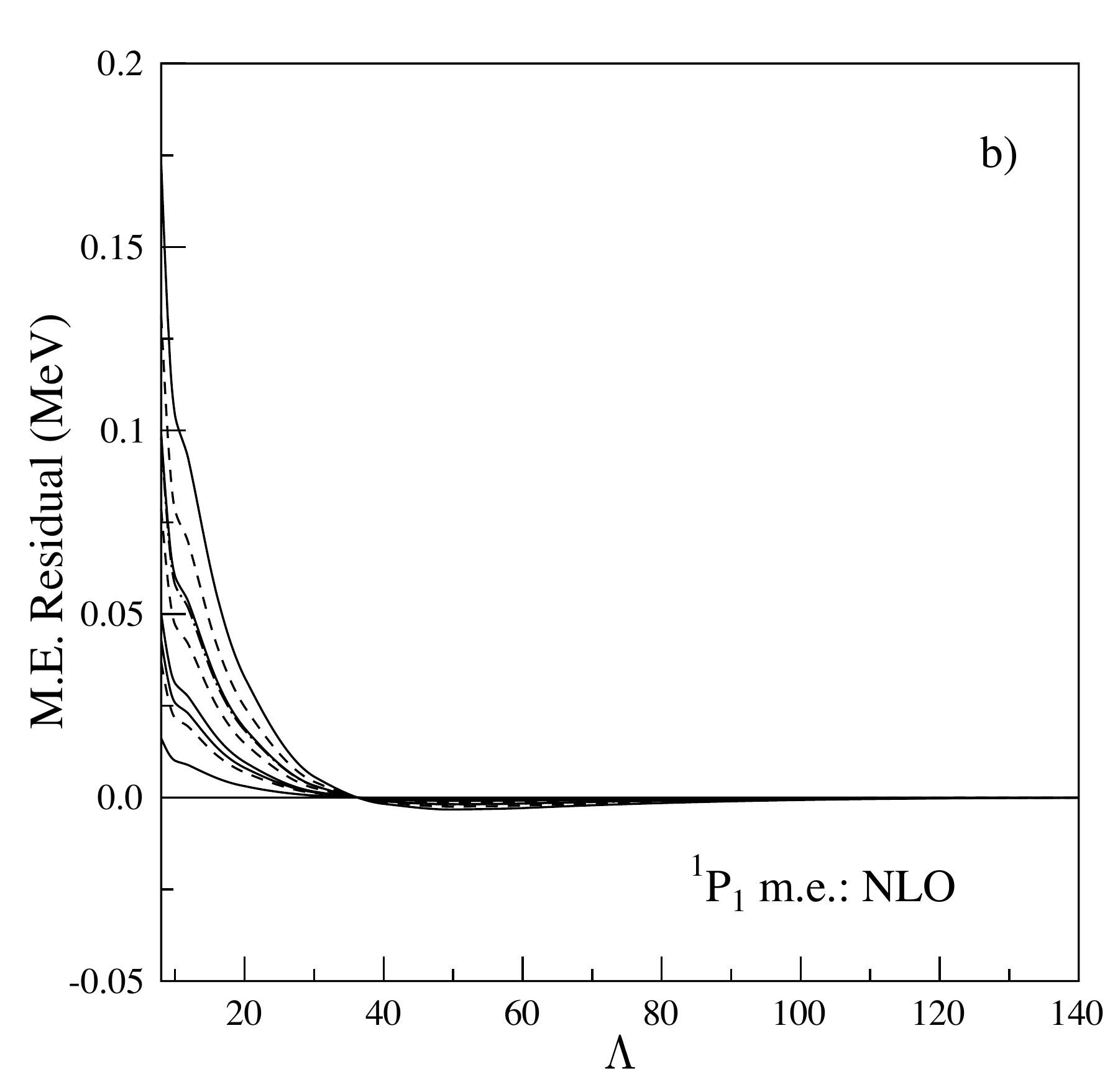}
\end{center}
\caption{As in Fig. \ref{fig_3s1},  but for the $^1P_1$ channel. }
\label{fig_1p1}
\end{minipage}%
\begin{minipage}{0.5\linewidth}
\begin{center}
\includegraphics[width=8cm]{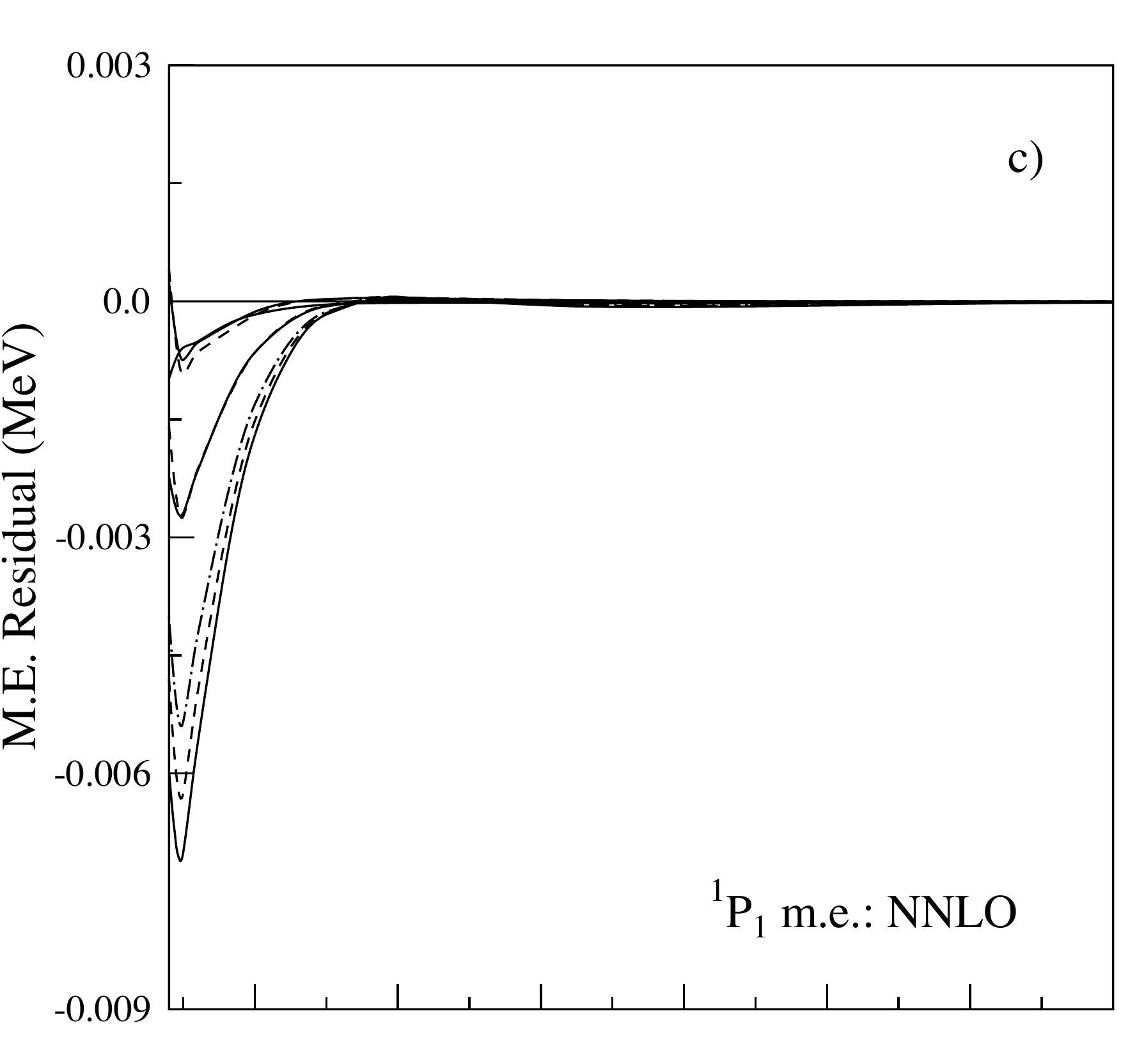}
\includegraphics[width=8cm]{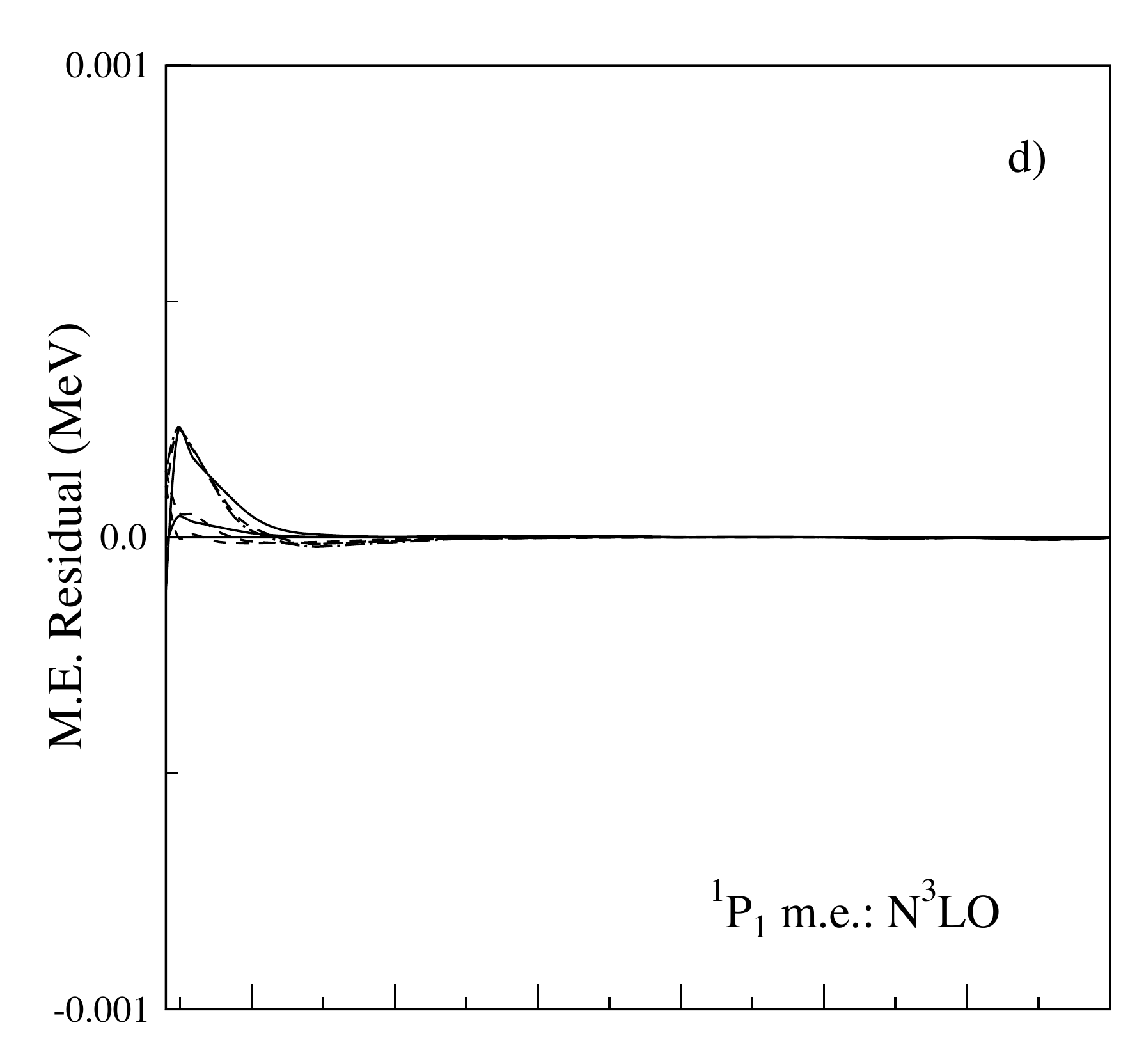}
\includegraphics[width=8cm]{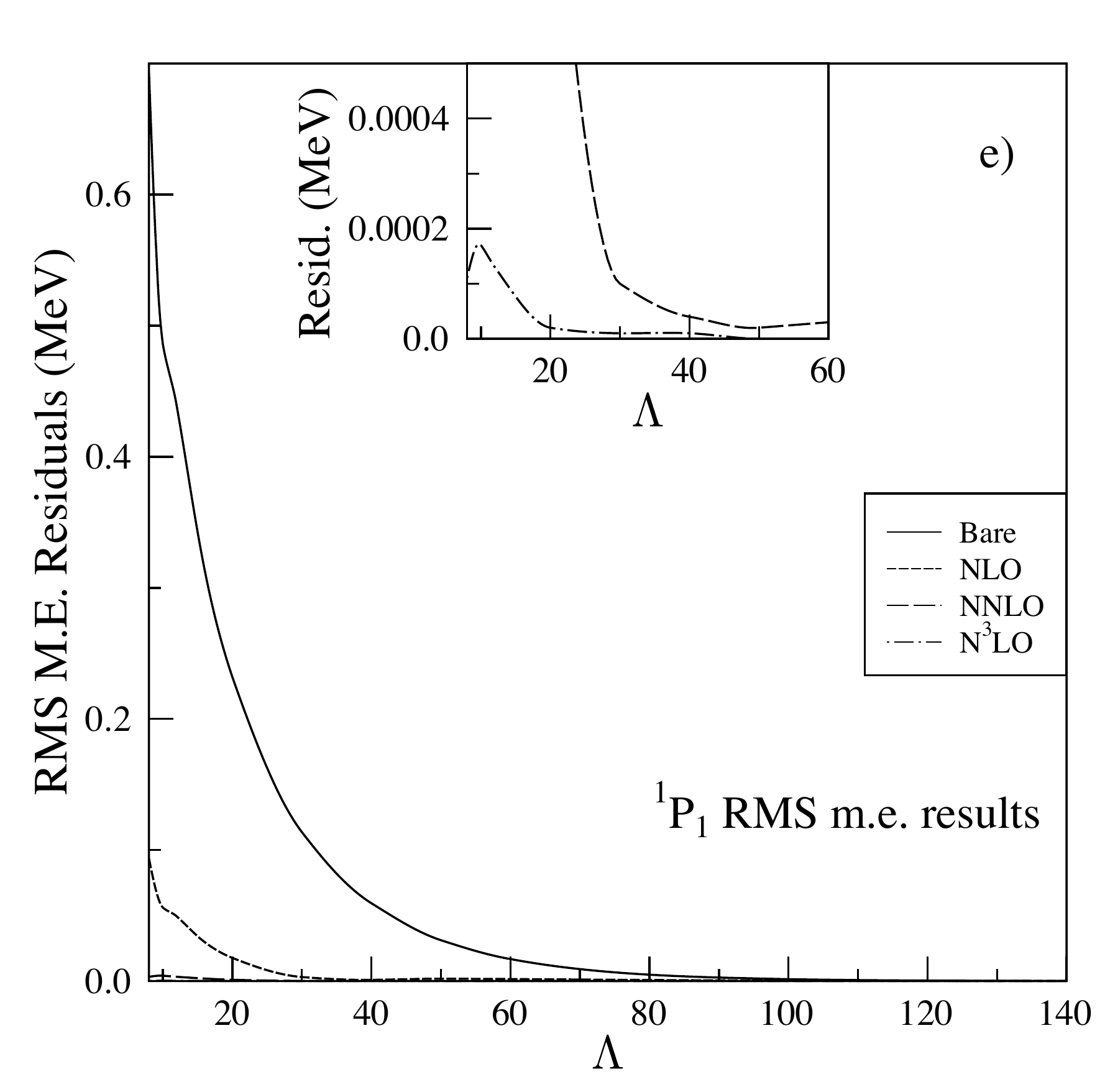}
\end{center}
\end{minipage}
\end{figure}

\begin{figure}
\begin{minipage}{0.5\linewidth}
\begin{center}
\includegraphics[width=8cm]{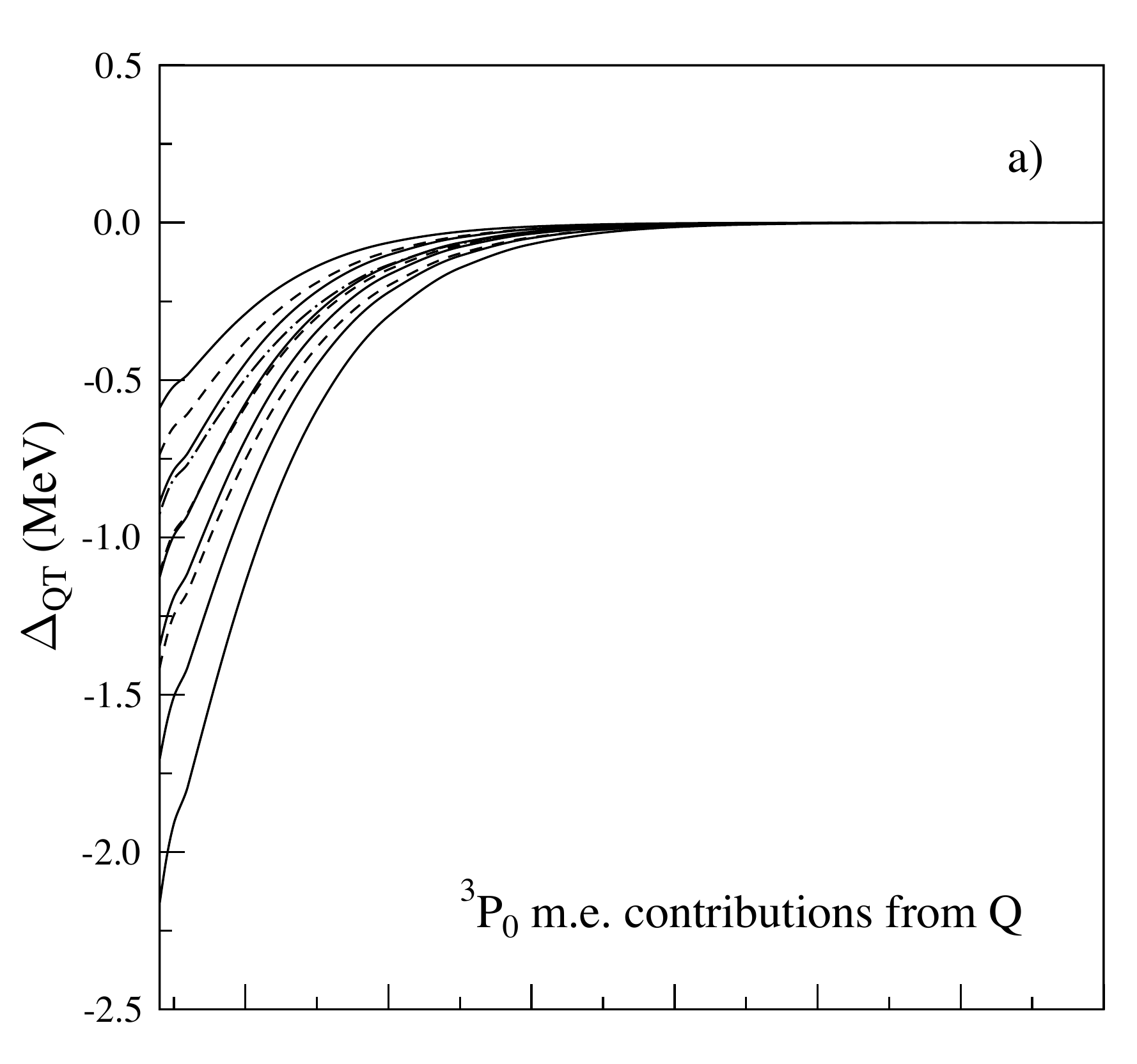}
\includegraphics[width=8cm]{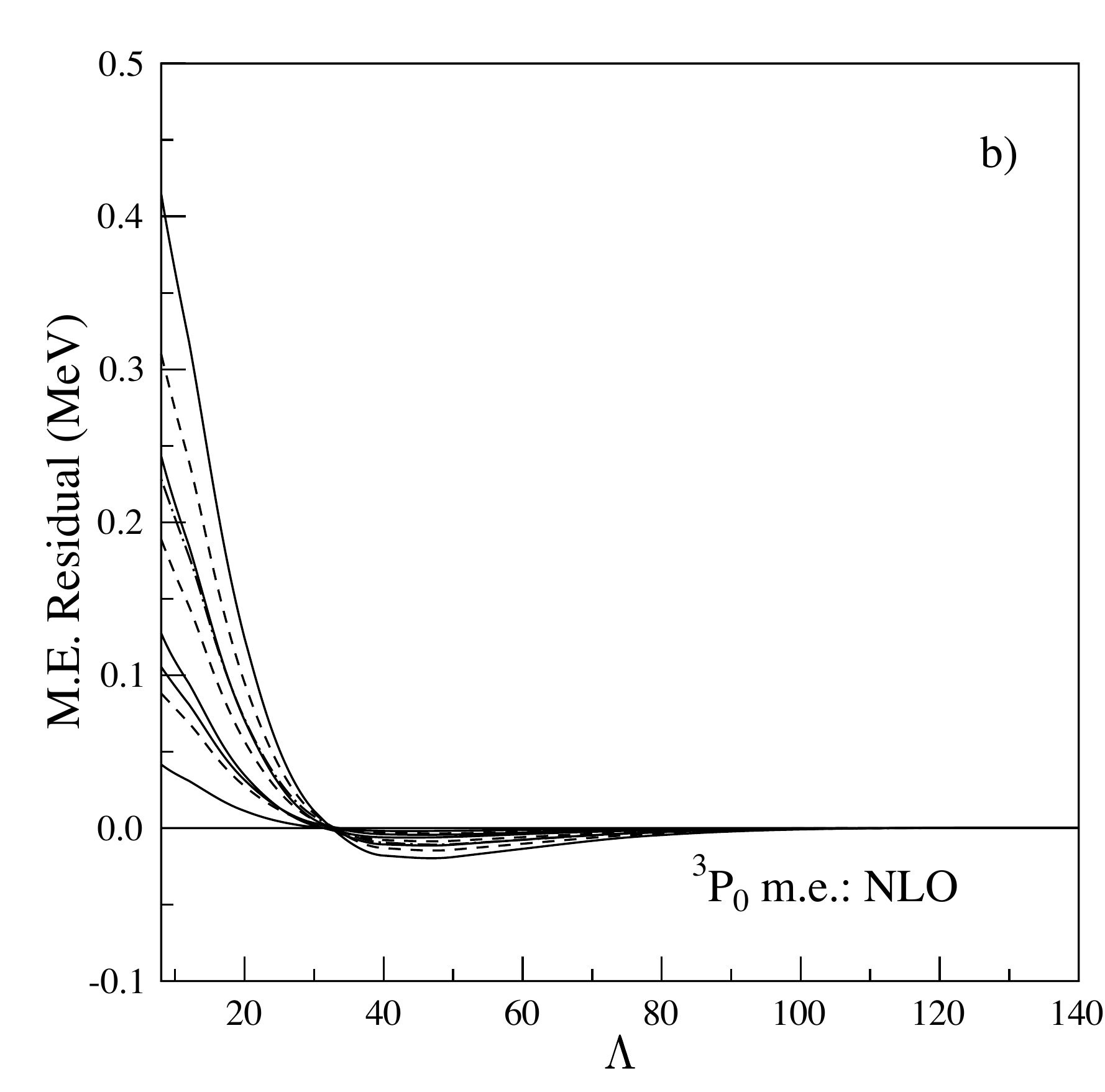}
\end{center}
\caption{As in Fig. \ref{fig_3s1},  but for the $^3P_0$ channel. }
\label{fig_3p0}
\end{minipage}%
\begin{minipage}{0.5\linewidth}
\begin{center}
\includegraphics[width=8cm]{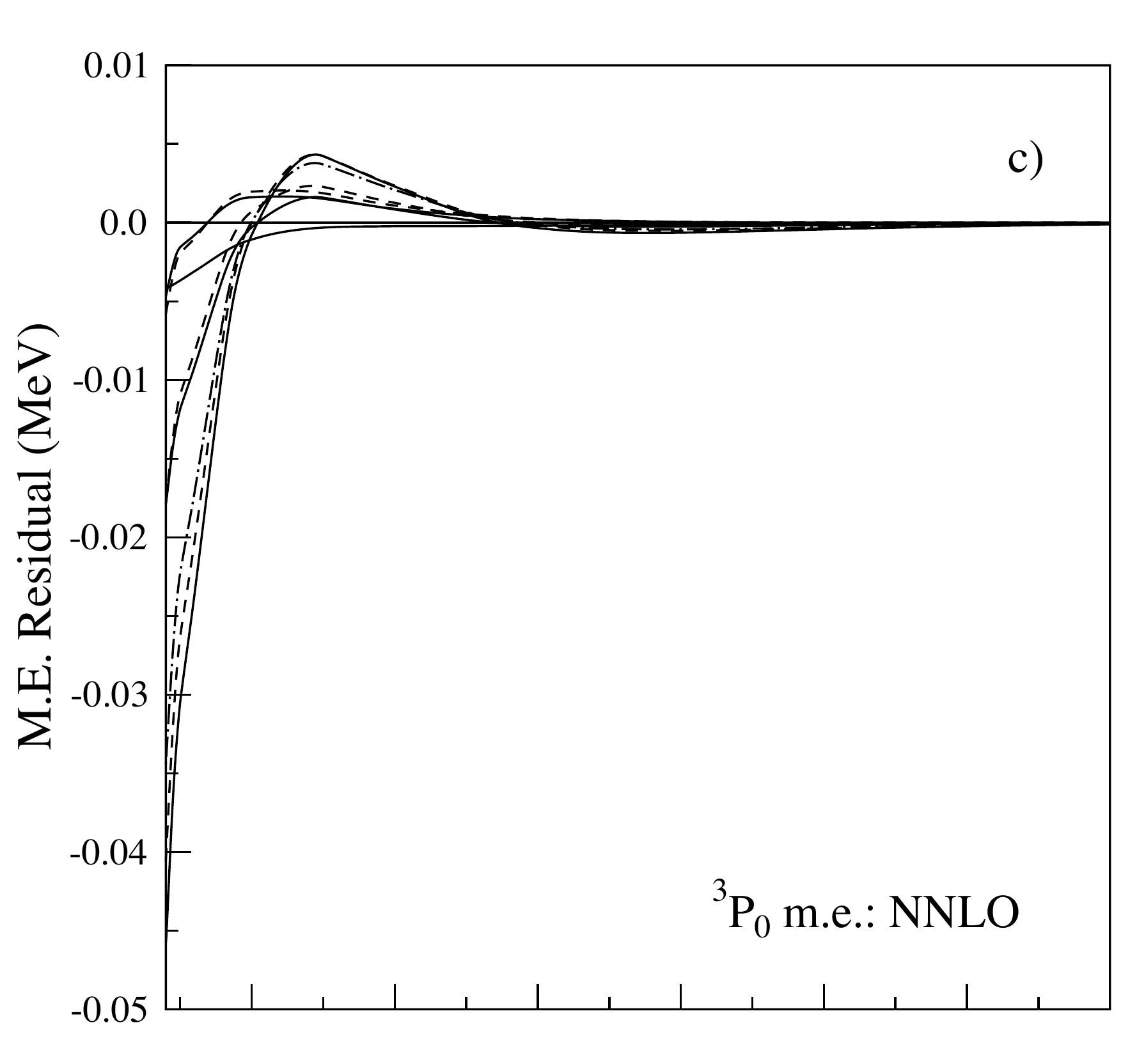}
\includegraphics[width=8cm]{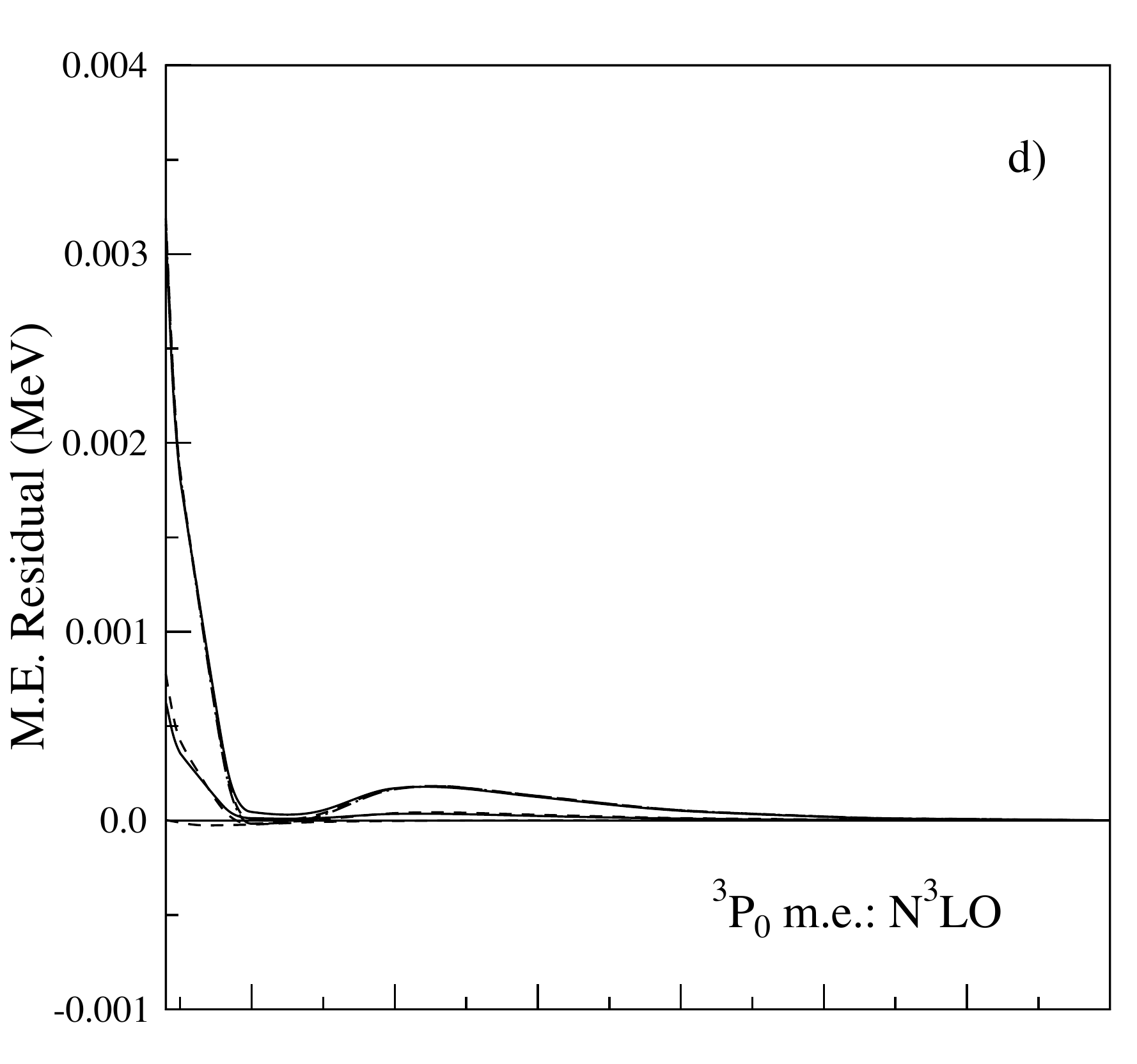}
\includegraphics[width=8cm]{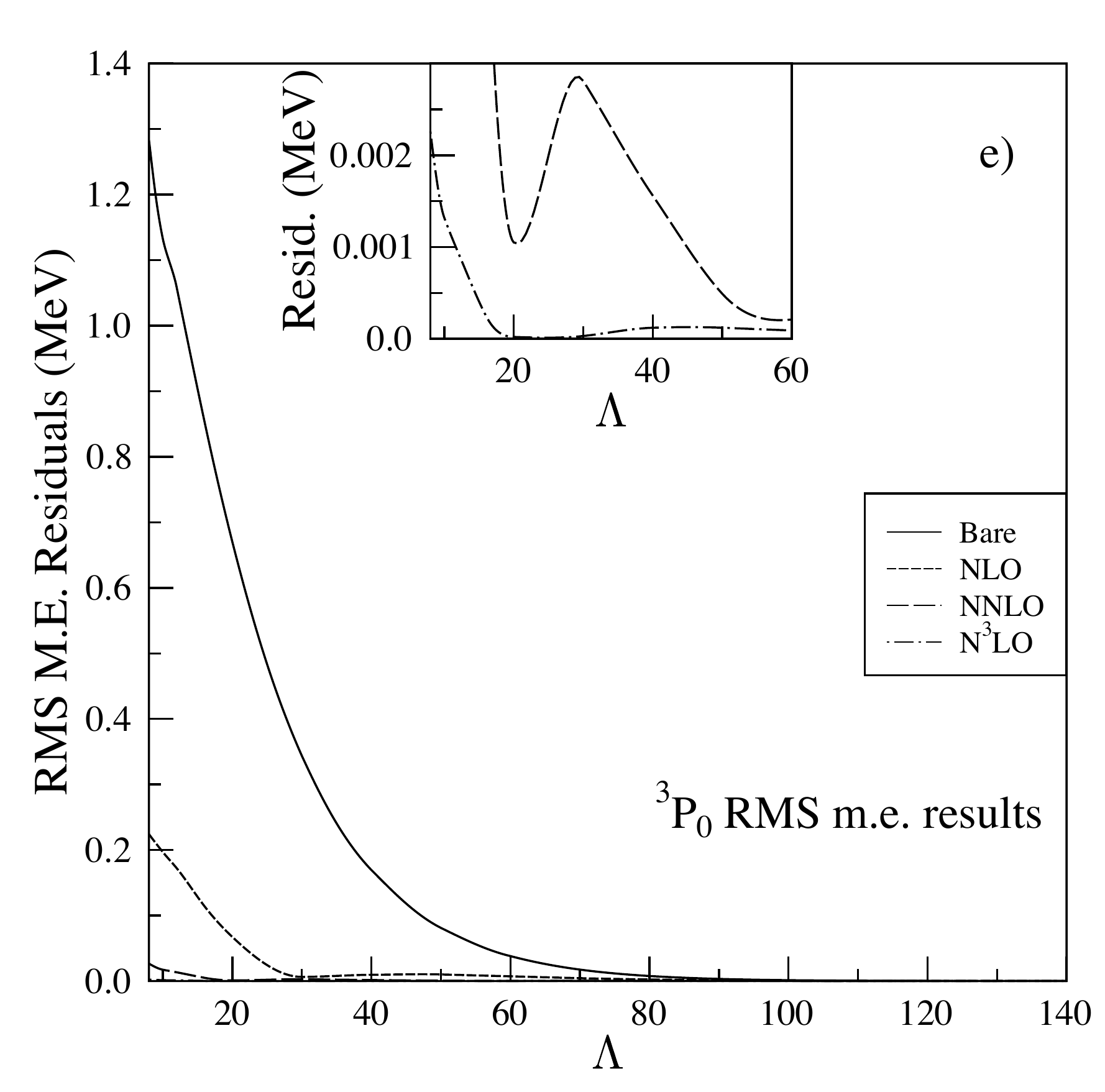}
\end{center}
\end{minipage}
\end{figure}

\begin{figure}
\begin{minipage}{0.5\linewidth}
\begin{center}
\includegraphics[width=8cm]{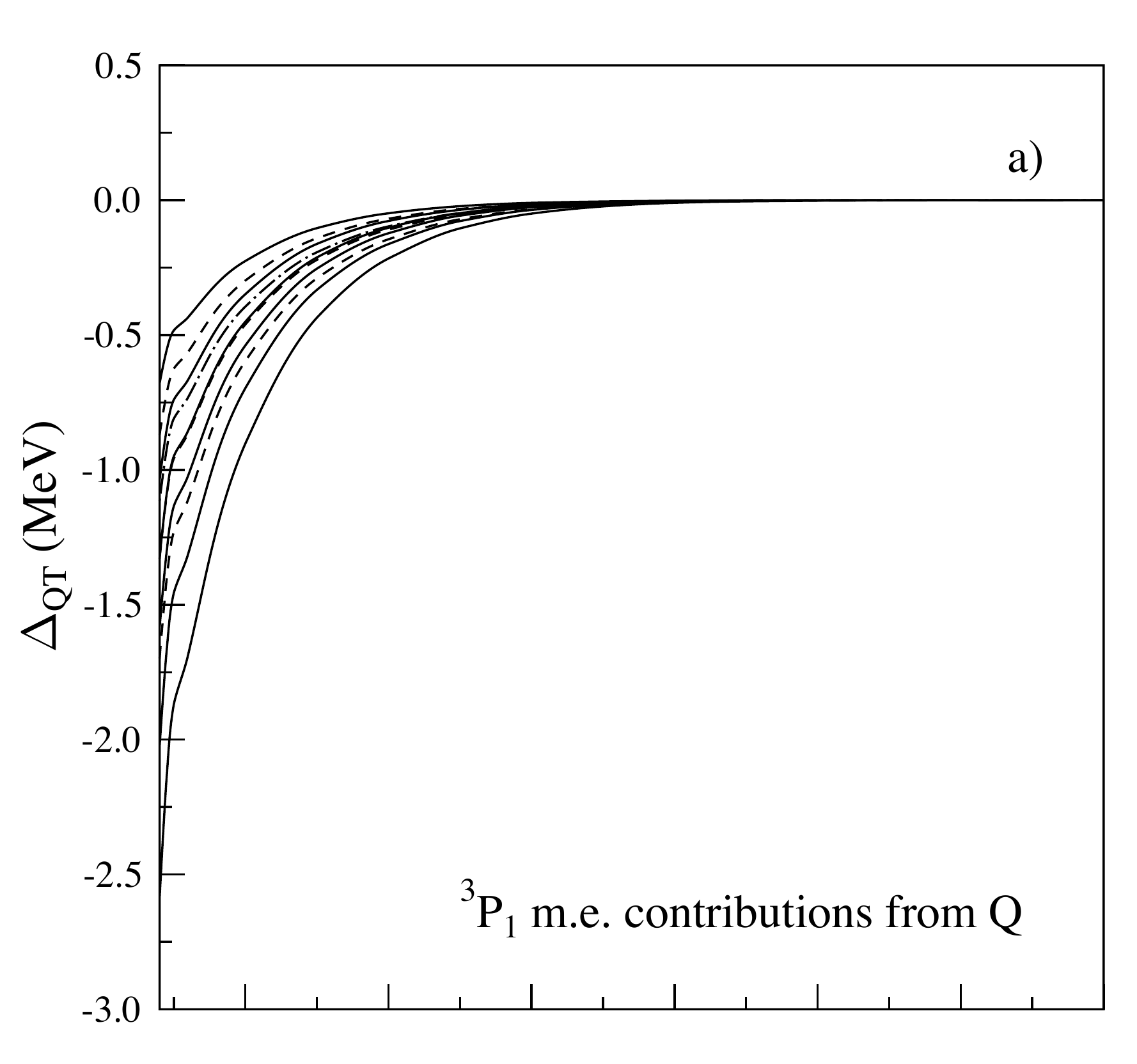}
\includegraphics[width=8cm]{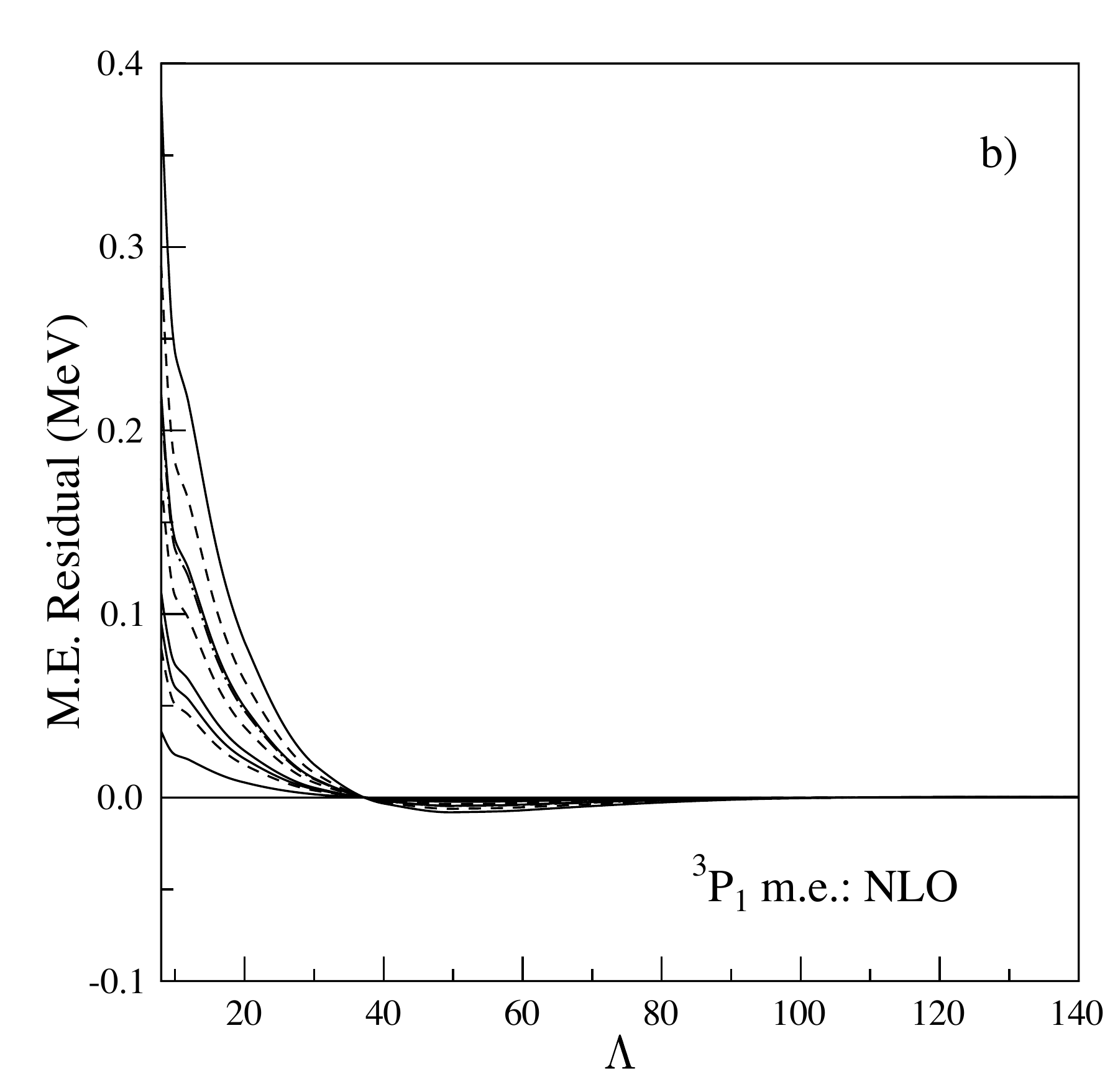}
\end{center}
\caption{As in Fig. \ref{fig_3s1},  but for the $^3P_1$ channel. }
\label{fig_3p1}
\end{minipage}%
\begin{minipage}{0.5\linewidth}
\begin{center}
\includegraphics[width=8cm]{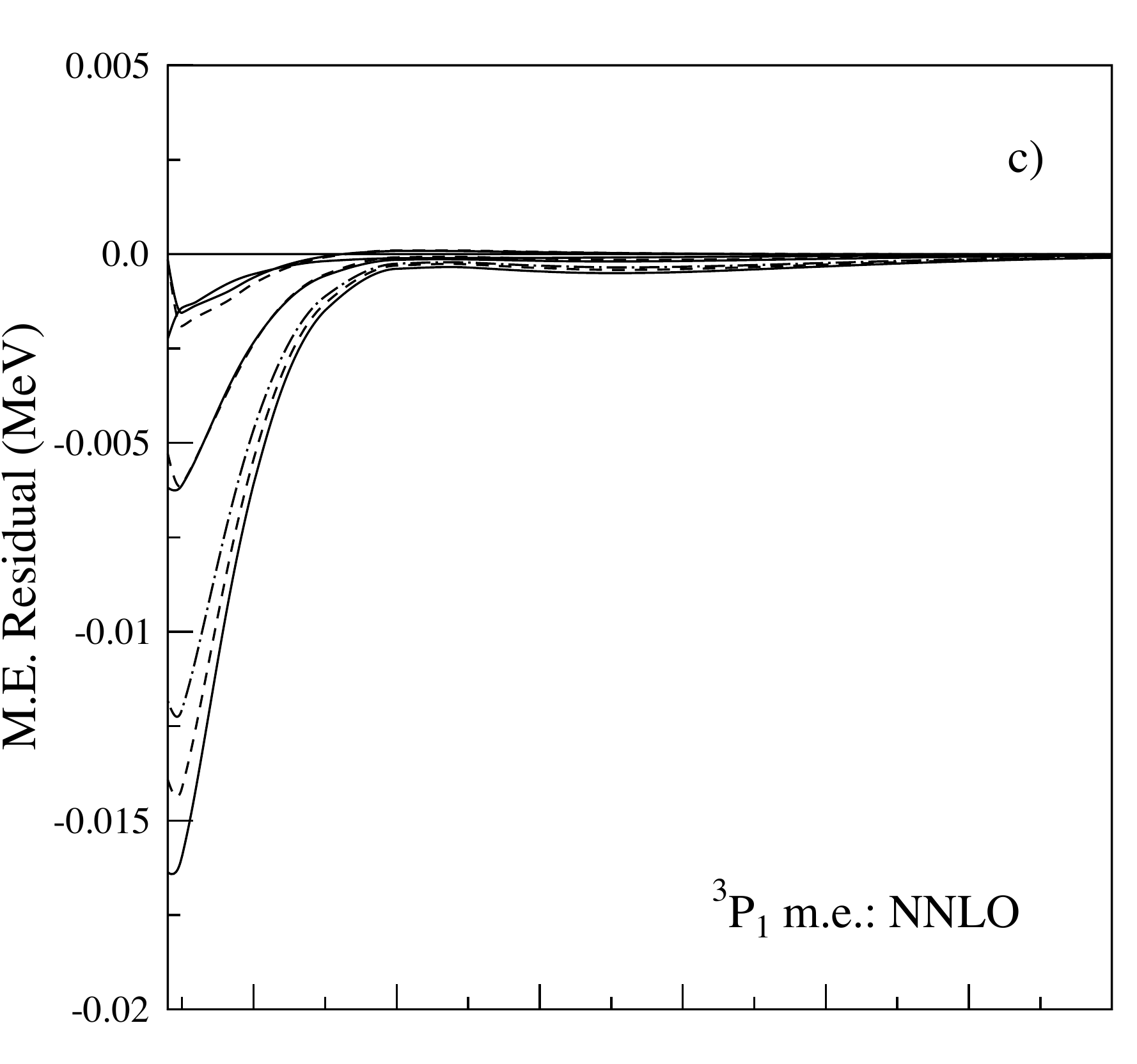}
\includegraphics[width=8cm]{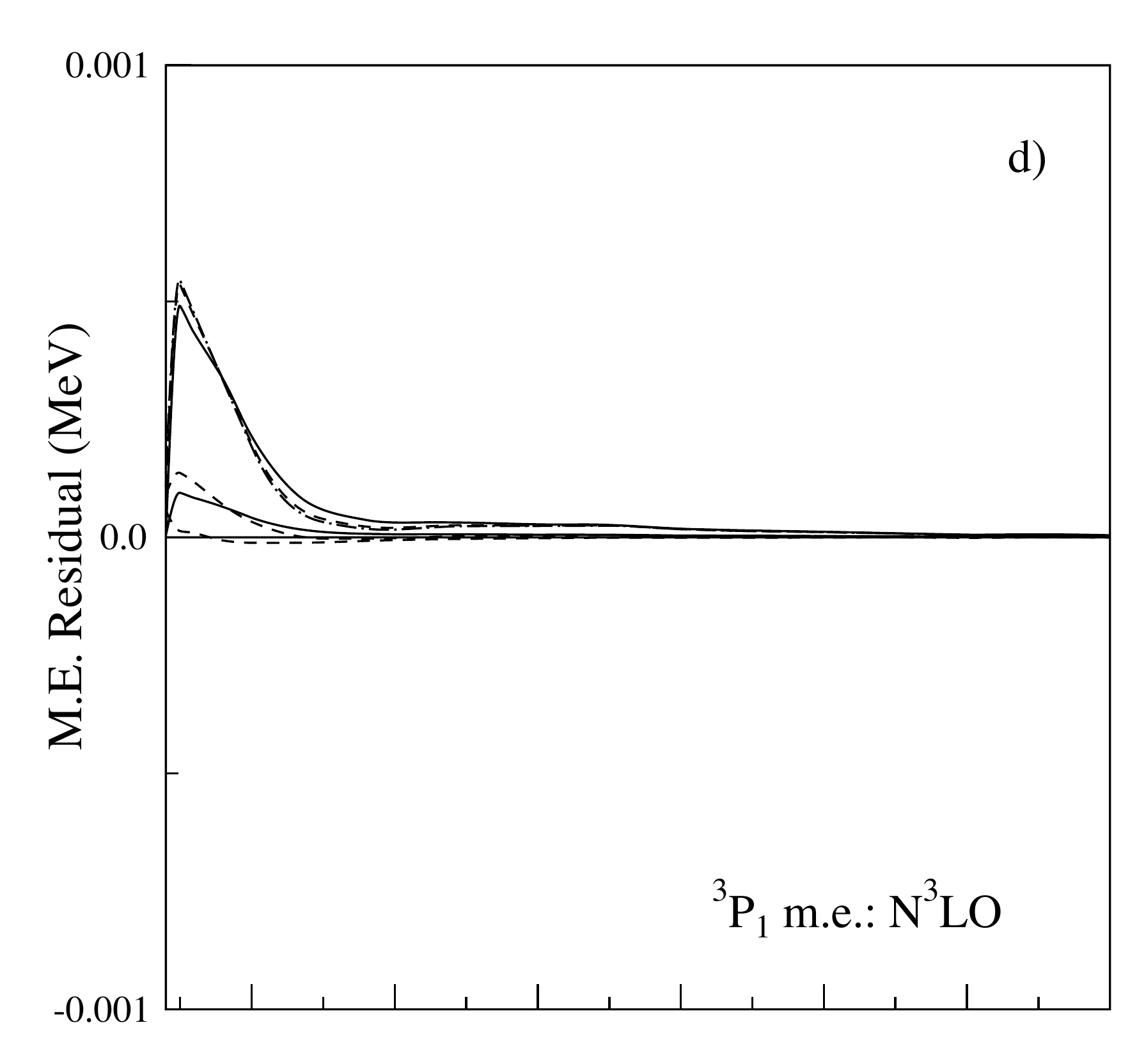}
\includegraphics[width=8cm]{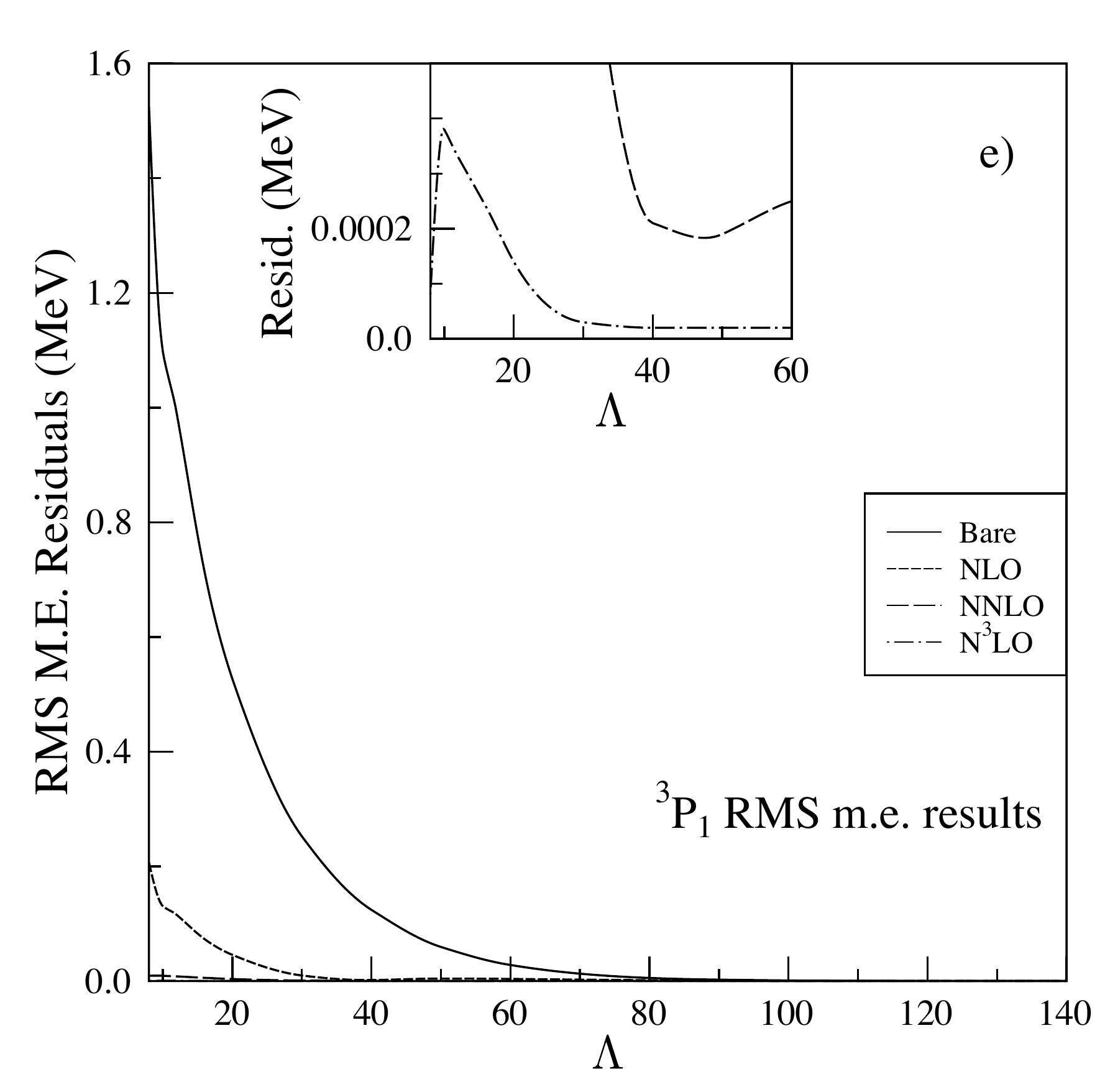}
\end{center}
\end{minipage}
\end{figure}

\begin{figure}
\begin{minipage}{0.5\linewidth}
\begin{center}
\includegraphics[width=8cm]{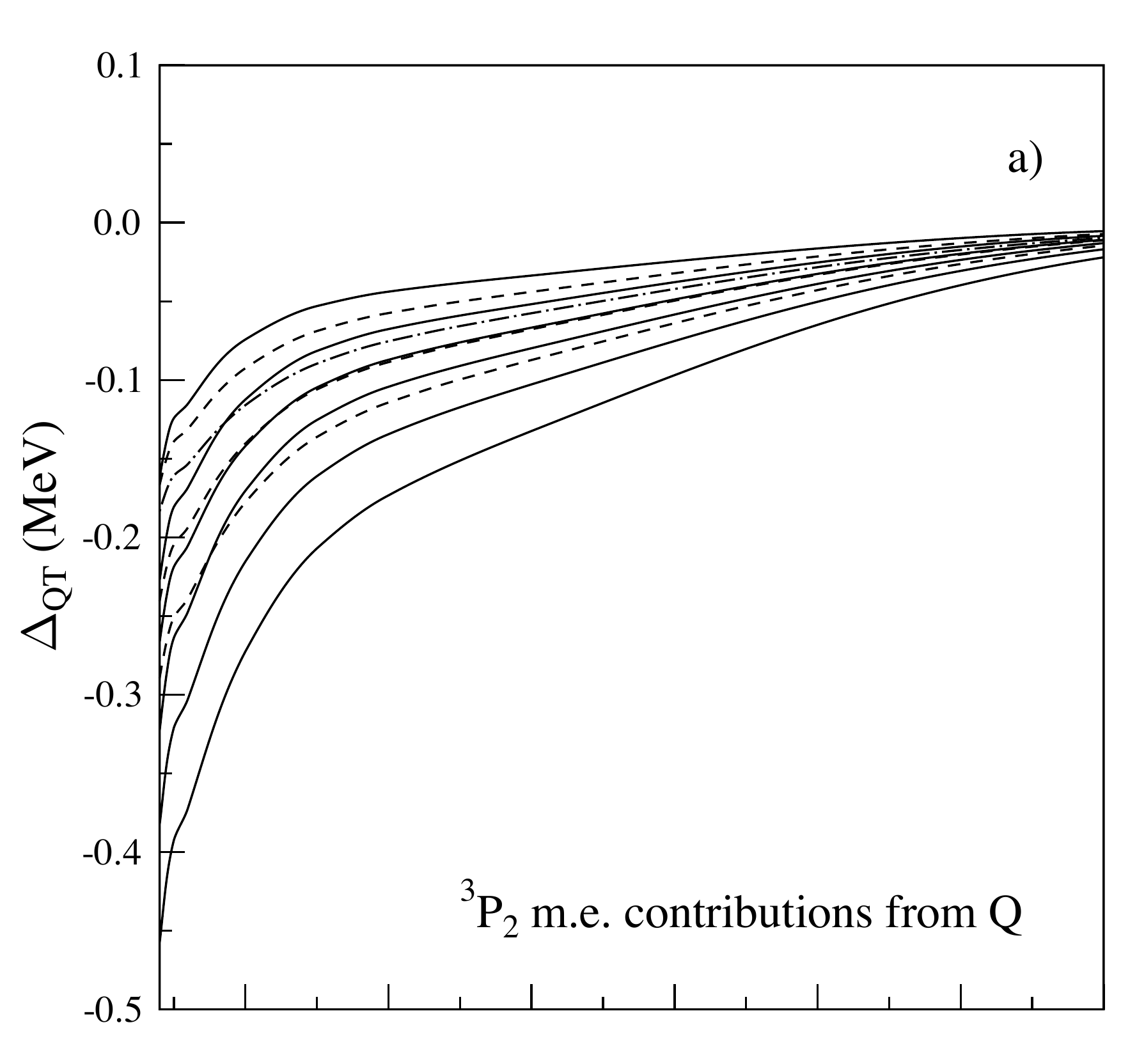}
\includegraphics[width=8cm]{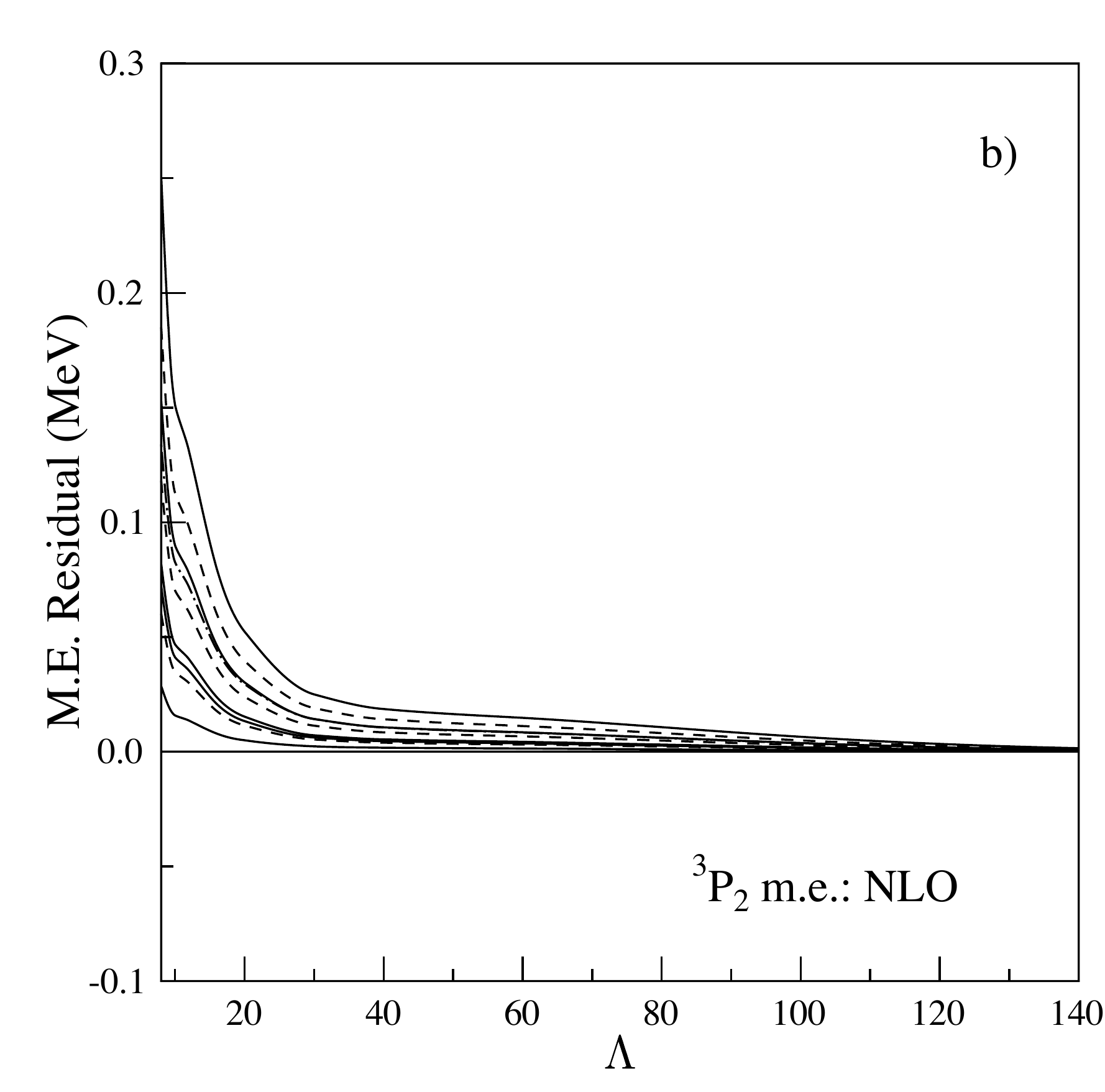}
\end{center}
\caption{As in Fig. \ref{fig_3s1},  but for the $^3P_2$ channel. }
\label{fig_3p2}
\end{minipage}%
\begin{minipage}{0.5\linewidth}
\begin{center}
\includegraphics[width=8cm]{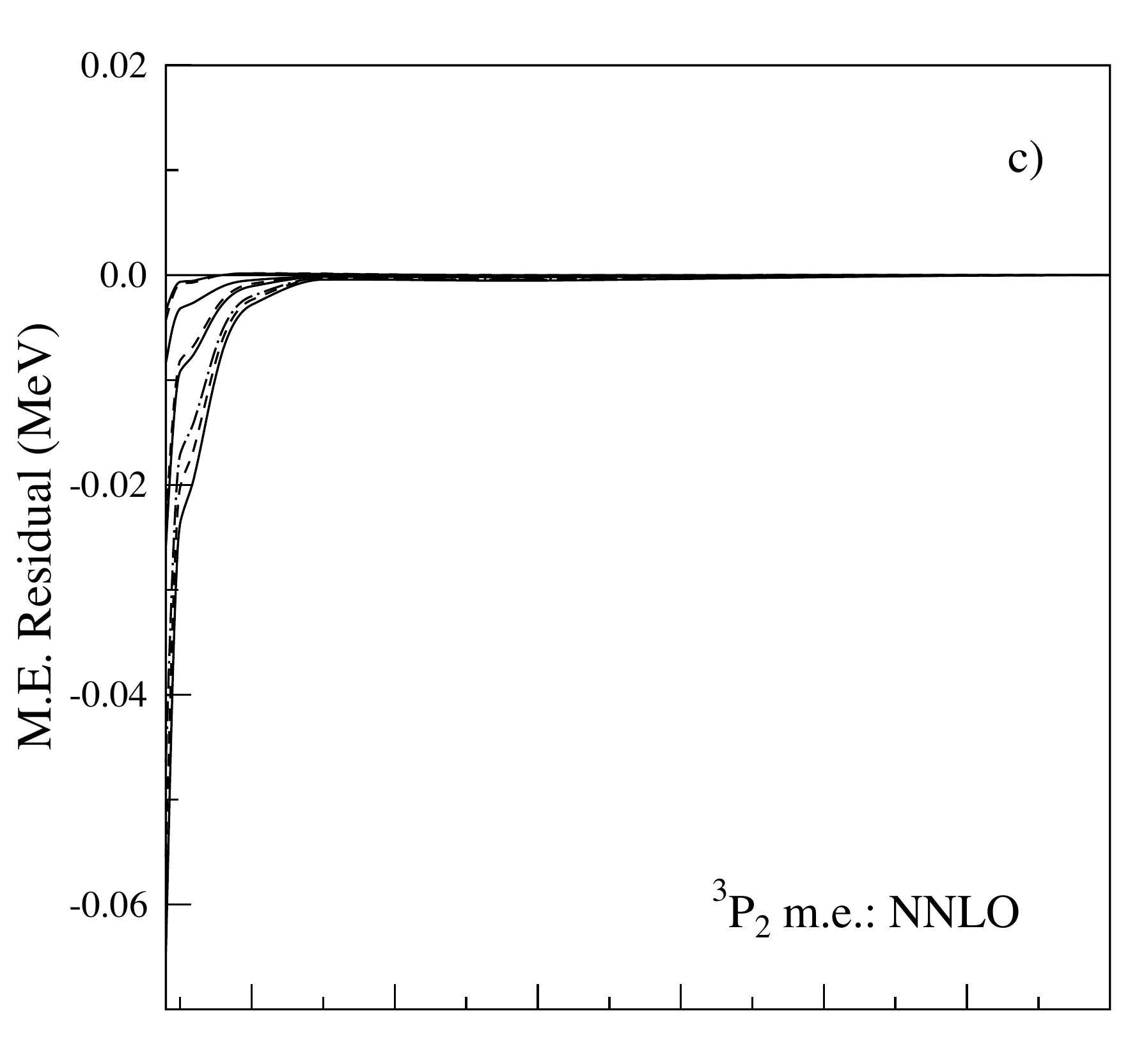}
\includegraphics[width=8cm]{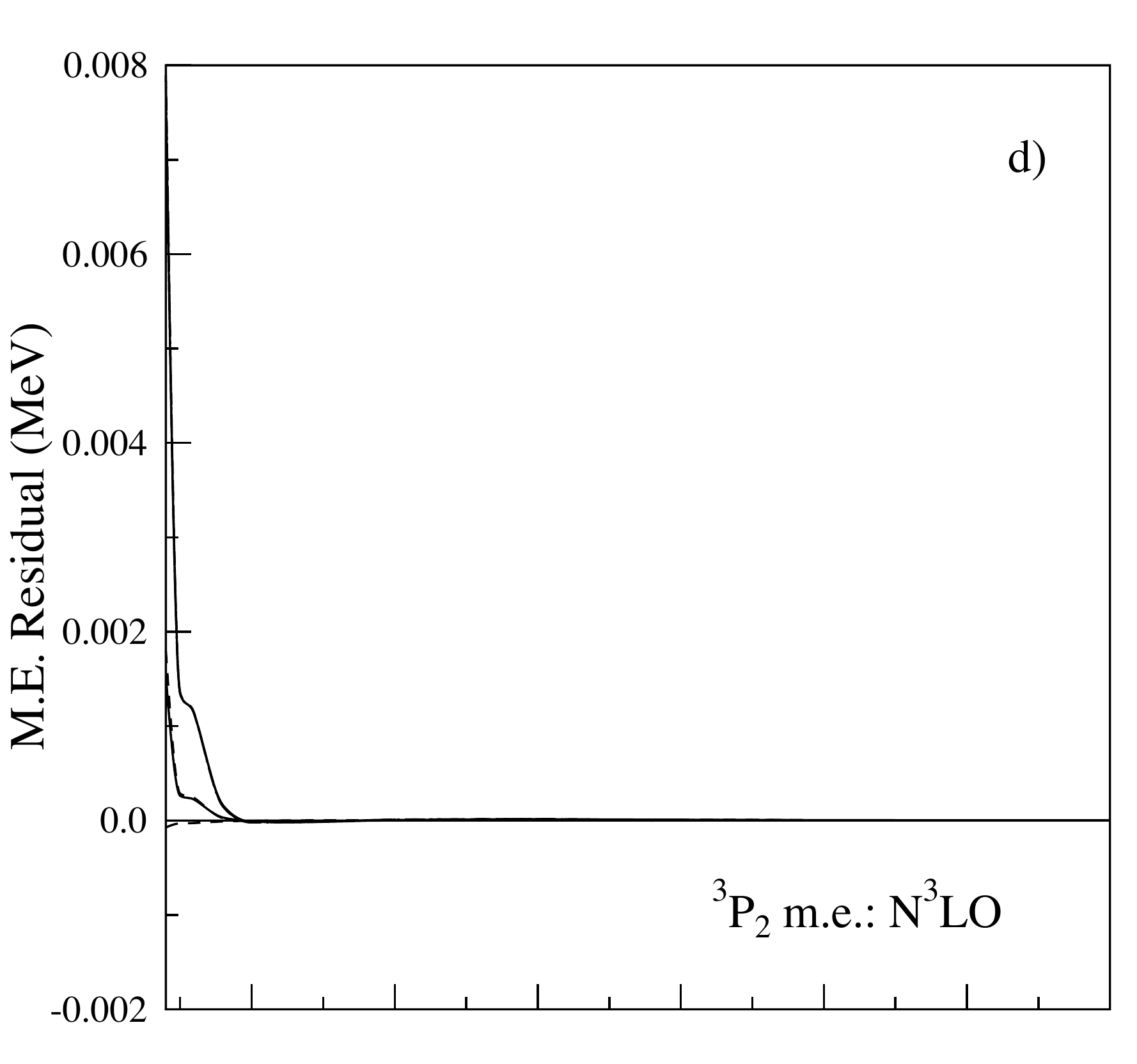}
\includegraphics[width=8cm]{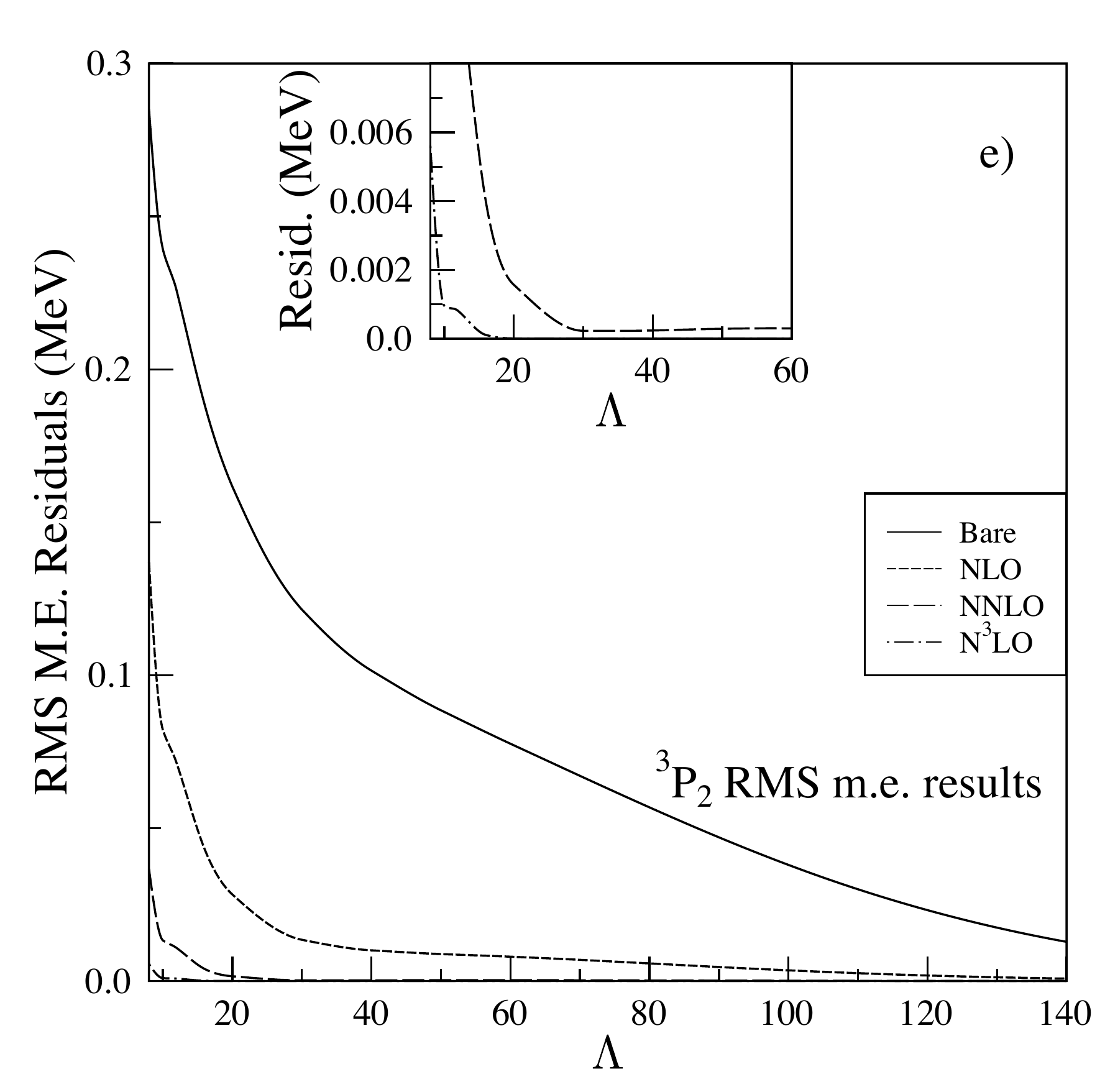}
\end{center}
\end{minipage}
\end{figure}

\begin{figure}
\begin{minipage}{0.5\linewidth}
\begin{center}
\includegraphics[width=8cm]{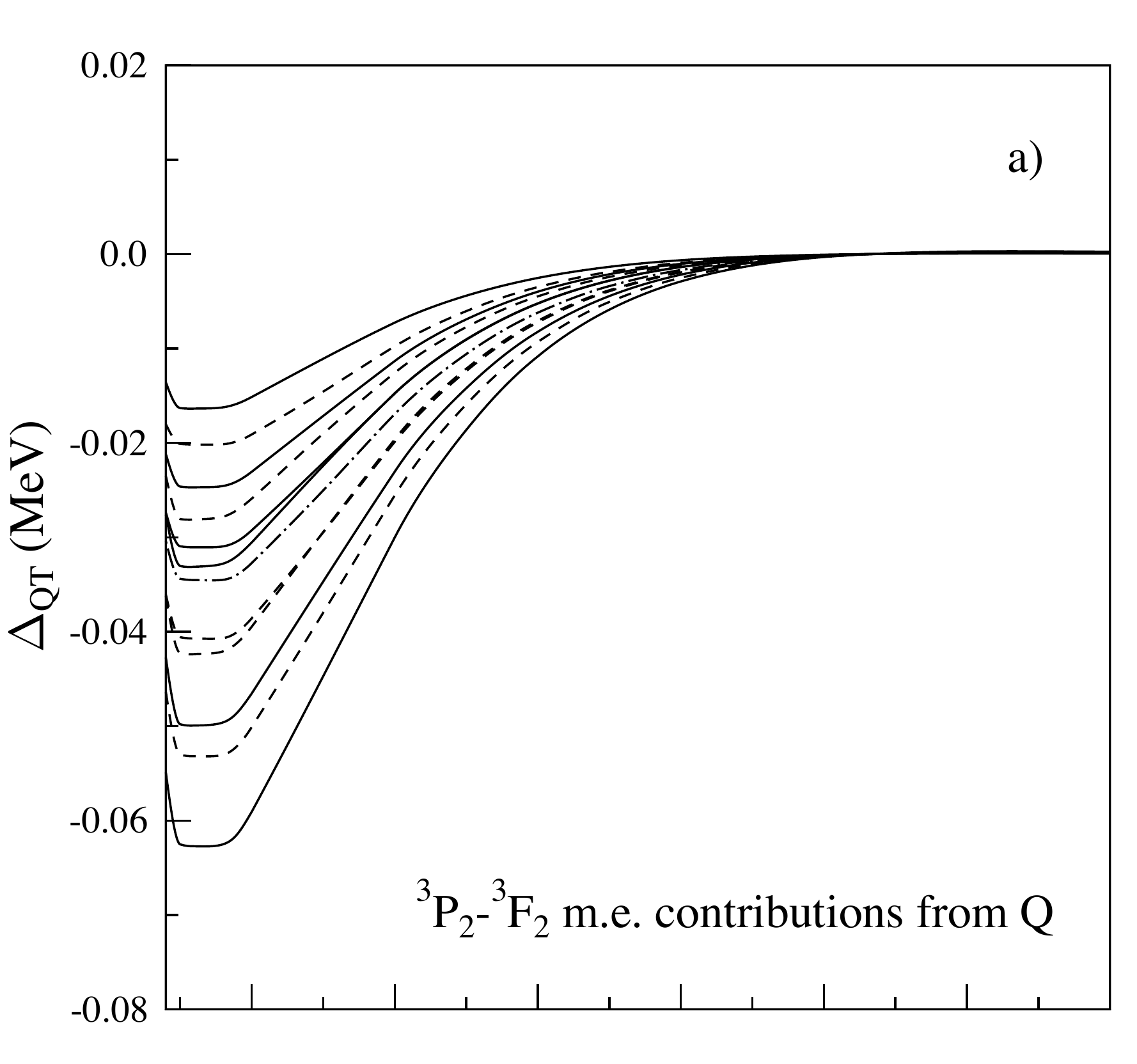}
\includegraphics[width=8cm]{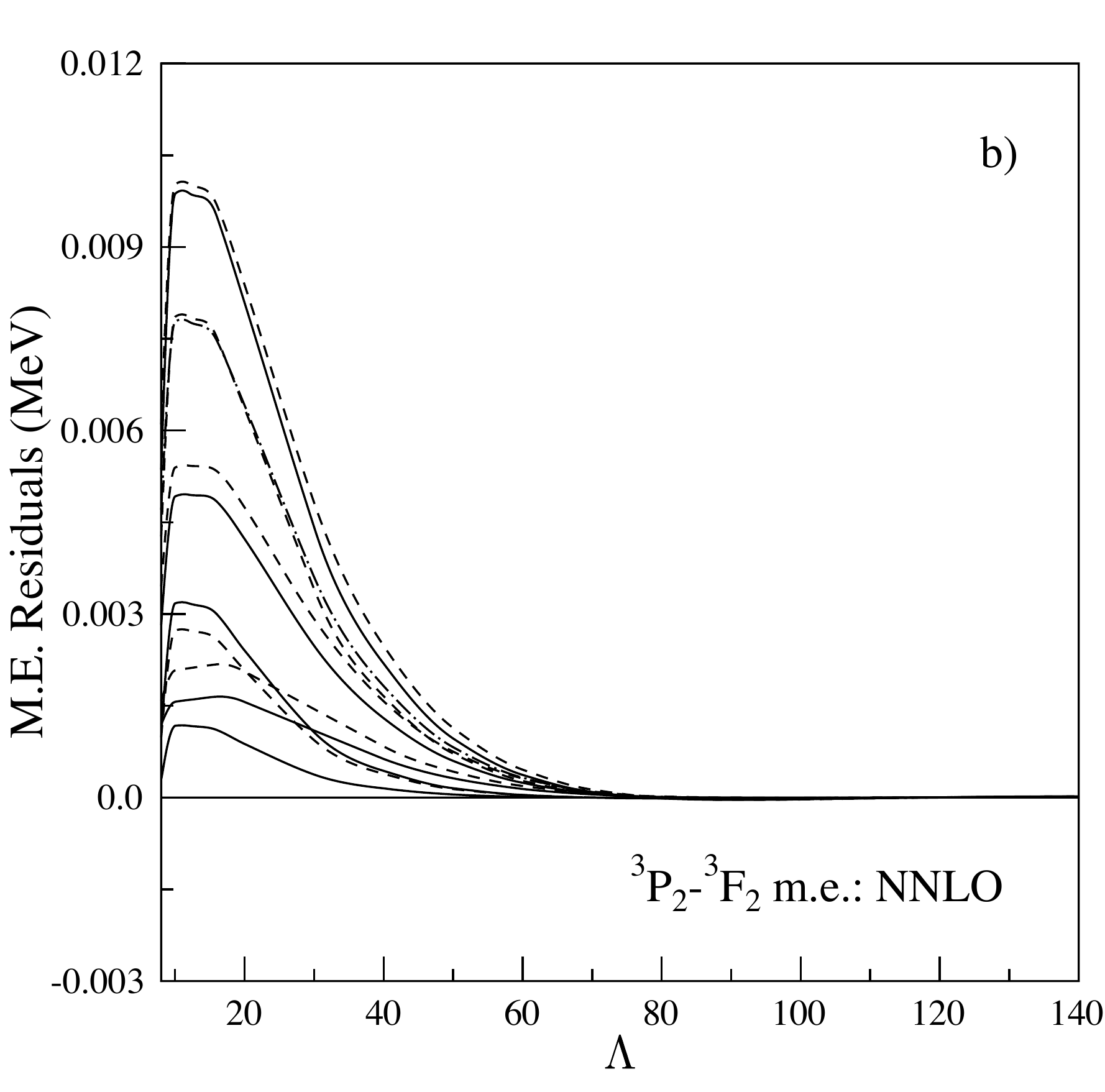}
\end{center}
\end{minipage}%
\begin{minipage}{0.5\linewidth}
\begin{center}
\includegraphics[width=8cm]{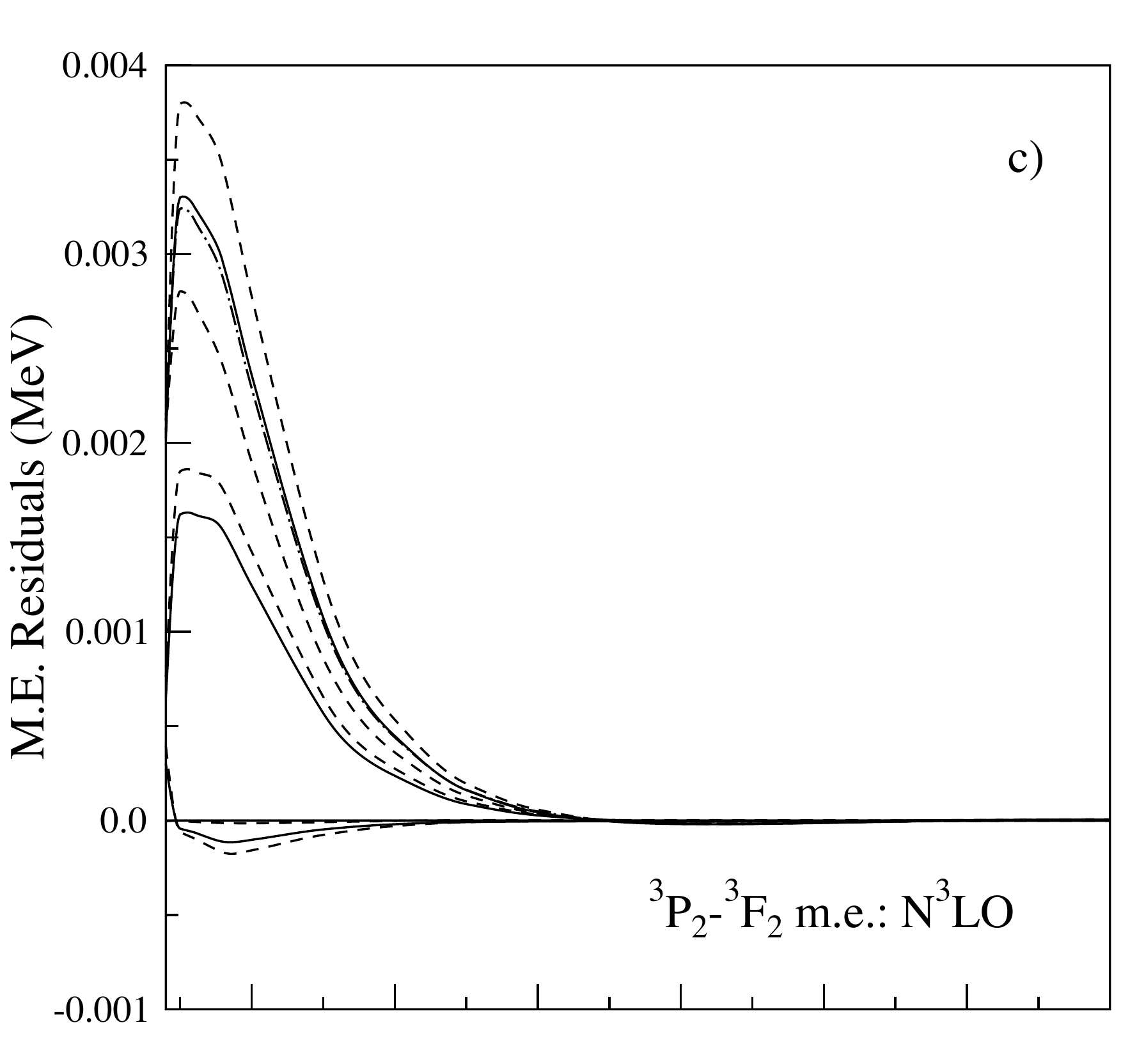}
\includegraphics[width=8cm]{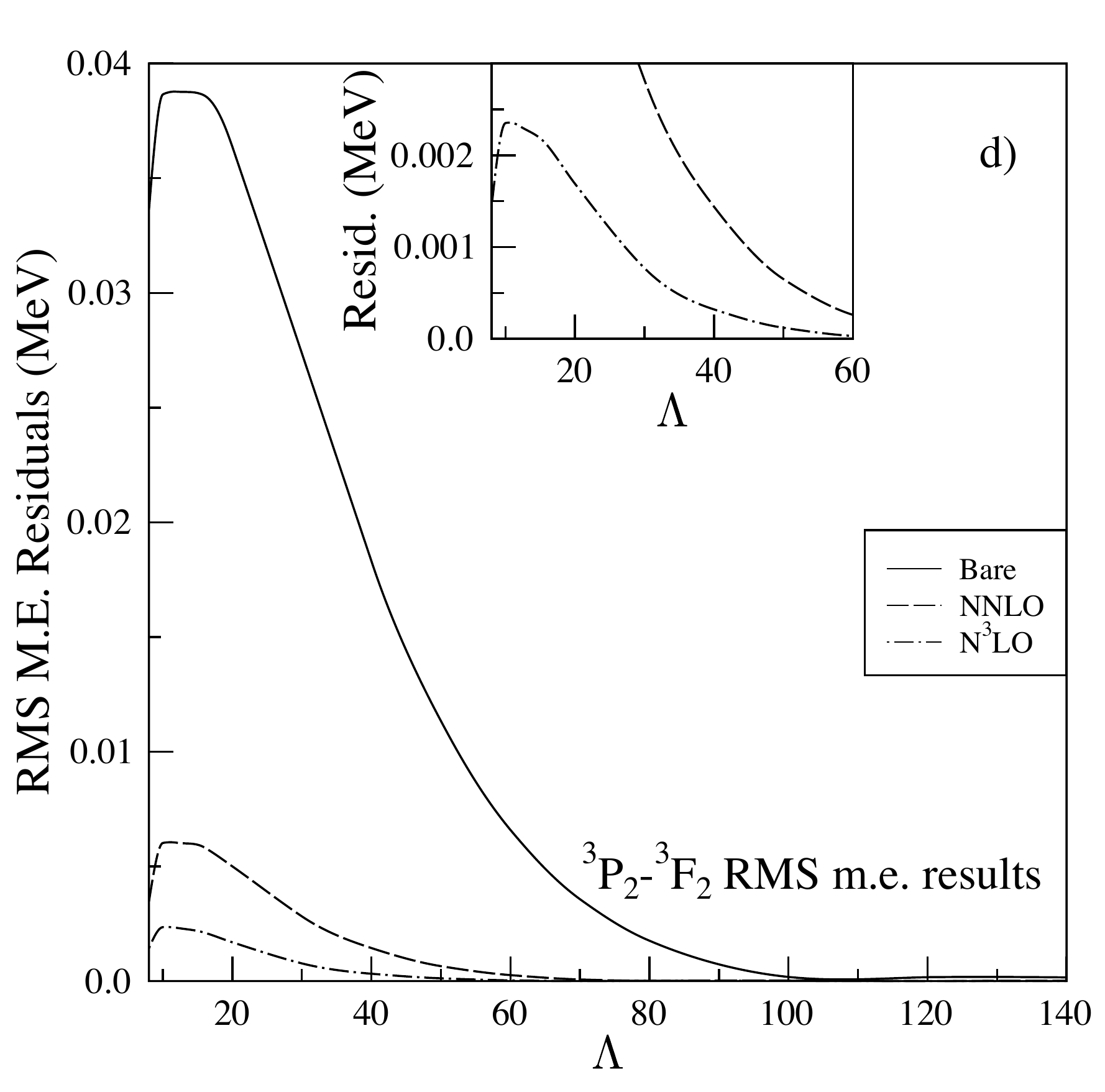}
\end{center}
\end{minipage}
\caption{As in Fig. \ref{fig_3s1},  but for the $^3P_2-{}^3F_2$ channel. }
\label{fig_pf}
\end{figure}

\begin{figure}
\begin{minipage}{0.5\linewidth}
\begin{center}
\includegraphics[width=8cm]{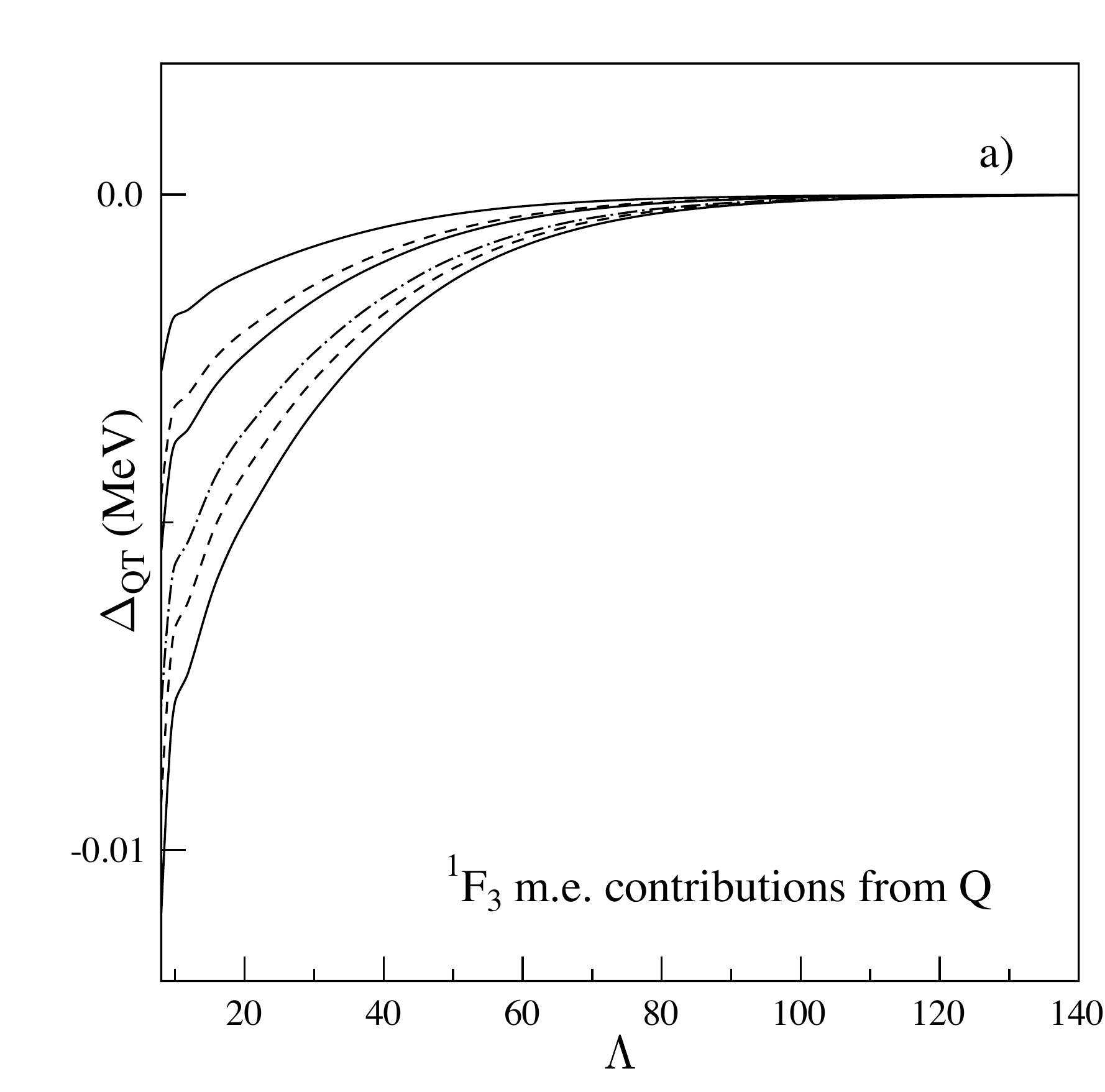}
\end{center}
\end{minipage}%
\begin{minipage}{0.5\linewidth}
\begin{center}
\includegraphics[width=8cm]{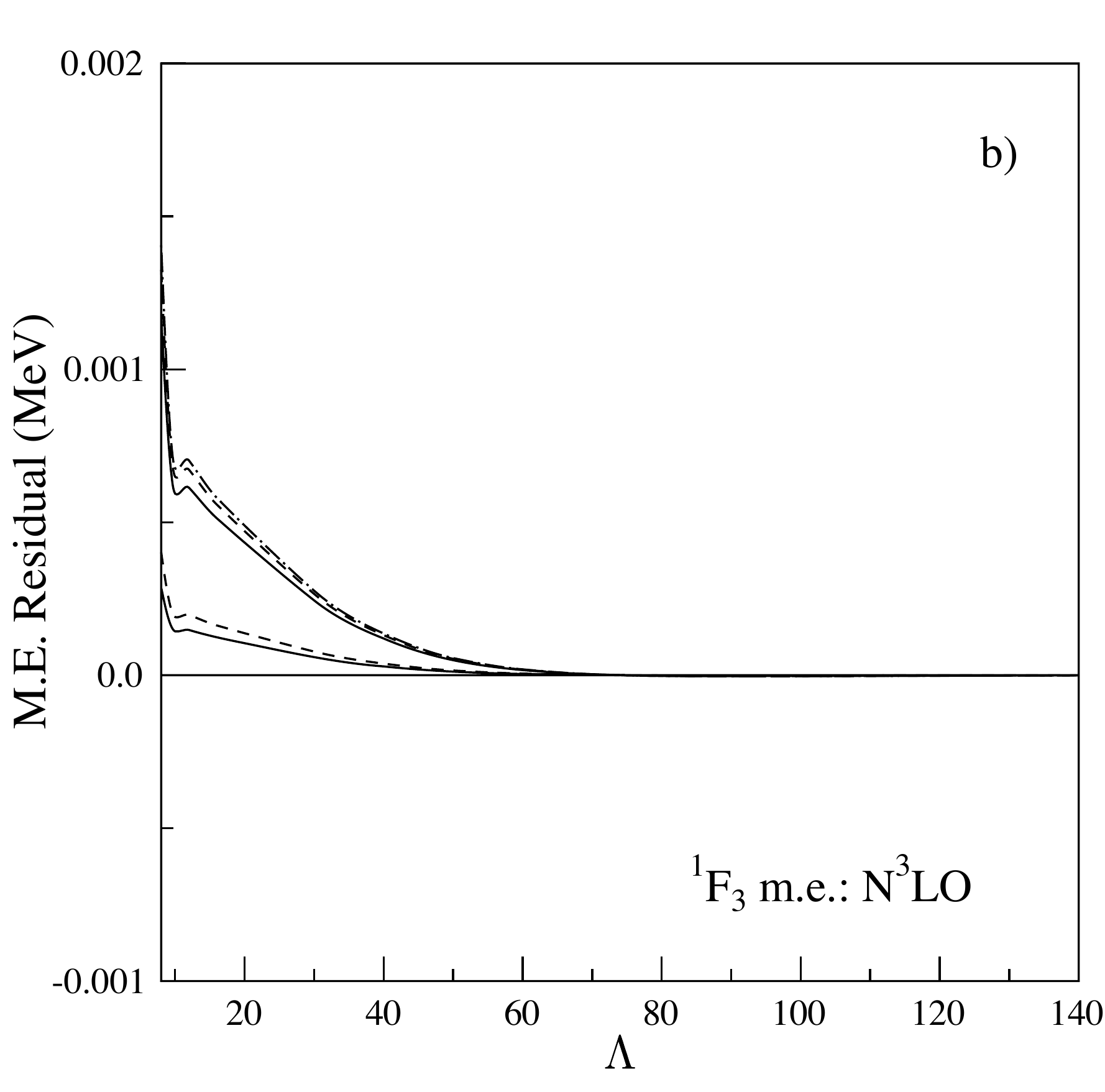}
\end{center}
\end{minipage}
\caption{The lowest contributing order to the $1F_3$ channel is N$^3$LO.  $\Delta_{QT}(\Lambda)$
and the N$^3$LO residuals for the five unconstrained matrix elements are shown. }
\label{fig_1f3}
\end{figure}

\begin{figure}
\begin{minipage}{0.5\linewidth}
\begin{center}
\includegraphics[width=8cm]{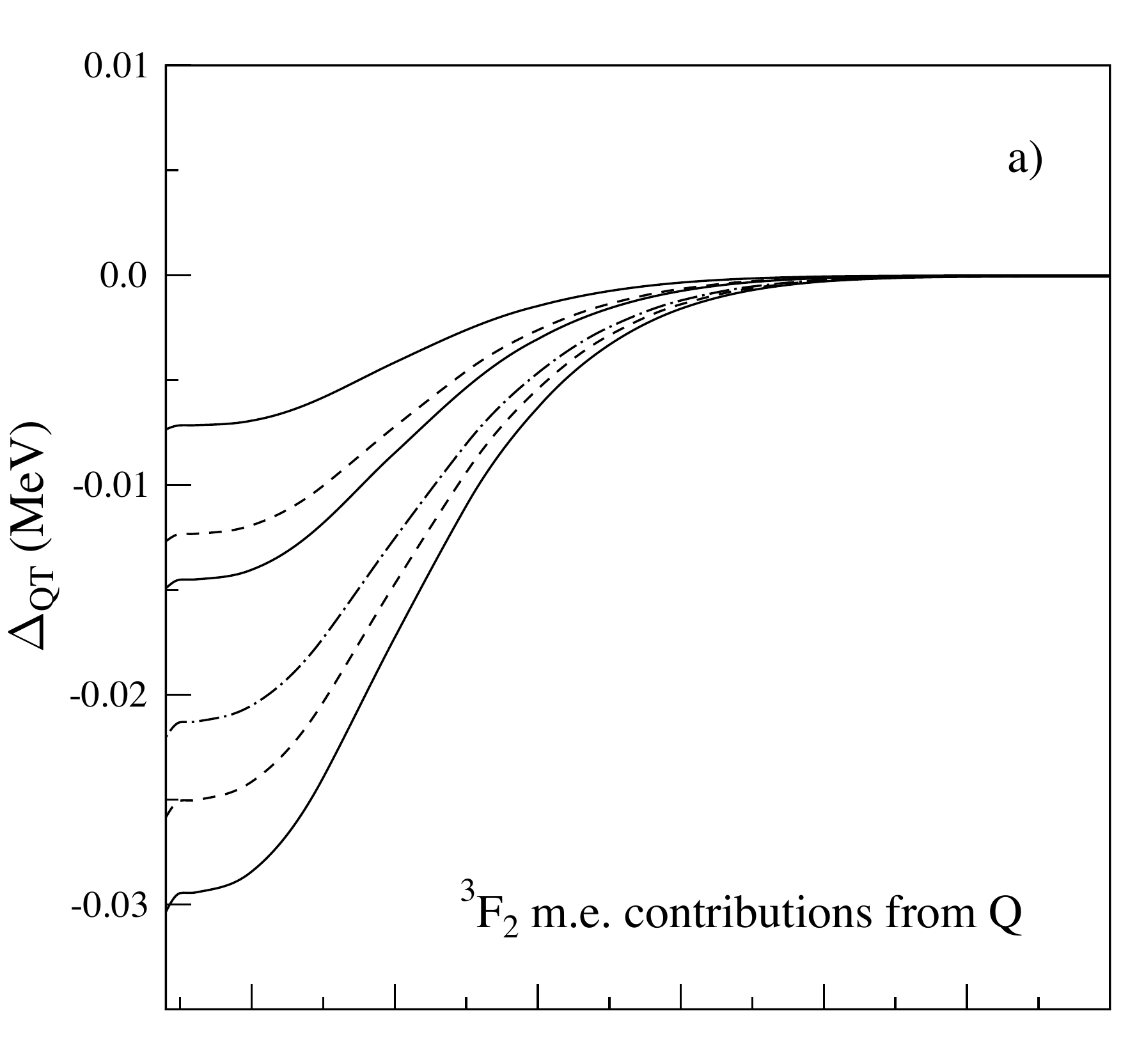}
\includegraphics[width=8cm]{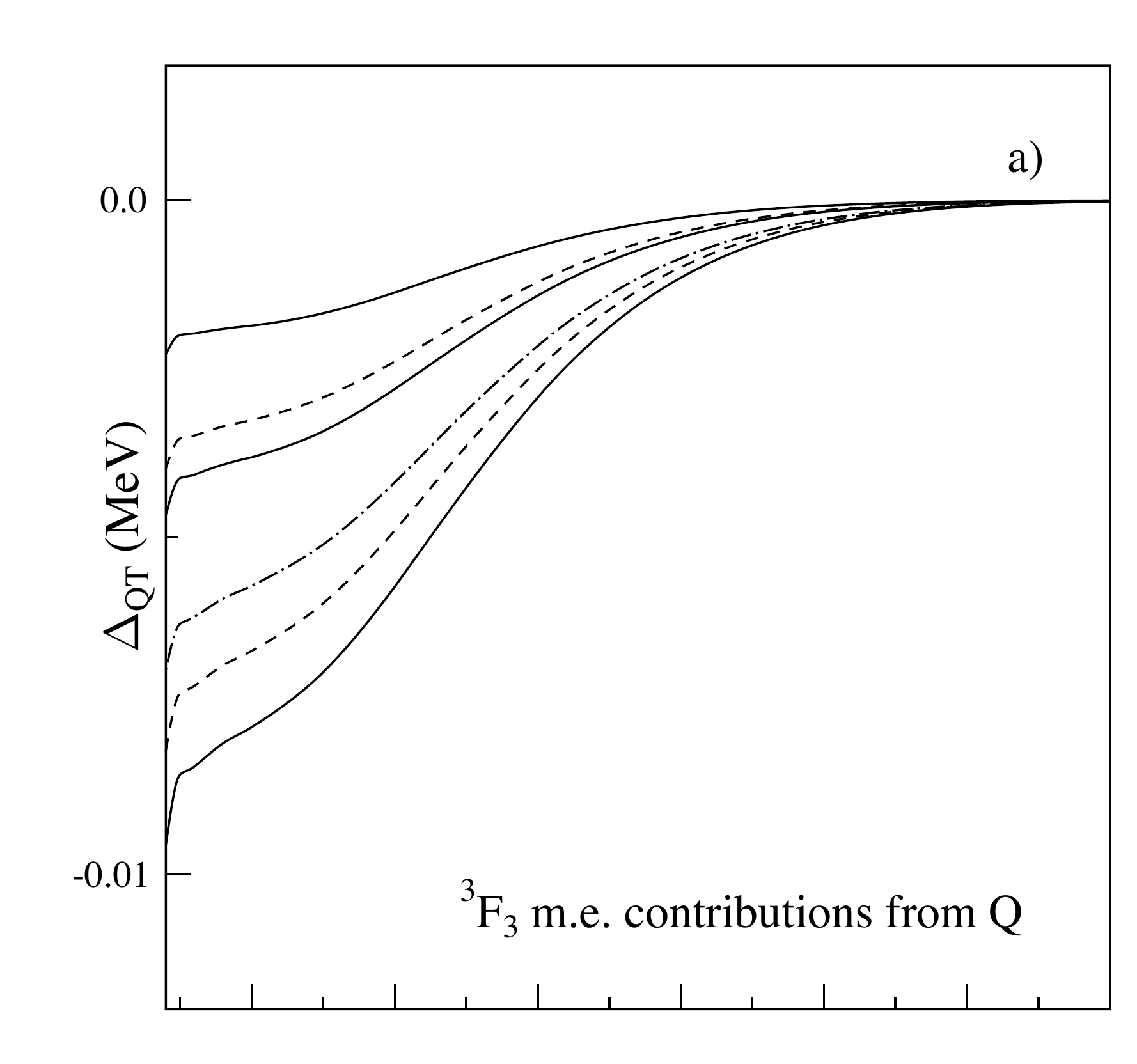}
\includegraphics[width=8cm]{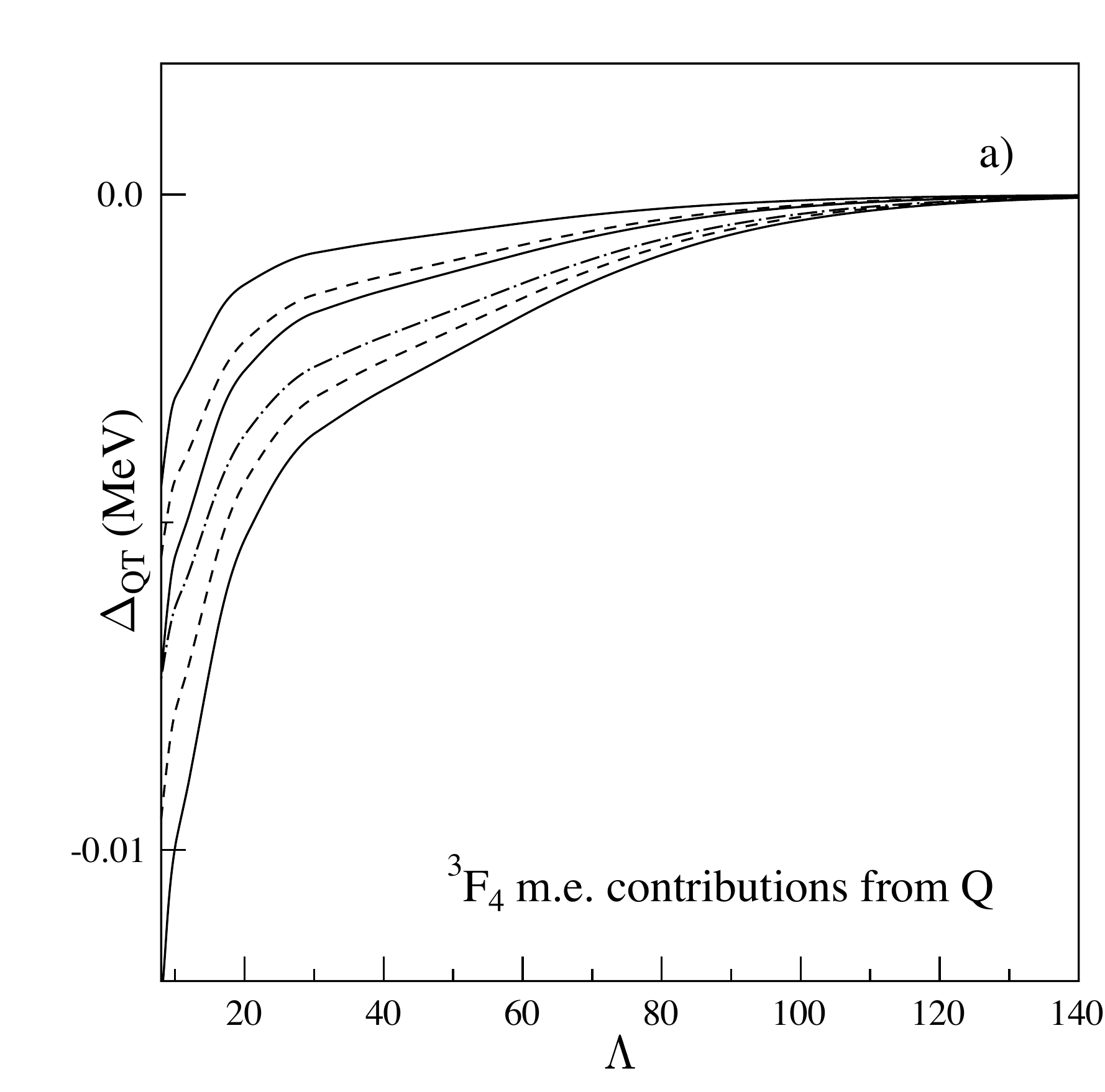}
\end{center}
\end{minipage}%
\begin{minipage}{0.5\linewidth}
\begin{center}
\includegraphics[width=8cm]{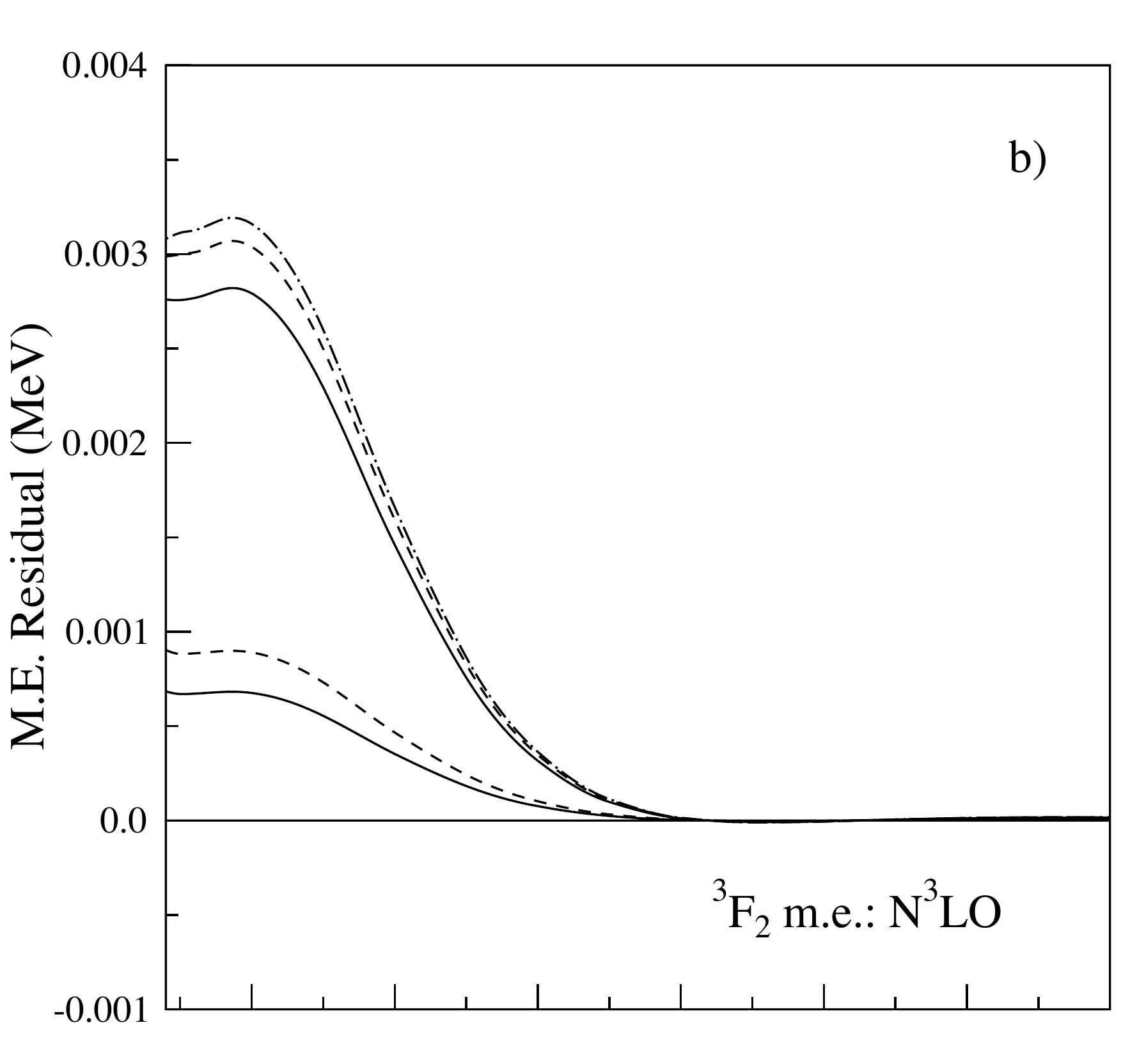}
\includegraphics[width=8cm]{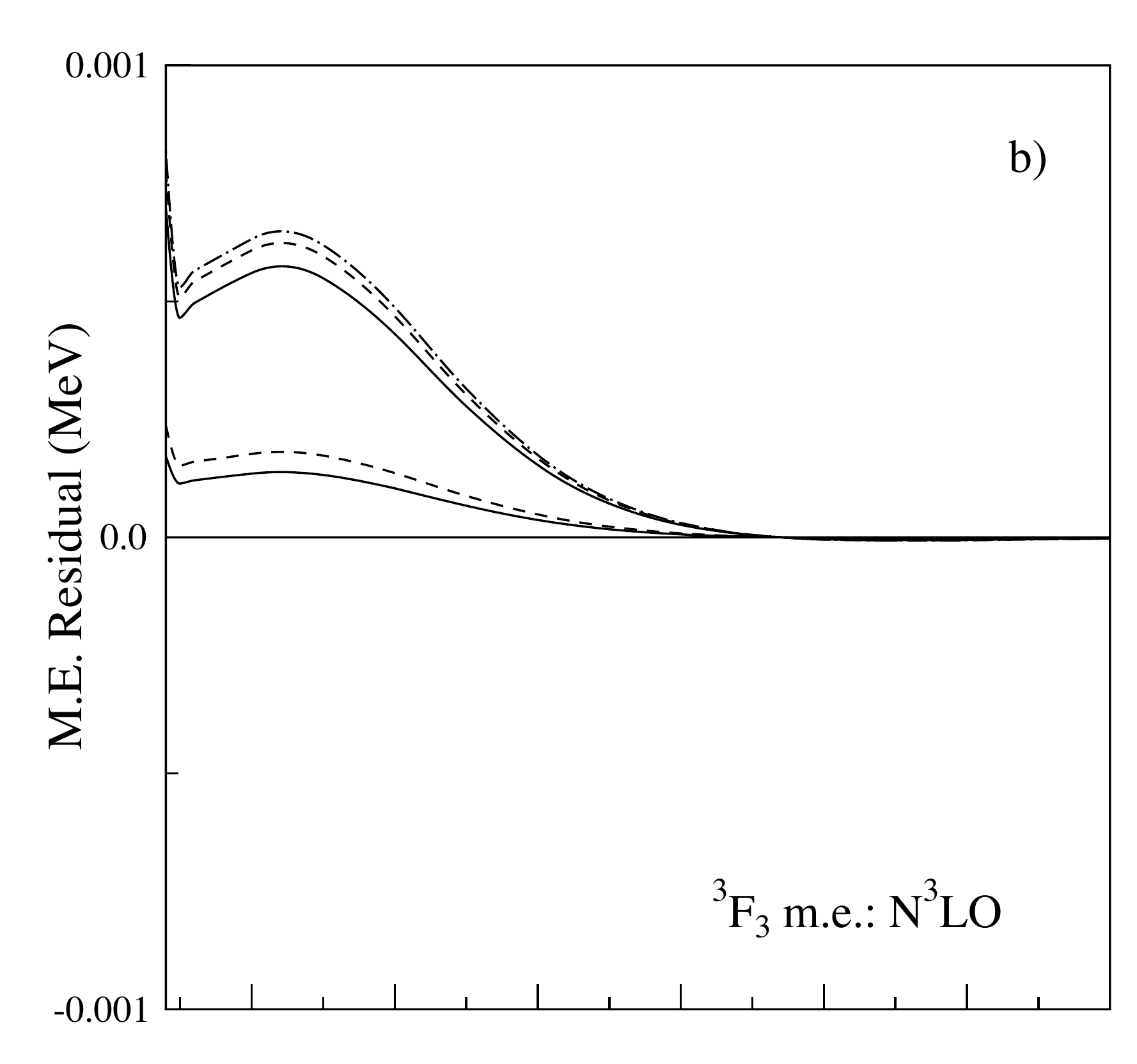}
\includegraphics[width=8cm]{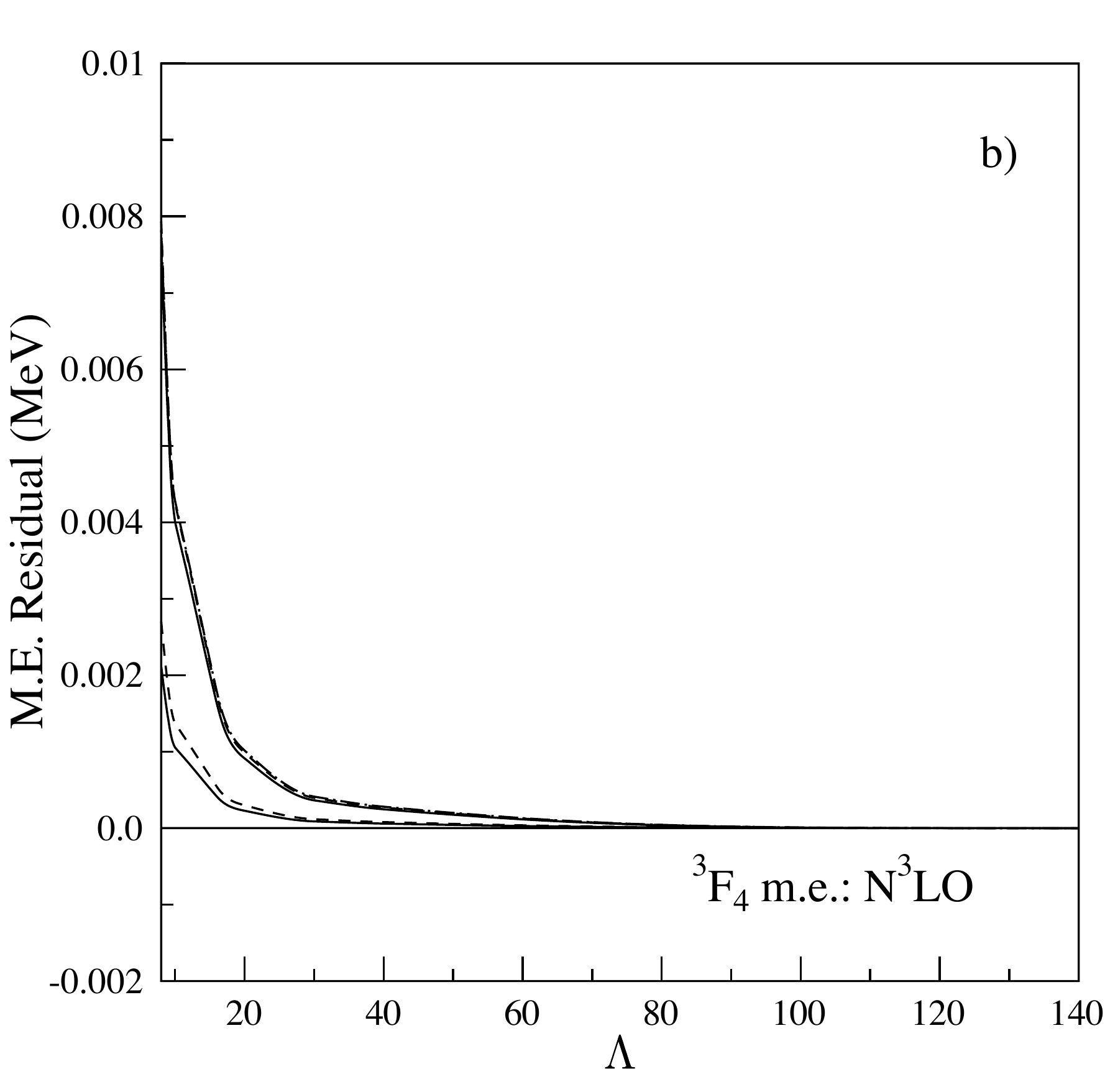}
\end{center}
\end{minipage}
\caption{As in Fig. \ref{fig_1f3},  but for the $^3F_J-{}^3F_J$ channels.  As has been noted in other
cases, the stretched $^3F_4$ case has the largest residual.}
\label{fig_3f}
\end{figure}

Figures \ref{fig_1p1} through \ref{fig_3f} show the convergence for the various channels involving
odd-parity states and contributing through N$^3$LO:
 $^1P_1-{}^1P_1$, $^3P_J-{}^3P_J$, $^3P_2-{}^3F_2$, $^1F_3-{}^1F_3$, and $^3F_J-{}^3F_J$.
While the spin-aligned channels show slightly large residuals, overall the RMS errors at
N$^3$LO are at the one-to-few keV level.  Thus a simple and essentially exact
representation for the effective interaction exists.\\

\noindent
{\it Expansion parameters, naturalness:}   The approach followed here differs from EFT,
where the formalism is based on an explicit expansion parameter, the ratio of the momentum
to a momentum cutoff.  The input into the present calculation is a set of numerical matrix
elements of an iterated, nonrelativistic potential operating in $Q$.  Potentials like $v_{18}$
are also effectively regulated at small $r$ by some assumed form, e.g., a Gaussian,
matched smoothly to the region in $r$ that is constrained by scattering data.   Thus
there are no singular potentials iterating in $Q$.

Intuitively it is clear that the convergence apparent in Table \ref{table:2} is connected
with the range of hard-core interactions (once edge states are transformed by summing $T$).
A handwaving argument can be made by assuming rescattering in $Q$ effectively
generates a potential of the form
\[ V_0 e^{-r_{12}^2/a^2}, \]
where $r_{12} =|\vec{r_1}-\vec{r_2}|$.
This ansatz is local, so there is some arbitrariness in mapping it onto contact-gradient
expansion coefficients, which correspond to the most general nonlocal potential.  But
a sensible prescription is to equate terms with equivalent
powers of $r^2$, in the bra and ket, when taking HO matrix of this potential.  Then one finds,
for S-wave channels
\begin{equation}
a(m',m) \equiv a_{N^{m'+m}LO}^{S,2m'2m} = {1 \over 4^{m'+m} m'! m!} {(2m'+2m+1)!! \over
(2m'+1)!! (2m+1)!!} V_0 \left[ {\pi a^2 \over a^2+ 2b^2} \right]^{3/2} \left[ a^2 \over a^2 + 2b^2 \right]^{m'+m}
\end{equation}
where $b$ is the oscillator parameter.  (The notation is such that, e.g., $a(m'=3,m=0)=
a_{N^3LO}^{S,60}$.)  The last term is thus the expansion parameter:
if the range of the hard-core physics residing in $Q$ is small compared to the natural nuclear
size scale $b$, then each additional order in the expansion should be suppressed by
$\sim (a/b)^2$.

One can use this crude ansatz to assess whether the convergence shown in Table \ref{table:2}
is natural, or within expectations.   The LO and NLO $^1S_0-{}^1S_0$ results effectively determine
$V_0$ and $a$; thus the  strengths of four NNLO and N$^3$LO potentials
can be predicted relative to that of $a_{LO}$ and $a_{NLO}$.  The predicted hierarchy
$a_{LO}:a_{NLO}:a_{NNLO}^{22}:a_{NNLO}^{40}:a_{N^3LO}^{42}:
a_{N^3LO}^{60}$ of 
\[ 1:6.3 \times 10^{-3}:6.7\times 10^{-5}:2.0\times 10^{-5}:3.0\times 10^{-7}:4.2\times 10^{-8} \]
matches the relative strengths of the couplings in the table quite well,  
\[1:6.3 \times 10^{-3}:6.4\times 10^{-5}:3.9\times 10^{-5}:2.1\times10^{-7}:5.7\times 10^{-8} ,\]
including qualitatively reproducing the ratios of the two NNLO and two N$^3$LO 
coefficients.  The parameters derived from $a_{LO}$ and $a_{NLO}$ are $a \sim 0.39$ fm
and $V_0 \sim -1.5$ GeV.  In the $^1S_0$ channel the bare Argonne $v_{18}$ potential at small $r$
can be approximated by a Gaussian with $a \sim 0.33$ fm and $V_0 \sim 3.0$ GeV.  So again
the crude estimates of range and even the strength are not unreasonable.  [Note that the signs of the
two $V_0$s are correct -- the $P$-space lacks the appropriate short-range repulsion and thus
samples the iterated bare potential at small $r$, a contribution that then must be subtracted
off when $H^{eff}$ is evaluated.]

\begin{figure}
\begin{minipage}{0.5\linewidth}
\begin{center}
\includegraphics[width=8cm]{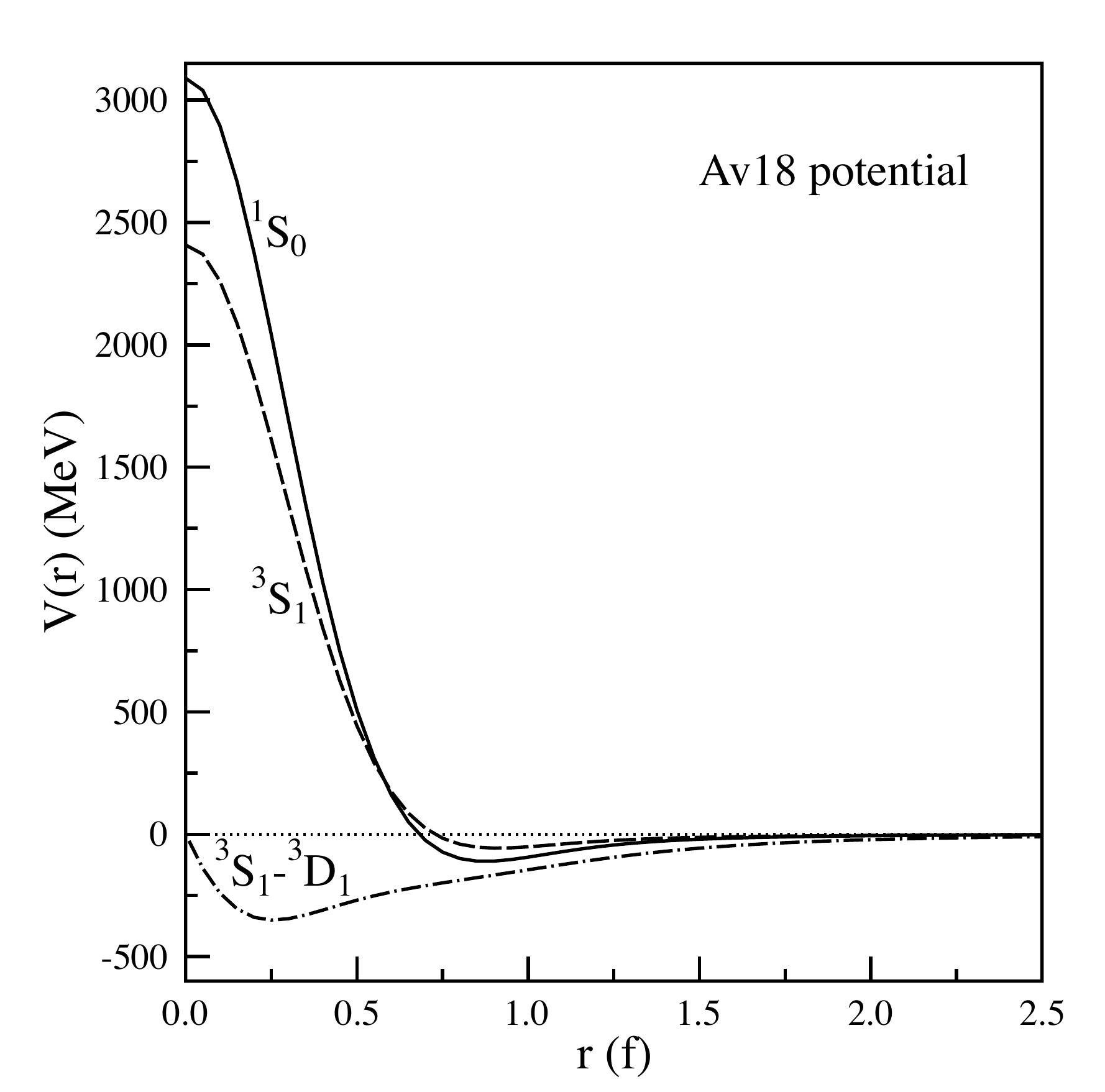}
\end{center}
\end{minipage}%
\begin{minipage}{0.5\linewidth}
\begin{center}
\includegraphics[width=8cm]{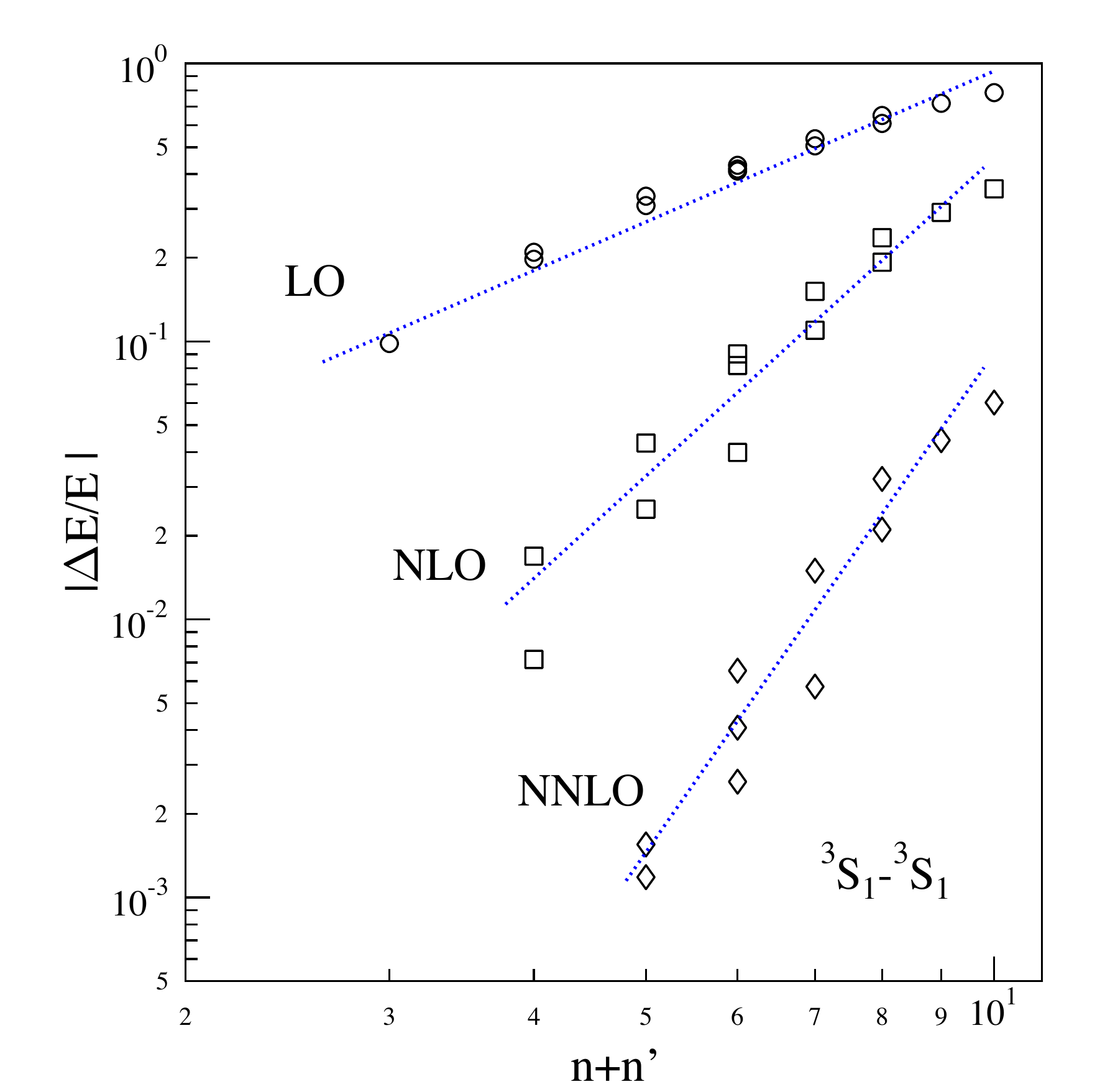}
\end{center}
\end{minipage}
\caption{(Color online) The left panel shows the radial dependence of the Argonne $v_{18}$ potential in the $^1S_0-{}^1S_0$,
$^3S_1-{}^3S_1$, and $^3S_1-{}^3D_1$ (tensor) channels.  The last is clearly more extended.
The right panel is a ``Lepage plot" displaying fractional errors as a function of the order of the
calculation, on log scales.  The steepening of the slope with order is the sign of a well behaved, converging effective theory.}
\label{fig_av18}
\end{figure}

A similar exercise in the $^3S_1$ channel yields the predicted hierarchy
\[ 1:2.2 \times 10^{-2}:8.3\times 10^{-4}:2.5\times 10^{-4}:13.1\times 10^{-6}:1.9\times 10^{-6} \]
which compares with the coupling ratios calculated from Table \ref{table:2}
\[ 1:2.2 \times 10^{-2}:8.8\times 10^{-4}:1.9\times 10^{-4}:9.3\times 10^{-6}:2.3\times 10^{-6}. \]
The convergence is very regular but slower: in this case the effective Gaussian parameter
needed to describe these trend is $a \sim 0.75$ fm.  The overall strength, $V_0 \sim -0.42$,
differs substantially from that found for the $^1S_0$ channel, though the underlying
$v_{18}$ potentials for $^3S_1-{}^3S_1$ and $^1S_0-{}^1S_0$ scattering are quite similar
(see Fig. \ref{fig_av18}).

The $^3S_1-{}^3S_1$ behavior is similar to that found in the other spin-aligned channels,
such as $^3D_3-{}^3D_3$ and $^3P_2-{}^3P_2$, where the scattering in $Q$ includes
contributions from the tensor force.  The tensor force contributes to the LO s-wave coupling
through intermediate $D$-states in $Q$, e.g., 
\[ \langle n' l'=0 | V_{SD}Q | n'' l=2 \rangle {1 \over \langle E \rangle} \langle n'' l=2 |Q V_{SD} |n l=0\rangle, \]
as the product of two tensor operators has an s-wave piece.  The radial dependence
of $V_{SD}$ for $v_{18}$, shown in Fig. \ref{fig_av18}, is significantly 
more extended than in central-force $^3S_1-{}^3S_1$ and $^1S_0-{}^1S_0$ cases.  This has the
consequences that (1) the mean excitation energy $\langle E \rangle$ for $^3S_1-{}^3D_1$ will
be lower (enhancing the importance of the tensor force) and (2) the 
$P$-space $\langle ^3S_1 | H^{eff} | ^3S_1 \rangle$ matrix element will reflect the extended
range.

Once this point is appreciated -- that the effective expansion parameter are naturally 
channel-dependent because of effects like the tensor force -- the results shown in Table
\ref{table:2} are very pleasing:
\begin{itemize}
\item  In each channel the deduced couplings $a_{LO},~a_{NLO},~a_{NNLO},~a_{N^3LO},...$
evolve in a very orderly, or natural, fashion: one can reliably predict the size of the
next omitted term.  The convergence appears related to an effective range characterizing
scattering in $Q$.
\item The convergence varies from channel to channel, but this variation reflects underlying
physics, such as role of the tensor force, governing the channel's range.
One does not find, nor perhaps should one expect to find,
some single parameter $p/\Lambda$ to characterize convergence independent of channel.
\item  The convergence is very satisfactory in all channels: the measure used in Table
\ref{table:2}, $\langle$Resid.$\rangle_{RMS}$, is an {\it exceedingly} conservative one,
as discussed below.  But even by by this standard, in only one channel ($^3D_3-{}^3D_3$)
do the RMS residual discrepancies among unconstrained matrix elements exceed
$\sim$ 10 keV.   Given the arguments above, it is perfectly sensible to work to order NNLO
in rapidly-converging channels like $^1S_0-{}^1S_0$ and N$^4$LO in slowly converging
channels like $^3D_3-{}^3D_3$.   As noted in the table, at N$^4$LO the residual in the
$^3D_3-{}^3D_3$ channel is reduced to 22 keV.
\end{itemize}

\noindent
{\it Convergence and the ``Lepage" plot:}   The procedure often followed in an
effective theory is to use information about the low-lying excitations to parameterize an
effective Hamiltonian, which is then used to predict properties of other states near the
ground state.  In contrast, the goal here has been to characterize the entire effective 
interaction to high accuracy.   As described below, the residual errors in the procedure
are typically dominated by matrix elements with the largest $n$ and $n^\prime$, corresponding
to minor components in the deuteron ground state, for example.   The difference in the
deuteron binding energy using exact matrix elements of $H^{eff}$ 
versus using the N$^3$LO expansion is quite small ($\sim$ 40 eV).

Order-by-order improvement should be governed by
nodal quantum numbers.  For example, in LO in the $^3S_1$ channel 
the omitted NLO term would be
\begin{equation}
-{8 a_{NLO}^{3S1} \over \pi^2} (n'+n-2) \left[ {\Gamma[n^\prime+1/2] \Gamma[n+1/2]
\over (n^\prime-1)! (n-1)!} \right]^{1/2}  \stackrel{n^\prime, n~\mathrm{large}}{\longrightarrow}
-{2a_{NLO}^{3S1} \over \pi^2} \left[(4 n^\prime-1)(4n-1)\right]^{1/2} (n^\prime+n-2)
\end{equation}
Thus the fractional error associated with the omission of the NLO terms relative to LO should
to be linear in the sum of the nodal
quantum numbers, if the expansion is capturing the correct physics.
That is, the expected
absolute (e.g., in keV) error for ($n^\prime,n$)=(5,5) would be about 16 times that for (1,2).

In higher orders this distinction between large and small $n$ grows.  At LO+NLO, the
expected fractional errors in 
matrix elements from omitted NNLO terms would be quadratic in $n$ and $n'$: the explicit functional
dependence is no longer simple as there are two NNLO operators, and one would not know
{\it a priori} the relevant quadratic combination of $n^\prime$ and $n$ governing the
error.   At NNLO the fractional error would be a cubic polynomial in 
$n$ and $n^\prime$.

While beyond LO the expected fractional errors have a dependence on both $n$ and $n^\prime$,
it is still helpful to display results as a 2D ``Lepage plot"  using $n+n^\prime$ -- proportional to the
average $\langle p^2 \rangle$ of bra and ket -- as the variable.  Such a plot makes clear whether 
improved fits in an effective theory are systematic -- that is, due to a correct description
of the underlying physics, not just additional parameters.
The use of a single parameter, $n+n^\prime$, of course maps multiple matrix
elements onto the same $x$ coordinate, when the ET indicates this is a bit too simple
beyond LO.  Nevertheless,  the right panel in Fig. \ref{fig_av18} still shows rather nicely 
that the nuclear 
effective interactions problem is a
very well behaved effective theory.   In LO the residual errors do map onto the
single parameter $n+n^\prime$ to very good accuracy,
and the residual error is linear.   The steepening of
the convergence with order is consistent with the expected progression from linear to
quadratic to cubic behavior in nodal quantum numbers. 
By NNLO, errors in unconstrained matrix elements for
small $n+n^\prime$ are tiny, compared to those with high $n+n^\prime$.  That is, the expansion
converges most rapidly for matrix elements between long-wavelength states, as it should.  
However, improvement is substantial and systematic everywhere, including at the 
largest $n+n^\prime$.

\section{Properties and Energy Dependence of the  Effective Interaction}
The results of the previous section demonstrate the existence of a simple systematic
operator expansion for the HOBET effective interaction.  Its behavior order-by-order
and in the Lepage plot indicates that the short-wavelength physics is being 
efficiently captured in the associated operator coefficients.

The error measure used in the  N$^3$LO fit is
dominated by the absolute errors in matrix elements involving
the highest nodal quantum numbers: these matrix elements are large even though they
may not play a major role in determining low-lying eigenvalues.
(It might have been better to use the fractional error
in matrix elements, a measure that would be roughly independent of $n^\prime$ and $n$.)
Other possible measures of error are the 
ground state energy; the first energy-moment of the effective interaction matrix (analogous
to the mean eigenvalue in the SM); the fluctuation between neighboring 
eigenvalues of that matrix (analogous to the level spacing in the SM); and the overlap of
the eigenfunctions of that matrix with the exact eigenfunctions (analogous to wave function
overlaps in the SM).  
The N$^3$LO interaction in the coupled $^3S_1-{}^3D_1$ channel
produces a ground state energy accurate to $\sim$ 40 eV;
a spectral first moment accurate to 1.81 keV; an RMS average deviation in the level spacing 
of 3.52 keV; and wave function overlaps that are unity to better than four significant digits.
As rescattering in $Q$ contributes  $\sim$ -10 MeV to eigenvalues, the accuracy of the
N$^3$LO representation of the effective interaction is, by these spectral measures, on the order
of 0.01\%.
As the best excited-state techniques in nuclear physics currently 
yield error bars of about 100 keV for the lightest nontrivial nuclei, this representation of the
two-body effective interaction is effectively exact \cite{argonne,nocore}.

The approach requires one to sum $QT$ to all orders, producing a result that
depends explicitly on $|E|$ -- which in this context should be measured relative to the
first breakup channel.  While the associated effects increase with decreasing $|E|$, it will
be shown later that the renormalization is substantial
even for well-bound
nuclear states.
The deuteron is definitely not an extreme case.   The effects are also sensitive to
the choice of $P$, through $b$, which
controls the mean momentum within $P$ -- a small $b$ reduces the missing
hard-core physics, but exacerbates the problems at long wavelengths, and
conversely.   Figure \ref{fig_1} suggests factor-of-two changes
in the $Q$-space contribution to the deuteron binding energy can result from
$\sim$ 20\% changes in $b$.   At the outset, the dependence on $|E|$ and $b$ seems
like a difficulty for 
nuclear physics, as modest changes in these parameters alter predictions.

One of the marvelous properties of the HO is that the $QT$ sum can be done.   
The two effects discussed above turn out to be
governed by a single parameter, $\kappa$.  The associated effects are nonperturbative
in {\it both} $QT$ and $QV$.  In the case of $QT$ an explicit sum to all orders is
done.  The effects are also implicitly
nonperturbative in $QV$, because of the dependence on $|E|$.  This is why the BH
approach is so powerful: because $|E|$ is determined self-consistently, it is simple
to incorporate this physics directly into the iterative process (which has been shown to
converge very rapidly in the HOBET test cases  A=2 and 3).
When this is done, one finds that $\kappa$ affects results in three ways:
\begin{itemize}
\item the rescattering of $QT$ to all orders, $T (E-QT)^{-1} QT$, is absorbed into a
new ``bare" matrix element
$\langle \alpha | T | \widetilde{\beta}(\kappa) \rangle$;
\item the new ``bare" matrix element $\langle \widetilde{\alpha}(\kappa) | V | \widetilde{\beta}(\kappa) \rangle$
captures the effects of $QT$ in all orders on the contribution first-order in $V$; and
\item the matrix elements of the short-range operators $\bar{O}$, which contain all the multiple scattering of $QV$,  are similarly modified, $\langle \widetilde{\alpha}(\kappa) | \bar{O} | \widetilde{\beta}(\kappa) \rangle$.
\end{itemize}

So far the discussion has focused on the problem of a single bound state of fixed
binding energy $|E|$, the deuteron ground state.  No discussion has occurred
of expectations for problems in which multiple
bound states, each with a different $H^{eff}(|E|)$, might arise.   But 
1) the dependence of $H^{eff}(|E|)$ on $\kappa$ arises already in the single-state case, which
was not {\it a priori} obvious; and 2) state dependence (energy dependence in the
case of BH) must arise in the case of multiple states, as this is the source of the
required nonorthogonality of states when restricted to $P$, a requirement for a
proper effective theory.  So a question clearly arises about the connection between
the explicit $\kappa$ dependence found for fixed $|E|$, and the additional 
energy dependence that might occur for a spectrum of states.

Because other techniques, like Lee-Suzuki, have been used to address problem 2),
it is appropriate to first stress the relationship between $\kappa$ and the strong interaction
parameters provided in Table \ref{table:2}.  The choice $\Lambda_P$=8 is helpful, as
it shows there is no relation.  Every short-range coefficient arising through order N$^3$LO
was determined from nonedge matrix elements: the fitting procedure matches the
coefficients to the set of matrix elements with $n^\prime+n \leq 5$, and there are no edge
states satisfying this constraint.  Nothing in the treatment of the strong interaction ``knows"
about edge states.  This then makes clear how efficiently $\kappa$ captures
the remaining missing physics.  Without $\kappa$ one would have, in the contact-gradient
expansion to
N$^3$LO, a total of 78 poorly reproduced edge-state matrix elements, 10 of which
would be $S$-state matrix elements with errors typically of several MeV.  With $\kappa$ --
a parameter nature (and the choice of $b$) determines -- all of the 78 matrix elements are 
properly reproduced,
consistent with the general $\sim$ keV accuracy of the N$^3$LO description of $H^{eff}$. 

Suppose someone were to prefer an $H^{eff}$ free of any dependence on $|E|$, again in the 
context of an isolated state of energy $|E|$.  Could this be done?  Yes, but at the cost
of a cumbersome theory that obscures the remarkably simple physics behind the
proper description of the edge state matrix elements.  Suppose one wanted merely
to fix the five $^3S_1-{}^3S_1$ edge state matrix elements, those where 
$n^\prime=5$ couples to $n$=1, 2, 3,4, 
and 5.  One could introduce operators corresponding to the coefficients
\[ a_{N^4LO}^{S,80},~a_{N^5LO}^{S,82},~a_{N^6LO}^{S,84},~a_{N^7LO}^{S,86},~a_{N^8LO}^{S,88} \]
to correct these matrix elements.  It is clear all five couplings would be needed -- that's
the price one would pay for mocking up long-range physics (a long series of high-order
Talmi integrals) with a set of short-range operators of this sort. 

This would be a rather poorly motivated exercise: 
\begin{itemize}
\item  The problems in these matrix elements have nothing to do with high-order generalized
Talmi integrals of the strong potential, as was demonstrated
in the previous section.
\item  This approach does not  ``heal" the effective theory:  the poor running of 
matrix elements would remain.  There would be no systematic improvement, for all
matrix elements, as a function of $\Lambda$, as one progresses from LO, to NLO, etc.
The five parameters introduced above would remove the numerical discrepancies 
at $\Lambda_P$, but not fix the running as a function of $\Lambda$, even for just
the edge-state matrix elements.
\item  This approach amounts to parameter fitting, in contrast to the systematic
improvement demonstrated in the Lepage plot.   The parameter $a_{N^4LO}^{S,80}$
introduced to fix the $n=1$ to $n=5$ matrix element will not properly correct
the $n=2$ to $n=5$ matrix element, as the underlying physics has nothing to do
with the $r_1^8 r_2^0$-weighted Talmi integral of any potential.
\item  If $\Lambda_P$ is increased, the number of such edge-state matrix elements that will
need to be corrected by the fictitious potential increases.  
This contrasts with the approach where $|E|$ is
explicitly referenced: there the number of short-range coefficients needed to
characterize $Q$ will decrease (that is, the LO, NLO, ... expansion becomes more
rapidly convergent), while $\kappa$ remains the single parameter 
governing the renormalization of those coefficients for edge-state matrix elements.
\end{itemize}

While these reasons are probably sufficient to discard any such notion of building
a $\kappa$-independent $H^{eff}$, consider now the consequences of changing
$b$ -- which after all is an arbitrary choice.   The short-range coefficients in Table
\ref{table:2} will change: there is an underlying dependence on $QV(\vec{r}_{12}/b)$.  This governs
natural variations in the coefficients -- one could estimate those variations based on some
picture of the range of multiple scattering in $Q$, as was done in the ``naturalness"
discussion.  But there would be additional changes in the ratios of
edge to nonedge matrix elements, reflecting the changes in $\kappa$.  This would 
induce in any $\kappa$-independent potential unnatural evolution in $b$.  That is, the
fake potential would look fake, as $b$ is changed.

The arguments above apply equally well to the case of the
state-dependence associated with techniques like Lee-Suzuki.  To an accuracy of about
95\%, the $\kappa$-dependence isolated in $H^{eff}$ is {\it also} the state-dependence 
that one encounters when $|E|$ is changed.  This is a lovely result: the natural 
$\kappa$-dependence that is already present in the case of a short-range
expansion of $H^{eff}$ for a fixed state, also gives us ``for free" the BH state-dependence.  The result 
is not at all surprising, physically: changes in $|E|$ will alter the balance between $QT$ and
$QV$, and that is precisely the physics that was disentangled by introducing $\kappa$.
Mathematically, it is also not surprising: changing $b$ at fixed $|E|$ alters $\kappa$,
just as changes in $|E|$ for fixed $b$ would.  Thus all 
of the $QT$ effects identified above, in considering a single state, must also arise 
when one considers spectral properties.

This argument depends on showing that other, implicit energy dependence in $H^{eff}$
is small compared in the explicit dependence captured in $\kappa$.   Such implicit
dependence can reside in only one place, the fitted short-range coefficients.

\subsection{Energy Dependence}
The usual procedure for solving the BH equation,
\[ H^{eff} = H + HQ {1 \over E-QH} QH ,\]
 involves steps to ensure self-consistency.  As
the energy appearing in the Green's function is the energy of the state being calculated,
self-consistency requires iteration on this energy until convergence
is achieved: an initial guess for $E$ yields an $H^{eff}(E)$ and thus an eigenvalue $E^\prime$,
which then can be used in a new calculation of the interaction $H^{eff}(E^\prime)$.  
This procedure is iterated
until the eigenvalue coresponds to the energy used in calculating $H^{eff}$.  In practice,
the convergence is achieved quite rapidly,
typically after about five cycles.

As the BH procedure produces a Hermitian $H^{eff}$, this energy dependence is 
essential in building into the formalism the correct relationship between the $P$-space
and full-space wave functions, that the former are the restrictions of the latter (and
thus cannot form an orthonormal set).   This relationship
allows the wave function to evolve smoothly to the
exact result, in form and in normalization, as $\Lambda_P \rightarrow \infty$. 

Generally this energy dependence remains implicit because the BH equation is solved
numerically: one obtains distinct sets of matrix elements  $\langle \alpha | H^{eff}(E_i) | \beta \rangle$ 
for each state $i$, but the functional dependence on $E_i$ is not immediately apparent.  
But that is not the case in the present treatment, where an
analytic representation for the effective interaction has been obtained.  

While significant energy-dependent effects governed by $\kappa$ have been isolated,
additional sources remain in the case of a spectrum of bound states.  The identified energy-dependent terms are
\begin{itemize}
\item $\langle \alpha | T + {E \over E-QT} QT | \beta \rangle = \langle \alpha |T|\widetilde{\beta}(\kappa) \rangle$;
\item $\langle \alpha | {E \over E-TQ} V {E \over E-QT} | \beta \rangle = \langle \widetilde{\alpha}(\kappa) |V|\widetilde{\beta}(\kappa) \rangle$; and
\item $\langle \alpha | {E \over E-TQ} \bar{O} {E \over E-QT} | \beta \rangle = \langle \widetilde{\alpha}(\kappa) |\bar{O}|\widetilde{\beta}(\kappa) \rangle$
\end{itemize}
The implicit energy dependence not yet isolated resides in the coefficients of the contact-gradient
expansion,
\begin{itemize}
\item $\langle \alpha | V {E \over E-QH} QV | \beta \rangle = \langle \alpha | \bar{O}(E)|\beta \rangle$.
\end{itemize}
To isolate this dependence, one must repeat the program that was executed for the deuteron ground state at a variety
of energies, treating $H^{eff}(|E|)$ as a function of $|E|$.  The resulting variations in the extracted coefficients
will then determine the size of the implicit energy dependence.  Of course, all of the explicit energy
dependence is treated as before, using the appropriate $\kappa.$

\begin{figure}
\begin{minipage}{0.5\linewidth}
\begin{center}
\includegraphics[width=9cm]{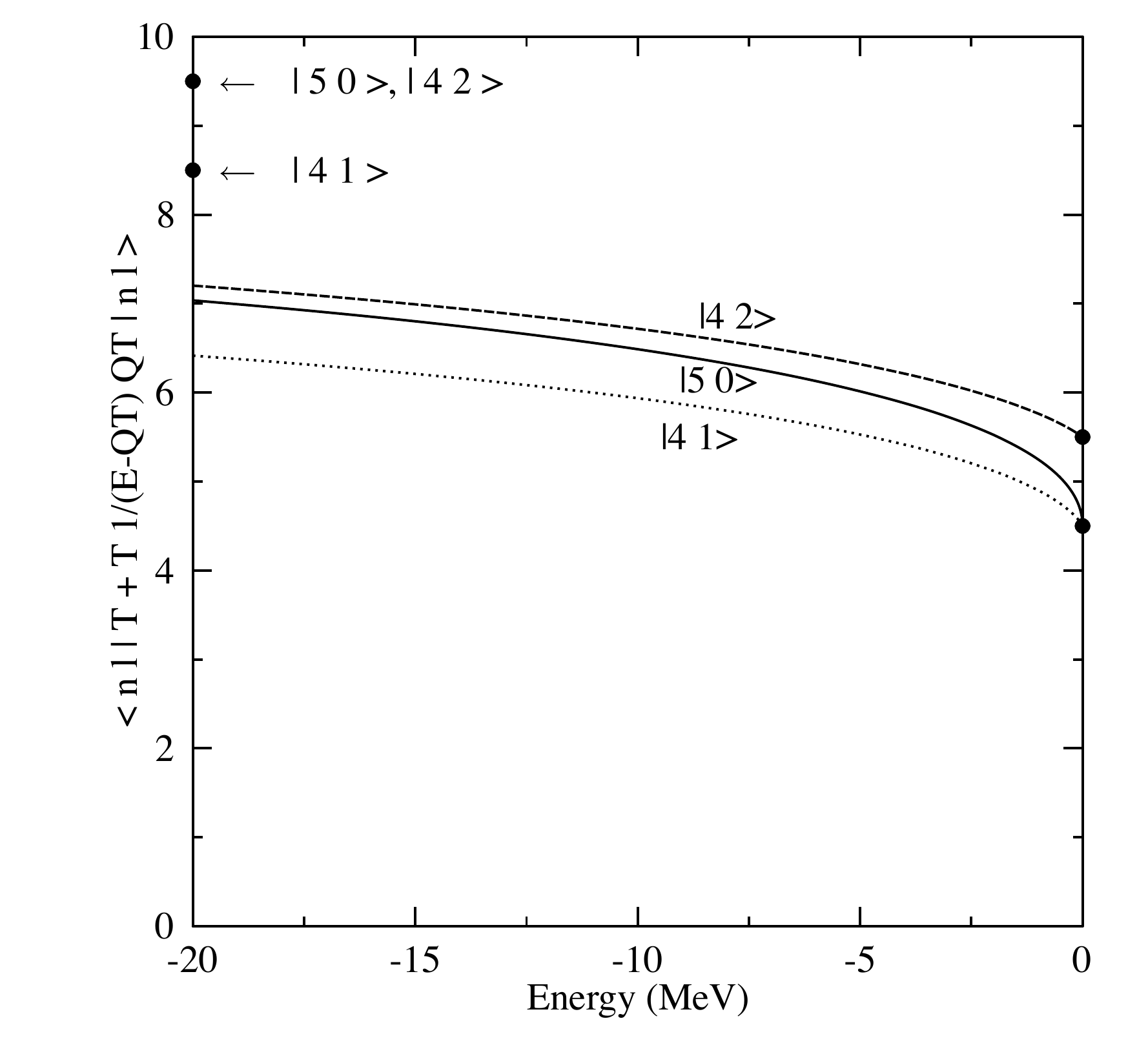}
\includegraphics[width=9cm]{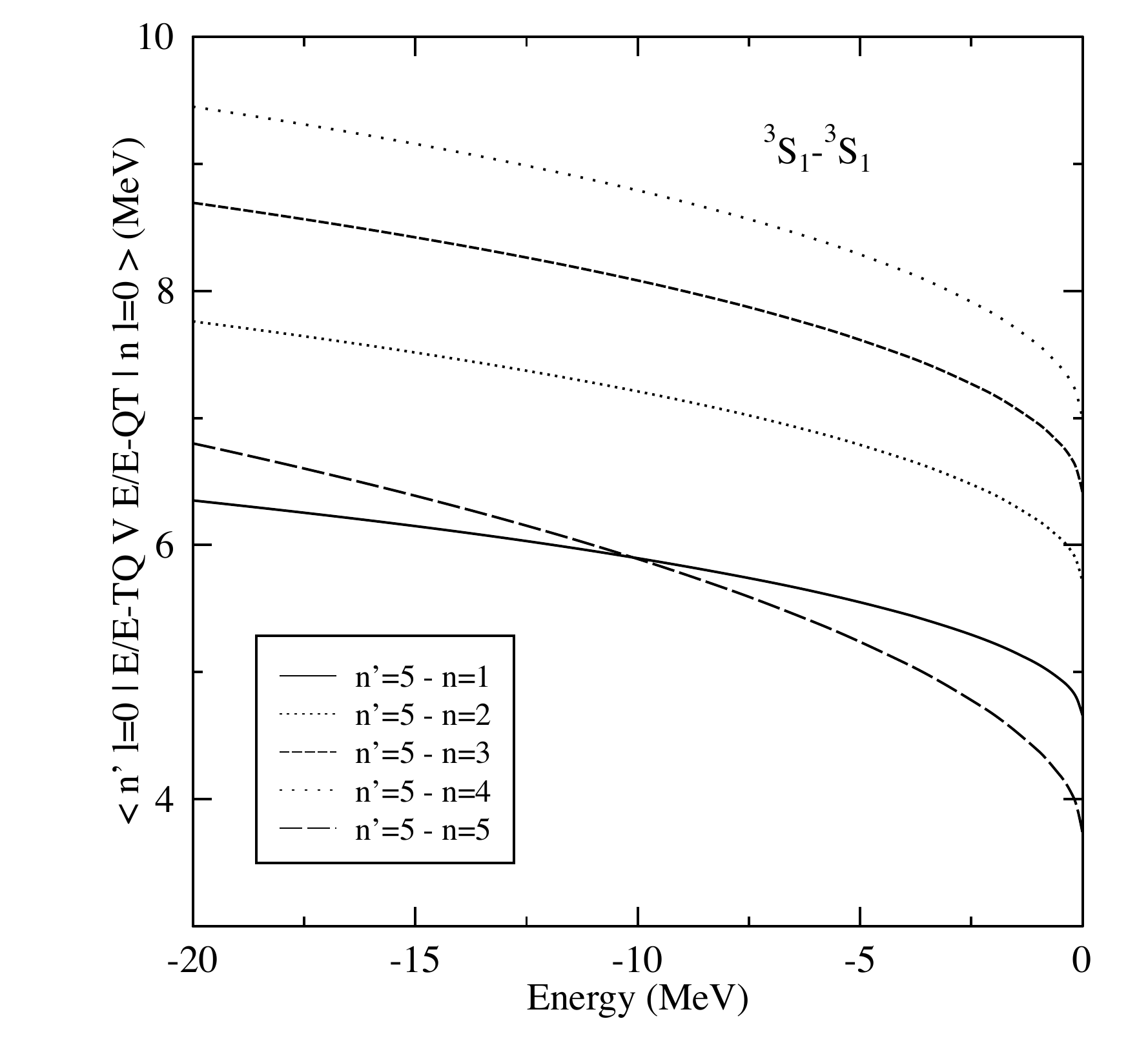}
\end{center}
\end{minipage}%
\begin{minipage}{0.5\linewidth}
\begin{center}
\includegraphics[width=9cm]{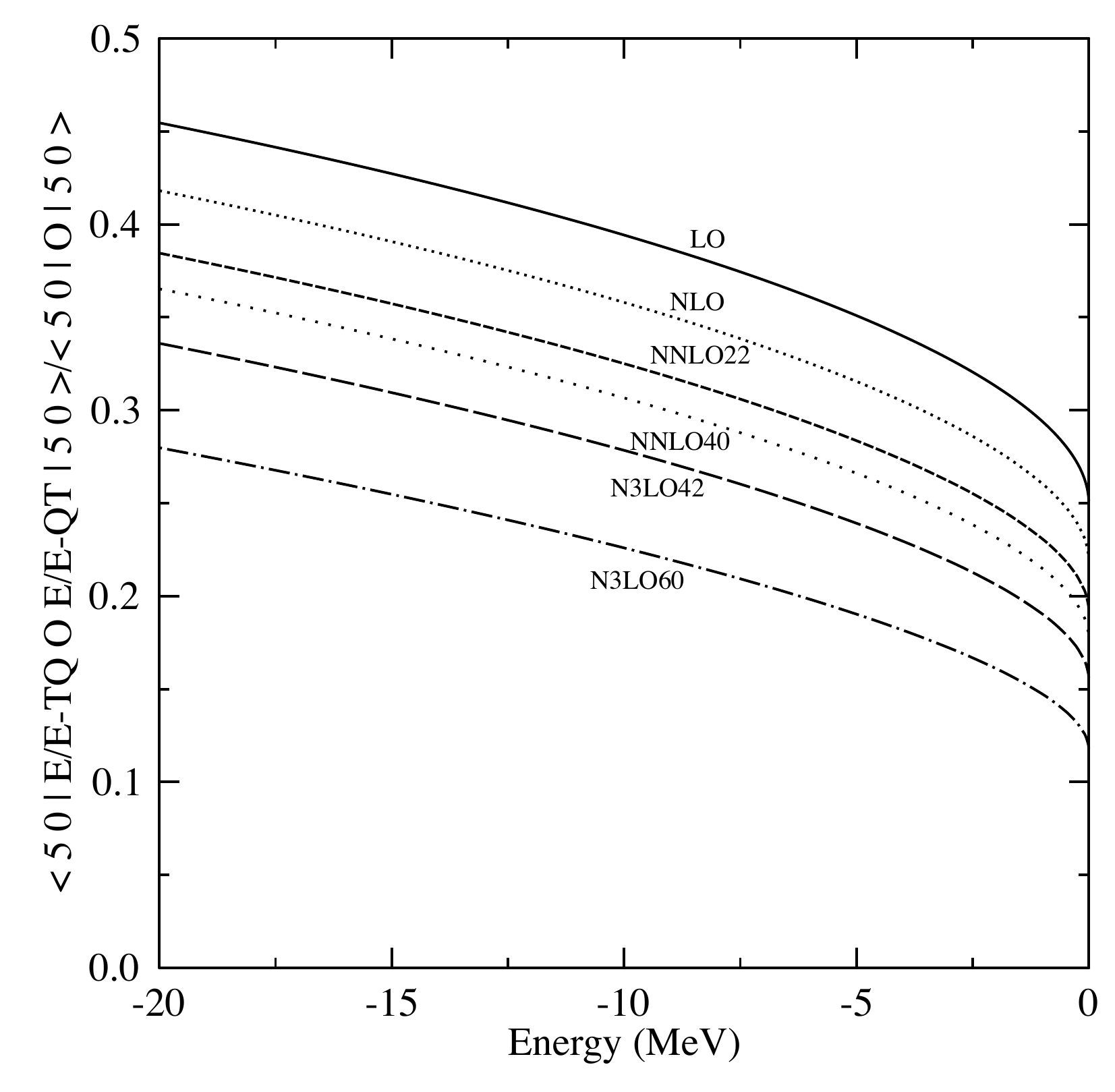}
\end{center}
\caption{Contributions to $H^{eff}$ with explicit energy dependence, for $P$ defined
by $\Lambda_P=8$ and $b=1.7$ fm.
The upper left panel shows the diagonal ``bare" kinetic energy term $\langle \alpha | T |
\widetilde{\alpha} \rangle$ for the edge states $| \alpha \rangle = | n=5~ l=0 \rangle$, 
$| n=4 ~l=1 \rangle$, and $| n=4~ l=2 \rangle$.  The dots indicate the limiting values
for very large and very small binding energies.  The kinetic energy plotted is dimensionless, 
given in terms of  $\hbar \omega/2$.  The lower left panel gives the matrix elements
of the bare potential $V$ between $^3S_1$ edge states, as a function of $E$.
The  upper right panel shows the evolution of the
quantities $a_{LO}^\prime(E;n',l',n,l)/a_{LO}(n',l',n,l)$, $a_{NLO}^\prime(E;n',l',n,l)/
a_{NLO}(n',l',n,l)$, etc., through N$^3$LO for the diagonal matrix element with $| n=5~ l=0 \rangle.$
The general softening of such matrix elements is apparent, for small binding energy -- repeated
scattering by $T$ through high-energy oscillator states in $Q$ 
spreads the wave function and thus reduces the effects of the strong
potential at short range.  This effect is carried by the edge states, because their renormalization
is affected by the missing long-range physics.  See the text for further discussion.}
\label{fig_edep}
\end{minipage}
\end{figure}

The simplest of the explicit terms is the ``bare" kinetic energy \[ \langle n^\prime l |  T | \widetilde{n} \widetilde{ l}(\kappa) \rangle \equiv \langle n^\prime l | T + T {1 \over E-QT} QT |n l \rangle = \langle n^\prime l | T | n l  \rangle + {\hbar \omega \over 2} \delta_{n^\prime n} \sqrt{n(n+l+1/2)} ~\widetilde{g}_1(-\kappa^2;n,l). \]
where effects only arise in the double-edge-state case.   Two limits define the range of
variation.  As $|E| \rightarrow \infty$, $\widetilde{g}_1 \rightarrow 0$, so
the edge-state matrix element takes on its bare value, $(2n+l-1/2) \hbar \omega/2$.
Similarly one can show $\widetilde{g}_1(-\kappa^2;n,l) \rightarrow n$ as the binding energy $|E|$
approaches zero.  Thus for small binding, the matrix element approaches $(n+l-1/2) \hbar \omega/2$.
Thus the range is a broad one, $n \hbar \omega/2$, about 35 MeV for the parameters used in
this paper.  The behavior between these limits
can be calculated.  The results over 20 MeV in binding are shown in the upper left
panel of  Fig. \ref{fig_edep} for $S$, $P$, and $D$ states.  One finds that even deeply
bound (E=-20 MeV) states have very significant corrections due to $QT$: the scattering in $Q$
reduces the edge-state kinetic energy matrix elements by (2-3) $\hbar \omega/2$, which
serves to lower the energy of the bound state.  The kinetic energy decreases monitonically
as $|E| \rightarrow 0$.

The second $\kappa$-dependent term, the ``bare" potential energy 
$\langle \widetilde{\alpha}(\kappa) |V|\widetilde{\beta}(\kappa) \rangle$, 
is displayed over the same range in the lower left
panel of Fig. \ref{fig_edep} for the five $^3S_1-{}^3S_1$ edge-state matrix elements.
These matrix elements are again quite sensitive to $|E|$, varying by 2-3 MeV over the
20 MeV range displayed in the figure.

The third $\kappa$-dependence is the renormalization of the contact-gradient
coefficients for edge states,
\begin{equation}
\langle n^\prime ~l^\prime | {E \over E-TQ} \bar{O} {E \over E-QT}  | n~l\rangle =
\sum_{i,j=0} \widetilde{g}_j(-\kappa^2;n',l') \widetilde{g}_i(-\kappa^2;n,l)~\langle n^\prime+j~ l^\prime |~ \bar{O}~ | n+i~l \rangle
\label{ren}
\end{equation}
Here $\bar{O}$ is fixed, while the explicit energy dependence carried by the 
$\widetilde{g}_i$  (i.e., the effects of the interplay between $QT$ and $QV$) is evaluated.
The upper left panel in Fig. \ref{fig_edep} gives the result for the diagonal
edge-state matrix element, $| n^\prime~l^\prime \rangle=|n~l \rangle = |5~0 \rangle$.
As has been seen in other cases, the reduction due to the $QT-QV$ interplay is substantial 
throughout the  illustrated 20 MeV range.  Thus the large effects
observed for the deuteron, a relatively weakly bound state, are in fact generic.
But weakly bound states are more strongly affected, with the differences between the
corrections for the double-edge states changing by a factor of nearly two between
$|E|$=20 MeV and $|E| \sim 0$ MeV.
The results for 
single-edge-state matrix elements are similar, but  the
changes are smaller by a factor of two.

In doing these calculations, some care is needed in going to the limit of very small binding energies.  One can show for edge states
\begin{equation}
\widetilde{g}_i(-\kappa^2;n,l) \stackrel{\mathrm{small}~\kappa}{\longrightarrow} (-1)^i
\left[ {\Gamma(n+l+1/2) (n-1+i)! \over \Gamma(n+l+1/2+i) (n-1)! }\right]^{1/2} 
\end{equation}
If one uses this in Eq. (\ref{ren}) with $\kappa \equiv 0$, one finds that
\[ {\sum_{i=0} \widetilde{g}_i(0;n,l) \langle \vec{r}=0~ |~n+i~l \rangle \over  \langle \vec{r}=0~ |~n~l \rangle} \]
oscillates (for an edge state) between 0 and 1, with every increment in $i$.  However,
a nonzero $\kappa^2$ acts as a convergence factor.  If it is quite small, but
not zero, the ratio then goes smoothly to 1/2.  Consequently, as Fig. \ref{fig_edep} shows,
$a_{LO}^\prime/a_{LO} \rightarrow$ 1/4 in the limit of small, but nonzero $\kappa$.
 
The effects illustrated in Fig. \ref{fig_edep} -- the three effects explicitly governed
by $\kappa$ -- are associated with the coupling between $P$ and $Q$ generated
by $T$.   Because this operator connects states with $\Delta n = \pm 1$,
there is no large energy scale associated with excitations.  
As the effects are encoded into a subset of the matrix elements,
the overall scale of the $\kappa$ dependence on spectral properties is, at this
point, still not obvious.   

This leaves us with one remaining term that, qualitatively, seems quite different,
\begin{equation}
V {1 \over E-QH} QV \leftrightarrow \{a_{LO}(|E|),~ a_{NLO}(|E|),~ a_{NNLO}(|E)|,~a_{N^3LO}(|E|),...\}.
\label{constante}
\end{equation}
Here the energy dependence is implicit, encoded in the parameters fitted to the
lowest energy matrix elements of $H^{eff}$.    The underlying potentials are dominated
by strong, short-ranged potentials, much larger than nuclear binding energies.  Thus
the implicit ratio governing this energy dependence -- $|E|$ vs. the strength of
the hard-core potential --  is a small parameter.  For this reason one  
anticipates that the resulting energy dependence might be  gentler than in the
cases just explored.

\begin{figure}
\begin{minipage}{0.5\linewidth}
\begin{center}
\includegraphics[width=9cm]{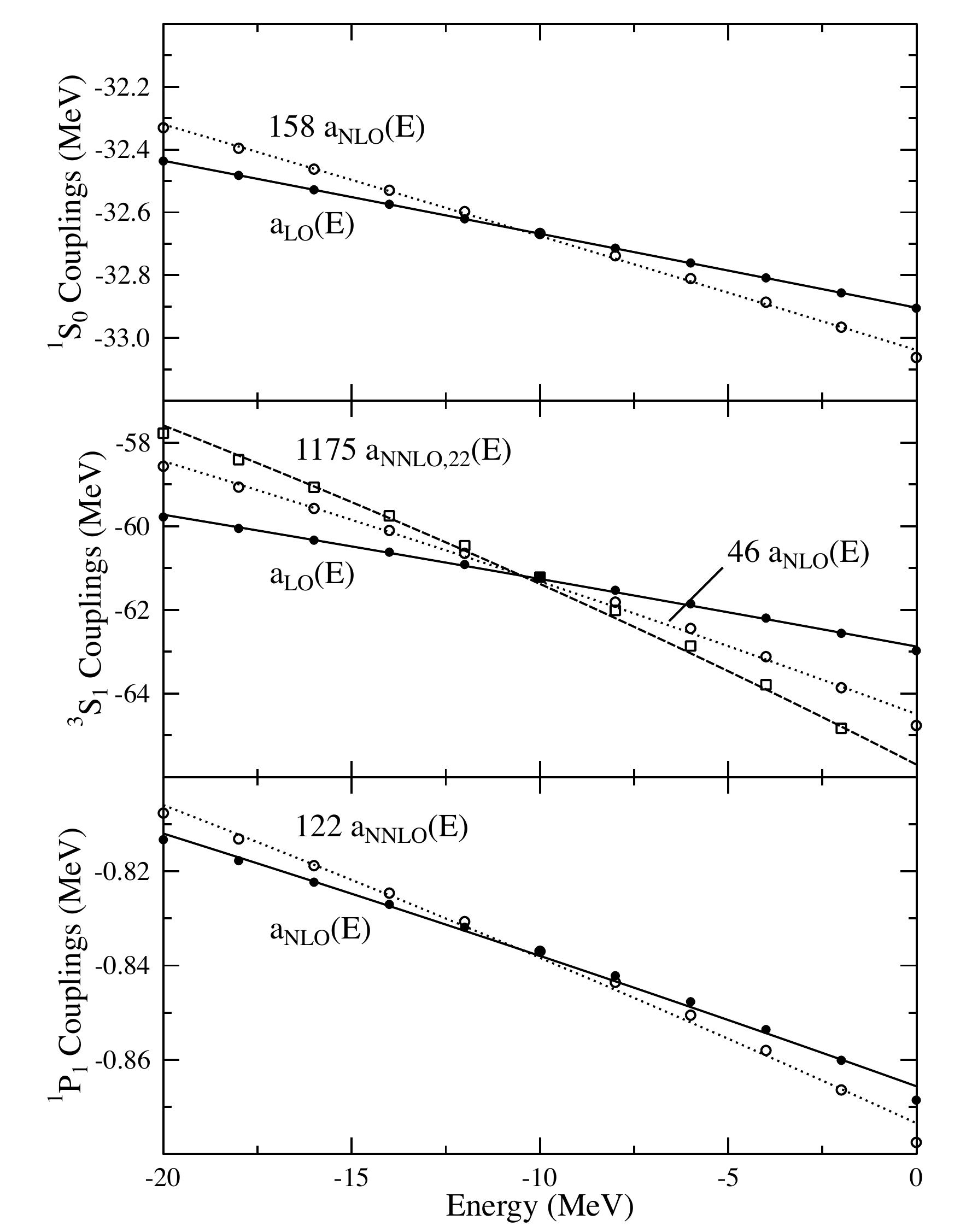}
\end{center}
\end{minipage}%
\begin{minipage}{0.5\linewidth}
\begin{center}
\includegraphics[width=9cm]{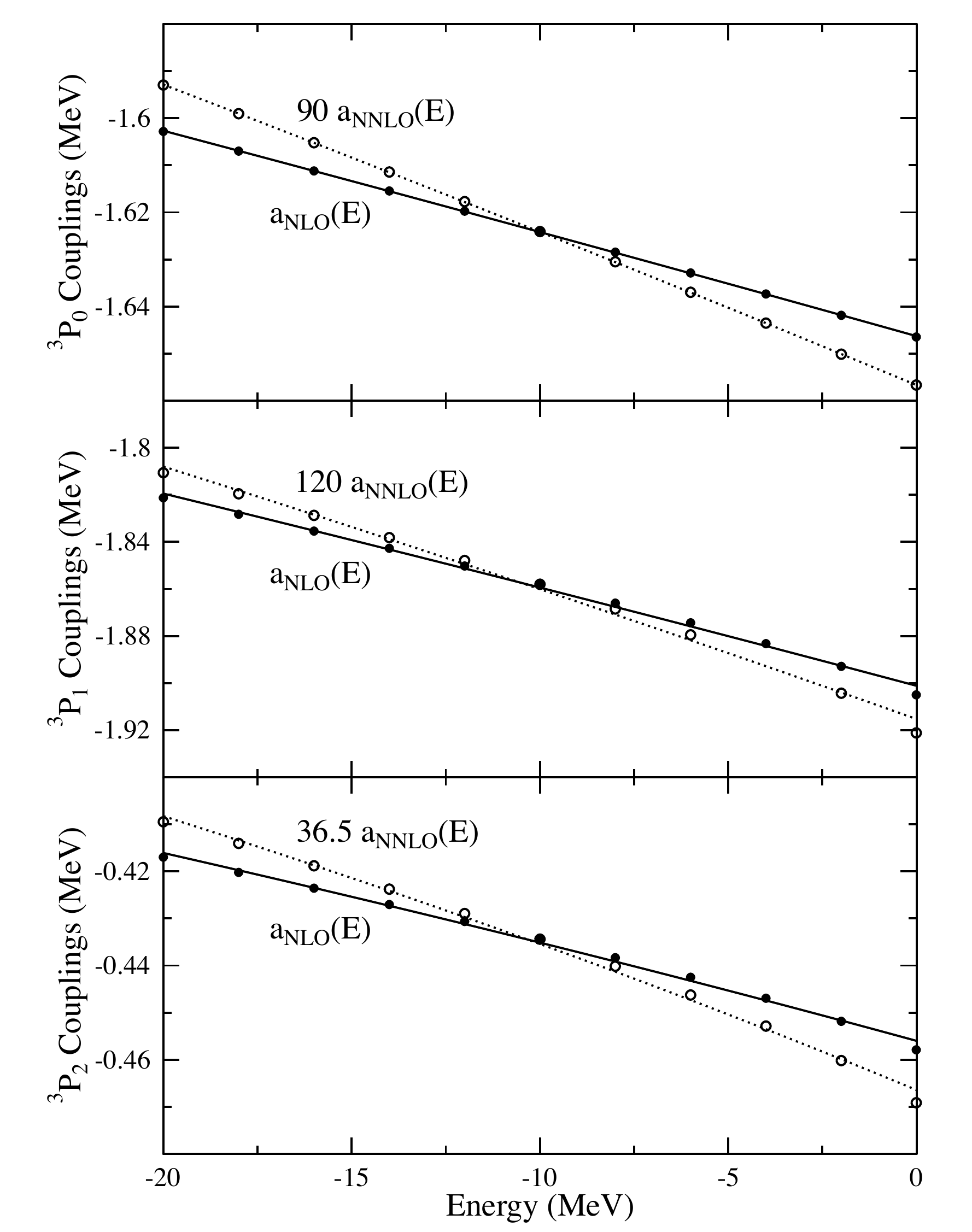}
\end{center}
\end{minipage}
\caption{The calculated energy dependence of derived coefficients for the 
contact-gradient expansion are indicated by the markers, for the various $S-S$ 
and $P-P$ channels.  Over a 10 MeV interval typical of bound-state nuclear
spectra, variations are typically at the few percent level.  The continuous lines represent
simple linear fits, $a$(10 MeV)/$a(E)$  = 1 + $\alpha |E|$, to the results.  The fit
is generally excellent.}
\label{fig_aofe}
\end{figure}

After repeating the fitting procedure over a range of energies, one obtains the results
shown in Fig. \ref{fig_aofe}.    Because the energy variation is quite small, results are
provided only for the channels that contribute in low order,  
$^1S_0$, $^3S_1$, $^1P_1$ and $^3P_J$.  The variation is very modest
and regular, varying inversely with $|E|$ and
well fit by the assumption (motivated by the form of $V (E-QH)^{-1} QV$)
\[ a(E) = {a(10 MeV) \over 1 + \alpha |E|}. \]
The variation is typically at the level of a few percent, over 20 MeV.
The progression in the slopes within each channel, order by order, correspond to
expectation: the lowest order terms, which account for the hardest part of the scattering
in $Q$, have the weakest dependence on $|E|$.  Comparisons between channels also
reflect expectations.  In the earlier discussion of naturalness, the 
rapid convergence in the $^1S_0$ channel, order by order, was consistent with
very short range interactions in $Q$.  Accordingly, $a_{LO}^{1S0}$
varies by just 0.72\% over a 10 MeV interval, and $a_{NLO}^{1S0}$ by 1.10\%.
This channel  contrasts with the $^3S_1$ channel, where convergence
in the contact-gradient expansion is slower, consistent with somewhat longer range
interactions in $Q$.  For the $^3S_1$ case one finds 2.64\% variations in $a_{LO}(^3S_1)$ and
5.17\% variations in $a_{NLO}(^3S_1)$ per 10 MeV interval.

Are such variations of any numerical significance, compared to the explicit
variations isolated in $\kappa$?  That is, if one were to determine
a HOBET interaction directly from bound-state properties of light nuclei, would 
the neglect of this implicit energy dependence lead to significant errors in binding
energies?  One can envision doing such a fit over bound-state data spanning
$\sim$ 10 MeV, finding the couplings as a function of $|E|$, so that the error induced by
using average energy-independent couplings $a_{LO}(|\bar{E}|)$ can be
assessed.  These errors would reflect variations in the matrix
elements to which these couplings are fit, following the procedures previously described.
Such a study showed that only two channels exhibited
drifts $\Delta$ in excess of 15 keV over 10 MeV,
\[ a_{LO}^{1S0}: \Delta \sim \pm 21 \mathrm{~keV}~~~a_{LO}^{3S1}: \Delta \sim \pm 148 
\mathrm{~keV}~~~
a_{NLO}^{3S1}: \Delta \sim \pm 32 \mathrm{~keV} \]
One concludes that the $^3S_1$ channel is, by a large factor, the dominant source of
implicit energy dependence in the HOBET interaction.

This allows one to do a more quantitative calculation that focuses on the most difficult
channel ($^3S_1$) and compares the relative sizes of the $\kappa$-dependent and
implicit energy dependences, as reflected in changes in the matrix $H^{eff}(|E|)$.
Thus this matrix is constructed at
$|E|=10$ MeV and at $|E| \sim 0$ MeV (including the coupling to $^3D_1$), and 
changes in global quantities
of that matrix over 10 MeV are
examined: shifts in the first moment (the average eigenvalue) , the RMS
shifts of levels relative to the first moment (related to the stability of level splittings),
and eigenvalue overlaps.   The four energy-dependent effects discussed here are
separately turned on and off.
Thus this exercise should provide a good test of the relative
importance of these effects.  The results are shown in Table \ref{table:3}.

\begin{table}
\begin{center}
\caption{Spectral property variations in $H^{eff}(E)$ over 10 MeV}
\label{table:3}
\begin{tabular}{|l|c|c|c|c|}
        \hline
Term & Parameter & 1st Moment Shift (MeV) & RMS Level Variation (MeV) & Wave Function Overlaps \\
\hline \hline
 $\langle \alpha |T | \widetilde{\beta} \rangle$ & $\kappa$ & 2.554 & 1.107 & 95.75-99.74\% \\
 $\langle \widetilde{\alpha} | V | \widetilde{\beta} \rangle$ & $\kappa$ & 0.272 & 0.901 & 99.35-99.82\% \\
 $\langle \widetilde{\alpha} | \bar{O} | \widetilde{\beta} \rangle$ & $\kappa$ & -0.239 & 0.957 & 99.51-99.99\% \\
 $\langle \alpha | \bar{O}(E) | \beta \rangle$ & implicit & 0.135 & 0.107 & 99.95-100\% \\
 \hline  \hline
\end{tabular}
\end{center}
\end{table}

Despite the selection of the worst channel, $^3S_1$,
the implicit energy dependence is small, intrinsically and in comparison 
with the implicit energy dependence embedded in $\kappa$.   The implicit
dependence in the first moment -- a quantity important to absolute binding energies -- is
5\% that of the explicit dependence in $\langle \alpha |T | \widetilde{\beta} \rangle$.   The
RMS shifts in levels relative the first moment are at the $\sim$ 1 MeV level for each of
the implicit terms, but $\sim$ 100 keV for the implicit term.  Eigenfunction overlaps show
almost no dependence on the implicit term,
exceeding 99.95\% in all cases: variations 10-100 times larger arise
from the analytical terms in $\kappa$.

Thus a simple representation of the HOBET effective interaction exists:
\begin{itemize}  
\item The requirements for a state of fixed $|E|$ are a series of short-range 
coefficients and a single parameter $\kappa$ that governs long-range corrections 
residing in $Q$,
including certain terms that couple $QV$ and $QT$.   By various measures explored here, an
N$^3$LO expansion is accurate to about a few keV 
\item  The $\kappa$ dependence found for a state of definite energy $|E|$ 
also captures almost all of the energy
dependence resulting from varying $|E|$, the state-dependence in BH.
Even in the most troublesome channel, calculations show that $\sim$ 95\%
of the energy dependence associated with changes in $|E|$ is explicit.   
It appears that neglect of the implicit energy dependence would induce errors of $\lsim$ 100 keV,
for a spectrum spanning $\pm$ 10 MeV.   This kind of error would be within the uncertainties
of the best {\it ab initio} excited-state methods for light (p-shell) nuclei, 
such as Green's function Monte Carlo \cite{argonne}
or large-basis no-core SM diagonalizations \cite{nocore}.
\item  If better results are desired, the program described here can be extended to include
the implicit energy dependence.  The expansion around an average energy $E_0$
\[ V{1 \over E-QH} QV = V \left[ {1 \over E_0-QH} - {1 \over E_0-QH} (E-E_0) {1 \over E_0-QH} + \cdots \right] QV \]
generates the correction linear in $E$ that is seen numerically.  This second term is
clearly quite small, explicitly suppressed by the ratio of scales discussed above.
But, in any troublesome channel, the second term could be represented by
contact-gradient operators of low order, with the contribution suppressed by an
overall factor of $(E-E_0)$.
\end{itemize}

\section{Discussion and Summary}
One of the important motivations for trying to formulate an effective theory for nuclei in
a harmonic oscillator basis is the prospect of incorporating into the approach some
of the impressive numerical technology of approaches like the SM.
Numerical techniques could be used to solve significant $P$-space problems, 
formulated in spaces, such as completed $N \hbar \omega$ bases, that preserve
the problem's translational invariance.  

My collaborators and I made an initial effort to construct a HOBET some years ago, using
a contact-gradient expansion modeled after EFT.  We performed a shell-by-shell integration,
in the spirit of 
a discrete renormalization group method, but encountered several abnormalities connected 
with the running of the coefficients of the expansion.   
Subsequent numerical work in which we studied
individual matrix elements revealed the problems illustrated in Fig. \ref{fig_simple}.
These problems -- the difficulty of representing $Q$-space contributions that are
both long range and short range -- are not only important for HOBET, but also are
responsible for the lack of convergence of perturbative expansions of the effective
interaction.  Fig. \ref{fig_1} provides one example.  In other work \cite{luu} we have shown that
convergent expansions in the bare interaction (deuteron) or g-matrix ($^3$He/$^3$H)
for $H^{eff}$ do exist, if the long-range part of this problem is first solved, as we have done here.

Thus the current paper returns to the problem of constructing a contact-gradient
expansion for the effective interaction, taking into account what has been learned since
the first, less successful effort.
This paper introduces a form for that expansion that eliminates operator mixing,
simplifying the fitting of coefficients and guaranteeing that coefficients determined in
a given order remain fixed when higher-order terms are added.  Thus the N$^3$LO
results presented here contain the results of all lower orders.  The expansion is 
one in nodal quantum numbers, and is directly connected with traditional Talmi
integral expansions, generalized for nonlocal interactions. 

Convergence does vary from channel to channel, but in each channel the order-by-order
convergence is very regular.   Each new order brings down the
scale $\Lambda$ at which deviations appear, and in each new order the Lepage plot 
steepens, showing that the omitted physics does have the expected
dependence on higher-order polynomials in $(n^\prime, n)$.
The channel-by-channel variations in convergence reflect similar
behavior seen in EFT approaches, where the need for alternative power counting 
schemes has been noted to 
account for this behavior.   From a practical standpoint, however,
the N$^3$LO results are effectively exact:  in the most important difficult channel, $^3S_1$,
measures of the quality of the matrix $H^{eff}$ yielded results on the order of (1-3) keV.

The summation done over $QT$ yields a simple result, but still one that is quite remarkable
in that long-range physics is governed by a single parameter $\kappa$, that depends
on the ratio of $|E|$ and $\hbar \omega$.   Despite all of the attractive analytic properties
of the HO as a basis for bound states, its unphysical binding at large $r$ has been viewed
as a shortcoming.  But the ladder properties of the HO in fact
allow an exact summation of $QT$.  It seems unlikely that any other bound-state
basis would allow the coupling of $P$ and $Q$ by $T$ to be exactly removed.  That is,
the HO basis may be the only one that allows the long-range physics in $Q$ to be
fully isolated, and thus subtracted systematically.  In this sense it may be the optimal basis
for correctly describing the asymptotic behavior of the wave function.  Note, in particular,
that the right answer is not going to result from using ``improved" single-particle
bases: $\kappa$ depends of $|E|$, not on single-particle energies of some mean field.

The effects associated with $\kappa$ are large, typically shifting edge-state matrix elements
by several MeV, and altering spectral measures, like the first energy moment of $H^{eff}$,
by similar amounts.  This dependence, if not isolated, destroys the
systematic order-by-order improvement important to HOBET, as Fig. \ref{fig_simple}
clearly illustrates.

The explicit energy dependence captured in $\kappa$ accounts for almost
all of the energy dependence of $H^{eff}(|E|)$.  In more complicated calculations this
dependence, in the
BH formulation used here, generates the state-dependence that allows ET wave functions
to have the proper relationship to the exact wave functions, namely that the former are the
$P$-space restrictions of the latter.  
While in principle additional energy dependence important to this
evolution resides in
$V (E-QH)^{-1} QV$ and thus in the coefficients of the contact-gradient expansion, in
practice this residual implicit energy dependence was found to be very weak.  This
dependence was examined channel by channel, and its impact on global properties of
$H^{eff}(|E|)$ was determined for the most troublesome channel, $^3S_1$.    Even in this channel, the
impact of the remaining implicit energy dependence on $H^{eff}(|E|)$ spectral properties such 
as the
first moment, eigenvalue spacing, and eigenvalue overlaps, was found to
be quite small compared to the explicit dependence isolated in $\kappa$.  
This is physically very reasonable:
$QT$ generates nearest-shell couplings between $P$ and $Q$, so that excitation scales
are comparable to typical nuclear binding energies.  Thus this physics, extracted
and expressed as a function of $\kappa$, should be sensitive to binding energies.
In contrast, $V (E-QH)^{-1} QV$ 
involves large scales associated with the hard core, and thus should be relatively insensitive
to variations in $|E|$.  In the $^3S_1$ channel, the explicit dependence 
captured in $\kappa$ is about 20 times larger than the implicit energy dependence buried
in the contact-gradient coefficients.  Numerically, the latter could cause drifts on the
order of 100 keV over 10 MeV intervals.  Thus, to an excellent approximation, one could
treat these coefficients as constants in fitting the properties of low-lying spectra.  
Alternatively, the HOBET procedure for accounting for this implicit energy dependence
has been described, and could be used in any troublesome channel.

The weakness of the implicit energy dependence will certainly
simplify future HOBET efforts to determine 
$H^{eff}(E,b,\Lambda_P)$ directly from data (rather than from an NN potential like $v_{18}$).
Indeed, such an effort will be the next step in the program. The
approach outlined here is an attractive starting point, as it can be shown that
the states  $|\widetilde{\alpha}(\kappa) \rangle$ become asymptotic plane-wave states, when
$E$ is positive.  Thus the formalism relates bound and continuum states through a 
common set of strong-interaction coefficients operating in a finite orthogonal basis.

The relationship between current work and some
more traditional treatments of the $H^{eff}$ for model-based approaches, like the SM,
should be mentioned.
Efforts like those of Kuo and Brown are often based on the division $H=H_0+(V-V_0)$,
where $H_0$ is the HO Hamiltonian \cite{kuobrown}.   Such a division would allow the same BH
reorganization done here: $QT$ and $Q(H_0-V_0)$ are clearly equivalent.   But in
practice terms are, instead, organized in perturbation theory according to $H_0$, i.e., so that
Green's functions involve single-particle energies.  This would co-mingle the long-
and short-range physics is a very complicated way.  In addition, often the definitions of $Q$
and $P$ used in numerical calculations are not those of the HO: instead, a plane-wave
momentum cut is often used, which simplifies the calculations but introduces
uncontrolled errors.  Either this approximation (plane waves
are diagonal in $T$) or the use of perturbation theory (because of the co-mingling) would
appear to make it impossible to separate long- and short-range physics correctly, as has
been done here.

Another example is $V_{low~k}$, in which a softer two-nucleon potential is derived
by integration over high-momentum states \cite{achim}.   This is a simpler description of $Q$ than
arises in bound-state bases problems, like those considered here:  the division between
$P$ and $Q$ is a specified momentum, and $T$ is diagonal.  There would be no
analog of the $\kappa$-dependence
found in HOBET.   However, HOBET and $V_{low~k}$ may have an interesting relationship.
Effective operators for HOBET and for EFT approaches (which also employ a momentum 
cutoff) agree in lowest contributing order.  When there are differences in higher order, it would
seem that these differences must vanish by taking the appropriate limit, namely
the limit of the HOBET $Q(b,\Lambda_P)$ where $b \rightarrow \infty$
while $\Lambda_P/b$ is kept fixed.  This keeps the average  $\langle p^2 \rangle$ of the
last included shell fixed,  while forcing the number of shells to infinity and the shell splitting to zero.  
Numerically it would be sufficient to approach this limit, so that $Q$ resembles the plane-wave
limit over a distance characteristic of the nuclear size.
It is a reasonable conjecture that $V_{low~k}$ would emerge from such
a limit of the HOBET $H^{eff}(b,\Lambda_P)$.  It would follow that all $\kappa$
dependence should vanish in that limit.  It would be interesting to try to verify these
conjectures in future work, and to study the evolution of the HOBET effective interaction
coefficients as this limit is taken

The state-dependence of effective interactions is sometimes treated
in nuclear physics by the method of Lee and Suzuki.  One form of Lee-Suzuki 
produces a Hermitian energy-independent interaction.  While it is always possible to
find such an $H$ to reproduce eigenvalues, it is clear that basic wave function requirements of an 
effective theory -- that the included wave functions correspond to restrictions of the
true wave functions to $P$ -- are not consistent with such an $H$.  

Another form produces
an energy-independent but nonHermitian $H$.  This can be done consistently in an 
effective theory.
However, the results presented here make it difficult to motivate such a transformation.
It appears that the state-dependence is almost entirely attributable to the interplay between
$QT$ and $QV$, removed here analytically in terms of a function
of one parameter $\kappa$, which relates the bound-state momentum scale (in $\hbar \omega$)
to a state's energy.   There is no obvious benefit in obscuring this simple dependence
in a numerical transformation of the potential, given that the Lee-Suzuki method is not
easy to implement.  The physics is far more transparent in the BH formulation, and the 
self-consistency required in BH makes the use of an energy-dependent potential as
easy as an energy-independent one.  More to the point, the necessary $\kappa$ dependence
is already encoded in the potential for a single state of definite energy $|E|$ -- thus 
no additional complexity is posed by the state-dependence of the potential.    

I thank Tom Luu and Martin Savage for helpful discussions.  This work was supported in
part by the Office of Nuclear Physics and SciDAC, US Department of Energy.

\renewcommand{\theequation}{A-\arabic{equation}}
\setcounter{equation}{0}  
 \section{APPENDIX: Effective Interaction Matrix Elements}  
 
 This appendix provides details on the evaluation of the modified HO states  \begin{equation}
 {E \over E-QT} |n~l \rangle.
 \label{eq:start}
 \end{equation}
 for nodal quantum numbers $n=1,2,3,...$, and for corresponding contact-gradient effective interaction
 matrix elements between such states.  
Closed-form expressions allow these matrix elements to be evaluated quickly to any order.
Here two alternative evaluations are provided, one  
based on a harmonic oscillator expansion and one on the free Green's function.\\
 
 \noindent
 {\it Harmonic Oscillator Expansion:}  The harmonic oscillator Green's function expansion is
 \begin{equation}
  {E \over E-QT} |n~l \rangle = \sum_{i=0}^\infty \widetilde{g}_i(-\kappa^2; n,l) |n+i~l\rangle
 \end{equation} 
 where the $\widetilde{g}_i $ are determined by a set of continued fractions generated from the ladder
 properties of the operator $QT$.  In practice the sum can be truncated: numerical
 convergence is discussed below, in comparison with the Green's function approach.  
 While this paper focuses
 on the simple example of the deuteron, the approach is more general : the relationship
 between the relative kinetic energy and the three-dimensional harmonic oscillator can be extended to 
 the n-dimensional harmonic oscillator, with the hyperspherical harmonics replacing the spherical harmonics as eigenfunctions of the kinetic energy.   The corresponding ladder properties for the harmonic oscillator in hyperspherical
 coordinates are known \cite{hyper}: this is the essential requirement for the expansion.
 
 As discussed in the text, it is convenient to modify the usual contact-gradient expansion, 
 to remove operator mixing and create an expansion in nodal quantum numbers.  Each term $O$
 in the usual contact-gradient expansion is replaced by
 \begin{equation}
 O \rightarrow \bar{O} \equiv e^{r^2/2}Oe^{r^2/2}
 \end{equation}
 where $r$ is the dimensionless Jacobi coordinate $|\vec{r}_1-\vec{r}_2|/(b \sqrt{2})$.  Gradients appearing in $O$ are also defined in terms of this dimensionless coordinate.
 
 In each partial wave the lowest contributing operators are based on gradients, maximally
 coupled, acting on wave functions, with the result evaluated at $\vec{r}=0$.  For example,
 for $S$, $P$, and $D$ states
 \begin{eqnarray}
 e^{r^2/2} R_{nl}(r) Y_{l0}(\Omega_r) \big|_{\vec{r}=0} &=& \delta_{l0} {1 \over \pi} \left[ {2 \Gamma(n+1/2) \over (n-1)!} \right]^{1/2}\nonumber \\
 \overrightarrow{\nabla}_{10}~ e^{r^2/2} R_{nl}(r) Y_{l0}(\Omega_r) \big|_{\vec{r}=0} &=& \delta_{l1} \sqrt{{2(2n + 1) \over 3}}  R_{n0}(r)Y_{00}(\Omega_r) \big|_{r=0} =\delta_{l 1} ~2 \left[ {1! \over 3!!} \right]^{1/2} {1 \over \pi} \left[ {2~ \Gamma[n+3/2] \over (n-1)!} \right]^{1/2}\nonumber \\
 (\overrightarrow{\nabla} \otimes \vec{\nabla})_{20} e^{r^2/2} R_{nl}(r) Y_{l0}(\Omega_r) \big|_{\vec{r}=0} &=& \delta_{l2} \sqrt{{8(2n+1)(2n+3) \over 15}} R_{n0}(r) Y_{00}(\Omega_r) \big|_{\vec{r}=0} =\delta_{l 2} ~2^2 \left[ {2! \over 5!!} \right]^{1/2} {1 \over \pi} \left[ {2~ \Gamma[n+5/2] \over (n-1)!} \right]^{1/2} 
 \nonumber \\
 \label{eq:ex}
 \end{eqnarray}
As the gradients are maximally coupled, all coupling schemes are equivalent.  If one defines by
$(\overrightarrow{\nabla}^q)_{q0}$ the expressions with $q$ gradients maximally coupled, the results of Eq. (\ref{eq:ex})
are examples of the more general formula
\begin{equation}
({\overrightarrow{\nabla}}^q)_{q0}~e^{r^2/2} R_{nl}(r) Y_{l0}(\Omega_r) \big|_{\vec{r}=0} = \delta_{l q} ~2^l \left[ {l! \over (2l+1)!!} \right]^{1/2} {1 \over \pi} \left[ {2~ \Gamma[n+l+1/2] \over (n-1)!} \right]^{1/2}.
\end{equation}
A form of this equation that will be used below is
\begin{eqnarray}
&&(\overrightarrow{\nabla}^q)_{q0}~e^{r^2/2} R_{n-p~l}(r) Y_{l0}(\Omega_r) \big|_{\vec{r}=0} = \delta_{l q} ~2^l \left[ {l! \over (2l+1)!!} \right]^{1/2} {1 \over \pi} \left[ {2 ~\Gamma[n+l+1/2-p] \over (n-1-p)!} \right]^{1/2} \nonumber \\
&&~~~~~~= \left[ {(n-1)! ~\Gamma [n+l+1/2-p] \over (n-1-p)! ~\Gamma[n+l+1/2] } \right]^{1/2} \delta_{l q}~ 2^l \left[ {l! \over (2l+1)!!}\right]^{1/2} {1 \over \pi} \left[ {2 ~\Gamma [n+l+1/2] \over (n-1)!}\right]^{1/2}
\label{eq:form}
\end{eqnarray}

Contact-gradient operators beyond the lowest contributing order involve $\overrightarrow{\nabla}^2$ acting on
harmonic oscillator wave functions.  One can quickly verify
\begin{equation}
\overrightarrow{\nabla}^2 e^{r^2/2} R_{nl}(r) Y_{lm}(\Omega_r) = - 4 \sqrt{(n-1)(n+l-1/2)} e^{r^2/2} R_{n-1~l}(r) Y_{lm}(\Omega_r)
\label{laplacian}
\end{equation}
so that 
\begin{equation}
(\overrightarrow{\nabla}^2)^p e^{r^2/2} R_{nl}(r)Y_{lm}(\Omega_r) = (-4)^p \left[{(n-1)!~\Gamma [n+l+1/2] \over
(n-1-p)!~\Gamma[n+l+1/2-p]} \right]^{1/2} e^{r^2/2} R_{n-p~l}(r) Y_{lm}(\Omega_r).
\label{eq:higher}
\end{equation}
Thus by first using  Eq. (\ref{eq:higher}) and then applying Eq. (\ref{eq:form}), one finds
the general expression for a contact-gradient operator of arbitrary order acting on a harmonic oscillator state
\begin{equation}
(\overrightarrow{\nabla}^2)^p (\overrightarrow{\nabla}^q)_{q0} e^{r^2/2} R_{nl}(r) Y_{lm}(\Omega_r) \big|_{\vec{r}=0} =
\delta_{lq}~(-4)^p {(n-1)! \over (n-1-p)!}  \left( 2^l \left[{l! \over (2l+1)!!} \right]^{1/2} {1 \over \pi}
\left[{ 2~\Gamma[n+l+1/2] \over (n-1)!} \right]^{1/2} \right)
\label{eq:finalHO}
\end{equation}

Equation (\ref{eq:finalHO}) defines the needed matrix elements, evaluated below 
for each partial-wave channel contributing through  N$^3$LO.  [Note that if one
wanted to write a potential, as opposed to partial-wave matrix elements of that potential 
that are given here, suitable projection
operators could be inserted as needed.  For example, the $l=0$ triplet and singlet
channels could be distinguished by introducing the projection operators $P(^3S_1)=(3 + \vec{\sigma}_1
\cdot \vec{\sigma}_2)/4=\vec{S}^2/2$ and $(1 - \vec{\sigma}_1 \cdot \vec{\sigma}_2)/4
= 1 - \vec{S}^2/2$, respectively;  the three triplet $l=1$ channels could be distinguished from
the singlet $l=1$ channel and from each other by the projectors
\begin{eqnarray}
P(^3P_0) &=& {(-1 + (\vec{l} \cdot \vec{S})^2)~ \vec{S}^2 \over 6} \nonumber \\
P(^3P_1) &=& {(2 - \vec{l} \cdot \vec{S} - (\vec{l} \cdot \vec{S})^2)~ \vec{S}^2 \over 4}\nonumber \\
P(^3P_2) &=& {(2+ 3~ \vec{l} \cdot \vec{S} + (\vec{l} \cdot \vec{S})^2)~ \vec{S}^2 \over 12}
\end{eqnarray}
and so on.]   The matrix elements of Table \ref{table:1}, which all are scalar products
of spin-spatial tensor operators, are of two types.  One is diagonal in $l$, where
$O= \overleftarrow{O}^L_l \cdot \delta(\vec{r})~ \overrightarrow{O}^R_l$, with $\overleftarrow{O}^L_l$
and $\overrightarrow{O}^R_l$ spatial tensors.  In this case
\begin{equation}
\langle n^\prime (lS)JM TM_T | O | n(lS)JMTM_T\rangle =
\langle n^\prime l m_l=0 | \overleftarrow{O}^L_{l0}~\delta({\vec{r}})~\overrightarrow{O}^R_{l0} | n l m_l=0\rangle
\label{eq:diagonal}
\end{equation}
The second type of operator,
 $O= \overleftarrow{O}^L_l \cdot \delta(\vec{r})~ [\overrightarrow{O}^R_{l+2}  \otimes [\sigma(1)
 \otimes \sigma(2)]_2]_l$, enters for the off-diagonal triplet $S-D$,$P-F$, and $D-G$ cases,
\begin{eqnarray}
&&\langle n^\prime (lS=1)J=l+1~M TM_T |O | n(l+2~S=1)J=l+1~MTM_T\rangle = \nonumber \\
&&~~~~~~2 \sqrt{{2 l+1 \over 2 l +3}} ~\langle n^\prime~ l ~m_l=0 | \overleftarrow{O}^L_{l~0}~\delta(\vec{r})~ \overrightarrow{O}^R_{l+2~0}| n ~{l+2} ~m_l=0\rangle
\label{eq:nondiagonal}
\end{eqnarray}

As both of these expressions reduce the matrix element to a product of terms like
Eq. (\ref{eq:finalHO}), the needed N$^3$LO matrix elements follow directly.
For the $^3S_1 \leftrightarrow {}^3S_1$ or $^1S_0\leftrightarrow {}^1S_0$  channels,
\begin{eqnarray}
\lefteqn{\langle n^\prime (l=0S)JM TM_T | e^{r^2/2} \biggl[ a_{LO}^{3S1} \delta(\vec{r}) + a_{NLO}^{3S1}(\overleftarrow{\nabla}^2 \delta(\vec{r}) + \delta(\vec{r}) \overrightarrow{\nabla}^2)
+ a_{NNLO}^{3S1,22} \overleftarrow{\nabla}^2 \delta(\vec{r}) \overrightarrow{\nabla}^2
+ a_{NNLO}^{3S1,40} (\overleftarrow{\nabla}^4 \delta(\vec{r}) + \delta(\vec{r}) \overrightarrow{\nabla}^4)}  \nonumber \\
&&+a_{N^3LO}^{3S1,42} (\overleftarrow{\nabla}^4 \delta(\vec{r}) \overrightarrow{\nabla}^2+ \overleftarrow{\nabla}^2\delta(\vec{r}) \overrightarrow{\nabla}^4)
+ a_{N^3LO}^{3S1,60}(\overleftarrow{\nabla}^6 \delta(\vec{r}) + \delta(\vec{r}) \overrightarrow{\nabla}^6) \biggr] e^{r^2/2} | n(l=0S)JMTM_T\rangle = \hspace{1.2in} \nonumber \\
&&\hspace{0.3in} {2 \over \pi^2}
\left[{ \Gamma[n^\prime+1/2] \Gamma[n+1/2] \over (n^\prime-1)!(n-1)!} \right]^{1/2}~
\biggl(a_{LO}^{3S1} -4((n^\prime-1)+(n-1)) a_{NLO}^{3S1} + \nonumber \\
&&\hspace{0.3in}+16\bigl[(n^\prime-1)(n-1)a_{NNLO}^{3S1,22} +  
((n^\prime-1)(n^\prime-2)+(n-1)(n-2)) a_{NNLO}^{3S1,40} \bigr]
-64 \bigl[ ((n^\prime-1)(n^\prime-2)(n-1)+ \nonumber \\
&& \hspace{0.3in}(n^\prime-1)(n-1)(n-2))a_{N3LO}^{3S1,42}+((n^\prime-1)(n^\prime-2)(n^\prime-3)+(n-1)(n-2)(n-3))a_{N3LO}^{3S1,60}\bigr] \biggr);
 \label{firstho}
\end{eqnarray}
For the $^3S_1 \leftrightarrow {}^3D_1$ channel, recalling $\overrightarrow{D}^0_0 \equiv
((\overrightarrow{\nabla} \otimes \overrightarrow{\nabla})_2 \otimes (\sigma(1) \otimes \sigma(2))_2)_{00}$,
\begin{eqnarray}
\lefteqn{\langle n^\prime (l=0S=1)J=1M TM_T |e^{r^2/2} \biggl[ a_{NLO}^{SD}(\delta(\vec{r}) \overrightarrow{D}^0_0+\overleftarrow{D}^0_0 \delta(\vec{r})  )
+ a_{NNLO}^{SD,22} (\overleftarrow{\nabla}^2 \delta(\vec{r}) \overrightarrow{D}^0_0
+\overleftarrow{D}^0_0 \delta(\vec{r}) \overrightarrow{\nabla}^2)+}\nonumber \\
&&a_{NNLO}^{SD,04} (\delta(\vec{r})\overrightarrow{\nabla}^2\overrightarrow {D}^0_0+
\overleftarrow{D}^0_0 \overleftarrow{\nabla}^2\delta(\vec{r})) + a_{N^3LO}^{SD,42} (\overleftarrow{\nabla}^4 \delta(\vec{r}) \overrightarrow{D}^0_0+ \overleftarrow{D}^0_0\delta(\vec{r}) \overrightarrow{\nabla}^4)
+ \nonumber \\
&&a_{N^3LO}^{SD,24}(\overleftarrow{\nabla}^2 \delta(\vec{r})\overrightarrow{\nabla}^2 \overrightarrow{D}_0^0 + \overleftarrow{D}^0_0 \overleftarrow{\nabla}^2 \delta(\vec{r}) \overrightarrow{\nabla}^2)+
a_{N^3LO}^{SD,06} ( \delta(\vec{r})\overrightarrow{\nabla}^4 \overrightarrow{D}_0^0 + \overleftarrow{D}^0_0 \overleftarrow{\nabla}^4 \delta(\vec{r})) \biggr] e^{r^2/2} | n(l=2S=1)J=1MTM_T\rangle = \nonumber \\
&& \hspace{0.3in} {8 \over 15} \sqrt{10} {2 \over \pi^2}
\left[{ \Gamma[n^\prime+1/2] \Gamma[n+5/2] \over (n^\prime-1)!(n-1)!} \right]^{1/2}
\biggl(a_{NLO}^{SD}-4\bigl[(n^\prime-1) a_{NNLO}^{SD,22} + (n-1)a_{NNLO}^{SD,04} \bigr]
+ \nonumber \\
&& \hspace{0.3in} 16\bigl[(n^\prime-1)(n^\prime-2)a_{N^3LO}^{SD,42} + 
(n^\prime-1)(n-1) a_{N^3LO}^{SD,24} +
(n-1)(n-2)a_{N^3LO}^{SD,06} \bigr] \biggr);
\end{eqnarray}
For the $^1D_2 \leftrightarrow {}^1D_2$ or $^3D_J \leftrightarrow {}^3D_J$ channels, recalling
$\overrightarrow{D}^2_M \equiv
(\overrightarrow{\nabla} \otimes \overrightarrow{\nabla})_{2M}$,
\begin{eqnarray}
\lefteqn{\langle n^\prime (l=2S)J=1M TM_T |e^{r^2/2} \biggl[ a_{NNLO}^{1D2}\overleftarrow{D}^2 \cdot  \delta(\vec{r}) \overrightarrow{D}^2+
a_{N^3LO}^{1D2} (\overleftarrow{D}^2 \overleftarrow{\nabla}^2 \cdot  \delta(\vec{r}) \overrightarrow{D}^2+\overleftarrow{D}^2 \cdot  \delta(\vec{r})\overrightarrow{\nabla}^2 \overrightarrow{D}^2)
 \biggr] e^{r^2/2} } \nonumber \\
&& | n(l=2S)J=1MTM_T\rangle={32 \over 15} {2 \over \pi^2}
\left[{ \Gamma[n^\prime+5/2] \Gamma[n+5/2] \over (n^\prime-1)!(n-1)!} \right]^{1/2}
\biggl(a_{NNLO}^{1D2}-4\bigl[(n^\prime-1)+(n-1)\bigr] a_{N^3LO}^{1D2}  \biggr); \hspace{0.53in}
\end{eqnarray}
For the $^3D_3 \leftrightarrow {}^3G_3$ channel, recalling $\overrightarrow{G}^2_M \equiv
\left[((\overrightarrow{\nabla} \otimes \overrightarrow{\nabla})_2 \otimes (\overrightarrow{\nabla} \otimes \overrightarrow{\nabla})_2)_4 \otimes (\sigma(1) \otimes \sigma(2))_2 \right]_{2M}$,
\begin{eqnarray}
\lefteqn{\langle n^\prime (l=2S=1)J=3M TM_T |e^{r^2/2}  a_{N^3LO}^{DF}(\overleftarrow{D}^2 \cdot  \delta(\vec{r}) \overrightarrow{G}^2
+\overleftarrow{G}^2 \cdot  \delta(\vec{r~~}) \overrightarrow{D}^2)
 e^{r^2/2} | n(l=4S=1)J=3MTM_T\rangle=}\nonumber \\
&&\hspace{0.3in} {512 \over 315} \sqrt{15} {2 \over \pi^2}
\left[{ \Gamma[n^\prime+5/2] \Gamma[n+9/2] \over (n^\prime-1)!(n-1)!} \right]^{1/2}
a_{N^3LO}^{DF}; \hspace{3.6in}
\end{eqnarray}
For the $^1P_1 \leftrightarrow {}^1P_1$ or $^3P_J \leftrightarrow {}^3P_J$  channels,
\begin{eqnarray}
\lefteqn{\langle n^\prime (l=1S)JM TM_T | e^{r^2/2} \biggl[ a_{NLO}^{1P1} \overleftarrow{\nabla} \cdot \delta(\vec{r}) \overrightarrow{\nabla} + a_{NNLO}^{1P1}(\overleftarrow{\nabla} \overleftarrow{\nabla}^2 \cdot \delta(\vec{r})\overrightarrow{\nabla} + \overleftarrow{\nabla} \cdot \delta(\vec{r}) \overrightarrow{\nabla}^2 \overrightarrow{\nabla}) +}\nonumber \\
&&a_{N^3LO}^{1P1,33} \overleftarrow{\nabla}\overleftarrow{\nabla}^2 \cdot \delta(\vec{r})\overrightarrow{\nabla}^2\overrightarrow{\nabla} 
+ a_{N^3LO}^{1P1,51} (\overleftarrow{\nabla}\overleftarrow{\nabla}^4 \cdot \delta(\vec{r})\overrightarrow{\nabla} + \overleftarrow{\nabla} \cdot \delta(\vec{r}) \overrightarrow{\nabla}^4 \overrightarrow{\nabla}) \biggr] e^{r^2/2} | n(l=1S)JMTM_T \rangle = \nonumber \\
&&\hspace{0.3in}{4 \over 3} {2 \over \pi^2}
\left[{ \Gamma[n^\prime+3/2] \Gamma[n+3/2] \over (n^\prime-1)!(n-1)!} \right]^{1/2}~
\biggl(a_{NLO}^{1P1} -4((n^\prime-1)+(n-1)) a_{NNLO}^{1P1} + \hspace{1.9in} \nonumber \\
&&\hspace{0.3in}+16\bigl[(n^\prime-1)(n-1)a_{N^3LO}^{1P1,33} +  
((n^\prime-1)(n^\prime-2)+(n-1)(n-2)) a_{N^3LO}^{1P1,51} \bigr]
 \biggr);
\end{eqnarray}
For the $^3P_2 \leftrightarrow {}^3F_2$ channel, recalling $\overrightarrow{F}^1_M \equiv
\left[(\overrightarrow{\nabla} \otimes (\overrightarrow{\nabla} \otimes \overrightarrow{\nabla})_2)_3 \otimes (\sigma(1) \otimes \sigma(2))_2 \right]_{1M}$,
\begin{eqnarray}
\lefteqn{\langle n^\prime (l=1S=1)J=2M TM_T | e^{r^2/2} \biggl[ a_{NNLO}^{PF}(\overleftarrow{\nabla}\cdot \delta(\vec{r})\overrightarrow{F}^1 + \overleftarrow{F}^1 \cdot \delta(\vec{r})  \overrightarrow{\nabla}) + 
a_{N^3LO}^{PF,33} (\overleftarrow{\nabla}\overleftarrow{\nabla}^2 \cdot \delta(\vec{r})\overrightarrow{F}^1 +}\nonumber \\
&& \overleftarrow{F}^1 \cdot \delta(\vec{r}) \overrightarrow{\nabla}^2 \overrightarrow{\nabla} )
+ a_{N^3LO}^{PF,15} (\overleftarrow{\nabla} \cdot \delta(\vec{r}) \overrightarrow{\nabla}^2 \overrightarrow{F}^1 + \overleftarrow{F}^1 \overleftarrow{\nabla}^2 \cdot \delta(\vec{r}) \overrightarrow{\nabla} ) \biggr] e^{r^2/2} | n(l=3S=1)J=2MTM_T \rangle = \nonumber \\
&& \hspace{0.3in} {32 \over 35} \sqrt{14} {2 \over \pi^2}
\left[{ \Gamma[n^\prime+3/2] \Gamma[n+7/2] \over (n^\prime-1)!(n-1)!} \right]^{1/2}~
\biggl(a_{NNLO}^{PF} -4\bigl[(n^\prime-1) a_{N^3LO}^{PF,33} +(n-1) a_{N^3LO}^{PF,15}\bigr]
 \biggr); \hspace{1.3in}
\end{eqnarray}
And finally, for the $^1F_3 \leftrightarrow {}^1F_3$ or $^3F_J \leftrightarrow {}^3F_J$  channels,
recalling $\overrightarrow{F}^3_M \equiv
(\overrightarrow{\nabla} \otimes (\overrightarrow{\nabla} \otimes \overrightarrow{\nabla})_2)_{3M}$,
\begin{eqnarray}
\lefteqn{\langle n^\prime (l=3S)JM TM_T | e^{r^2/2} \biggl[ a_{N^3LO}^{1F3}\overleftarrow{F}^3\cdot \delta(\vec{r})\overrightarrow{F}^3 \biggr] e^{r^2/2} | n(l=3S)JMTM_T \rangle}  \nonumber \\
&& \hspace{0.3in} {128 \over 35}  {2 \over \pi^2}
\left[{ \Gamma[n^\prime+7/2] \Gamma[n+7/2] \over (n^\prime-1)!(n-1)!} \right]^{1/2}~
a_{N^3LO}^{PF}. \hspace{3.9in}
\label{lastho}
\end{eqnarray}

In each case the general matrix element of the effective interaction, for edge amd nonedge states,
can then be expanded in terms of Eqs. (\ref{firstho}-\ref{lastho}),
\begin{eqnarray}
\lefteqn{\langle n^\prime (l^\prime S)JM TM_T | {E \over E-TQ} \bar{O} {E \over E-QT}  | n(lS)JMTM_T \rangle =} \nonumber \\
&&\hspace{0.3in} \sum_{i,j=0} \widetilde{g}_j(-\kappa^2;n',l') \widetilde{g}_i(-\kappa^2;n,l)~\langle n^\prime+j~ (l^\prime S)JM TM_T |~ \bar{O}~ | n+i~(lS)JMTM_T \rangle~. \hspace{1.5in}
\end{eqnarray}
 
\noindent
{\it Free  Green's Function Results:}
An equivalent set of results can be derived by making use of Eq. (\ref{eq:partial-wave})
\begin{eqnarray}
\lefteqn{\langle \vec{r} | {E \over E-QT} | n~l ~m_l \rangle \equiv \langle \vec{r} | \widetilde{\alpha}_{nlm_l} \rangle = -Y_{lm_l}(\Omega_r) \biggl[ {1 \over \sqrt{r}}~ I_{l+1/2}(\kappa r) \int_r^\infty {r^\prime}^2 d r^\prime {1 \over \sqrt{r^\prime}} K_{l+1/2}(\kappa r^\prime)~\langle \vec{r}^{~\prime}|\alpha_{nlm_l}>} \nonumber \\
&&\hspace{0.3in}+ {1 \over \sqrt{r}} ~K_{l+1/2}(\kappa r) \int_0^r  {r^\prime}^2  dr^\prime {1 \over \sqrt{r^\prime}} I_{l+1/2}(\kappa r^\prime)~\langle \vec{r}^{~\prime}|\alpha_{nlm_l}> \equiv
\widetilde{R}^\alpha_{nl}(r) Y_{lm_l} (\Omega_r) \hspace{2.0in}
\label{eq:GF}
\end{eqnarray}
where the source term in the Green's function is
\begin{eqnarray}
\lefteqn{|\alpha_{nlm_l} \rangle \equiv \bigl[-\kappa^2 - (2n+l-1/2) - \widetilde{g}_1(-\kappa^2;n,l) \sqrt{n(n+l+1/2)} ~\bigr] |nlm_l>
-  \sqrt{(n-1)(n+l-1/2)} |n-1lm_l\rangle}  \nonumber \\
&&\hspace{0.3in}  - \sqrt{n(n+l+1/2)} P[n+1,l] |n+1l m_l\rangle~. \hspace{4.1in}
\end{eqnarray}
The inclusion of the last term makes this equation valid for non-edge as well as edge states:
$P[n+1,l]=1$ if and only if $|n+1~l\rangle$ belongs to $P$.  The reproduction of simple HO
nonedge states is a helpful numerical check. Here
$\langle \vec{r} | \alpha_{nlm_l} \rangle \equiv R^\alpha_{nl}(r) Y_{lm_l} (\Omega_r)$.

The first task is to evaluate $(\overrightarrow{\nabla}^2)^p e^{r^2/2}$ on this wave function.  The
inclusion of $e^{r^2/2}$ -- which eliminates mixing among HO states and simplifies other
HO expressions -- is a bit of an annoyance in the Green's function case,
generating a series of surface terms.
One finds the generic result, analogous to Eq. (\ref{laplacian}),
\begin{eqnarray}
\lefteqn{(\overrightarrow{\nabla}^2)^p~e^{r^2/2}  \widetilde{R}^\alpha_{nl}(r) Y_{lm_l} (\Omega_r) = e^{r^2/2} Y_{lm_l} (\Omega_r) \times} \nonumber \\
 &&\hspace{0.3in} \biggl\{ \biggl[-f_I^p(\kappa^2,r^2) {1 \over \sqrt{r}} I_{l+1/2}(\kappa r) -   f_{I^\prime}^p(\kappa^2,r^2)
2r {d \over dr} \biggl(( {1 \over \sqrt{r}} I_{l+1/2}(\kappa r)\biggr) \biggr]
 \int_r^\infty x^2 dx {1 \over \sqrt{x}} K_{l+1/2}(\kappa x) R^\alpha_{nl}(x) \nonumber \\
&&\hspace{0.3in} +\biggl[-f_K^p(\kappa^2,r^2) {1 \over \sqrt{r}} K_{l+1/2}(\kappa r) -  f_{K^\prime}^p(\kappa^2,r^2)
2r {d \over dr} \biggl(( {1 \over \sqrt{r}} K_{l+1/2}(\kappa r)\biggr) \biggr] \int_0^r x^2 dx {1 \over \sqrt{x}} I_{l+1/2}(\kappa x) R^\alpha_{nl}(x) \biggr\} \hspace{0.5in} \nonumber \\
&&\hspace{0.4in} +f_\alpha^p(\kappa^2,r^2) e^{r^2/2} R^\alpha_{nl}(r) Y_{lm_l} (\Omega_r)+ f_{\alpha^\prime}^p (\kappa^2,r^2) 2r {d \over dr} \bigl( e^{r^2/2} R^\alpha_{nl}(r)  Y_{lm_l} (\Omega_r)\bigr) \nonumber \\
&&\hspace{0.4in} +f_{\alpha^{\prime \prime}}^p(\kappa^2,r^2) \overrightarrow{\nabla}^2 \bigl( e^{r^2/2} R^\alpha_{nl}(r) Y_{lm_l} (\Omega_r) \bigr)+ f_{\alpha^{3 \prime}}^p (\kappa^2,r^2) 2r {d \over dr} \overrightarrow{\nabla}^2 \bigl( e^{r^2/2} R^\alpha_{nl}(r)  Y_{lm_l} (\Omega_r) \bigr) \nonumber \\
&&\hspace{0.4in} +f_{\alpha^{4 \prime}}^p(\kappa^2,r^2) \overrightarrow{\nabla}^4 \bigl(e^{r^2/2} R^\alpha_{nl}(r) Y_{lm_l} (\Omega_r)\bigr)+ f_{\alpha^{5 \prime}}^p (\kappa^2,r^2) 2r {d \over dr}  \overrightarrow{\nabla}^4 \bigl(e^{r^2/2} R^\alpha_{nl}(r)  Y_{lm_l} (\Omega_r)\bigr)  ~~+ ...\biggr\}\hspace{0.8in}
\end{eqnarray}
where each $f(\kappa^2,r^2)$ is a polynomial that can be evaluated using standard
gradient formulas for spherical harmonics.  At N$^3$LO $f_{\alpha^{4\prime}}^p$ is the
highest contributing surface term.  This form allows one to use Eq. (\ref{laplacian}) to
evaluate repeated operations of $\overrightarrow{\nabla}^2$.
One can show, by expanding this expression around $r=0$, that the leading order terms,
which come from the second line above and from the surface terms in line four and the following
lines, are proportional to the solid harmonics
$r^l Y_{lm}(\Omega_r)$, with corrections involving additional powers of $r^2$.  As the lowest order contributing operators, $(\overrightarrow{\nabla}^q)_{q0}$, can annihilate only the former at 
the origin, in fact this expression effectively simplifies for contact-gradient purposes
\begin{eqnarray}
\lefteqn{(\overrightarrow{\nabla}^2)^p~e^{r^2/2}  \widetilde{R}^\alpha_{nl}(r) Y_{lm_l} (\Omega_r) \stackrel{eff.}{\longrightarrow} - ~{r^l  Y_{lm_l}(\Omega_r) \over (2l+1)!!}  \times} \nonumber \\
 &&\hspace{0.1in} \biggl\{ \biggl[f_I^p(\kappa^2,0) +2 l f_{I^\prime}^p(\kappa^2,0) \biggr]
\sqrt{{2 \over \pi}}~ \biggl( \kappa^{l+1/2} \int_0^\infty x^2 dx {1 \over \sqrt{x}} K_{l+1/2}(\kappa x) R^\alpha_{nl}(x) \biggr)  +
~2^{l+1}\left[{2 \Gamma[n+l+1/2] \over \pi (n-1)!} \right] \times \hspace{0.5in} \nonumber \\
&&\hspace{0.1in} \biggl[ \biggl[ f_\alpha^p(\kappa^2,0) + 2 l f_{\alpha^\prime}^p (\kappa^2,0) \biggr]  \biggl(\kappa^2 +3n+l-3/2+ \widetilde{g}_1(-\kappa^2;n,l) \sqrt{n(n+l+1/2)}+(n+l+1/2) P[n+1,l] \biggr) \nonumber \\
&&\hspace{0.1in} -4 \biggl[ f_{\alpha^{2\prime}}^p(\kappa^2,0) + 2 l f_{\alpha^{3\prime}}^p (\kappa^2,0) \biggr] \biggl((n-1) \bigl[\kappa^2 +3n+l-5/2+ \widetilde{g}_1(-\kappa^2;n,l) \sqrt{n(n+l+1/2)}~\bigr]\nonumber \\
&& \hspace{3.0in} + n(n+l+1/2) P[n+1,l] \biggr) \nonumber \\
&& +16 \biggl[ f_{\alpha^{4\prime}}^p(\kappa^2,0) + 2 l f_{\alpha^{5\prime}}^p (\kappa^2,0) \biggr] (n-1)(n-2) \bigl[\kappa^2 +3n+l-7/2+ \widetilde{g}_1(-\kappa^2;n,l) \sqrt{n(n+l+1/2)}~\bigr]\nonumber \\
&& \hspace{3.0in} + n(n-1)(n+l+1/2) P[n+1,l]  \biggr] \biggr\}
\end{eqnarray}
For the cases of interest here (through N$^3$LO), the required nonzero polynomials are
\[ f_I^p(\kappa^2,0) +2 l f_{I^\prime}^p(\kappa^2,0)= \left\{ \begin{array}{ll}
1 & \mbox{~~$p$=0} \\
3 + \kappa^2 + 2l & \mbox{~~$p$=1} \\
15 + 10 \kappa^2 + \kappa^4 + 4l(l+1) + 2 l (6+2 \kappa^2)& \mbox{~~$p$=2} \\
105 + 105\kappa^2+21\kappa^4+\kappa^6+4 l(l+1)(13+3\kappa^2)+2l(45+30\kappa^2+3\kappa^4+4l(l+1)) & \mbox{~~$p$=3} \end{array} \right. \]
\[ f_\alpha^{p=1}(\kappa^2,0) + 2 l f_{\alpha^\prime}^{p=1} (\kappa^2,0)=f_{\alpha^{2\prime}}^{p=2}(\kappa^2,0) + 2 l f_{\alpha^{3\prime}}^{p=2} (\kappa^2,0)=f_{\alpha^{4\prime}}^{p=3}(\kappa^2,0) + 2 l f_{\alpha^{5\prime}}^{p=3} (\kappa^2,0)=1 \hspace{1.5in} \]
\[ f_\alpha^{p=2}(\kappa^2,0) + 2 l f_{\alpha^\prime}^{p=2} (\kappa^2,0)=7 + \kappa^2 + 2l ~~~~~~~~f_{\alpha^{2\prime}}^{p=3}(\kappa^2,0) + 2 l f_{\alpha^{3\prime}}^{p=3} (\kappa^2,0)=
11 + \kappa^2 + 2l \hspace{2.2in} \]
\[ ~f_\alpha^{p=3}(\kappa^2,0) ~+ ~2 l f_{\alpha^\prime}^{p=3} (\kappa^2,0)=
57 + 18\kappa^2+ \kappa^4+4l(l+1)+4l(6+2\kappa^2) \hspace{3.4in} \]

The analog of Eq. (\ref{eq:finalHO}) then becomes
\begin{eqnarray}
\lefteqn{(\overrightarrow{\nabla}^2)^p (\overrightarrow{\nabla}^q)_{q0}~e^{r^2/2}  R^\alpha_{nl}(r) Y_{lm_l} (\Omega_r)\big|_{\vec{r}=0} = - \delta_{lq} ~\sqrt{l! \over  4 \pi (2l+1)!!}  ~\times} \nonumber \\
 &&\hspace{0.1in} \biggl\{ \biggl[f_I^p(\kappa^2,0) +2 l f_{I^\prime}^p(\kappa^2,0) \biggr]
\biggl( \sqrt{{2 \over \pi}}~ \kappa^{l+1/2} \int_0^\infty x^2 dx {1 \over \sqrt{x}} K_{l+1/2}(\kappa x) R^\alpha_{nl}(x) \biggr)  +
~2^{l+1}\left[{2 \Gamma[n+l+1/2] \over \pi (n-1)!} \right] \times \hspace{0.4in} \nonumber \\
&&\hspace{0.1in} \biggl[ \biggl[ f_\alpha^p(\kappa^2,0) + 2 l f_{\alpha^\prime}^p (\kappa^2,0) \biggr] \biggl(\kappa^2 +3n+l-3/2+ \widetilde{g}_1(-\kappa^2;n,l) \sqrt{n(n+l+1/2)}+(n+l+1/2) P[n+1,l] \biggr) \nonumber \\
&&\hspace{0.1in} -4 \biggl[ f_{\alpha^{2\prime}}^p(\kappa^2,0) + 2 l f_{\alpha^{3\prime}}^p (\kappa^2,0) \biggr] \biggl((n-1) \bigl[\kappa^2 +3n+l-5/2+ \widetilde{g}_1(-\kappa^2;n,l) \sqrt{n(n+l+1/2)}~\bigr]\nonumber \\
&& \hspace{3.0in} + n(n+l+1/2) P[n+1,l] \biggr) \nonumber \\
&& +16 \biggl[ f_{\alpha^{4\prime}}^p(\kappa^2,0) + 2 l f_{\alpha^{5\prime}}^p (\kappa^2,0) \biggr] (n-1)(n-2) \bigl[\kappa^2 +3n+l-7/2+ \widetilde{g}_1(-\kappa^2;n,l) \sqrt{n(n+l+1/2)}~\bigr]\nonumber \\
&& \hspace{3.0in} + n(n-1)(n+l+1/2) P[n+1,l]  \biggr] \biggr\}
\label{eq:finalB}
\end{eqnarray}
where
\begin{equation}
(\overrightarrow{\nabla}^q)_{q0}{r^l  Y_{lm_l}(\Omega_r) \over (2l+1)!!}\bigg|_{\vec{r}=0} = \delta_{lq} ~\sqrt{l! \over  4 \pi (2l+1)!!}
\end{equation}
has been used, and where the remaining integral can be evaluated using
\begin{eqnarray}
 \lefteqn{\sqrt{{2 \over \pi}}~ \kappa^{l+1/2} \int_0^\infty x^2 dx {1 \over \sqrt{x}} K_{l+1/2}(\kappa x) R^\alpha_{nl}(x)= -\sqrt{2 (n-1)! \Gamma[n+l+1/2]} \sum_{m=0}^n {(-2)^m \over m! (n-m)! \Gamma[l+3/2+m]}}\nonumber \\
 && \hspace{0.1in}  \times \biggl[ (\kappa^2+3n + l-m -3/2 +\widetilde{g}_1(-\kappa^2;nl) \sqrt{n(n+l+1/2} )(n-m) +n(n+l+1/2) P[n+1,l] \biggr]  \hspace{1.1
 in} \nonumber \\
 && \hspace{0.1in} \times \sum_{i=0}^l {(\sqrt{2} \kappa)^{l-i} \over 2^i} {(l+i)! \over i! (l-i)!}\biggl[ -\sqrt{2} \kappa \Gamma[m+3/2+(l-i)/2]~{}_1F_1[m+3/2+ (l-i)/2; 3/2; \kappa^2/2] \nonumber \\
 && \hspace{1.8in}+  \Gamma[m+1+(l-i)/2]~{}_1F_1[m+1+(l-i)/2;1/2;\kappa^2/2] \biggr].
\end{eqnarray}
With the effects of contact-gradient operators on wave functions of the form of Eq. (\ref{eq:start})
thus determined, these results can be plugged into Eqs. (\ref{eq:diagonal}) and (\ref{eq:nondiagonal}) to
produce the needed expressions for matrix elements.

The contact-gradient matrix elements needed through N$^3$LO have been evaluated using both
the HO expansion and the free Green's function.   The resulting agreement is a nice check.
The sum over HO excitations is truncated at some $N$: at N$^3$LO the choice $N$=400
will give results accurate to at least 0.01\%  for the most sensitive N$^3$LO operators, which depend on higher derivatives at the origin.  The Mathematica script for this summation is simple and efficient, so there is no practical limit to the $N$s that can be handled.
Similarly, the Green's function expressions can be evaluated 
very easily: this is the recommended scheme for evaluating edge-state matrix elements
for contact-gradient operators.  
 
Eq. (\ref{eq:GF}) is also an efficient way to generate the wave function at all values of $r$, which is needed for evaluations of edge-state matrix elements
of the bare $V$.  One finds
\begin{eqnarray}
\lefteqn{\widetilde{R}_{nl}(r) = \sqrt{2 (n-1)!  \Gamma[n+l+1/2]} \sum_{m=0}^n {(-1)^m \over
m! (n-m)! \Gamma[l+3/2+m]} } \nonumber \\
&& \times \biggl[ \bigl(n-m \bigr) \biggl( \kappa^2 +3n-m+l-3/2+\widetilde{g}_1(-\kappa^2; nl) \sqrt{n(n+l+1/2)}\biggr)+n(n+l+1/2)) P[n+1,l] \biggr]  \hspace{1.0in} \nonumber \\
&& \times \left[ {1 \over \sqrt{r}} I_{l+1/2}(\kappa r) G_1[\kappa,l,m,r]+ {1\over \sqrt{r}} K_{l+1/2}(\kappa r)
G_2[\kappa,l,m,r] \right]
\end{eqnarray}
where
\begin{eqnarray}
\lefteqn{G_1[\kappa,l,m,r] = e^{-\kappa r -r^2/2} \sqrt{\pi} ~
\sum_{i=0}^l {(l+i)! \over i! (l-i)!} {2^{m+(l-i)/2} \over (2 \kappa)^{i+1/2}} ~~\sum_{j=0}^{2m+1+l-i} {(2m+1+l-i)! \over
(2m+1+l-i-j)! j!} ({r \over \sqrt{2}})^{2m+1+l-i-j}} \nonumber \\
&&\times \left[ -\sqrt{2} (\kappa+r) \Gamma[1+j/2]~ {}_1F_1[1+j/2;3/2;(\kappa+r)^2/2]
+ \Gamma[(1+j)/2]~ {}_1F_1[(1+j)/2;1/2;(\kappa+r)^2/2] \right] \hspace{0.7in}
\end{eqnarray}
and
\begin{eqnarray}
\lefteqn{G_2[\kappa,l,m,r]={1 \over \sqrt{\pi}}~ \sum_{i=0}^l ~{(l-i)! \over i! (l-i)!}~ {2^{m+(l-i)/2}  \over (2 \kappa)^{i+1/2}} }\nonumber \\
&&\hspace{0.6in} \biggl[ ((-1)^i+(-1)^l) \kappa \sqrt{2}~ \Gamma[m+3/2+(l-i)/2]~{}_1F_1[m+3/2+(l-i)/2;3/2;\kappa^2/2]
\hspace{1.2in} \nonumber \\
&& \hspace{0.6in} + ((-1)^i-(-1)^l) \Gamma[m+1+(l-i)/2]~{}_1F_1[m+1+(l-i)/2;1/2;\kappa^2/2] \nonumber \\
&&\hspace{0.6in} - e^{-r^2/2}~ \sum_{j=0}^{2m+1+l-i} ~{(2m+1+l-i)! \over (2m+1+l-i-j)! j!} ({r \over \sqrt{2}})^{2m+1+l-i-j} \nonumber \\
&&\hspace{0.8in} \biggl( (-1)^i e^{\kappa r} \sqrt{2}~ (\kappa-r) ~\Gamma[1+j/2]~{}_1F_1[1+j/2;3/2;(\kappa-r)^2/2]
\nonumber \\
&&\hspace{0.8in} + (-1)^l e^{-\kappa r} \sqrt{2}~ (\kappa+r)~ \Gamma[1+j/2]~{}_1F_1[1+j/2;3/2;(\kappa+r)^2/2] \nonumber \\
&&\hspace{0.8in} +  (-1)^i e^{\kappa r}~ \Gamma[(1+j)/2]~{}_1F_1[(1+j)/2;1/2;(\kappa-r)^2/2] \nonumber \\
&&\hspace{0.8in} - (-1)^l e^{-\kappa r}~ \Gamma[(1+j)/2]~{}_1F_1[(1+j)/2;1/2;(\kappa+r)^2/2] \biggr) \biggr].
\end{eqnarray}

\end{document}